\documentclass[%
reprint,
superscriptaddress,
nofootinbib,
nobibnotes,
 amsmath,amssymb,
 aps,
 prd,
 longbibliography,
 twocolumn,
]{revtex4-2}% 4-2 has problem with bibtex

\usepackage{siunitx}

\DeclareSIUnit \mm {\milli\meter}
\DeclareSIUnit \cm {\centi\meter}
\DeclareSIUnit \us {\micro\second}
\DeclareSIUnit \ms {\milli\second}
\DeclareSIUnit \pA {\pico\ampere}
\DeclareSIUnit \pC {\pico\coulomb}
\DeclareSIUnit \fC {\femto\coulomb}
\DeclareSIUnit \fF {\femto\farrad}
\DeclareSIUnit \pF {\pico\farrad}
\DeclareSIUnit \mV {\milli\volt}
\DeclareSIUnit \kV {\kilo\volt}
\DeclareSIUnit \V {\volt}
\DeclareSIUnit \GOhm {\giga\ohm}
\DeclareSIUnit \MOhm {\mega\ohm}
\DeclareSIUnit \ton {\tonne}
\DeclareSIUnit \kton {\kilo\tonne}
\DeclareSIUnit \kt {\kilo\tonne}
\DeclareSIUnit \Mt {\mega\tonne}
\DeclareSIUnit \eV {\electronvolt}
\DeclareSIUnit \keV {\kilo\electronvolt}
\DeclareSIUnit \MeV {\mega\electronvolt}
\DeclareSIUnit \GeV {\giga\electronvolt}
\DeclareSIUnit \km {\kilo\meter}
\DeclareSIUnit \kW {\kilo\watt}
\DeclareSIUnit \MW {\mega\watt}
\DeclareSIUnit \MHz {\mega\hertz}
\DeclareSIUnit \kHz {\kilo\hertz}
\DeclareSIUnit \mrad {\milli\radian}
\DeclareSIUnit \year {year}
\DeclareSIUnit \POT {POT}
\DeclareSIUnit \sig {$\sigma$}
\DeclareSIUnit\parsec{pc}
\DeclareSIUnit\lightyear{ly}
\DeclareSIUnit\foot{ft}
\DeclareSIUnit\ft{ft}
\usepackage[printwatermark]{xwatermark}
\usepackage{graphicx}% Include figure files
\usepackage{dcolumn}% Align table columns on decimal point
\usepackage{bm}% bold math
\usepackage{xcolor}
\usepackage[hidelinks]{hyperref}
\hypersetup{backref=true,       
    pagebackref=true,               
    hyperindex=true,                
    colorlinks=true,                
    breaklinks=true,                
    urlcolor= black,                
    linkcolor= red,                
    bookmarks=true,                 
    bookmarksopen=false,
    filecolor=black,
    citecolor=blue,
    linkbordercolor=blue
}
\usepackage{lineno}
\usepackage{mathtools}
\usepackage{amsmath}
\usepackage{breqn}
\usepackage{comment}
\usepackage{multirow}
\usepackage{caption}
\usepackage{subcaption}
\usepackage{float}
\usepackage{makecell}
\usepackage{overpic}
\usepackage{breqn}
\usepackage[labelformat=simple]{subcaption}
\usepackage{appendix}

\newlength{\figwidth}
\newlength{\fighalfwidth}
\newcommand{\dqdx}{$dQ/dx$}

\newcommand{\nue}{$\nu_{e}$}
\newcommand{\nueCC}{$\nu_{e}$\,CC}
\newcommand{\numu}{$\nu_{\mu}$}
\newcommand{\numuCC}{$\nu_{\mu}$\,CC}
\newcommand{\pizero}{$\pi^{0}$}
\newcommand\brabarb{\scalebox{.3}{(}\raisebox{-1.7pt}[0pt][0pt]{$-$}\scalebox{.3}{)}}

\newcommand{\SM}{eLEE$_{x=0}$}
\newcommand{\LEE}{eLEE$_{x=1}$}

\begin{document}
\raggedbottom

%\title{A Search for $\nu_e$ Low-Energy Excess with Wire-Cell Event Reconstruction in the MicroBooNE Experiment}
\title{Search for an anomalous excess of inclusive charged-current $\nu_e$ interactions \\
in the MicroBooNE experiment using Wire-Cell reconstruction}
%Search for an anomalous excess with inclusive charged-current $\nu_e$ interactions \\
%in the MicroBooNE experiment using the Wire-Cell Event Reconstruction}
%\title{Search for anomalous events in the MicroBooNE experiment \\
%with inclusive charged-current $\nu_e$ interactions from the Booster Neutrino Beamline \\
%using the Wire-Cell Event Reconstruction}

%\linenumbers

% List of institutions in command form:
\newcommand{\Bern}{Universit{\"a}t Bern, Bern CH-3012, Switzerland}
\newcommand{\BNL}{Brookhaven National Laboratory (BNL), Upton, NY, 11973, USA}
\newcommand{\UCSB}{University of California, Santa Barbara, CA, 93106, USA}
\newcommand{\Cambridge}{University of Cambridge, Cambridge CB3 0HE, United Kingdom}
\newcommand{\CIEMAT}{Centro de Investigaciones Energ\'{e}ticas, Medioambientales y Tecnol\'{o}gicas (CIEMAT), Madrid E-28040, Spain}
\newcommand{\Chicago}{University of Chicago, Chicago, IL, 60637, USA}
\newcommand{\Cincinnati}{University of Cincinnati, Cincinnati, OH, 45221, USA}
\newcommand{\CSU}{Colorado State University, Fort Collins, CO, 80523, USA}
\newcommand{\Columbia}{Columbia University, New York, NY, 10027, USA}
\newcommand{\Edinburgh}{University of Edinburgh, Edinburgh EH9 3FD, United Kingdom}
\newcommand{\FNAL}{Fermi National Accelerator Laboratory (FNAL), Batavia, IL 60510, USA}
\newcommand{\Granada}{Universidad de Granada, Granada E-18071, Spain}
\newcommand{\Harvard}{Harvard University, Cambridge, MA 02138, USA}
\newcommand{\IIT}{Illinois Institute of Technology (IIT), Chicago, IL 60616, USA}
\newcommand{\KSU}{Kansas State University (KSU), Manhattan, KS, 66506, USA}
\newcommand{\Lancaster}{Lancaster University, Lancaster LA1 4YW, United Kingdom}
\newcommand{\LANL}{Los Alamos National Laboratory (LANL), Los Alamos, NM, 87545, USA}
\newcommand{\Manchester}{The University of Manchester, Manchester M13 9PL, United Kingdom}
\newcommand{\MIT}{Massachusetts Institute of Technology (MIT), Cambridge, MA, 02139, USA}
\newcommand{\Michigan}{University of Michigan, Ann Arbor, MI, 48109, USA}
\newcommand{\Minnesota}{University of Minnesota, Minneapolis, MN, 55455, USA}
\newcommand{\NMSU}{New Mexico State University (NMSU), Las Cruces, NM, 88003, USA}
\newcommand{\Oxford}{University of Oxford, Oxford OX1 3RH, United Kingdom}
\newcommand{\Pitt}{University of Pittsburgh, Pittsburgh, PA, 15260, USA}
\newcommand{\Rutgers}{Rutgers University, Piscataway, NJ, 08854, USA}
\newcommand{\SLAC}{SLAC National Accelerator Laboratory, Menlo Park, CA, 94025, USA}
\newcommand{\SDSMT}{South Dakota School of Mines and Technology (SDSMT), Rapid City, SD, 57701, USA}
\newcommand{\Maine}{University of Southern Maine, Portland, ME, 04104, USA}
\newcommand{\Syracuse}{Syracuse University, Syracuse, NY, 13244, USA}
\newcommand{\TelAviv}{Tel Aviv University, Tel Aviv, Israel, 69978}
\newcommand{\Tennessee}{University of Tennessee, Knoxville, TN, 37996, USA}
\newcommand{\UTA}{University of Texas, Arlington, TX, 76019, USA}
\newcommand{\Tufts}{Tufts University, Medford, MA, 02155, USA}
\newcommand{\VTech}{Center for Neutrino Physics, Virginia Tech, Blacksburg, VA, 24061, USA}
\newcommand{\Warwick}{University of Warwick, Coventry CV4 7AL, United Kingdom}
\newcommand{\Yale}{Wright Laboratory, Department of Physics, Yale University, New Haven, CT, 06520, USA}
%%\newcommand{\listerThanks}{Now at University of Wisconsin, Madison}

% So that institutions appear in alphabetical order:
\affiliation{\Bern}
\affiliation{\BNL}
\affiliation{\UCSB}
\affiliation{\Cambridge}
\affiliation{\CIEMAT}
\affiliation{\Chicago}
\affiliation{\Cincinnati}
\affiliation{\CSU}
\affiliation{\Columbia}
\affiliation{\Edinburgh}
\affiliation{\FNAL}
\affiliation{\Granada}
\affiliation{\Harvard}
\affiliation{\IIT}
\affiliation{\KSU}
\affiliation{\Lancaster}
\affiliation{\LANL}
\affiliation{\Manchester}
\affiliation{\MIT}
\affiliation{\Michigan}
\affiliation{\Minnesota}
\affiliation{\NMSU}
\affiliation{\Oxford}
\affiliation{\Pitt}
\affiliation{\Rutgers}
\affiliation{\SLAC}
\affiliation{\SDSMT}
\affiliation{\Maine}
\affiliation{\Syracuse}
\affiliation{\TelAviv}
\affiliation{\Tennessee}
\affiliation{\UTA}
\affiliation{\Tufts}
\affiliation{\VTech}
\affiliation{\Warwick}
\affiliation{\Yale}

% Authors in alphabetical order
%\author{P.~Abratenko} \affiliation{\Michigan} % for CC incl and any papers using MCS
\author{P.~Abratenko} \affiliation{\Tufts} 
\author{R.~An} \affiliation{\IIT}
\author{J.~Anthony} \affiliation{\Cambridge}
\author{L.~Arellano} \affiliation{\Manchester}
\author{J.~Asaadi} \affiliation{\UTA}
\author{A.~Ashkenazi}\affiliation{\TelAviv}
\author{S.~Balasubramanian}\affiliation{\FNAL}
\author{B.~Baller} \affiliation{\FNAL}
\author{C.~Barnes} \affiliation{\Michigan}
\author{G.~Barr} \affiliation{\Oxford}
% ...

\collaboration{The MicroBooNE Collaboration}
\thanks{microboone\_info@fnal.gov}\noaffiliation

\date{\today}% It is always \today, today,
%\preprint{APS/123-QED}
\begin{abstract}
We report a search for an anomalous excess of inclusive charged-current (CC) \nue\ interactions using the Wire-Cell event reconstruction package in the MicroBooNE experiment, which is motivated by the previous observation of a low-energy excess (LEE) of electromagnetic events from the MiniBooNE experiment.
With a single liquid argon time projection chamber detector, the measurements of \numuCC\ interactions as well as \pizero\ interactions are used to constrain signal and background predictions of \nueCC\ interactions. A data set collected from February 2016 to July 2018 corresponding to an exposure of 6.369~$\times$~10$^{20}$ protons on target from the Booster Neutrino Beam at FNAL is analyzed.
With $x$ representing an overall normalization factor and referred to as the LEE strength parameter, we select 56 fully contained \nueCC\ candidates while expecting 69.6~$\pm$~8.0~(stat.)~$\pm$~5.0~(sys.) and 103.8~$\pm$~9.0~(stat.)~$\pm$~7.4~(sys.) candidates after constraints for the absence (\SM) of the median signal strength derived from the MiniBooNE observation and the presence (\LEE) of that signal strength, respectively.
Under a nested hypothesis test using both rate and shape information in all available channels, the best-fit $x$ is determined to be 0 (\SM) with a 95.5\% confidence level upper limit of $x$ at 0.502. 
Under a simple-vs-simple hypotheses test, the \LEE\ hypothesis is rejected at 3.75$\sigma$, while the \SM\ hypothesis is shown to be consistent with the observation at 0.45$\sigma$.
In the context of the eLEE model, the estimated 68.3\% confidence interval of the \nueCC\ hypothesis 
to explain the LEE observed in the MiniBooNE experiment is disfavored at a significance level of more than 2.6$\sigma$ (3.0$\sigma$) considering MiniBooNE's full (statistical)
uncertainties.
\end{abstract}

\maketitle

%\tableofcontents

\section{Introduction}~\label{sec:introduction}
%% What is the low-energy excess and its relation to sterile neutrino and other new physics interpretations
Neutrino flavor oscillation is one of the few observations in particle physics for evidence of physics beyond the standard model. While the majority of neutrino oscillation data can be successfully explained by a three-neutrino framework~\cite{Zyla:2020zbs}, the exact mechanism for neutrinos to acquire their masses remains a puzzle. In addition, the fact that the mass of the electron neutrino is at least five orders of magnitude smaller than that of the electron~\cite{KATRIN:2019yun} is an interesting mystery. The possible existence of additional neutrino flavors may provide a natural explanation of the small neutrino mass~\cite{King:2003jb}. Constrained by precision electroweak measurements~\cite{ALEPH:2005ab}, these additional neutrinos are expected to be sterile~\cite{Pontecorvo:1967fh} not participating in any fundamental interaction of the standard model. Despite the fact that there is no known mechanism to detect them directly, they may be indirectly observed in neutrino oscillation experiments where sterile neutrinos could mix with the three active neutrinos and affect the way they oscillate.

Because of strong theoretical motivations, there are many dedicated programs searching
for sterile neutrinos. While most of the results are consistent with 
the three-neutrino framework without sterile neutrinos (see Ref.~\cite{MINOS:2020iqj,PROSPECT:2020sxr} among others), 
there are several experimental anomalies suggesting the existence of an eV-mass-scale 
sterile neutrino:
%In addition to the theoretical motivations to search for sterile neutrinos,
%light sterile neutrinos at the eV mass scale may explain several experimental 
%anomalies, while many direct searches yield null results. 
 i) the observation that calibrated $\nu_e$ sources ($^{51}$Cr for GALLEX~\cite{Kaether:2010ag} and BEST~\cite{Barinov:2021asz}, $^{51}$Cr and $^{37}$Ar for SAGE~\cite{Abdurashitov:2009tn}) produced lower rates of measured  $\nu_e$ than expected in the three-neutrino framework, which could be explained by $\nu_e$ disappearance considering light sterile neutrinos; ii) the reactor anti-neutrino anomaly~\cite{Mention:2011rk}, which suggests that the observed deficit in the measured $\bar\nu_e$ events relative to the expectation based on the recent reactor anti-neutrino flux calculations~\cite{Huber:2011wv,Mueller:2011nm} could be explained by $\bar{\nu}_e$ disappearance considering light sterile neutrinos, although there are recent experimental measurements~\cite{DayaBay:2017jkb,DayaBay:2021dqj} and 
 improved flux calculations~\cite{Estienne:2019ujo,Giunti:2021kab} that disfavor this explanation; 
 iii) the Neutrino-4~\cite{Serebrov:2020kmd} anomaly, which suggests reactor $\bar{\nu}_e$ 
 oscillation at a few meters; and iv) the LSND~\cite{Aguilar:2001ty} and MiniBooNE~\cite{Aguilar-Arevalo:2013pmq,Aguilar-Arevalo:2020nvw} anomalies, which suggest $\nu_{e}$ appearance from $\nu_{\mu}$ to $\nu_e$ oscillations considering light sterile neutrinos. However, there are significant challenges in explaining 
 all available experimental results with a sterile neutrino oscillation model
 in a global fit~\cite{Giunti:2019aiy}.
 
The MiniBooNE experiment observes an anomalous excess of electromagnetic (electron- or photon-like) events in the data above the prediction for reconstructed neutrino energies below 800 MeV, which is commonly referred to as the MiniBooNE low-energy excess (LEE).
%% Why use LArTPC to do search ??
The interpretation of the MiniBooNE LEE observation is limited by the capability of its detector technology, which primarily uses the pattern of Cherenkov rings to differentiate between muons and electrons. However, the Cherenkov technology does not perform well in separating an electron from a photon leading to the possibility that the LEE may come from an excess of photon background rather than from the \nue\ charged-current (CC) interactions. The Liquid Argon Time Projection Chamber (LArTPC)~\cite{willis74, rubbia77,Chen:1976pp,Nygren:1976fe} has superb imaging capabilities through the combination of both tracking and calorimetry in a fully active volume. 
Throughout the years since its inception, many LArTPCs ranging in size from hundreds of liters 
to hundreds of cubic meters have been constructed and operated for neutrino experiments~\cite{ICARUS:2004wqc,Anderson:2012vc,CAPTAIN:2013irr,Hahn:2016tia,Cavanna:2014iqa,Acciarri:2016smi,DUNE:2021hwx}.
Although electrons and photons with energy greater than tens of MeV both induce electromagnetic (EM) showers in LArTPCs, electrons can be differentiated from photons in neutrino interactions through identifying a gap between neutrino and EM shower vertices, topological pattern recognition, and reconstructed ionization energy loss per unit length ($dE/dx$) measurement~\cite{Cavanna:2018yfk}. This improves the identification of \nueCC\ interactions and will enable sensitive measurements such as searches for $\overset{\brabarb}{\nu}_{\mu}\rightarrow\overset{\brabarb}{\nu}_{e}$ oscillations. LArTPCs can also reconstruct and identify hadrons at lower energies than Cherenkov detectors, which can only detect charged particles with energies above their respective Cherenkov thresholds. The extended capability of LArTPCs to detect these lower energy hadrons enables the study of various exclusive final states in the detector.

%% What is the overall search strategy ??
The MicroBooNE experiment was designed primarily to explore the nature of the MiniBooNE LEE observation using an 85-ton active volume LArTPC~\cite{Acciarri:2016smi} through its excellent electron versus photon separation.
The MicroBooNE detector is located at the Booster Neutrino Beam (BNB) ~\cite{AguilarArevalo:2008yp} at the Fermi National Accelerator Laboratory in Batavia, IL, USA. It is in the same beam line with a similar distance to the neutrino 
source as the MiniBooNE detector. 
While MicroBooNE is capable of searching for exotic signatures, such as electron-positron pairs~\cite{uB_higgs}, the MicroBooNE experiment's main goal is to search for an LEE signal in \nueCC\ interactions with an electron in the final state (referred to as the eLEE search) as well as in neutral-current (NC) interactions with a single photon in the final state. In these LEE searches, the MicroBooNE experiment has developed four analyses: i) a photon-based-search focusing on the $\Delta$-decay hypothesis~\cite{uB_gLEE}; ii) an exclusive search using the \nue\ charged-current quasi-elastic (CCQE) channel~\cite{uB_DL}; iii) a semi-inclusive search for events with one electron and no pions in the final state~\cite{uB_PeLEE}; and iv) a fully inclusive search for events with one electron and any final state, which is what this article describes. The results of the three eLEE searches are individually quantified with an 
empirical model constructed based on MiniBooNE data (see Sec.~\ref{sec:overview}) 
instead of performing 
a sterile neutrino oscillation model fit and summarized together in Ref.~\cite{uB_eLEE_PRL}. Looking ahead, a dedicated search of sterile 
neutrino(s) oscillations through comparison of multiple detectors at different baselines is the goal of the upcoming Short-Baseline Neutrino 
(SBN) program~\cite{Antonello:2015lea}.

%% Introduction for the rest of the paper 

This paper is organized as follows. An overview of the search strategy is presented in Sec.~\ref{sec:overview}. The MicroBooNE experiment and the Wire-Cell event reconstruction package is introduced in Sec.~\ref{sec:uboone}. The event selections including \numuCC\ and \nueCC\ selections\footnote{In this paper, $\nu_{x}$ refers to both $\nu_{x}$ and its counterpart $\bar{\nu}_{x}$.} are described in Sec.~\ref{sec:event_select}. Section~\ref{sec:systematics} describes the systematic uncertainties considered in this work. To search for new physics, validations of the overall model including the predictions of both signal and background are crucial. Various examinations of the model including i) signal prediction, ii) background prediction, and iii) modeling of neutrino energy reconstruction are summarized in Sec.~\ref{sec:validation}. The physics sensitivities and the final result of the eLEE search are presented in Sec.~\ref{sec:results} followed by the summary in Sec.~\ref{sec:summary}. 

\section{Overview of the analysis}\label{sec:overview}
In this paper, we report the results of the eLEE search using the Wire-Cell event reconstruction package~\cite{Qian:2018qbv,Abratenko:2020hpp} after analyzing the MicroBooNE data set collected from February 2016 to July 2018. The search for an eLEE signal is equivalent to testing the null hypothesis, which corresponds to the nominal prediction without an eLEE signal. This prediction is formed based on the state-of-the-art understanding of the BNB flux~\cite{AguilarArevalo:2008yp, uBpublic_flux}, 
the neutrino-argon interaction cross section~\cite{Andreopoulos:2009rq, GENIE:2021npt}, the detector
simulation~\cite{Adams:2018dra,Abratenko:2020bbx,Adams:2019ssg}, and data-driven constraints on \nueCC\ 
signal and background predictions. 
An alternative hypothesis is formed based on an eLEE model which 
represents an anomalous enhancement in the rate of intrinsic \nueCC\ events at true neutrino energies 
between 200~MeV and 800~MeV with a fixed spectral shape extracted from the MiniBooNE experiment. 
This alternative hypothesis has several advantages over a dedicated sterile neutrino oscillation model~\cite{Conrad:2016sve, Conrad:2013mka}. 
Firstly, this model is constructed from the MiniBooNE data with a minimum set of assumptions
and is agnostic of sterile neutrino oscillation model parameters. Secondly, this model contains many fewer
free parameters, which is beneficial for the single detector configuration of MicroBooNE. Thirdly, this model avoids complications introduced by simultaneous modifications in multiple channels including $\nu_\mu$ disappearance, $\nu_\mu$ to $\nu_e$ appearance, and $\nu_e$ disappearance in data. 
Finally, it allows for a quantitative comparison of results between MicroBooNE and MiniBooNE, particularly given the difficulty in explaining the shape of the MiniBooNE LEE in global fits with a sterile neutrino model~\cite{Giunti:2019aiy}.

%MiniBooNE 
%and MicroBooNE in searching for a $\nu_e$ LEE.

This eLEE model used as the alternative hypothesis is obtained by unfolding the observed
excess of electron-like events in MiniBooNE to true neutrino energy under a CCQE hypothesis
and applying the excess-to-intrinsic $\nu_e$ ratios directly to the flux of intrinsic
$\nu_e$ expected in the MicroBooNE detector.
The MiniBooNE data and simulation results reported in 2018~\cite{MiniBooNE:2018esg} were used as an input to the D'Agostini's multi-dimensional unfolding
procedure~\cite{DAgostini:1994fjx}. 
We note that this eLEE model is constructed using the true neutrino energy and is 
limited below 800 MeV, beyond which the unfolded LEE signal is negligible. 
Furthermore, the additional information regarding the lepton kinematics reported by MiniBooNE in 2020~\cite{Aguilar-Arevalo:2020nvw} is not taken into account. 
In addition, a 68.3\% confidence interval of the eLEE strength parameter of $1\pm0.21$ ($1\pm0.08$) considering both statistical and systematic (only statistical) uncertainties is estimated from the reported 4.8$\sigma$ (12.2$\sigma$) LEE significance from the latest MiniBooNE 2020 result~\cite{Aguilar-Arevalo:2020nvw} considering both neutrino and anti-neutrino data taking. 
If only neutrino data taking is considered, the LEE significance would slightly reduce to 4.7$\sigma$ (11.7$\sigma$) considering full (statistical) uncertainty, respectively.   

\begin{figure}[t]
  \centering
  \includegraphics[width=0.48\textwidth]{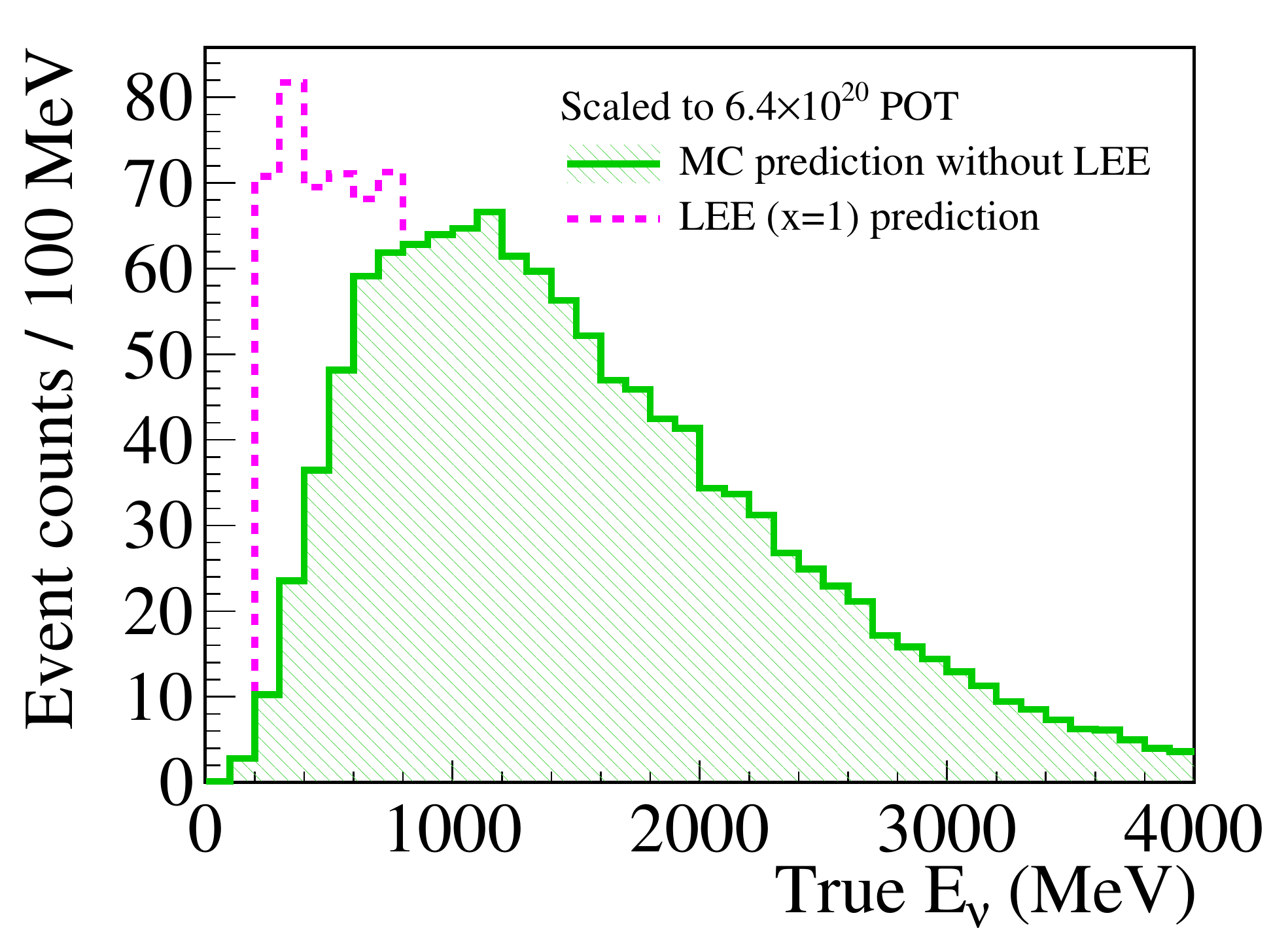}
  \put(-120,173){MicroBooNE Simulation}
  \caption{Expected \nueCC\ events in the MicroBooNE detector for \LEE~hypothesis and \SM~hypothesis as a function of true neutrino energy at 6.4$\times$10$^{20}$ proton-on-target (POT) exposure, assuming 100\% detection efficiency.}
  \label{fig:intro_LEE_illustration}
\end{figure}

In the hypothesis test, we allow the normalization of this eLEE model to float and define a non-negative strength parameter $x$, such that $x=0$ corresponds to the nominal prediction without an eLEE signal (\SM\ hypothesis) and $x=1$ corresponds to the prediction with an eLEE signal with magnitude equal to that of the unfolded median of the MiniBooNE LEE result (\LEE\ hypothesis). Figure~\ref{fig:intro_LEE_illustration} illustrates the prediction of the \LEE\ hypothesis with respect to that of the \SM\ hypothesis for \nueCC\ interactions in the TPC active volume in MicroBooNE assuming 100\% detection efficiency. 

%% What is Wire-Cell analysis strategy ??
We fit this eLEE model to our data by minimizing a combined Neyman-Pearson (CNP) $\chi^2$ test statistic~\cite{Ji:2019yca} that incorporates the MicroBooNE’s experimental uncertainties into a covariance matrix thereby obtaining a best fit value of $x=x_{\text{min}}$.  We compute the primary nested likelihood ratio test statistic, $\Delta \chi^2_{\rm nested} = \chi^2_{{\rm eLEEx}=x_0} - \chi^2_{{\rm eLEEx}=x_{\rm min}}$, for our data given a varying eLEE strength $x=x_0$ and obtain frequentist confidence intervals for the eLEE strength $x$ following the Feldman-Cousins procedure~\cite{Feldman:1997qc}. In addition to the $\Delta \chi^2_{\text{nested}}$, several other statistical tests are performed to provide additional information. They are i) goodness-of-fit (GoF) tests based on a Pearson $\chi^2$ to examine the compatibility between data measurement and Monte-Carlo (MC) prediction, especially with available constraints using the conditional covariance method~\cite{cond_cov}, and ii) a simple-vs-simple likelihood ratio test, $\Delta\chi^2_{\text{simple}} = \chi^2_{\text{eLEEx=1}} - \chi^2_{\text{eLEEx=0}}$ with the CNP $\chi^2$ to demonstrate the compatibility between data and the two signature hypotheses (\LEE\ and \SM). 

Without a near detector to measure the neutrino flux and neutrino-argon interaction cross section, the search for an eLEE in the \nueCC\ channel would suffer from large systematic uncertainties. In order to compensate for the lack of a near detector and to maximize the physics sensitivity of this search, \numuCC\ channel is used to constrain 
the \nueCC prediction. As to be elaborated in Sec.~\ref{sec:uboone_flux}, the prediction of $\nu_e$ and $\nu_\mu$ 
flux are strongly correlated, given the parent hadron species are mostly common. In addition, the \nueCC\ and \numuCC\ 
interaction cross section is also strongly correlated, given the lepton universality. Therefore, the measurement of 
\numuCC\ rates at different energies, which is proportional to the product of $\nu_\mu$ flux and $\nu_\mu$-Ar 
cross section, is efficient in constraining the prediction of \nueCC\ rates at different energies, which is 
proportional to the production of $\nu_e$ flux and $\nu_e$-Ar cross section.
In practice, a seven-channel fit strategy is adopted. The seven channels are: 
\begin{enumerate}
\item fully contained (FC) \nueCC,
\item partially contained (PC) \nueCC,
\item FC \numuCC,
\item PC \numuCC,
\item FC \numuCC\ with $\pi^0$ in the final state (FC CC$\pi^0$),
\item PC CC$\pi^0$, and
\item NC interactions with $\pi^0$ in the final state (NC$\pi^0$). 
\end{enumerate}
Here, the FC events are defined to be events with the reconstructed TPC activity
fully contained within the fiducial volume (3 cm inside the effective TPC boundary~\cite{Abratenko:2021bzb}, which is the corrected boundary that takes a space charge effect~\cite{Adams:2019qrr,Abratenko:2020bbx} into account). 
All non-FC events are defined as PC events.
FC events have many advantages such as higher signal-to-background ratio and better energy resolution when compared to the PC events. 
The primary channel that is sensitive to the eLEE search is FC \nueCC.
FC \numuCC\ and PC \numuCC\ channels are used to provide constraints on the prediction of \nueCC\ interaction rate and its systematic uncertainties. PC \nueCC\ channel, which is less sensitive to the eLEE search, also provides constraints to some extent. 
The other three channels, FC CC$\pi^0$, PC CC$\pi^0$, and NC$\pi^0$, are used to constrain/improve the background prediction since $\pi^0$ events are one of the major backgrounds in the \nueCC\ selection. 
These seven channels are designed to be orthogonal to each other. 
The \numuCC\ channel excludes \nueCC\ and CC$\pi^0$ candidates, and the CC$\pi^0$ channel excludes the \nueCC\ candidates.
%% blinding ...

To reduce bias, we adopted a blind analysis for this eLEE search. Two sidebands (near and far) are defined in addition to the signal region. The signal region is defined to contain the events with reconstructed neutrino energy $E^{rec}_{\nu}$ lower than 600~MeV and passing the \nueCC\ selection. The far sideband is defined to contain the events that fail a looser \nueCC\ selection or the events with $E^{rec}_{\nu}$ higher than 800~MeV. The far sideband essentially includes \numuCC\ events as well as high-energy \nueCC\ events, which are not sensitive to the eLEE signal.
The near sideband covers the remainder of the events between the far sideband and the signal region. A completely open data set with an exposure of 5.33$\times10^{19}$ proton-on-target (POT) is available for development of the event reconstruction, selection, and statistical analysis. This open data set is less than 10\% of the eventual data set of an exposure of $6.369\times10^{20}$ POT which is used in this analysis. The selection and analysis procedure were frozen after analyzing the open data set, and no changes were made after checking the consistency between the data measurement and MC prediction using the far sideband data set. 
The overall model including the central value predictions
and their associated uncertainties is quantitatively validated with extensive goodness-of-fit tests. The power of these 
validations are further enhanced through a conditional covariance matrix formalism, which allows for a comparison 
between data and a constrained model prediction. Constrained by the sideband data, the systematic uncertainties of the 
model prediction can be significantly reduced, allowing for a in-depth examination of the systematic uncertainties and estimated correlations between data samples.
The near sideband was openned subsequently for further examination and, only then, was the signal region unblinded. During the finalization of the analysis procedure, several fake data sets were generated based on different cross section models and injected with or without eLEE signals to test the robustness of the analysis procedure.
All details of the fake data sets were initially blind to the analysers, and the analysis results of these data sets were as expected, either extracting or excluding an eLEE signal as appropriate given the truth-level information released afterwards.

\section{The MicroBooNE Experiment}\label{sec:uboone}
\subsection{BNB Neutrino Flux}\label{sec:uboone_flux}
The booster neutrino beam line uses 8 GeV kinetic energy protons from the Booster accelerator to bombard a beryllium target. The primary hadrons produced in these interactions and the secondary hadrons further produced through additional interactions of primary hadrons and the surrounding materials are focused by a magnetic horn into a 50-m-long decay pipe. Decays of these hadrons produce a neutrino beam. The neutrino horn current polarity has been set to focus positive charged hadrons, resulting in a beam with a small antineutrino component for all data taking to date.

\begin{figure}[!htp]
\begin{overpic}[width=0.49\figwidth]{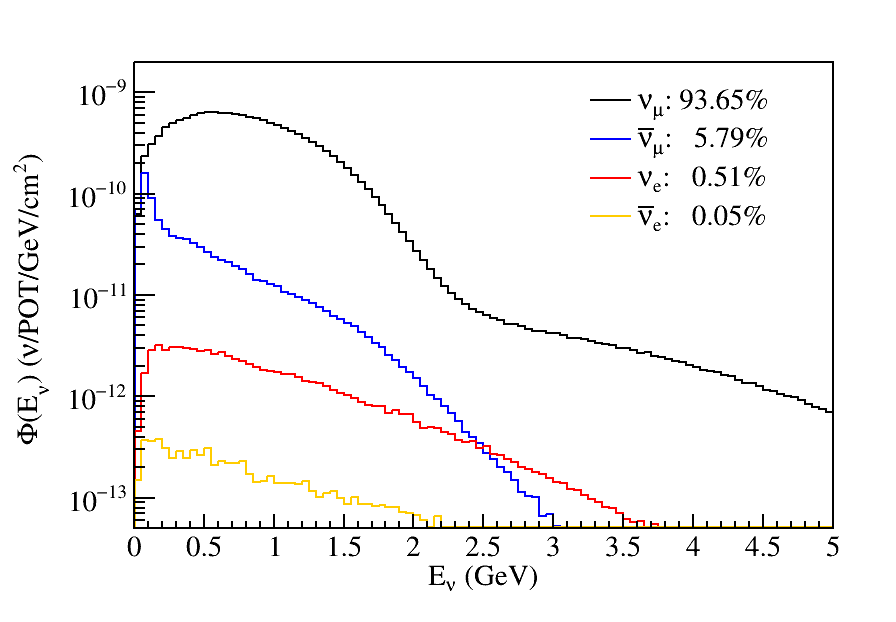}
\put(35,64){\text{(a)}\ \ \ \ \ \ \ \  MicroBooNE Simulation}
\end{overpic}    
\begin{overpic}[width=0.49\figwidth]{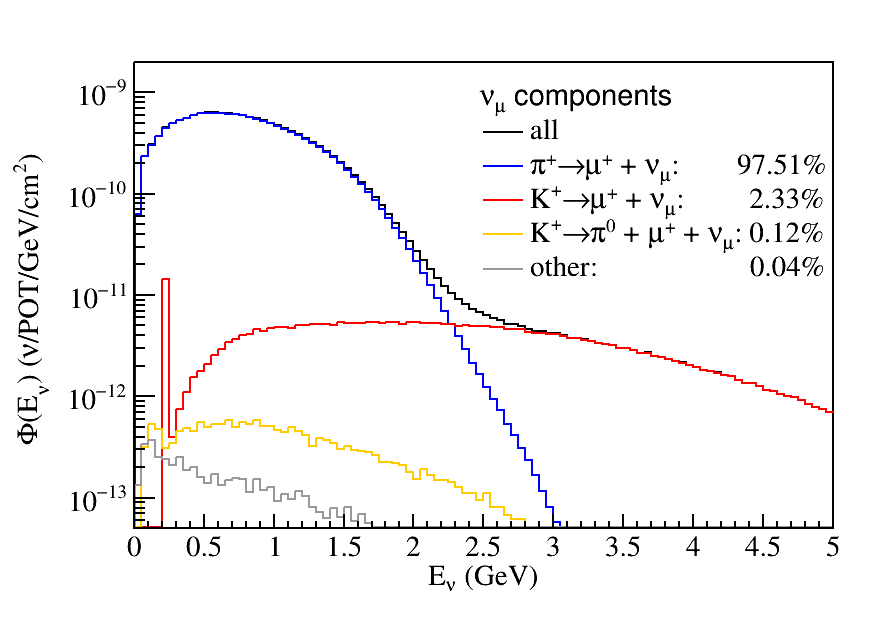}
\put(35,64){\text{(b)}\ \ \ \ \ \ \ \  MicroBooNE Simulation}
\end{overpic}    
\begin{overpic}[width=0.49\figwidth]{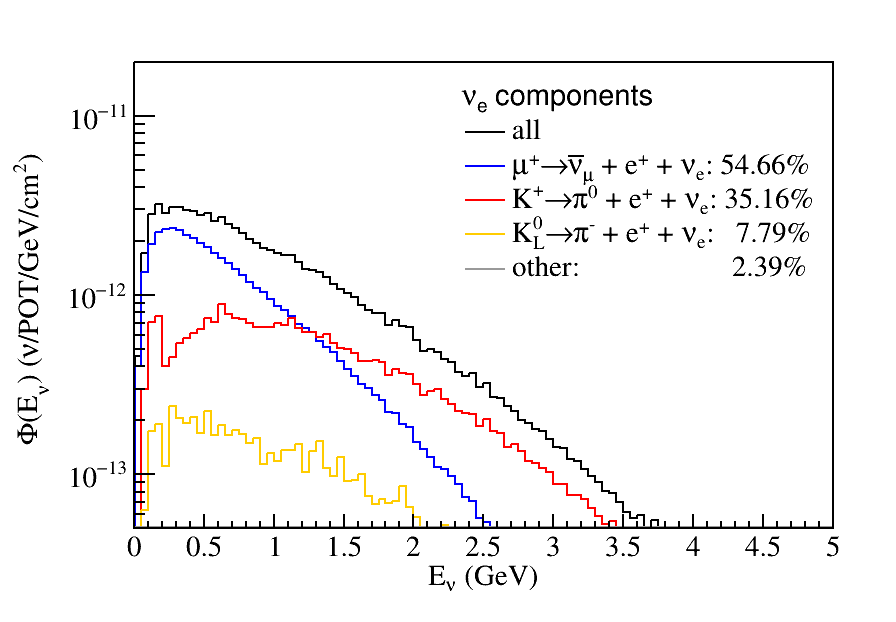}
\put(35,64){\text{(c)}\ \ \ \ \ \ \ \  MicroBooNE Simulation}
\end{overpic}    
\caption{Predicted neutrino flux from different decay modes at the MicroBooNE detector: (a) flux of different neutrino flavors, (b) $\nu_\mu$ flux of different decay modes (the peak at 236 MeV corresponds to Kaon decay at rest), (c) $\nu_e$ flux of different decay modes.}
  \label{fig:nu_flux}
\end{figure}

The BNB flux is simulated based on a \textsc{Geant4} framework~\cite{Agostinelli:2002hh} following the earlier work by the MiniBooNE collaboration~\cite{AguilarArevalo:2008yp}. The top panel of Fig.~\ref{fig:nu_flux} shows the composition of the BNB flux in terms of different neutrino flavors seen by the MicroBooNE detector, which is located at 468.5~m on axis from the target. The dominant neutrino species is the $\nu_\mu$, which is mostly produced by a $\pi^+$ two-body decay mode as shown in the middle panel in the same figure. For energies higher than $\sim$2.3 GeV, another two-body decay mode, $K^+ \rightarrow \mu^+ + \nu_\mu$, becomes the main mechanism to generate $\nu_\mu$. The $\nu_e$ flux is about 0.5\% of the overall neutrino flux. For neutrino energies below about 1.2 GeV, $\nu_e$'s are mostly produced by a three-body decay mode, $\mu^+ \rightarrow e^+ + \bar{\nu}_\mu + \nu_e$, with $\mu^+$ originating from the $\pi^+$ two-body decay mode as shown in the bottom panel of Fig.~\ref{fig:nu_flux}. For neutrino energies higher than 1.2 GeV, $\nu_e$s are mostly produced by another three-body decay mode, $K^+ \rightarrow \pi^0 + e^+ + \nu_e$. Since the parent hadron species are mostly common, the predictions of $\nu_\mu$ and $\nu_e$ flux are strongly correlated, which supports the overall strategy of using \numuCC\ events to provide constraints on the prediction of \nueCC\ events. 

\begin{figure*}[!htp]
	\includegraphics[width=0.4\figwidth]{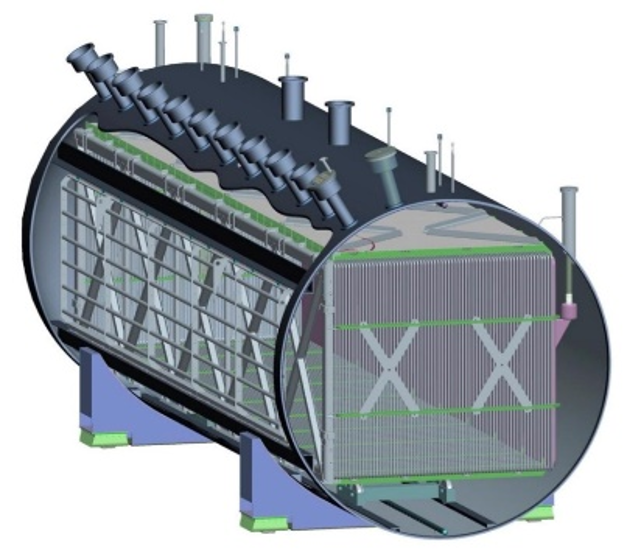}% Here is how to import EPS art
	\includegraphics[width=0.58\figwidth]{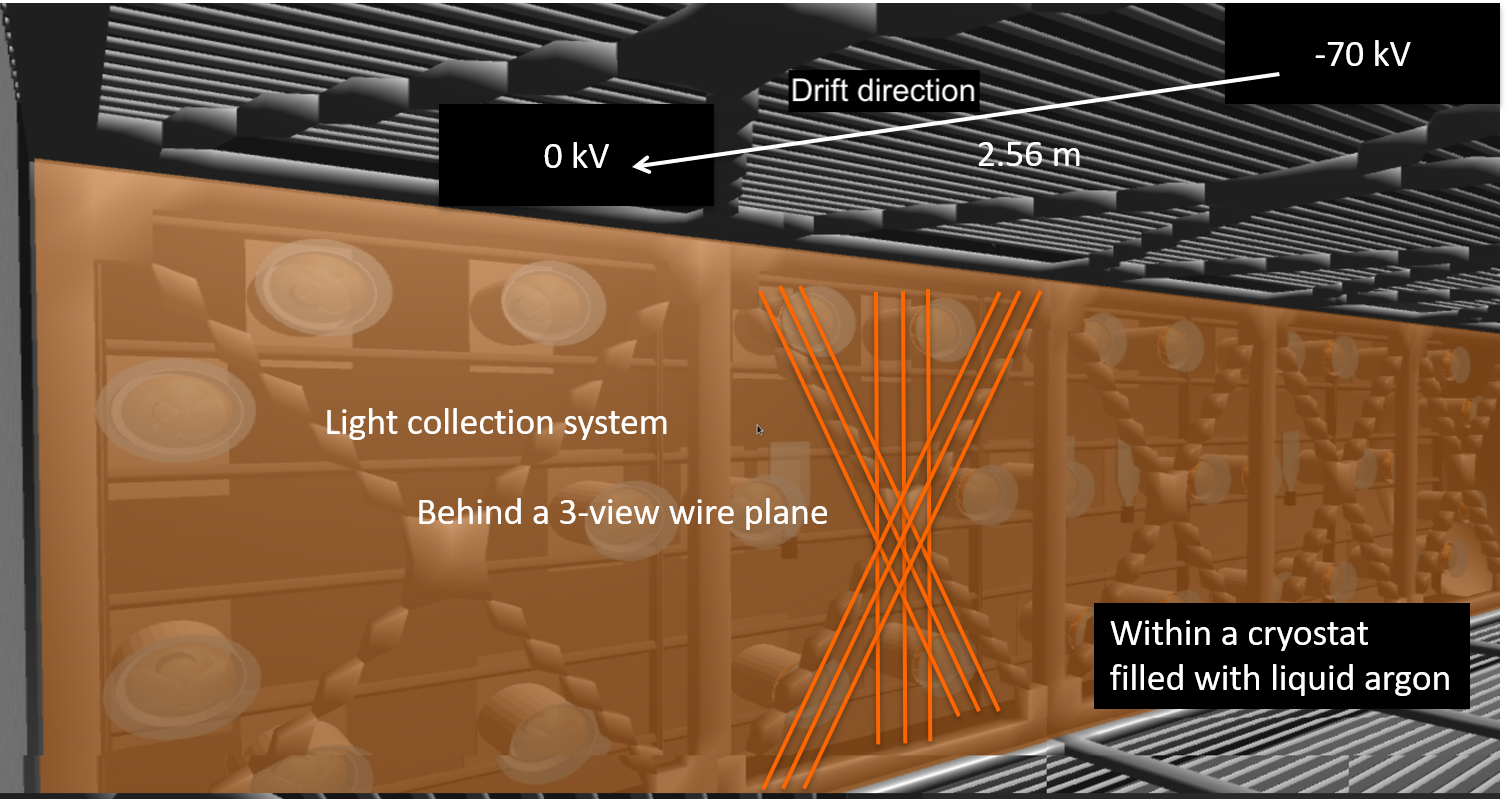}% Here is how to import EPS art
	\put(-360,-10){(a)}
	\put(-140,-10){(b)}
	\caption{\label{fig:uboone_det} Taken from Ref.~\cite{Abratenko:2021bzb}. (a) The MicroBooNE detector cryostat with the field cage shown inside. (b) Inside the cryostat of the MicroBooNE detector, visualized with the VENu software~\cite{DelTutto:2017vtk}. The maximum drift distance is 2.56~m with a drift electric field of 273~V/cm. The light-collection system, which consists of 32 PMTs, is located behind the three anode wire planes, which detect ionization charge. }
\end{figure*}

\subsection{MicroBooNE Detector}~\label{sec:uboone_det}
The MicroBooNE detector~\cite{Acciarri:2016smi} is a 10.4~m long, 2.6~m wide, and 2.3~m high LArTPC, located on-axis along the BNB at a distance of 468.5~m from the beryllium target, 72.5~m upstream of the MiniBooNE detector. It consists of approximately 85 metric tons of liquid argon in the active volume for ionization charge detection along with an array of 32 photomultiplier tubes (PMTs)~\cite{Briese:2013wua} for the scintillation light detection\footnote{One PMT stopped working after the run 1 data taking.}. Figure~\ref{fig:uboone_det}(a) shows the MicroBooNE TPC, which is housed in a foam-insulated evacuable cryostat vessel. The cathode-plane high voltage is set at -70~kV during normal operation, creating a drift field of 273~V/cm. The ionization electrons drift at a speed of 1.1~mm/$\mu$s~\cite{Li:2015rqa} in the drift field. This corresponds to 2.3~ms drift time for the maximum 2.56~m drift distance. 

As shown in Figure~\ref{fig:uboone_det}(b), there are three parallel wire readout planes at the anode side of the TPC. These planes are labeled as the ``U'', ``V'', and ``W'' planes\footnote{The ``W'' plane is sometimes also referred to as the ``Y'' plane.}, and each plane contains 2368, 2368, and 3456 wires, respectively. The wire spacing within a plane is 3~mm, and the planes are spaced 3~mm apart. The wires in the W plane are aligned vertically and the wires in the U and V planes are oriented at $\pm$60$^\circ$ with respect to the vertical direction. The different orientations of the wires allow for determination of the positions of the ionization electrons within the plane that is transverse to the drift direction. Bias voltages for the U, V, and W planes are -110\,V, 0\,V, and 230\,V, respectively, which satisfies the transparency condition that all drifting electrons pass through the U and V (induction) wire planes and are fully collected on the W (collection) plane. The induced current on each wire is amplified, shaped, and digitized through a custom designed front-end application-specific integrated circuit~\cite{Radeka:2011zz} operating at 89~K in the liquid argon. The direct implementation of readout electronics in the cold liquid significantly reduces electronics noise. The equivalent noise charge on each wire is generally below 400 electrons. A minimum ionizing particle usually produces in total 13,000 electrons at a single wire if the particle trajectory is perpendicular to the wire orientation~\cite{Acciarri:2017sde}.

% PMT system description ...
Figure~\ref{fig:uboone_det}(b) also shows the light-collection system behind the anode wire planes. This light-collection system is used to detect scintillation light from the liquid argon providing the precise timing of particle activity, which is crucial to rejection of the high-rate cosmic-ray background necessary in a surface-operating LArTPC detector like MicroBooNE. Thirty-two 8-inch Hamamatsu R5912-02MOD PMTs~\cite{Briese:2013wua} provide uniform coverage of the anode plane.  An acrylic plate coated with tetraphenyl butadiene is installed in front of each PMT to shift the ultraviolet argon scintillation light to the visible part of the spectrum to which the PMT is sensitive.  
The magnitude of the detected light on each PMT provides position information for time-isolated particle activities, which is compared with the light pattern predicted from the ionization charge signals in the TPC. A successful match~\cite{Abratenko:2020hpp} of charge and light signals determines the association between individual TPC activity and light detection and, therefore, the time of the corresponding TPC activity.

\subsection{Event Trigger and Readout}~\label{sec:uboone_readout}
The BNB delivers proton pulses at a rate of 5 Hz and approximately 4$\times$10$^{12}$ POT per pulse. Each pulse is called a beam spill and lasts $\sim$1.6~$\mu$s with 82 2-ns wide bunches of protons. The MicroBooNE detector is expected to have one neutrino interaction inside the TPC active volume per about 600 spills. Each beam spill initiates a hardware trigger in the MicroBooNE data acquisition (DAQ), which records 4.8~ms (digitized at 2~MHz) of TPC data of ionization charge signals and 6.4~ms (digitized at 64~MHz) PMT data of scintillation light signals. The PMT data contain two separate trigger streams, a forced trigger covering the beam spill which is referred to as the  beam discriminator and a self-trigger which is referred to as the  cosmic discriminator.  In each event, there are on average 26 cosmic-ray
muon induced activities observed in the 4.8~ms TPC readout window.

One such record of TPC and PMT data corresponds to one event which covers not only the beam spill but also the cosmic-ray muon activity in the proximity of the beam spill. 
The time window of an event accounts for the few-millisecond delay of the TPC readout signal coming from the relatively slow drift of ionization electrons.
Following the hardware trigger, a software trigger is applied to reduce the data size. It requires distinct PMT signals within the beam spill window. This results in a reduction of the event rate by a factor of 22 but a negligible efficiency loss for neutrino interactions.

	\begin{figure*}[htp!]
		\includegraphics[width=0.5\textwidth]{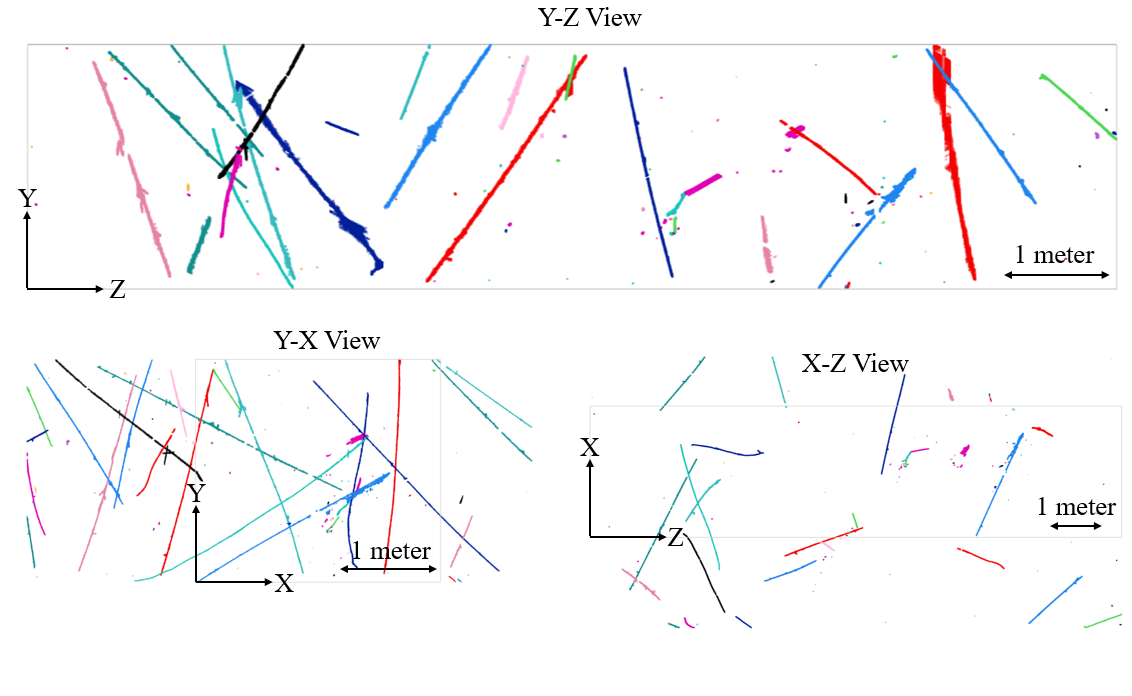}% Here is how to import EPS art
		\includegraphics[width=0.5\textwidth]{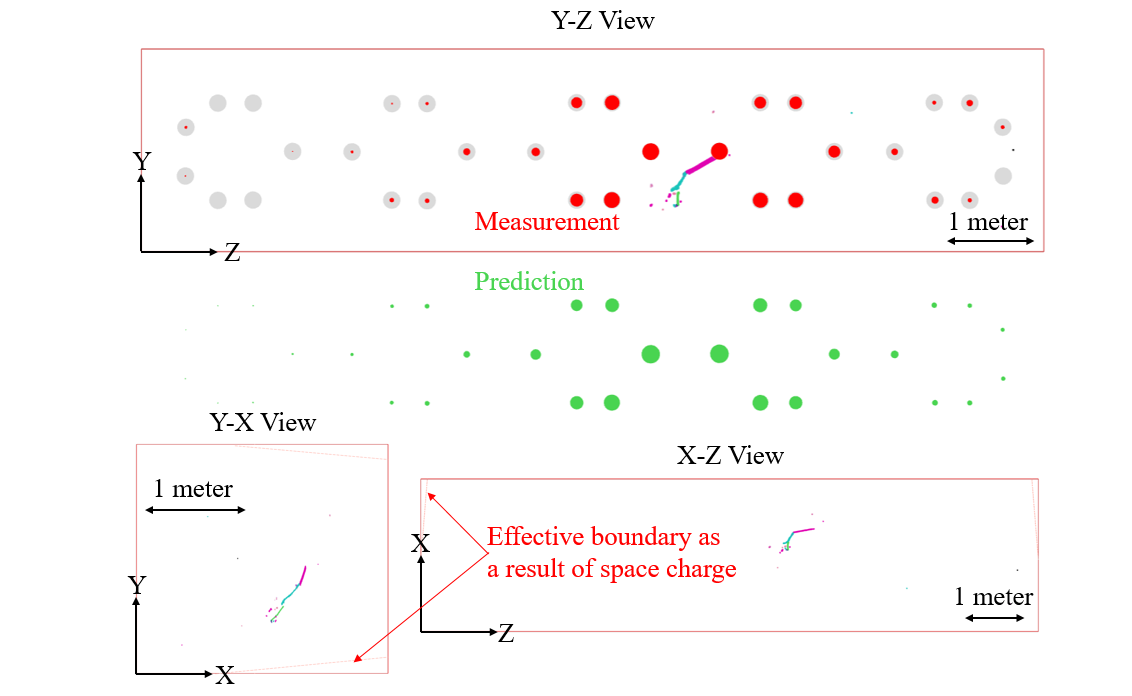}% Here is how to import EPS art
		\put(-330,160){MicroBooNE data run 6195 event 1020}
		\put(-360,-10){(a)}
		\put(-140,-10){(b)}
		\caption{\label{fig:matching} A \nueCC\ interaction candidate from MicroBooNE data. The X axis is the drift electric field direction from the TPC anode to the cathode. The Y axis is vertical up, and the Z axis is along the neutrino beam direction.
			%from run 6195, subrun 20, event 1020.
			Panel (a) shows three 2D projections of reconstructed 3D clusters in the full TPC readout window before charge-light matching. Each cluster is shown in a different color. The gray box represents the TPC active volume while the two ends along the X axis correspond to the trigger time and the maximum drift time relative to the trigger.
			Panel (b) shows the 2D projections of the \nueCC\ candidate cluster after applying the charge-light matching. The red (green) circles represent the observed (predicted) number of photoelectrons (PEs) at each PMT, where the area of the circle is proportional to the number of PEs.
			The effective detector boundary as a result of the space charge effect is indicated by the red dashed lines in the corner of the TPC active volume as shown in the ``Y-X view'' and ``X-Z view''.}
	\end{figure*}

\subsection{Monte Carlo Simulation}\label{sec:uboone_mc}
%GENIE event generator + GEANT4 simulation  
The simulated neutrino flux (introduced in Sec.~\ref{sec:uboone_flux}) is provided to the event generator \textsc{Genie} v3~\cite{Andreopoulos:2009rq, GENIE:2021npt} to generate neutrino-argon interactions inside\footnote{Here ``inside'' also includes the 11.1~mm thick stainless steel wall of the cryostat but excludes the foam insulation covering the cryostat outer surfaces.} and outside the cryostat. The latter case is also referred to as the {\it dirt} background. 
%% expanding on GENIE 
\textsc{Genie} v3.0.6, G18\_10a\_02\_11a, was used, which includes improvements on the usage of the Valencia model~\cite{Nieves:2011yp,Nieves:2004wx,Gran:2013kda} for the local Fermi gas nucleon momentum distributions, improvements in the CCQE and CC two-particles-two-holes (CC2p2h) interactions, and improvements in the treatments of final state interaction (FSI) and pion production, when compared to \textsc{Genie} v2\_12\_2 (used by previous MicroBooNE analyses~\cite{Abratenko:2019jqo,Abratenko:2020acr,Abratenko:2020sga,Abratenko:2021nlt}). In addition to the default configuration, the parameters governing the CCQE and CC2p2h models are adjusted according to the T2K CC0$\pi$ cross-section results~\cite{Abe:2016tmq}. Given that T2K has a neutrino flux in a similar energy range to that of MicroBooNE, this additional adjustment represents an improved model prediction and uncertainty treatment~\cite{uboone_genie_tune}, despite the fact that T2K data is on carbon-hydrogen and MicroBooNE's measurement is on argon. Corrections from carbon to argon targets are applied based on a smooth nuclear dependence (or $A$-dependence) that are determined in fits to inclusive and semi-inclusive electron scattering data.

The resulting final state particles of each simulated MC\footnote{Throughout this paper, we use the acronym ``MC'' to identify the output from our event generator simulation.} event are processed using the \textsc{LArSoft}~\cite{Snider:2017wjd} software framework, which is a toolkit to perform simulation, reconstruction, and analysis of LArTPC events. The final state particles are propagated through the detector using the \textsc{Geant4} toolkit~\cite{Agostinelli:2002hh} v4\_10\_3\_03c. The resulting energy depositions are further ported to dedicated detector simulation programs 
taking into account all detector effects to simulate the ionization charge and scintillation light signals after taking into account the space charge effect.
% recombination, space charge 
The space charge effect~\cite{Adams:2019qrr,Abratenko:2020bbx} is caused by the accumulation of positively charged ions inside the active volume. For on-surface LArTPC detectors such as MicroBooNE, cosmic-ray muons provide a constant source of positively charged ions, which distort the local drift electric field.
Consequently, the ionization electrons bend toward the detector center when drifting to the anode plane and the reconstructed positions of ionization electrons appear to be closer to the detector center compared to their true position making the effective detector boundary smaller than the actual active TPC boundary. In addition to the deformation of reconstructed positions, the distortion in the electric field also changes the amount of ionization electrons and scintillation photons through the charge recombination process~\cite{Jaskolski:2011qja,ArgoNeuT:2013kpa}. 
After comparing simulation (with these effects implemented) and calibration run data, the amount of ionization electrons is further scaled down in the simulation in order to include the effects from conversion of analog-to-digital converter values to the number of ionization electrons and position-dependent
energy calibrations~\cite{Adams:2019ssg}, which results in an average scaling factor of 0.86.

%TPC simulation
The position and number of ionization electrons modified by recombination and the space charge effect is ported to the TPC detector simulation~\cite{Adams:2018dra}, which takes into account the charge transportation and diffusion~\cite{Li:2015rqa}. The induced currents on the wires are simulated by convolving the ionization charge distribution at the wire plane with the position-dependent (at 1/10th of the wire pitch resolution) field response function, which are calculated by the dedicated \textsc{Garfield} simulation~\cite{Veenhof:1998tt} following Ramo's theorem~\cite{Ramo:1939vr}. The induced current is further convolved with the electronics response before adding the inherent electronics noise from data and is then digitized to generate the final waveform on each wire channel. 

%Light simulation
The optical detector simulation models the light emitted by charged particles interacting with the detector and produces signals in photomultiplier tubes. A full optical simulation implements the \textsc{Geant4} simulation of individual optical photons produced along the path of charged particles through both the scintillation and Cherenkov processes. These photons are stepped through several processes with the detector medium including Rayleigh scattering, reflection, and partial absorption in order to produce a realistic detector response to the light source. However, because of the vast number of photons typically produced in a neutrino physics event, the full optical simulation can take hours or days per event. Therefore, a fast optical simulation was developed to overcome this problem for routine simulation tasks. This mode utilizes a library of stored  data that represent the PMT acceptance of scintillation light signals\footnote{The contribution from Cherenkov light is insignificant at this stage.} to sample an expected detector response given an isotropic emission of light at a certain point in the volume. The PMT response is further convolved with the time distribution of these photons to generate the digitized waveform.

%Event overlay scheme
Compared to early MicroBooNE analyses~\cite{Abratenko:2019jqo,Abratenko:2020acr,Abratenko:2020sga,Abratenko:2021nlt}, the simulation used in this eLEE search adopts a scheme of overlaying the simulated neutrino interactions with dedicated beam-off data. These data are taken without the neutrino beam and are triggered by a random signal to mimic the neutrino beam gate. The simulated TPC and PMT waveforms from neutrino interactions are overlaid with the data waveform of a beam-off event. Such simulation is also referred to as overlay MC simulation. This scheme eliminates the systematic uncertainties in the simulation of excessive electronics noise and cosmic-ray muon activity. This scheme limits the statistics of the overlay MC sample because of the finite size of available beam-off data sample to be overlaid. Nevertheless, the statistics of the MC background 
sample is more than a factor of three larger than that of the data. As elaborated in Sec.~\ref{sec:syst_mcstat}, a Bayesian approach was developed to obtain an optimal estimation of MC statistical uncertainties, even when the predicted background is zero.

\subsection{Event Reconstruction}\label{sec:uboone_wirecell}
%% TPC Signal Processing
An end-to-end automated event reconstruction chain containing various fundamental reconstruction techniques was developed and implemented in this analysis. TPC signal processing, which mitigates the excess noise~\cite{Acciarri:2017sde} and deconvolves the detector response from the drift electric field and the electronics readout~\cite{Adams:2018dra, Adams:2018gbi}, provides the reconstructed ionization charge distributions for each wire to the subsequent calorimetry and topology reconstruction algorithms. A tomographic three-dimensional (3D) image reconstruction algorithm, Wire-Cell~\cite{Qian:2018qbv}, is used as the core algorithm of the reconstruction chain. Wire-Cell uses reconstructed ionization charge at different times and readout wire 1D positions to reconstruct the 3D images of ionization electrons without topology heuristic assumptions (e.g. tracks from muons/pions/hadrons or EM showers from electrons/photons) prior to the pattern recognition stage. Other algorithms such as clustering and de-ghosting~\cite{Abratenko:2020hpp} are implemented to further improve the quality of the 3D images particularly addressing the challenge that gaps occur in the 2D view of the charge signals because of the inefficiency of TPC signal processing or non-functional wires. 
%% 3-D clustering
The space points (representing the 3D voxels with non-zero reconstructed charge) of the reconstructed 3D image are grouped into {\it clusters}, each of which represents individual physical activity in the TPC from a cosmic-ray muon or a neutrino interaction. Figure~\ref{fig:matching}(a) shows each 2D projection view of an event's 3D image after imaging and clustering.  

\begin{figure*}[htp!]
  \centering
  \includegraphics[width=0.95\textwidth]{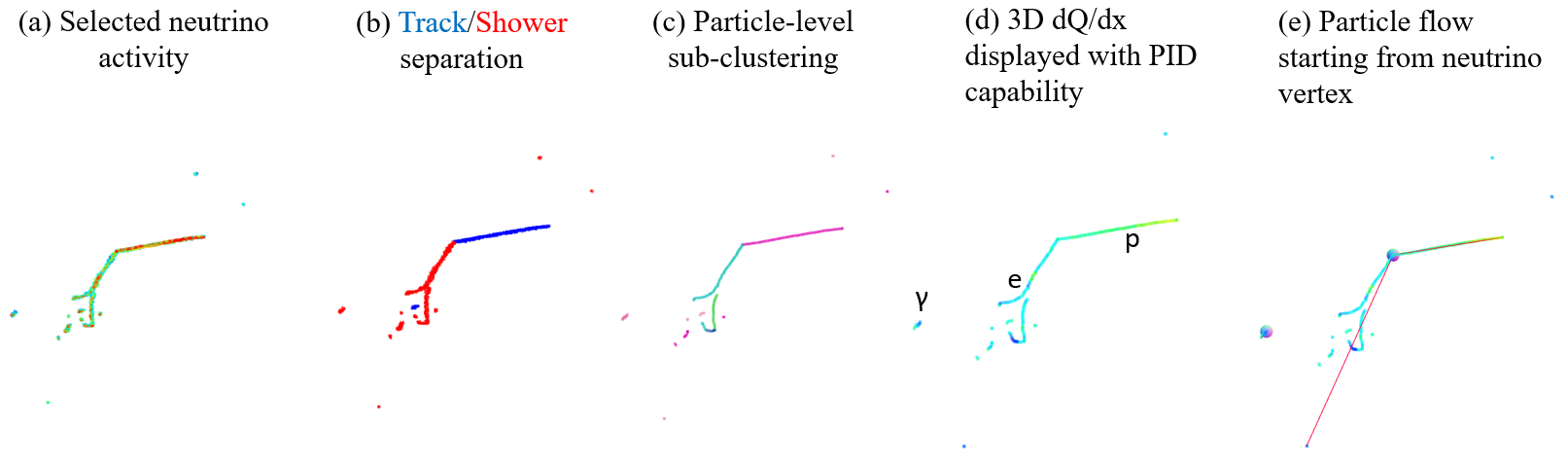}
  \put(-330,90){MicroBooNE data run 6195 event 1020}
  \caption{Displays of the Wire-Cell pattern recognition results at different stages.  (a) In-beam candidate neutrino cluster selected by the generic neutrino selection. The color scale represents the reconstructed charge associated with each space point, where blue and cyan are lower in charge and yellow and orange higher in charge. (b) Identified tracks and EM showers, which are displayed in blue and red, respectively. (c) Identified particles (or track segments), which are displayed in different colors. (d) Fitted $dQ/dx$ associated with each piece ($\sim$6~mm) along the trajectories. The blue, cyan, green, yellow, and red colors roughly correspond to 1/3, 1, 2, 3, and 4 times the $dQ/dx$ of a minimum ionizing particle (MIP). (e) Reconstructed particle flow starting from the primary neutrino interaction vertex, which is displayed in a rainbow-colored wheel.  %The original BEE weblink is \url{https://www.phy.bnl.gov/twister/bee/set/5a2e2fe7-e419-4cd2-8c3b-d9d53762f5a6/event/6/}.
  }
  \label{fig:illustration_pattern_recognition}
\end{figure*}

As previously discussed, the TPC ionization charge signal data, which provide the topology and calorimetry information, are collected separately from the PMT scintillation light signals, which provide the timing information, because of 
the longer drift time of ionization electrons relative to the light propagation. This results in a challenge, especially for surface-operating LArTPCs such as MicroBooNE, in identifying neutrino interactions from numerous cosmic-ray muon interactions~\cite{Acciarri:2017rnj, Adams:2018fud, Adams:2018sgn, Adams:2018lzd, Abratenko:2019jqo}.   
A many-to-many charge-light matching algorithm was developed to overcome this challenge by finding the corresponding light signals and providing the interaction time for each charge cluster~\cite{Abratenko:2020hpp}. About 70\% of the cosmic-ray muon events that pass the software trigger are rejected by requiring a charge cluster to have its start time coincide with the beam spill. These clusters are referred to as in-beam clusters.  
Figure~\ref{fig:matching}(b) shows an example from data, where the in-beam \nueCC\ candidate cluster is selected out of about 20 cosmic-ray muon clusters after charge-light matching.

% Generic neutrino selection
The majority of in-beam clusters still originate from cosmic-ray muons after charge-light matching for MicroBooNE events. Additional algorithms were developed to reject cosmic-ray backgrounds with 5-10\% efficiency loss for neutrino interactions~\cite{Abratenko:2020sxa}. The effective boundary of the TPC active volume considering the space charge effect~\cite{Adams:2019qrr, Abratenko:2020bbx} is used to define the fiducial volume in the cosmic-ray rejection as well as the subsequent neutrino selection. The fiducial boundary is defined as 3~cm inside the effective boundary, which leads to a fiducial volume of 94.2\% of the active TPC. The through-going muons, which traverse the TPC active volume, are rejected if the two ends of the track exit the fiducial boundary. The rejection of stopped muons, which enter the active volume and stop inside, is based on the identification of an increase of ionization charge loss per unit length ($dQ/dx$) near the end of the track (i.e.,~Bragg peak). This is obtained by a newly developed 3D track trajectory and $dQ/dx$ fitting procedure~\cite{Abratenko:2021bzb}, which is also an important ingredient in pattern recognition and particle identification. Note that the external cosmic-ray-tagger~\cite{Adams:2019bzt} system may provide additional rejection of cosmic-ray muon events, but is not included in this work as the system was not installed until late 2017. Using the above mentioned cosmic-ray rejection techniques including charge-light matching, the cosmic-ray background is reduced significantly resulting in less than 15\% cosmic-ray contamination in the selected neutrino candidate events while the original neutrino to cosmic-ray muon ratio was about 1:200 for the events passing the software trigger. The efficiency loss for CC neutrino interactions is 10-20\% for different flavors of neutrinos up to this stage~\cite{Abratenko:2020sxa, Abratenko:2021bzb}.

% Pattern recognition procedure ...
Pattern recognition is vital for the identification of different flavors of neutrinos, e.g. \nueCC\ and \numuCC, for a variety of physics analyses. The details of Wire-Cell pattern recognition can be found in Ref.~\cite{wire-cell-pr} and highlights are provided in the following. Wire-Cell pattern recognition starts by finding initial end points of track segments by searching for kinks and splits in the selected 3D in-beam cluster. Track segments and their end points are then determined by iterative multi-track trajectory and $dQ/dx$ fitting where linear algebra algorithms and graph theory operations are utilized to achieve a robust performance. Particle identification (PID) is performed based on the $dQ/dx$, topology information (direction, track or shower, etc.), and allowable particle flow relationships for each track segment. Candidate primary neutrino interaction vertices are concurrently identified as parts of the particle flow tree, which is a series of particles that starts from the neutrino interaction vertex and loops over all identified particles following the particle flow relationship. A deep neural network of \mbox{\textsc{SparseConvNet}}~\cite{scn1} is used to boost the performance of the neutrino vertex identification by predicting the distance from each 3D voxel to the neutrino vertex. It chooses from the neutrino vertex candidates, which are identified based on the above traditional algorithms, and determines the final reconstructed neutrino interaction vertex. The particle flow is then refined if needed. %The particles ($e^-$, $\mu^-$, $\pi^-$) and their respective counterparts ($e^+$, $\mu^+$, $\pi^+$) are in general indistinguishable in our pattern recognition. 
$\pi^0$ particles are reconstructed relying on the opening angle and topological information of the two decay $\gamma$'s and other $\pi^0$ decay modes like Dalitz decay plays a very minor role in this analysis. A proton ($p$) can be reconstructed if its length is $\gtrsim$1~cm which corresponds to a kinetic energy threshold of 35~MeV. Neutrons ($n$) are invisible in LArTPCs as they are neutral, but they could be suggested by off-vertex protons that result from the neutron scattering. Figure~\ref{fig:illustration_pattern_recognition} illustrates the results of pattern recognition at different stages. As reported in Ref.~\cite{wire-cell-pr}, this pattern recognition achieves 60-75\% efficiencies of good neutrino vertices (distance between reconstructed and true vertices below 1~cm) and subsequently a 80-90\% reconstruction efficiency for primary leptons
for charged-current neutrino interactions.   

\subsection{Neutrino Energy Reconstruction}\label{sec:uboone_nu_energy}
\begin{figure*}[htp!]
  \centering
  \begin{subfigure}[]{0.245\textwidth}
    \centering
    \includegraphics[width=\textwidth]{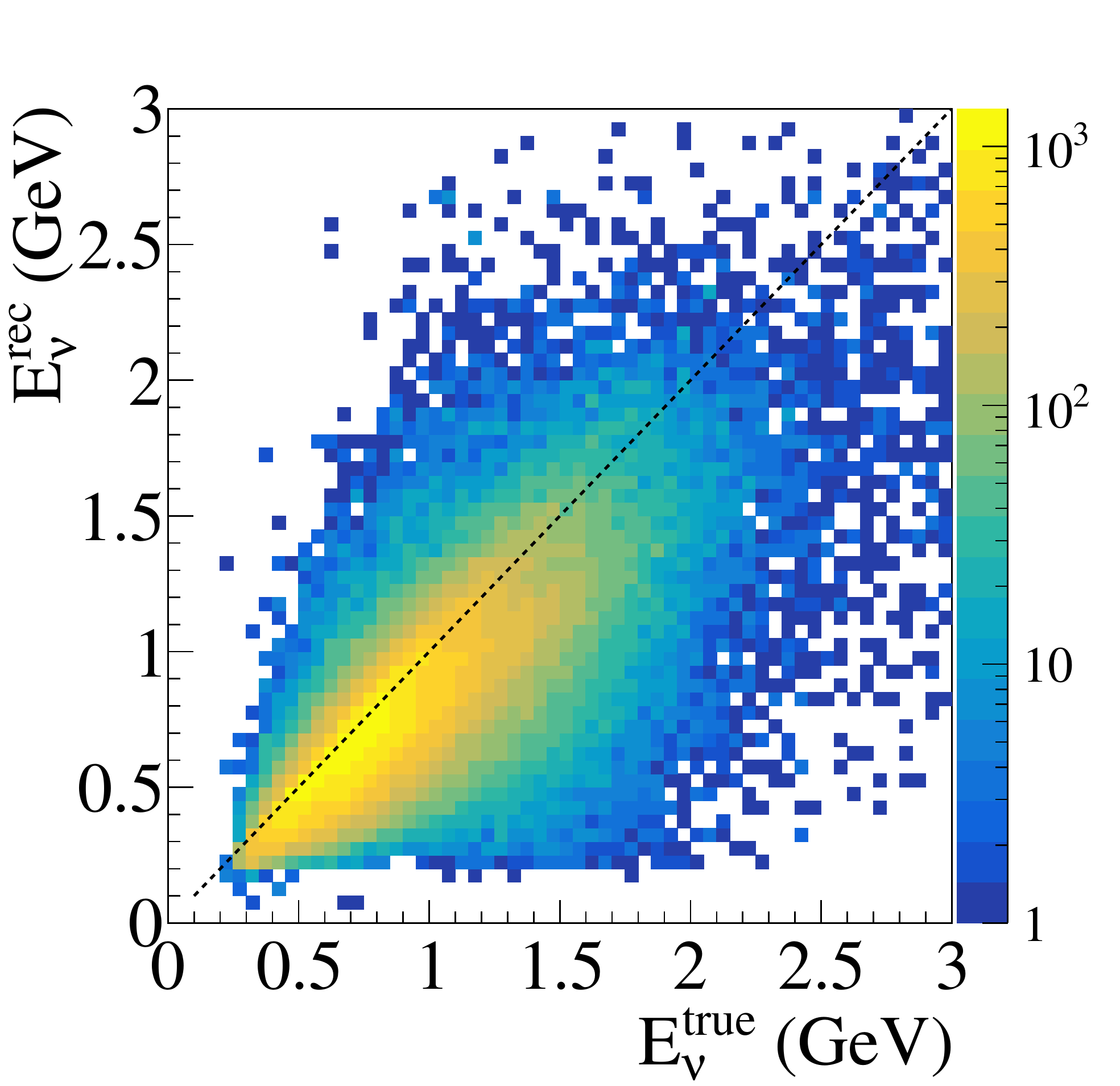}
    \put(-105,112){\scriptsize MicroBooNE Simulation}
    \caption{\numuCC\ candidates, FC}
  \end{subfigure}
  \begin{subfigure}[]{0.245\textwidth}
    \centering
    \includegraphics[width=\textwidth]{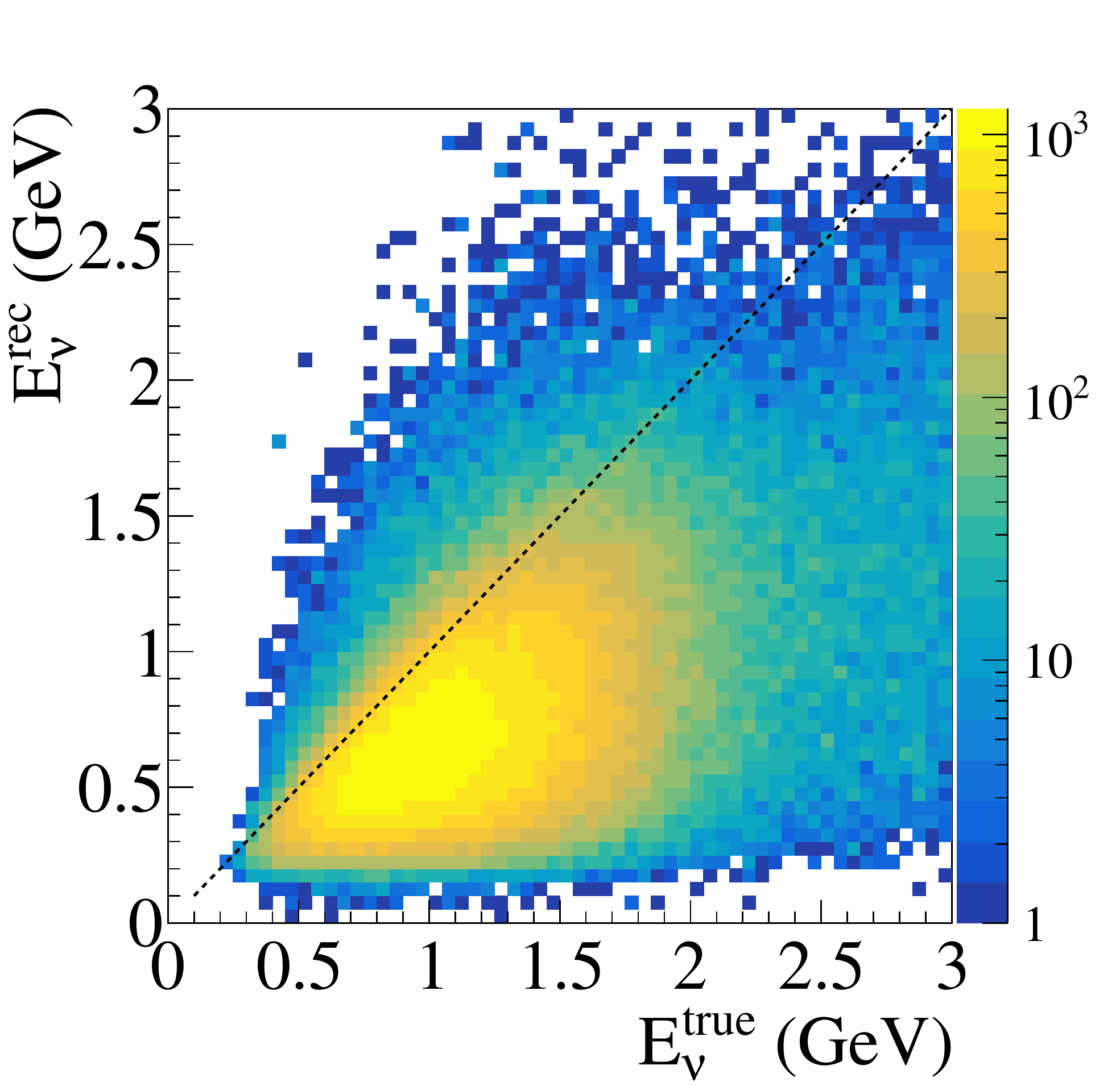}
    \put(-105,112){\scriptsize MicroBooNE Simulation}
    \caption{\numuCC\ candidates, PC}
  \end{subfigure}
  \begin{subfigure}[]{0.245\textwidth}
    \centering
    \includegraphics[width=\textwidth]{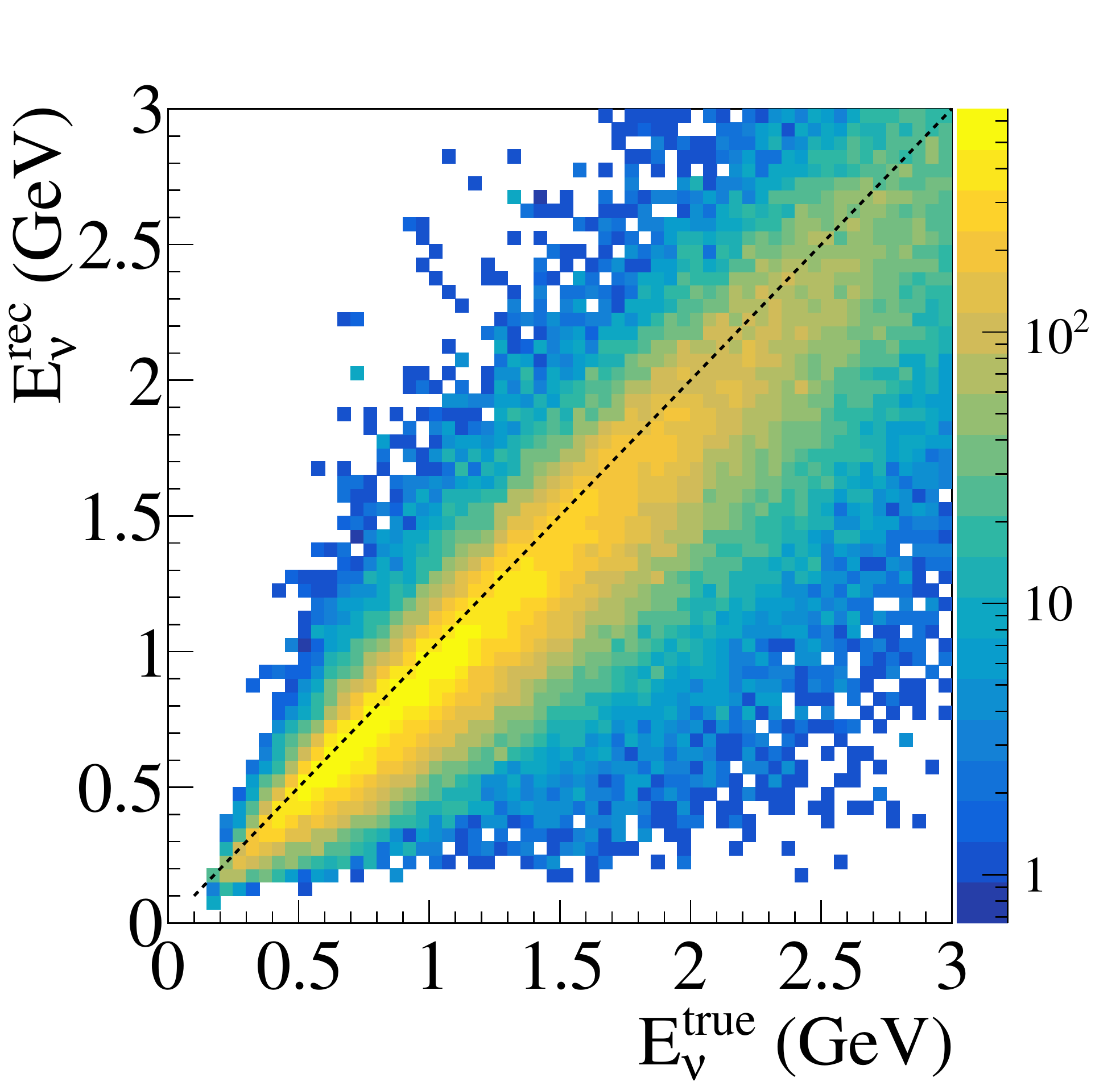}
    \put(-105,112){\scriptsize MicroBooNE Simulation}
    \caption{\nueCC\ candidates, FC}
  \end{subfigure}
  \begin{subfigure}[]{0.245\textwidth}
    \centering
    \includegraphics[width=\textwidth]{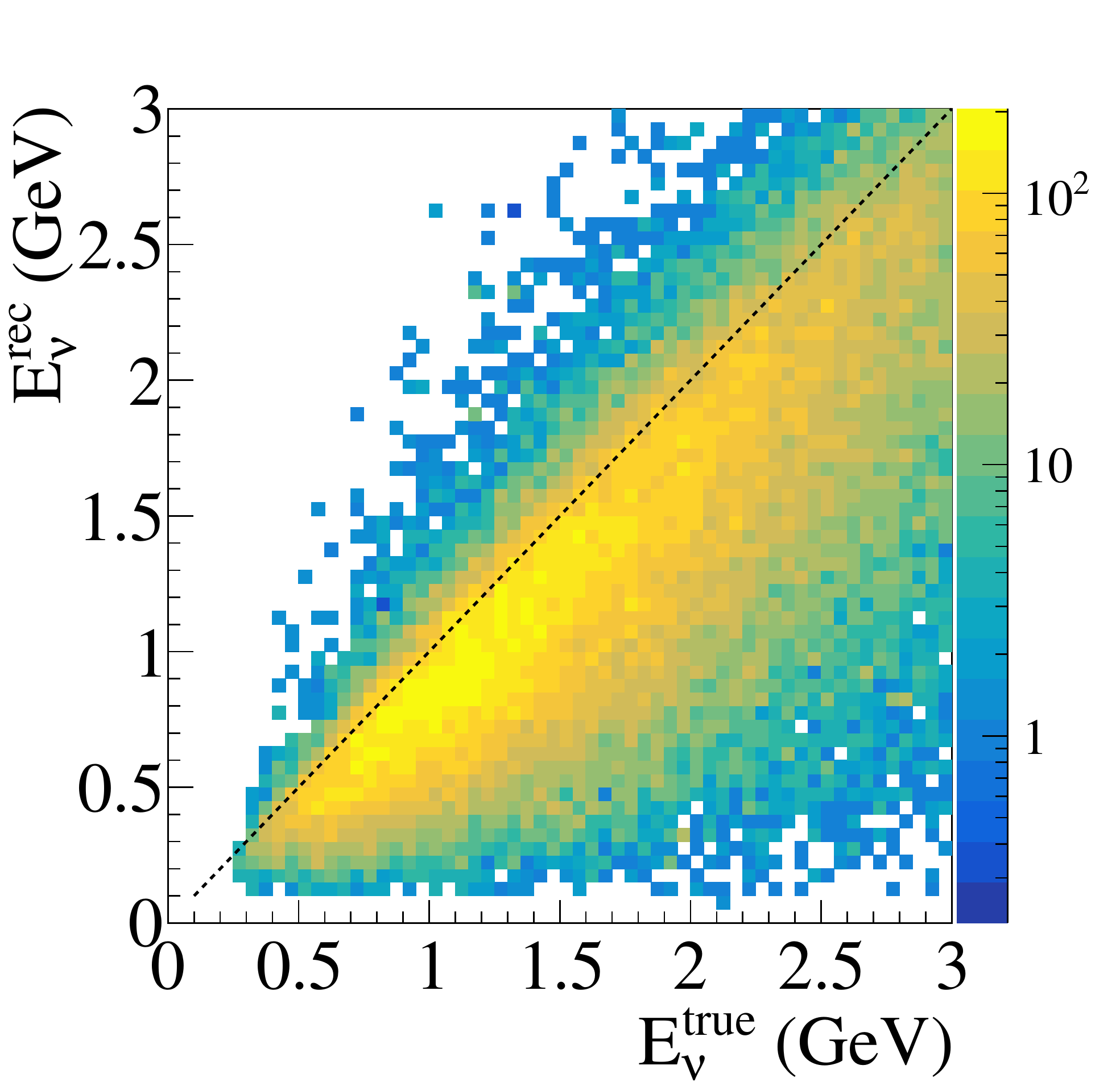}
    \put(-105,112){\scriptsize MicroBooNE Simulation}
    \caption{\nueCC\ candidates, PC}
  \end{subfigure}
  \begin{subfigure}[]{0.245\textwidth}
    \centering
    \includegraphics[width=\textwidth]{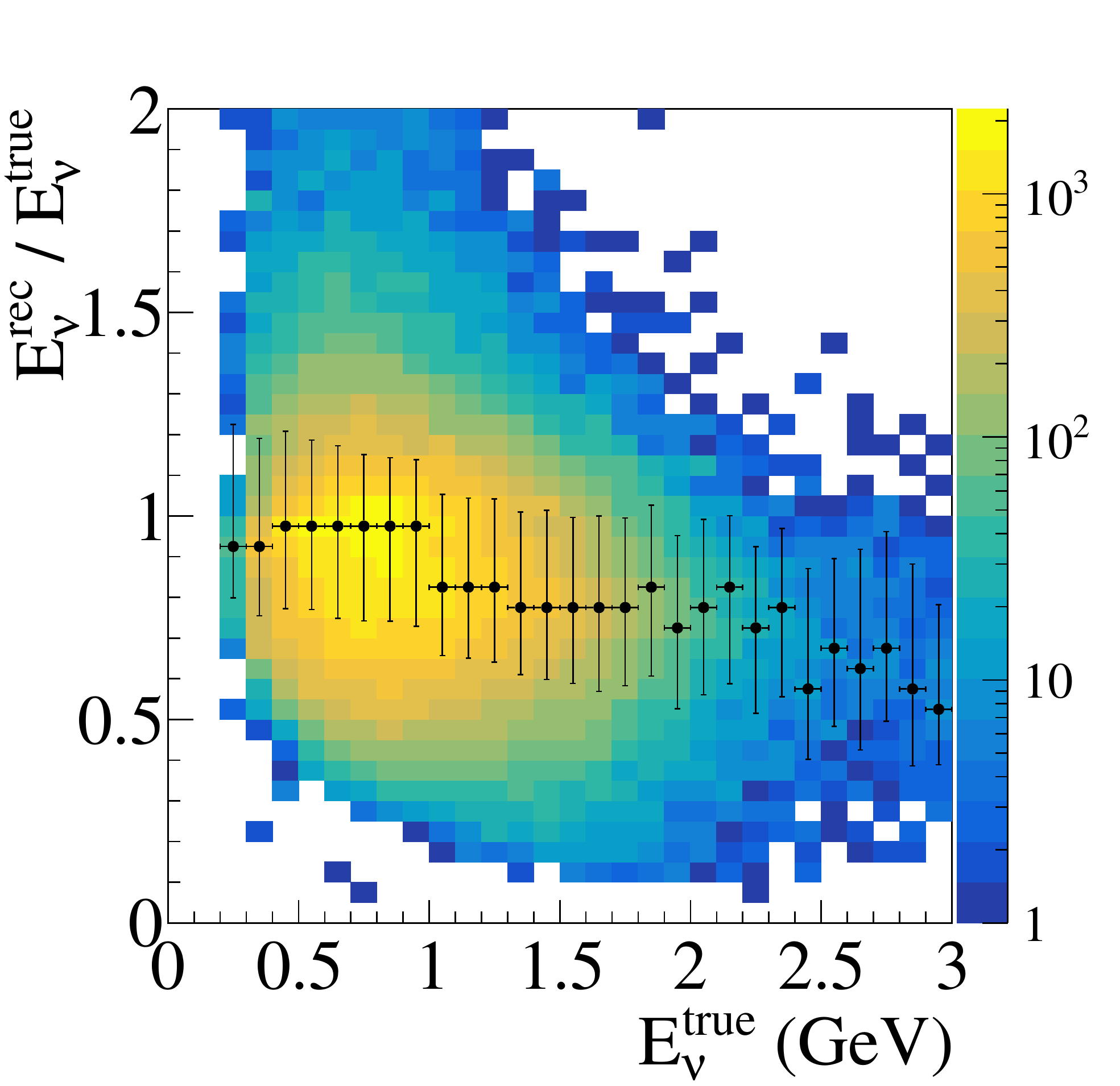}
    \put(-105,112){\scriptsize MicroBooNE Simulation}
    \caption{\numuCC\ candidates, FC}
  \end{subfigure}
  \begin{subfigure}[]{0.245\textwidth}
    \centering
    \includegraphics[width=\textwidth]{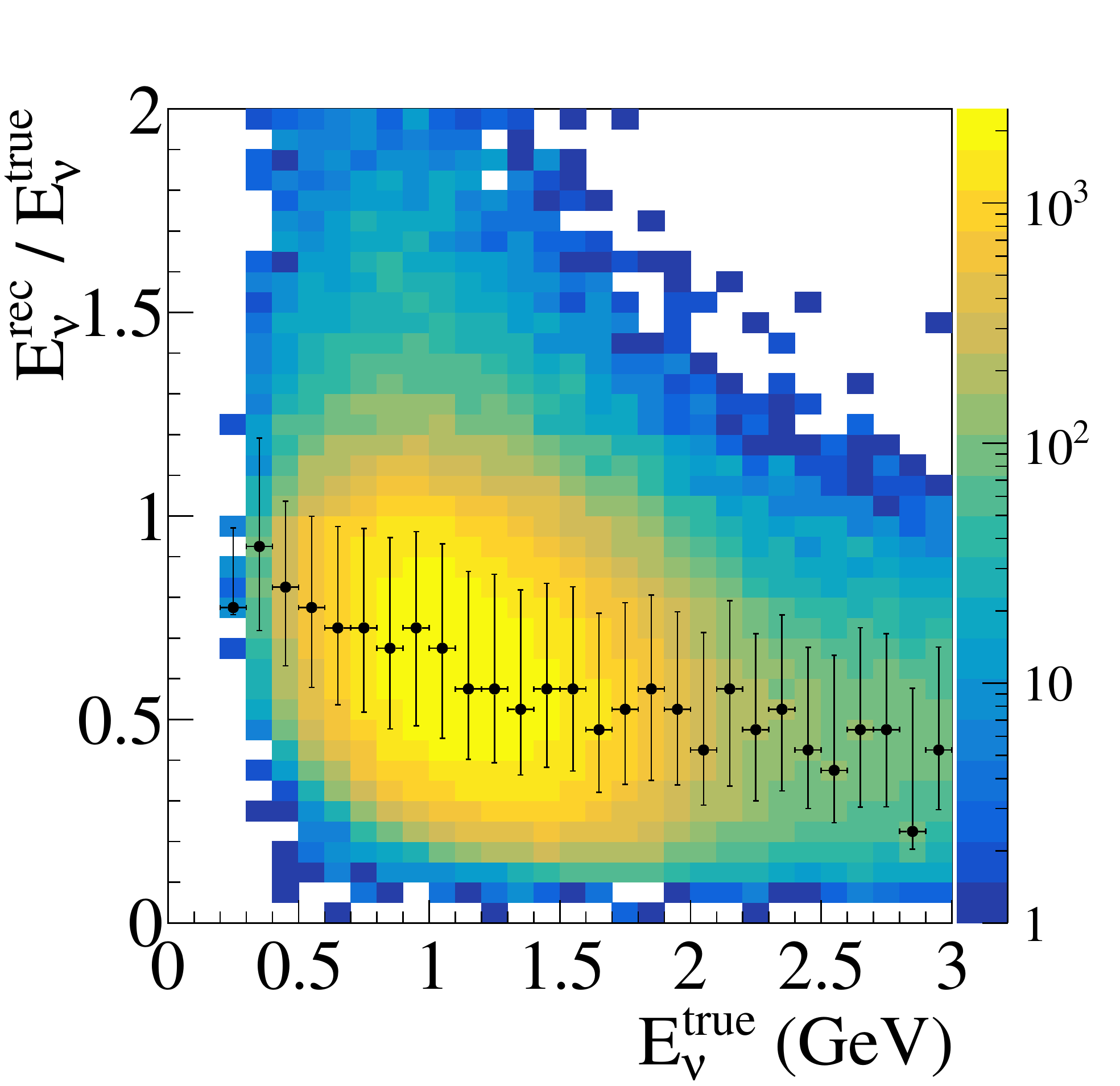}
    \put(-105,112){\scriptsize MicroBooNE Simulation}
    \caption{\numuCC\ candidates, PC}
  \end{subfigure}
  \begin{subfigure}[]{0.245\textwidth}
    \centering
    \includegraphics[width=\textwidth]{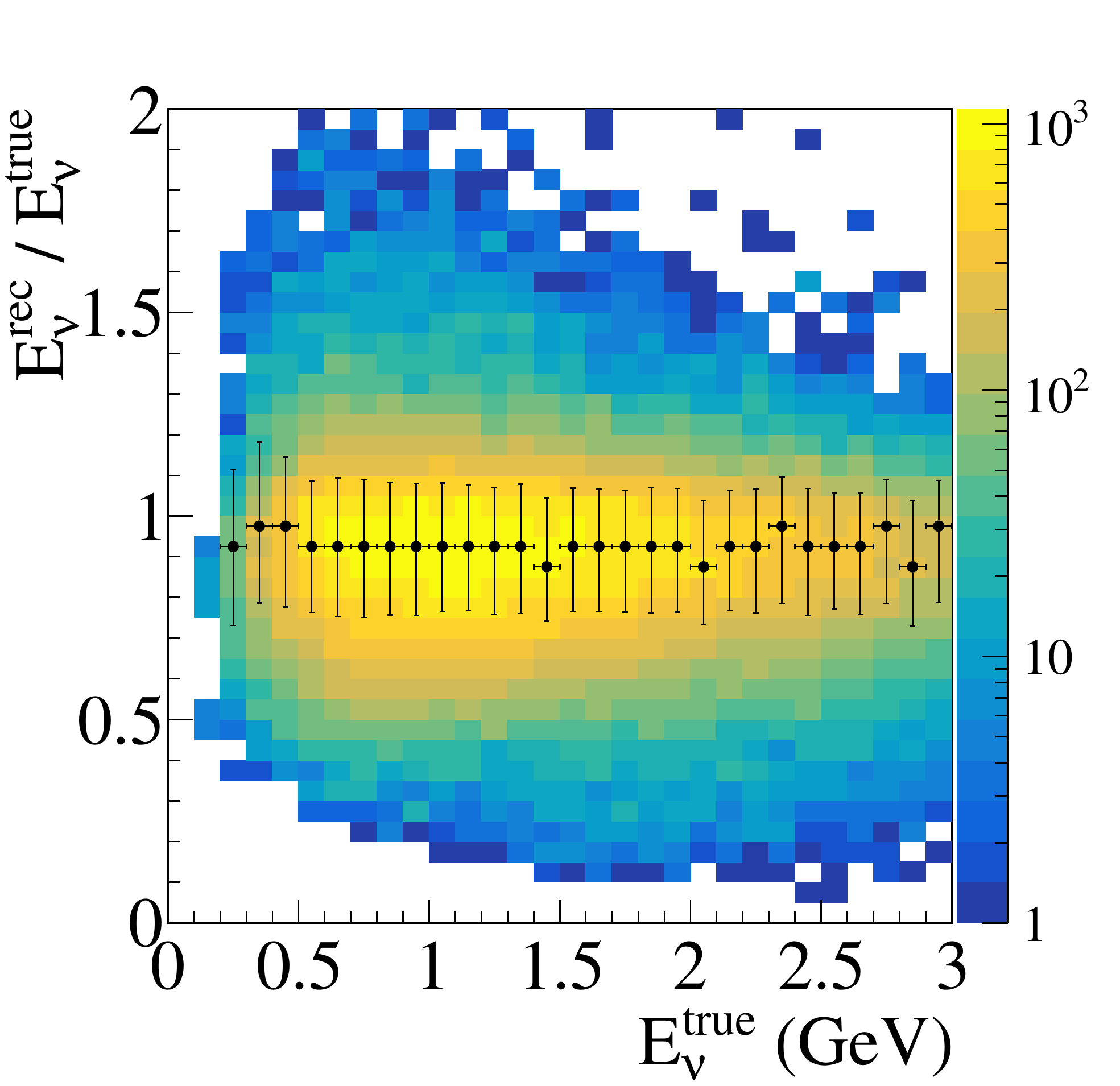}
    \put(-105,112){\scriptsize MicroBooNE Simulation}
    \caption{\nueCC\ candidates, FC}
  \end{subfigure}
  \begin{subfigure}[]{0.245\textwidth}
    \centering
    \includegraphics[width=\textwidth]{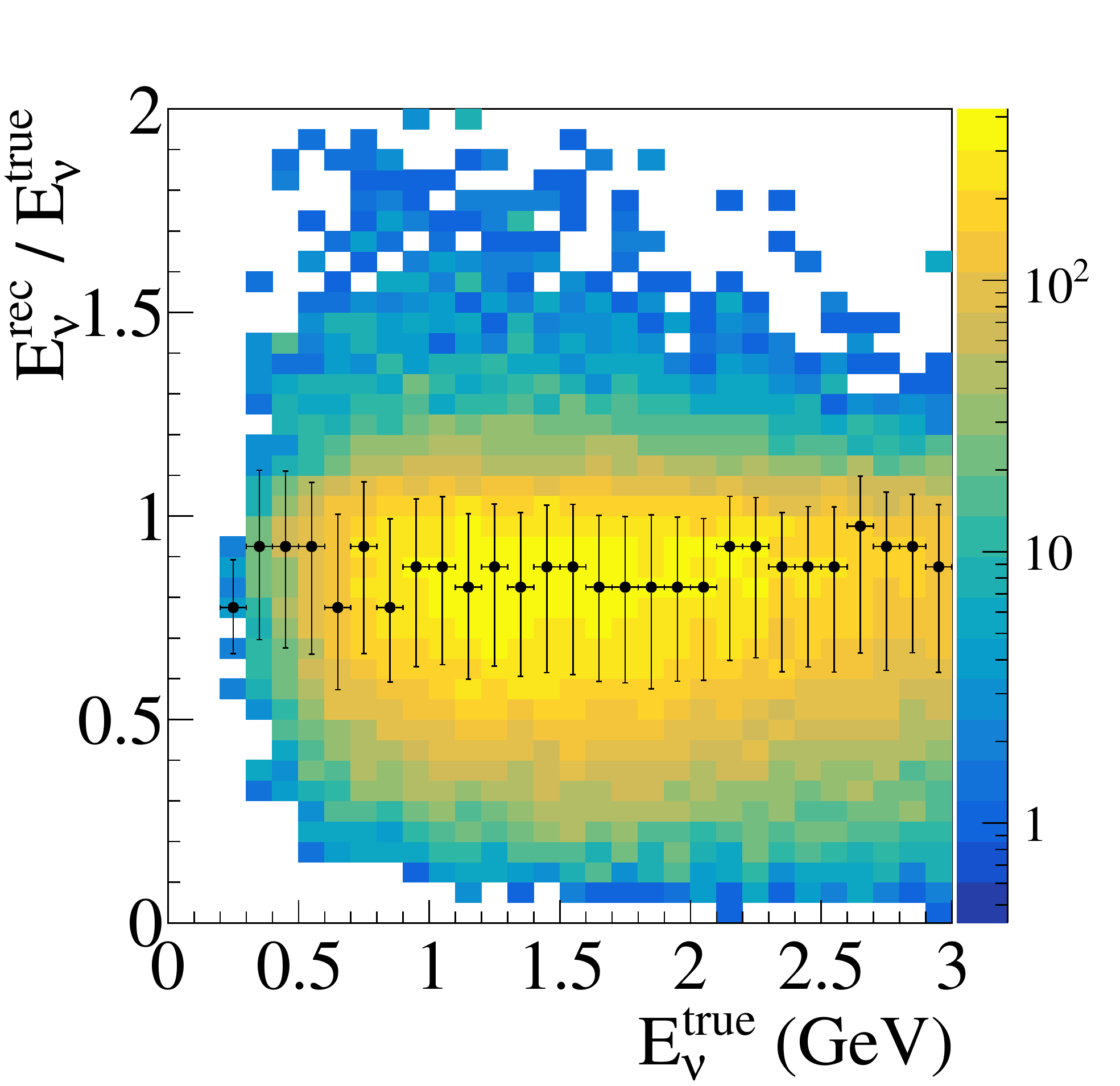}
    \put(-105,112){\scriptsize MicroBooNE Simulation}
    \caption{\nueCC\ candidates, PC}
  \end{subfigure}
  \caption{ Reconstructed neutrino energy vs. true neutrino energy [(a)-(d)] and reconstructed neutrino energy resolution and its bias [(e)-(h)] for FC and PC \numuCC\ and FC and PC \nueCC\ candidates. The black points in the energy resolution plots represent the peak positions for each bin indicating the typical bias, and the error bars represent 68.3\% quantiles from each bin's peak position.}
  \label{fig:energy_resolution}
\end{figure*}

Neutrino energy reconstruction for Wire-Cell inclusive neutrino selection adopts a calorimetric approach, essentially adding up the visible energy deposited in the TPC active volume. The reconstructed neutrino energy is given by the following formula:
\begin{equation}\label{eq:energy_nu_reco}
    E^{\text{rec}}_{\nu} = \sum_i \left( K^{\text{rec}}_{i} + m_{i} + B_{i}\right),
\end{equation}
where $i$ represents each identified particle in the reconstructed particle flow, $K^{\text{rec}}_{i}$ represents the reconstructed kinetic energy, $m_{i}$ represents the rest mass value from the PDG~\cite{Zyla:2020zbs}, and $B_{i}$ represents the average binding energy (8.6~MeV) per nucleon after which an argon-40 nucleus is completely disassembled into its constituent nucleons. $m_{i}$ is only added for particles reconstructed as $\mu^{\pm}$, $\pi^{\pm}$, and $e^{\pm}$ particles, and $B_{i}$ is added for each reconstructed proton in the particle flow, either a primary proton connected to the neutrino vertex or a secondary proton scattered off from an argon nucleus by a neutron.
There are three methods to calculate the kinetic energy, $K^{\text{rec}}_{i}$, for each particle:
\begin{itemize}
    \item Range: the range is used to calculate the kinetic energy of track-like particles $\mu^{\pm}$, $\pi^{\pm}$, and $p$ if they stop inside the active volume. This is based on the NIST PSTAR database~\cite{pstar} with a correction for different particle masses.
    \item Summation of $dE/dx$: for short ($<$4 cm) track-like particles that stop inside the active volume or any exiting particles, the (visible) kinetic energy is estimated by summing up $dE/dx$ for each piece ($\sim$6~mm) of the track. The energy loss per unit length $dE/dx$ is converted from the ionization charge per unit length $dQ/dx$ considering the recombination effect when the ionization electrons are produced. This method is also used to estimate the energy of long muons with significant delta-rays emitted along the trajectory. This avoids bias in calculating the range.
    Note that an {\it effective} recombination model that also takes into account the overall normalization difference between MicroBooNE data and simulation charge signals~\cite{Adams:2019ssg} is built by tuning the parameters of the modified box model from ArgoNeuT~\cite{ArgoNeuT:2013kpa}. This effective recombination model ($\alpha$=1.0 and $\beta$=0.255), as is used in current Wire-Cell energy reconstruction, has an improved agreement with MicroBooNE data; however, it still underestimates the energy by about 10\%. Given that the ionization charge in simulation has been scaled to match that in the data, this bias 
    from the effective recombination model appears in both data and simulation in the same way. The residual difference between 
    data and simulation is covered by the systematic uncertainties as discussed in Sec.~\ref{sec:systematics}.
    \item Overall charge-energy scaling: for EM showers (non-track topologies), 
    %from $e^{\pm}$ (primary electrons or from Compton scattering/pair production of $\gamma$'s, etc), 
    the energy is estimated by scaling the total reconstructed charge of the shower cluster by a factor of 2.50 after multiplying by the 23.6~eV per ionization pair~\cite{Shibamura:1975zz,Miyajima:1974zz}. 
    This factor is derived from the nominal simulation and takes into account the bias in the reconstructed charge with respect
    to the true charge, and the average recombination effect~\cite{Abratenko:2020hpp} which converts the deposited energy to the true charge. For data events, an additional scaling factor of 0.95, which is calibrated from the reconstructed $\pi^0$ invariant mass (Sec.~\ref{sec:pi0}), is applied.
\end{itemize}

Performance of this neutrino energy reconstruction is evaluated using MC samples by comparing the reconstructed values with the true values. The reconstructed neutrino energy versus the true neutrino energy as well as the energy resolution and bias relative to the true values are shown in Fig.~\ref{fig:energy_resolution}. These figures present the results for selected \numuCC\ and \nueCC\ candidates (see Sec.~\ref{sec:event_select}), respectively. The neutrino candidates are further divided into FC and PC samples (See Sec.~\ref{sec:overview}) based on the containment of the selected in-beam cluster. 
We should note that since an energetic muon track is much more extended than an electron shower with the same energy, 
the FC cut actually selects \numuCC\ events with higher energy transfer (thus larger missing hadronic energy on average) 
than the corresponding \nueCC\ events at the same neutrino energy.
For the FC samples, the reconstructed neutrino energy resolution is 15-20\% for \numuCC\ candidates with $\sim$10\% bias (towards lower energies) and 10-15\% for \nueCC\ candidates with $\sim$7\% bias (towards lower energies). The slightly worse energy resolution of \numuCC\ than that of \nueCC\ 
reflects the fact that \numuCC\ event selection can tolerate more imperfect event reconstruction given its higher 
signal to background ratio than that of \nueCC. 
Energy deposited outside the active volume by charged particles or carried away by neutral particles such as neutrons is missing in this calorimetric energy reconstruction. In the energy range of interest, below 800~MeV in true neutrino energy, we have achieved uniform performance in the resolution of reconstructed neutrino energy and its bias, especially in \nueCC\ FC signal event selection (Fig.~\ref{fig:energy_resolution}(g)).
More validation of this neutrino energy reconstruction in terms of the modeling of missing energy using \numuCC\ events and data-MC comparisons can be found in Sec.~\ref{sec:model_validation}.

\section{Event Selection}\label{sec:event_select}
The starting point of the neutrino event selection is the generic neutrino selection~\cite{Abratenko:2020sxa,Abratenko:2021bzb}, in which the cosmic-ray backgrounds are reduced resulting in an overall contamination below 15\%. After the generic neutrino selection, the efficiencies\footnote{In this article, the selection efficiency is defined as the number of selected events relative to the true neutrino interactions with vertices inside the fiducial volume.} for \numuCC\ and \nueCC\ events are approximately 80\% and 90\% with signal-to-background ratios of about 2:1 (purity of 66\%) and 1:250 (purity of 0.4\%), respectively.

For the search of the low-energy excess in the \nueCC\ channel, the event selections are designed to be as general as possible (i.e., inclusive \nueCC), so that more freedom in examining exclusive channels would be available at later stages of the analysis if an excess was to be observed. Since the \numuCC\ events are used to constrain the systematic uncertainties in neutrino flux, neutrino-argon interaction cross section, and detector effects, an inclusive \numuCC\ selection is also adopted.

\subsection{Charged-Current \nue\ Selection}\label{sec:nueCC}
The development of \nueCC\ selection involves two stages. The first stage is the categorization of non-\nueCC\ backgrounds, which is informed by hand scans of a small amount of background events.
Then, variables with signal-background discrimination capability which represent the characteristic features of each type of background in the first stage are used as input into boosted decision trees (BDTs) trained on large MC simulation samples. 
Events used in BDT training are removed in making predictions. %This evolution of selection philosophy is described in the Fig.~\ref{fig:nueCC_evolution}.

The basic selection of inclusive \nueCC\ events requires an EM shower with a reconstructed energy higher than 60~MeV connected to the (primary) neutrino vertex~\cite{wire-cell-pr}. The energy threshold is applied in order to exclude Michel electrons. When there are multiple reconstructed EM showers connected to the same neutrino vertex, the EM shower with the highest energy is taken as the primary electron candidate for further examinations. 
%Background taggers were developed by extracting features from a hand-scan effort.

The backgrounds are categorized into five major types. The first type focuses on primary electron identification, including the examination of the $dQ/dx$ profile at the first few centimeters of the shower (i.e. shower stem) and the identification of a gap between the shower and the neutrino vertex. This gap occurs in photon showers due to the photon conversion length of approximately 18~cm in liquid argon. The second type of background focuses on interactions with multiple EM showers in the final state, most likely from \pizero\ production. The third type focuses on muon-related misidentification as electrons. The fourth type focuses on more general background rejection using kinematic information, e.g. comparison of lepton kinematics between an electron candidate and its competing muon candidate in the same event\footnote{While this situation does not exist for true physics 
event, it can happen due to an imperfect event reconstruction.}. The last type focuses on interactions with poor pattern recognition,
which includes several different failure modes leading to incorrect pattern recognition. Beside these five major categorizations of background featuring in the \nueCC\ selection, there is a another set of dedicated taggers removing residual cosmic-ray muon induced backgrounds, which will be described in detail in the next section. 

The primary electron identification includes:
\begin{itemize}
    \item Gap cut: the beginning of the EM shower in each 2D projection view is examined to search for a gap to the
    identified neutrino interaction vertex. 
    \item Stem quality cut to remove backgrounds: the beginning of the shower is examined to ensure the quality of the shower stem. The checks include examinations of i) other minimum ionizing particle (MIP) tracks overlapping with the shower stem, and ii) possible track splitting, e.g. the splitting of pair-produced electron and positron instead of traveling in the same direction. 
    \item MIP $dQ/dx$ cut: we examine the $dQ/dx$ profile of the shower stem to ensure a MIP (electron-like) event. We calculate the length of the MIP-like track below a MIP threshold cut (i.e. 1.3 times the expected MIP $dQ/dx$). The calculation of the length also considers the possibility of a delta ray (i.e. a single sample with high dQ/dx). In addition, a high $dQ/dx$ value at the vertex because of additional vertex activities must be taken into account.
\end{itemize}

\begin{figure}[htb!]
\captionsetup[subfigure]{justification=centering}
  \centering
  \begin{subfigure}[]{0.99\columnwidth}
    \includegraphics[width=\textwidth]{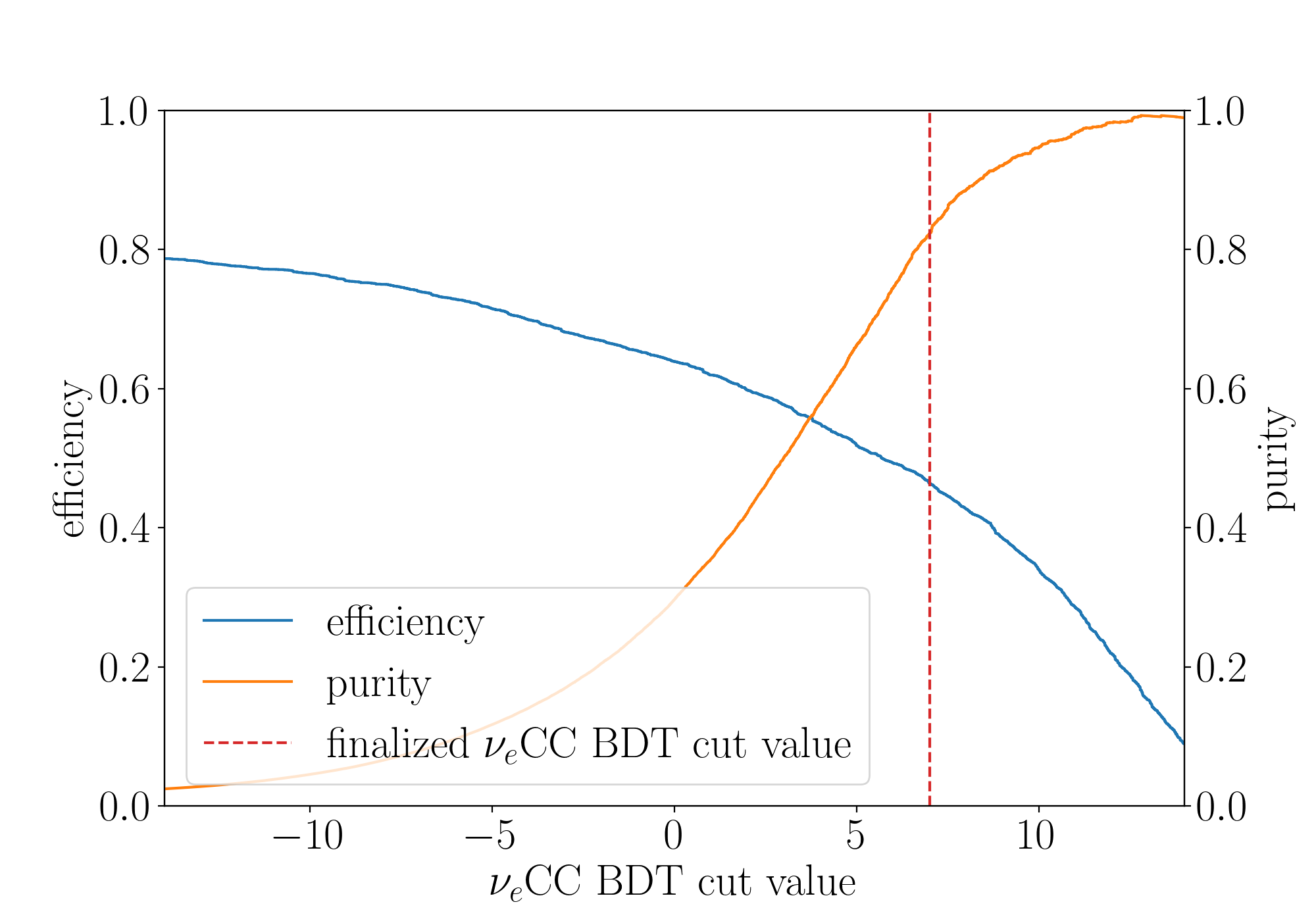}
    \put(-210,140){MicroBooNE Simulation}
    \caption{}
    \label{fig:nue_eff_pur}
  \end{subfigure}
  \begin{subfigure}[]{0.99\columnwidth}
    \includegraphics[width=\textwidth]{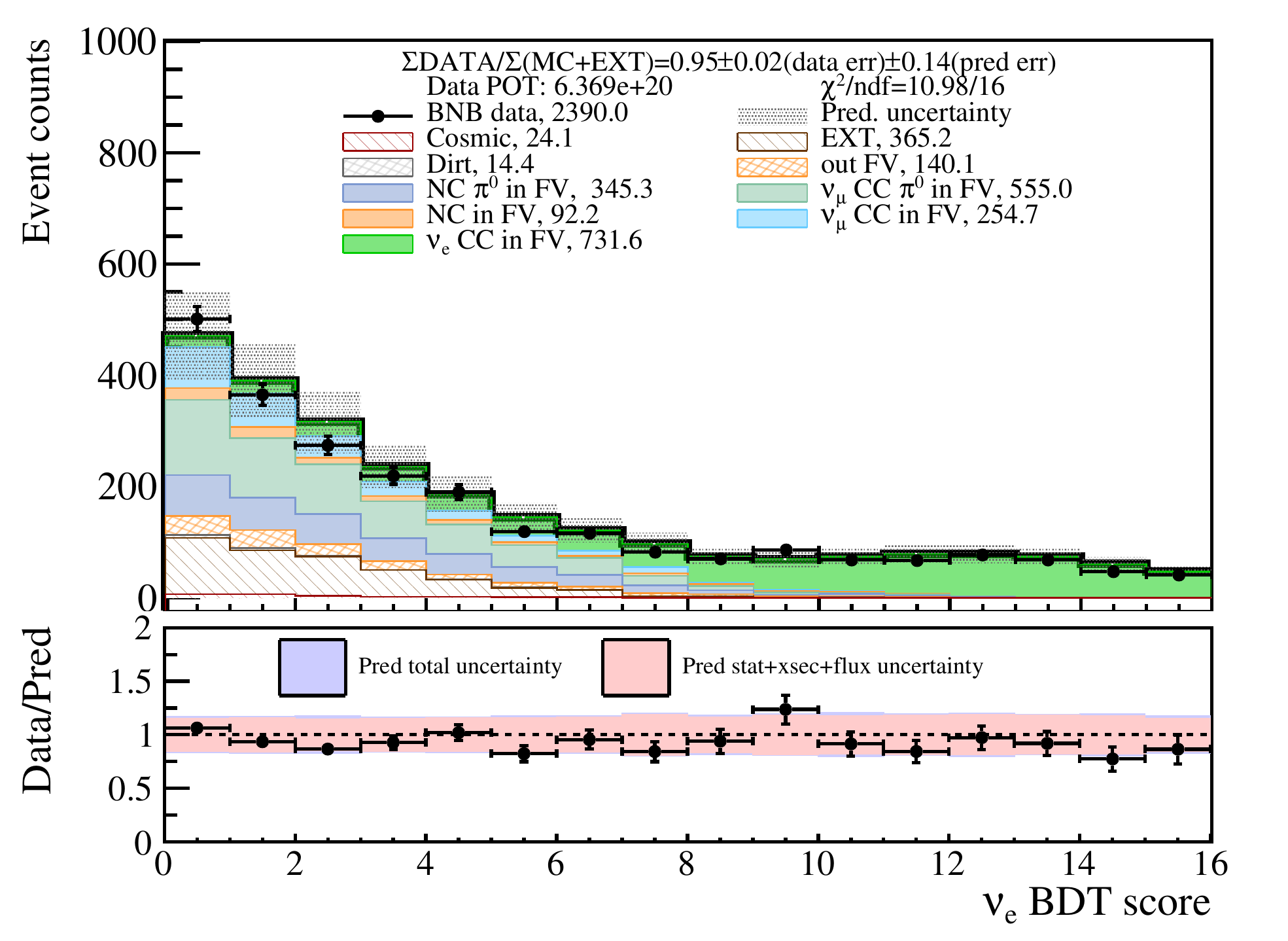}
    \put(-80, 100){MicroBooNE}
    \caption{}
    \label{fig:nue_bdtscore}
  \end{subfigure}
  \caption{(a) \nueCC\ selection efficiency and purity at different BDT cut values, with the finalized cut value of 7.0 indicated. (b) \nue\ BDT score distribution for events with BDT score$>$0.}
  \label{fig:nue_bdt}
\end{figure}

With the hand scanned features selected, we apply BDT techniques to high-statistics MC simulation samples to finalize the \nueCC\ selection. 
The usage of machine learning techniques mitigates the limitation of human learning when processing a large number of events. From among the many different machine learning tools, the BDT 
technique is chosen because it is more robust and approachable for general users compared to other multivariate 
analysis tools. The BDT package \textsc{XGBoost}~\cite{Chen:2016btl}, which provides fast and robust training through 
parallel tree boosting, is used. \textsc{XGBoost} also improves the model generalization and overcomes the issues of 
over-fitting in gradient boosting enabling the use of a large pool of variables (over 300 variables 
from all categorizations of backgrounds in this \nueCC\ selection) in a single model. To train the BDT, the true \nueCC\ events in the fiducial volume that pass the generic
neutrino selection and have at least one reconstructed electron EM shower are used to define the signal. 
In order to enhance the performance, the events with bad reconstruction, when the reconstructed neutrino and 
EM shower vertices are incorrectly reconstructed, are removed from the signal events.  The overlay MC simulation
after excluding true \nueCC\ events and the dedicated beam-off data are used as background in 
training the BDT. 
Figure~\ref{fig:nue_eff_pur} shows the signal efficiency and purity as a function of the \nueCC\ BDT score, 
and the distribution of \nue\ BDT score is shown in Fig.~\ref{fig:nue_bdtscore}. Here, the efficiency 
is defined with respect to all true \nueCC\ events with their neutrino interaction vertices inside the fiducial
volume. The cut value of 7.0 was chosen 
in order to maximize the \nue\ selection efficiency$\times$purity value. A final \nueCC\ selection with 46\% efficiency and 82\% purity is achieved.

\begin{figure*}[htp!]
\captionsetup[subfigure]{justification=centering}
  \centering
  \begin{subfigure}[]{0.47\textwidth}
    \includegraphics[width=\textwidth]{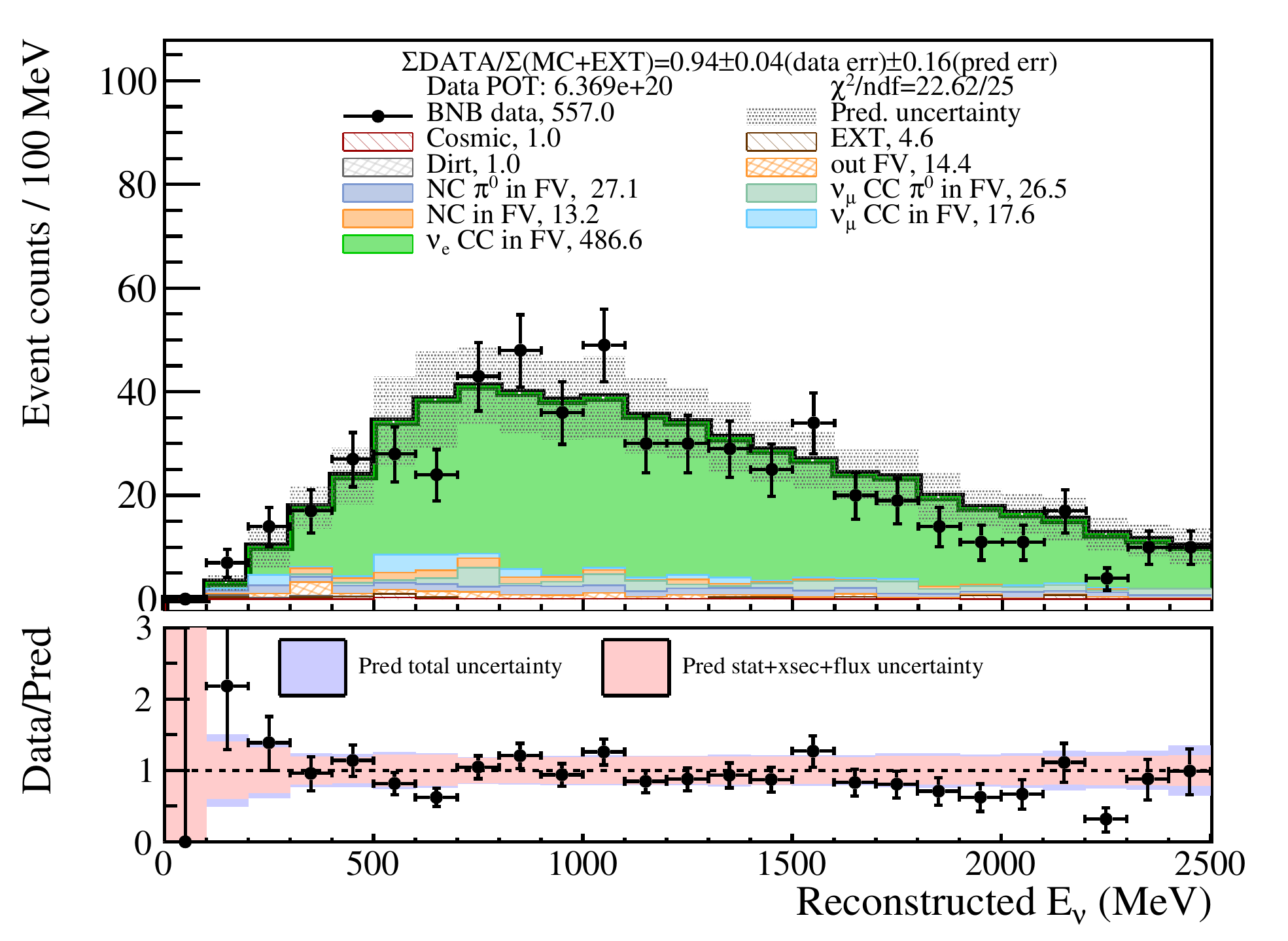}
    \put(-80,115){MicroBooNE}
    \put(-60,100){FC+PC}
    \caption{Neutrino energy, broken down with event types}
    \label{fig:nue_recoEnu}
  \end{subfigure}
  \begin{subfigure}[]{0.47\textwidth}
    \includegraphics[width=\textwidth]{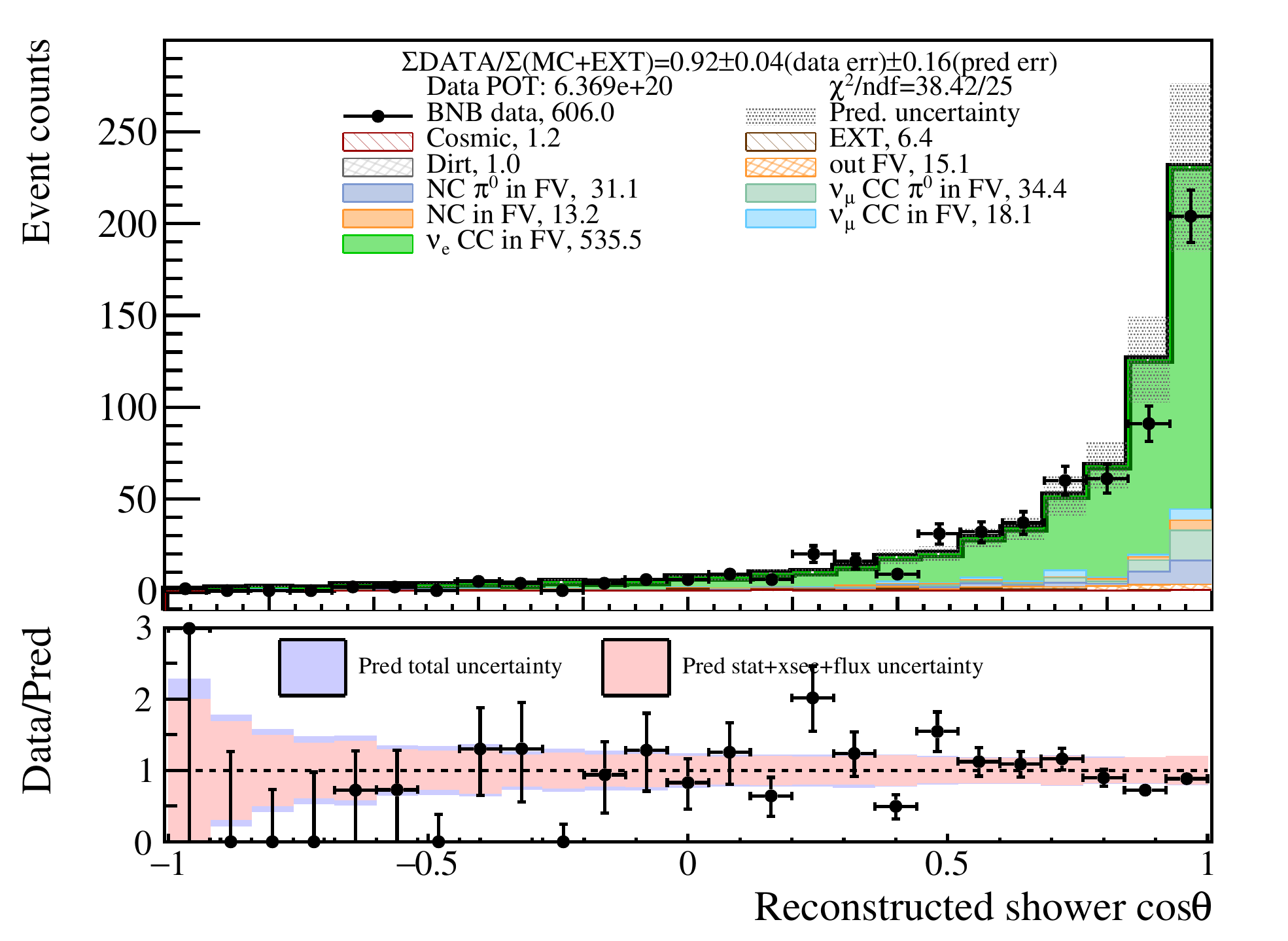}
    \put(-180,115){MicroBooNE}
    \put(-180,100){FC+PC}
    \caption{Shower cos$\theta$, broken down with event types}
    \label{fig:nue_costheta}
  \end{subfigure}
  \begin{subfigure}[]{0.47\textwidth}
    \includegraphics[width=\textwidth]{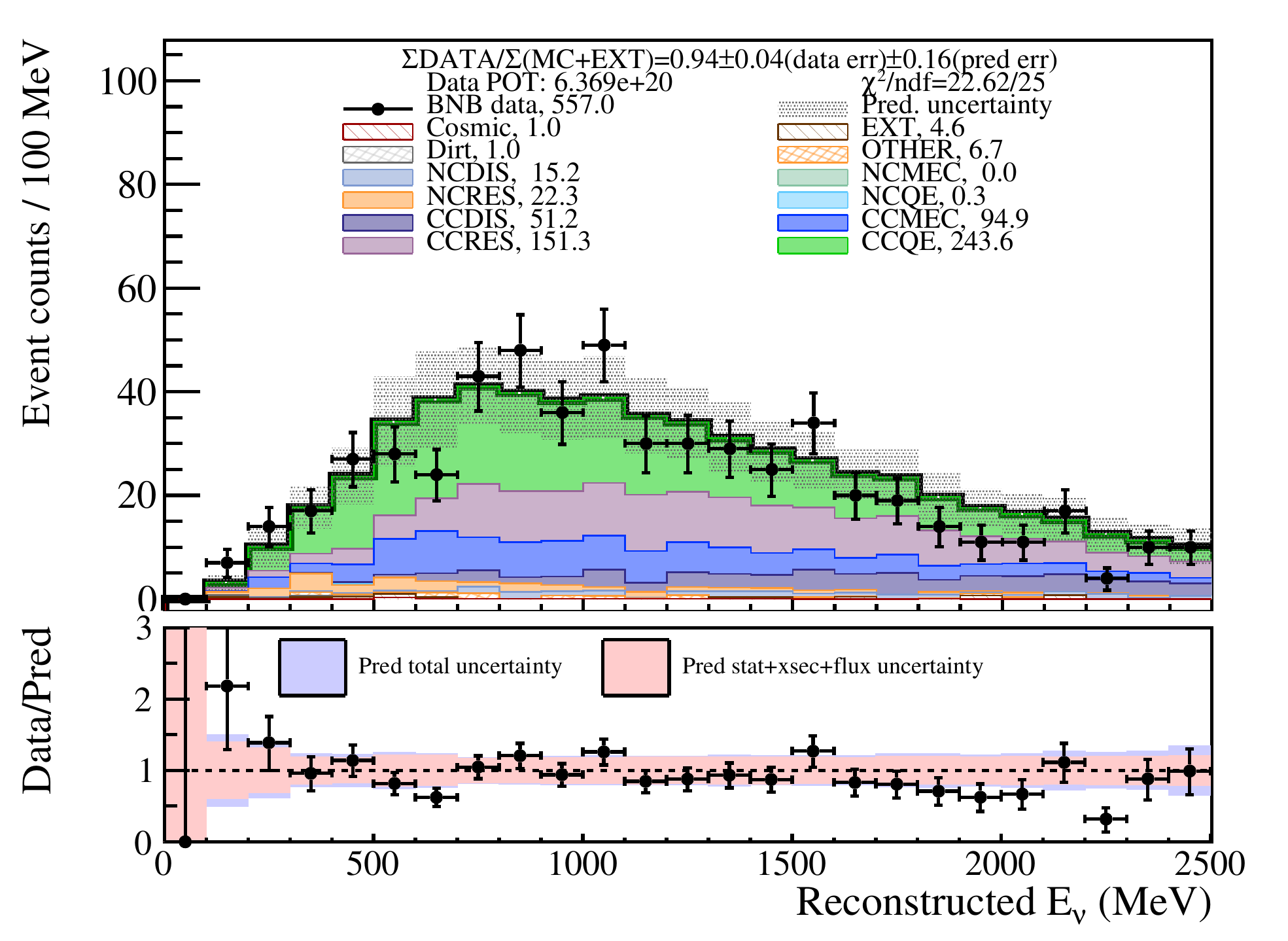}
    \put(-80,115){MicroBooNE}
    \put(-60,100){FC+PC}
    \caption{Neutrino energy, broken down with interaction types}
    \label{fig:nue_recoEnu_interactiontype}
  \end{subfigure}
  \begin{subfigure}[]{0.47\textwidth}
    \includegraphics[width=\textwidth]{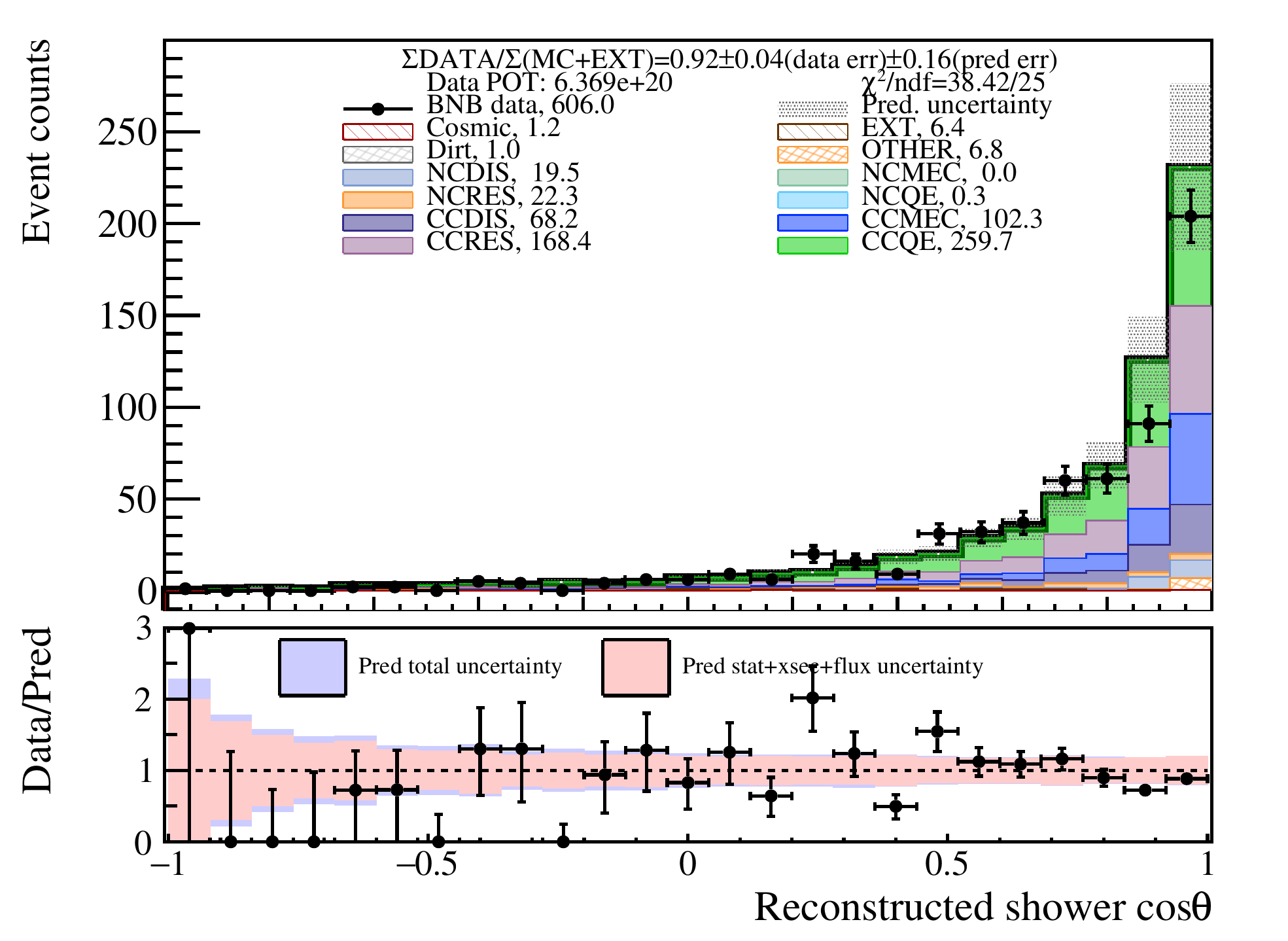}
    \put(-180,115){MicroBooNE}
    \put(-180,100){FC+PC}
    \caption{Shower cos$\theta$, broken down with interaction types}
    \label{fig:nue_costheta_interactiontype}
  \end{subfigure}
  \begin{subfigure}[]{0.47\textwidth}
    \includegraphics[width=\textwidth]{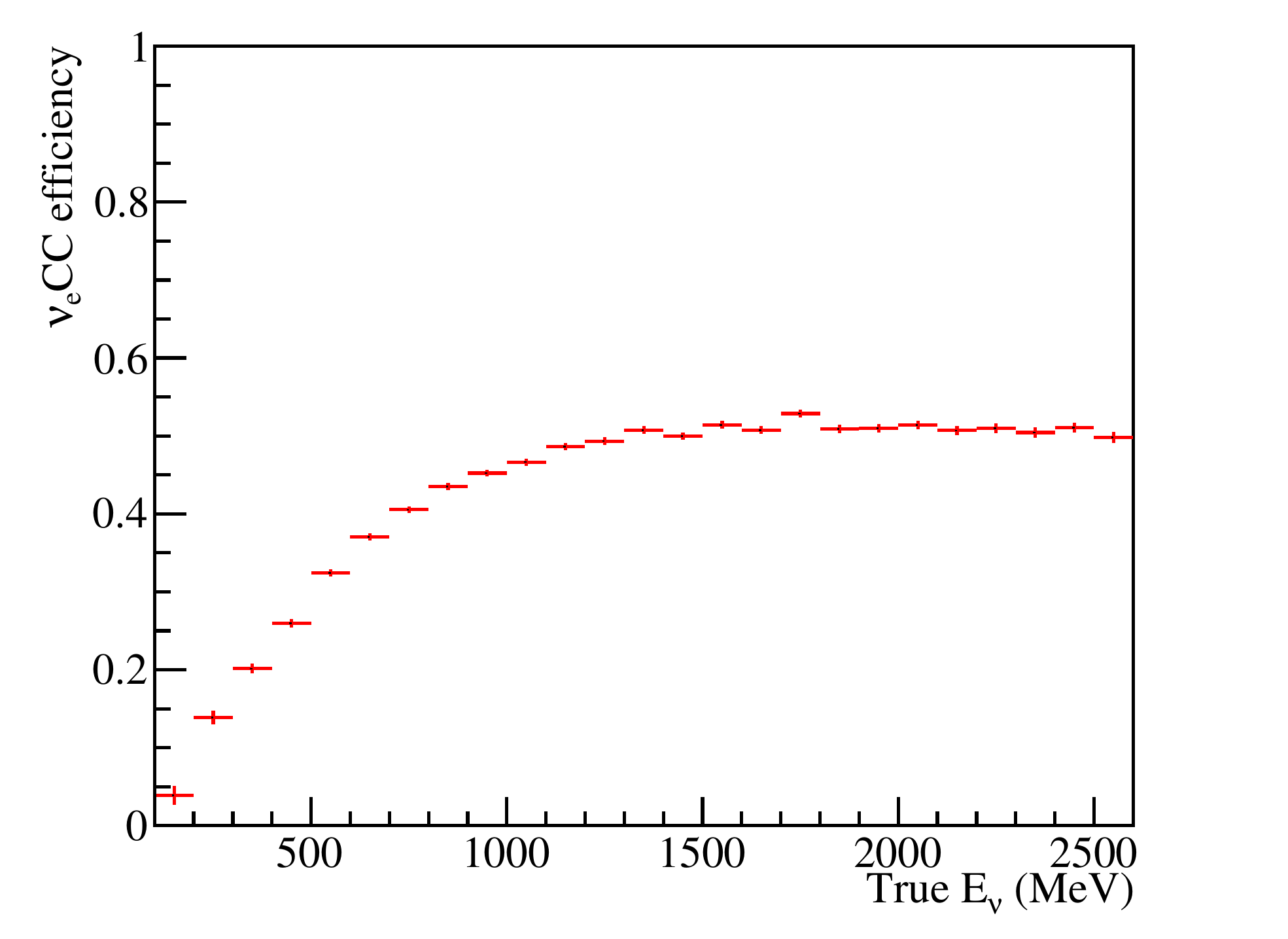}
    \put(-130,158){MicroBooNE Simulation}
    \caption{Efficiency as a function of true neutrino energy}
    \label{fig:nue_eff_recoEnu}
  \end{subfigure}
  \begin{subfigure}[]{0.47\textwidth}
    \includegraphics[width=\textwidth]{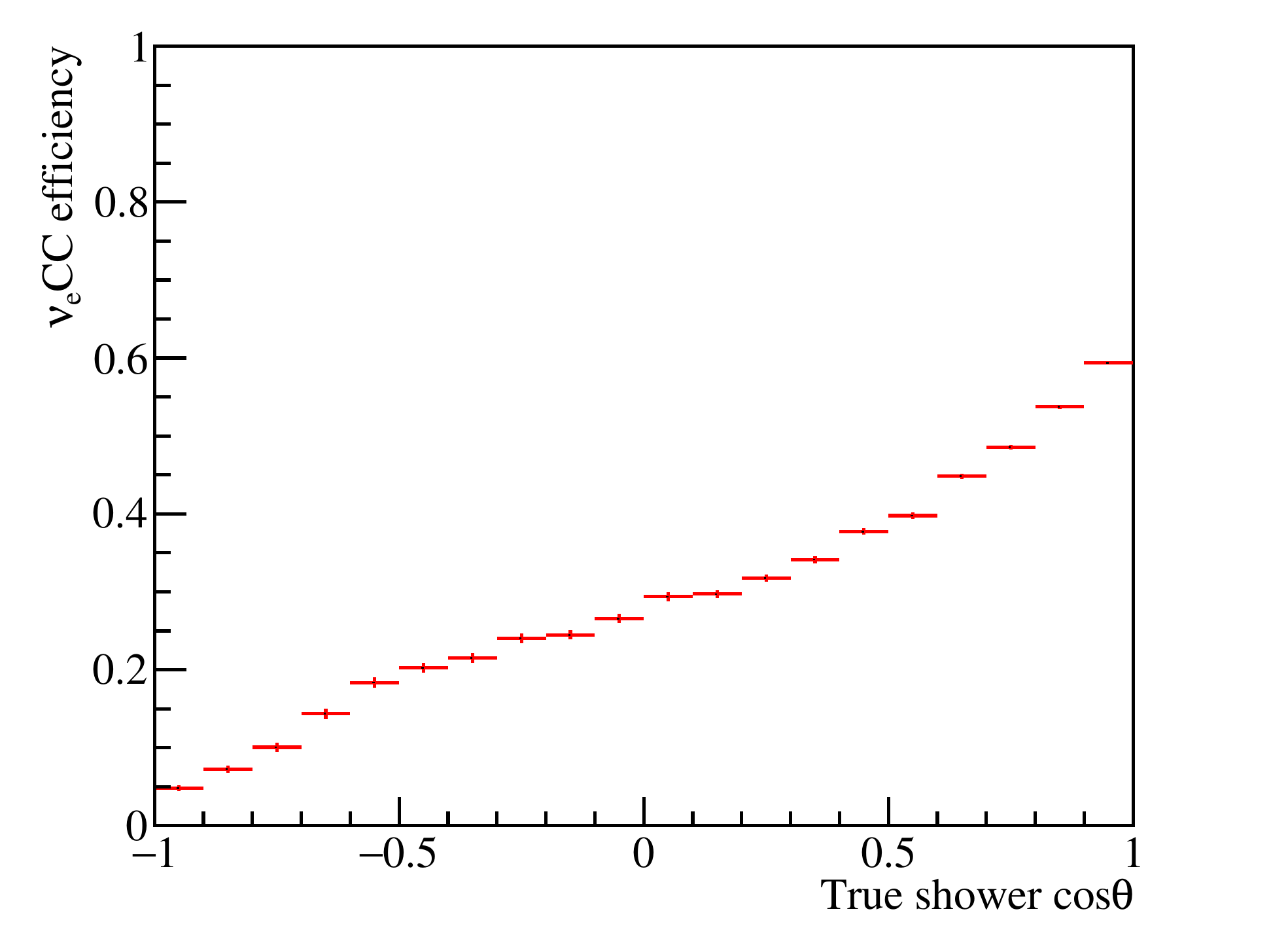}
    \put(-130,158){MicroBooNE Simulation}
    \caption{Efficiency as a function of true shower cos$\theta$}
    \label{fig:nue_eff_costheta}
  \end{subfigure}
  \caption{The final \nueCC\ selections as a function of reconstructed neutrino energy [(a) and (c)] and reconstructed shower cos$\theta$ [(b) and (d)]. (a) and (b) are categorized by event types, and (c) and (d) by interaction types. The number of events correspond to the range shown in the plot. The bottom sub-panels show both the statistical and systematic uncertainties. The pink band includes the statistical, cross section, and flux uncertainties. The purple band corresponds to the full uncertainty with the addition of detector systematic uncertainty. The selection efficiencies are shown as a function of (e) true neutrino energy and (f) true shower cos$\theta$ with only statistical uncertainty considered. The other dimensions are integrated in calculating
  these efficiencies.}
  \label{fig:nue_selection_recoEnu_theta}
\end{figure*}
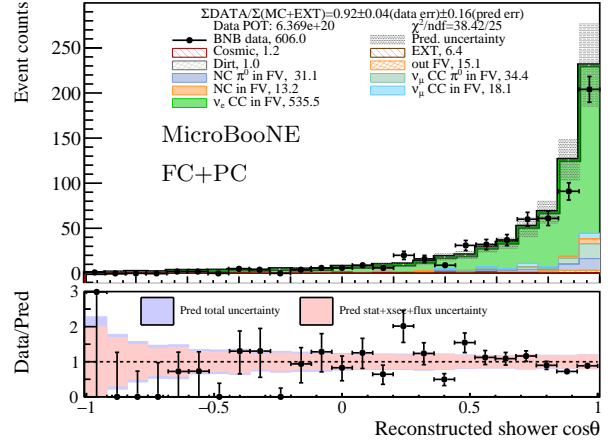
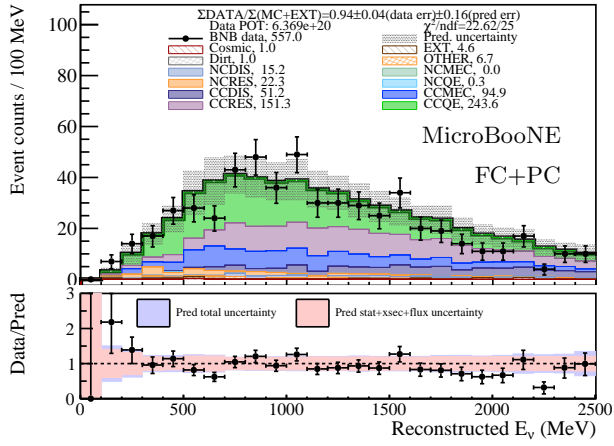
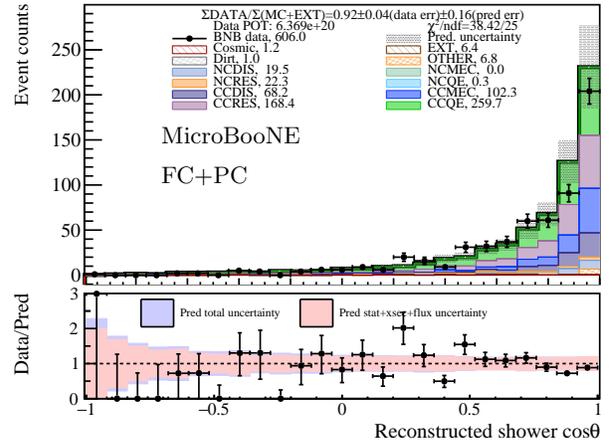
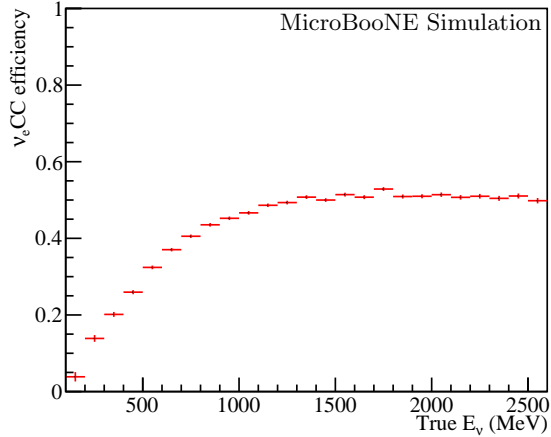
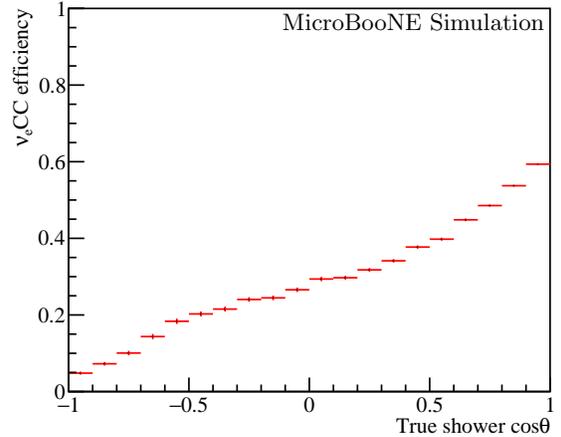

The selected events are categorized into different types which are determined by the truth information. Each category, which will be used throughout this paper, as shown in Fig.~\ref{fig:nue_recoEnu} and Fig.~\ref{fig:nue_costheta}, is defined as follows: (1) ``EXT'': cosmic-ray background from the beam-off data set that is external to the BNB data stream and triggered by cosmic-ray activity in coincidence with a fake beam spill, in which case the events have no BNB neutrino interactions; (2) ``Cosmic'':  mistakenly selected cosmic-ray background from BNB overlay MC simulation, in which case each event has a simulated BNB neutrino interaction overlaid with dedicated beam-off data; (3) ``Dirt'': neutrino interactions with their true neutrino interaction vertices outside the cryostat, as defined in Sec.~\ref{sec:uboone_mc}; (4) ``out FV'': neutrino-argon interactions with vertices outside the fiducial volume but within cryostat; (5) ``\numuCC\ $\pi^0$ in FV'': \numuCC\ interactions with vertices inside the fiducial volume and with at least one true $\pi^0$ in the final state; (6) ``\numuCC\ in FV'': \numuCC\ interactions with vertices inside the fiducial volume and with no $\pi^0$ in the final state; (7) ``NC $\pi^0$ in FV'': NC interactions with vertices inside the fiducial volume and with at least one true $\pi^0$ in the final state; (8) ``NC in FV'': NC interactions with vertices inside the fiducial volume and with no $\pi^0$ in the final state; (9) ``\nueCC\ in FV'': beam intrinsic \nueCC\ interactions with vertices in the fiducial volume; and (10) ``LEE'': all categories of excessive events originating from eLEE neutrino interactions, which accounts for the difference between the \LEE\ hypothesis and \SM\ hypothesis.
Except for ``EXT'', the other categories correspond to MC events which overlay simulated neutrino interactions with randomly triggered beam-off (cosmic) data. 
%``EXT'', ``Cosmic'', ``Dirt'' as shown in Fig.~\ref{fig:nue_recoEnu_interactiontype} and Fig.~\ref{fig:nue_costheta_interactiontype} share the same definitions as explained above. 
As shown in Fig.~\ref{fig:nue_recoEnu_interactiontype} and Fig.~\ref{fig:nue_costheta_interactiontype}, the other categories for the selected neutrino interactions correspond to different interaction types
obtained from the event generator. ``CC'' or ``NC'' represent charged-current or neutral-current. ``QE'', ``RES'', ``MEC'', and ``DIS'' represent quasi-elastic, resonance, meson exchange current, and deep inelastic scattering, respectively.
Figure~\ref{fig:nue_eff_recoEnu} and Figure~\ref{fig:nue_eff_costheta} show the \nueCC\ selection efficiency as a function of neutrino energy ($E_{\nu}$) and the cosine of the polar angle (cos$\theta$, relative to the BNB direction) of the electron EM shower, respectively.

Before applying the \nueCC\ selection to the BNB data stream, we validated its performance using 
the off-axis Neutrinos at the Main Injector (NuMI)~\cite{Adamson:2015dkw} neutrino beam at FNAL. 
The NuMI beam is created from
collisions of protons accelerated to an energy of 120 GeV with a graphite target. Similar to that of 
BNB, the charged hadrons are focused by a magnetic field into a 675~m long decay pipe. The distance 
between the NuMI target and the MicroBooNE detector is about 679~m. At an off-axis location of $\sim8^{\circ}$, the $\nu_e$'s with a sizable amount of $\bar{\nu}_e$'s above 200~MeV are mostly coming from the 3-body $K^+$ and unfocused $K^{0}_L$ decays.
Compared to that of the BNB, the percentage of \nue\ in the flux is an order of magnitude higher at $\sim$5 \%, which makes it ideal to validate the performance of the \nueCC\ selection. 
With 1.917$\times10^{20}$ POT exposure, we select 269 FC \nueCC\ and 162 PC \nueCC\ candidates with reconstructed neutrino energy below 2.5 GeV and with an overall efficiency of 42\% and purity of 91\%. 
The overall ratio between data and nominal NuMI MC prediction for FC and PC \nueCC\ without considering any anomalous enhancement
is 0.99$\pm$0.06 (stat.) and $1.08\pm0.08$ (stat.) indicating an overall good agreement.

\subsection{Charged-Current \numu\ Selection}\label{sec:numuCC}
The generic neutrino selection results in a 88.4\% selection efficiency and 65.0\% purity for \numuCC\ events~\cite{Abratenko:2020sxa,Abratenko:2021bzb}.  Here, the selection efficiency 
is defined with respect to all true \numuCC\ events with their neutrino interaction vertices inside the fiducial volume.
The achieved purity is limited by the residual cosmic-ray muon background, neutrino-induced background originating outside the fiducial volume, and NC interactions inside the fiducial volume. The precise reconstruction of the \numuCC\ neutrino vertex (resolution is less than 1~cm) and particle identification (an integrated efficiency of 90\% for primary muons) are leveraged to suppress these backgrounds. The reconstructed neutrino vertex is required to be inside the fiducial volume and the length of primary muons is required to be greater than 5~cm before applying the BDT selection.

%%%%%%%%%%%%%%%%%%%%%%%%%%%%%%%%%%%%%%%%%%%%%%%
% Residual cosmic rejection
In analogy to the \nueCC\ selection, human scans of the remaining backgrounds were performed to extract the main features of each type of background.
The residual cosmic-ray background is typically the result of incorrect charge-light matching where the TPC cluster is placed at an incorrect location along the electric field direction. A through-going muon could have only one track end reconstructed at the detector boundary instead of two track ends reconstructed at the detector boundary, mimicking a single muon starting inside the TPC and exiting the detector. A stopped muon might also appear to be fully contained and, alternatively, be reconstructed with the candidate neutrino vertex at the muon decay vertex connecting the muon and the Michele electron. These topological features are leveraged to do this background rejection.
For neutrino-induced background originating outside the fiducial volume, a charged hadron usually enters the detector and undergoes a hadronic interaction. For this kind of events, the neutrino vertex is typically reconstructed at the hadronic interaction point, and the event could then appear to originate inside the fiducial volume with a misidentified muon candidate. Note, with an exiting high-energy charged particle track, one may not achieve a reliable PID for MIP particles such as muons. 
The kinematics, especially the direction of the muon candidate, can be used to reject such background since most of the hadrons entering the detectors from outside of the detector are not as forward-going as expected.

\begin{figure}[!htp]
\captionsetup[subfigure]{justification=centering}
  \centering
  \begin{subfigure}[]{0.98\columnwidth}
    \includegraphics[width=\textwidth]{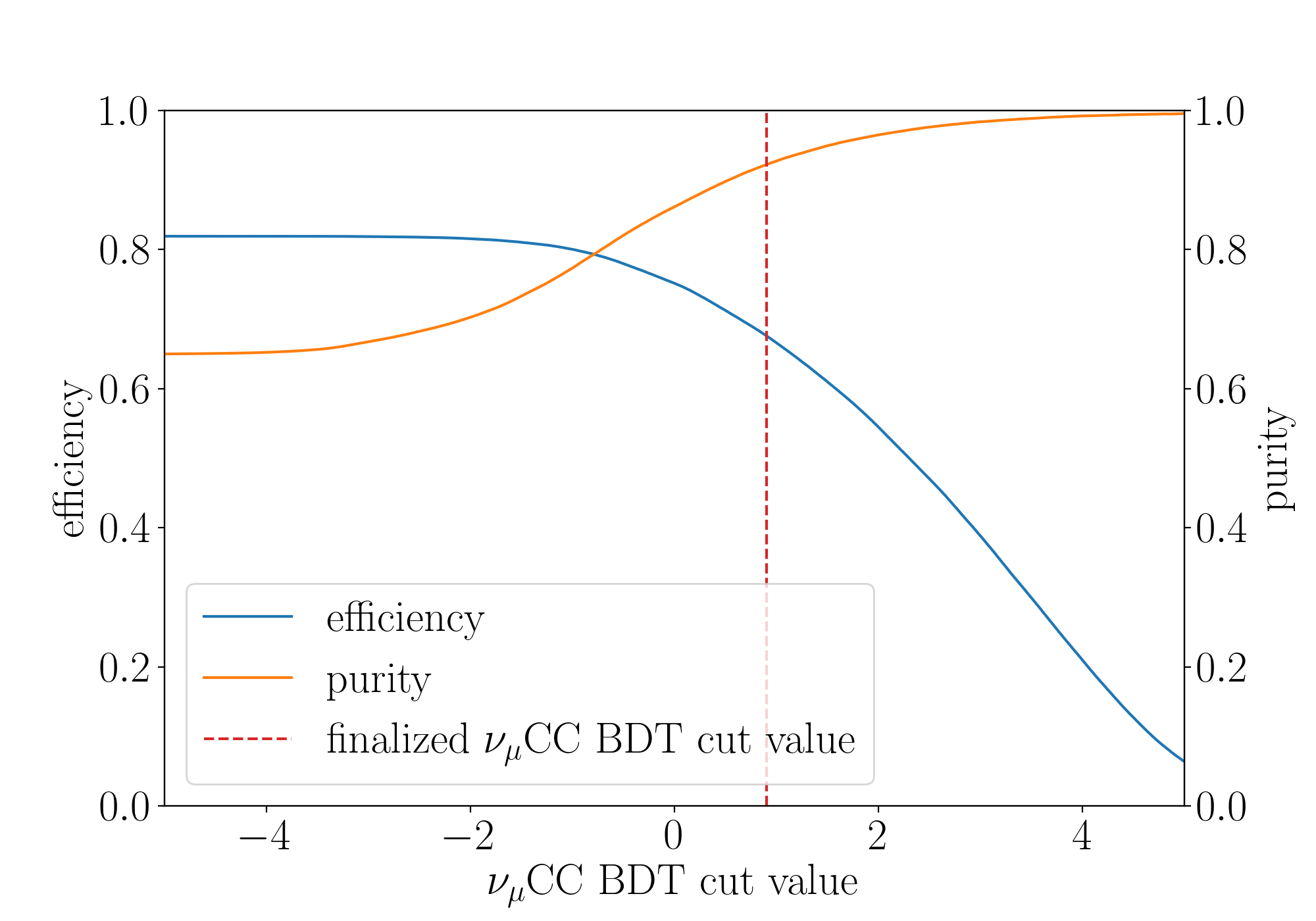}
    \put(-210,140){MicroBooNE Simulation}
    \caption{}
    \label{fig:numu_eff_pur}
  \end{subfigure}
  \begin{subfigure}[]{0.98\columnwidth}
    \includegraphics[width=\textwidth]{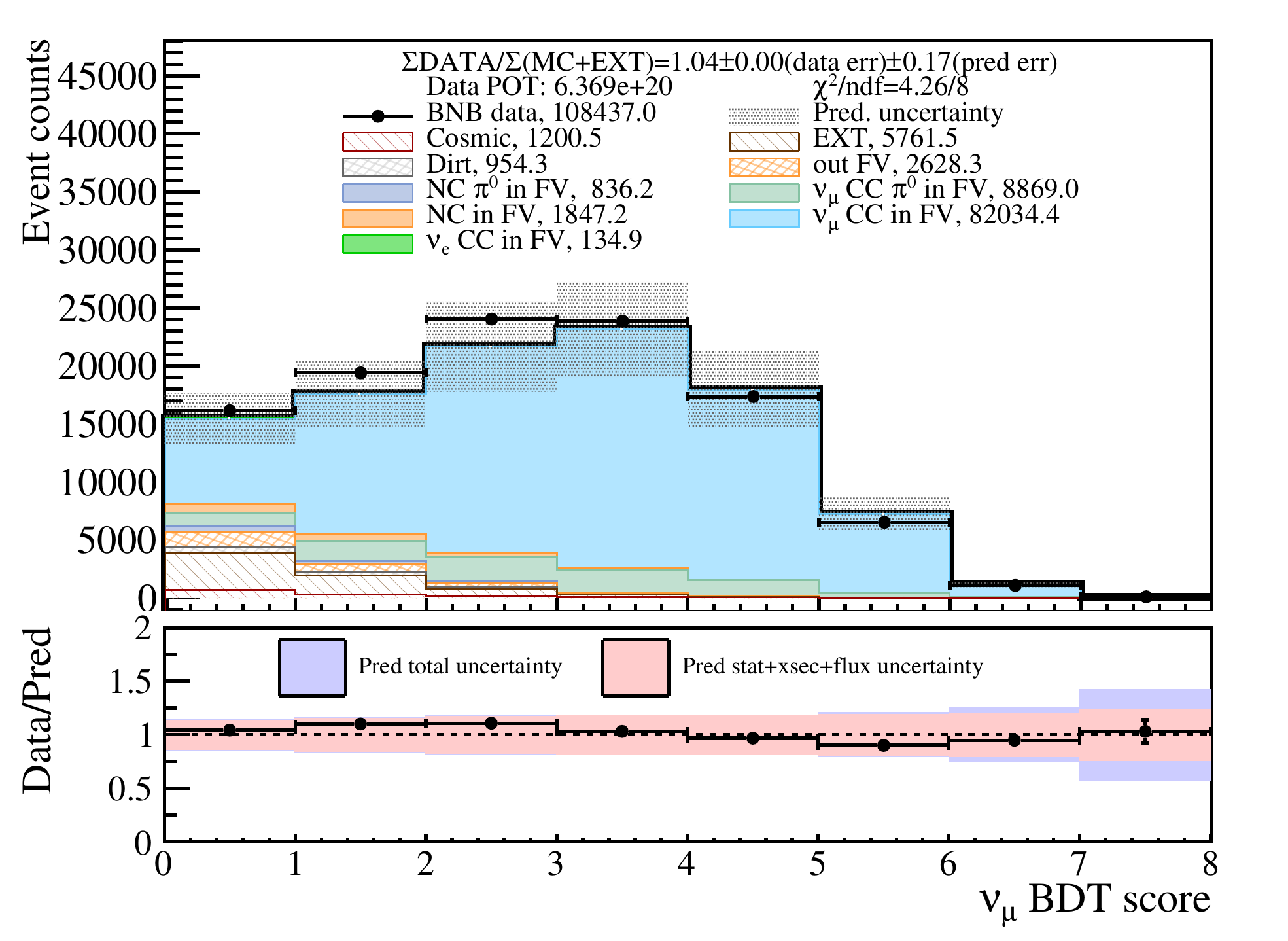}
    \put(-75, 100){MicroBooNE}
    \caption{}
    \label{fig:numu_bdtscore}
  \end{subfigure}
  \caption{(a) \numuCC\ selection efficiency and purity at different BDT cut values with the finalized cut value of 0.9 indicated. (b) \numu\ BDT score distribution for events with BDT score$>$0.}
  \label{fig:numu_bdt}
\end{figure}

%% Data/MC comparison discussion
\begin{figure*}[htp!]
\captionsetup[subfigure]{justification=centering}
  \centering
  \begin{subfigure}[]{0.48\textwidth}
    \includegraphics[width=\textwidth]{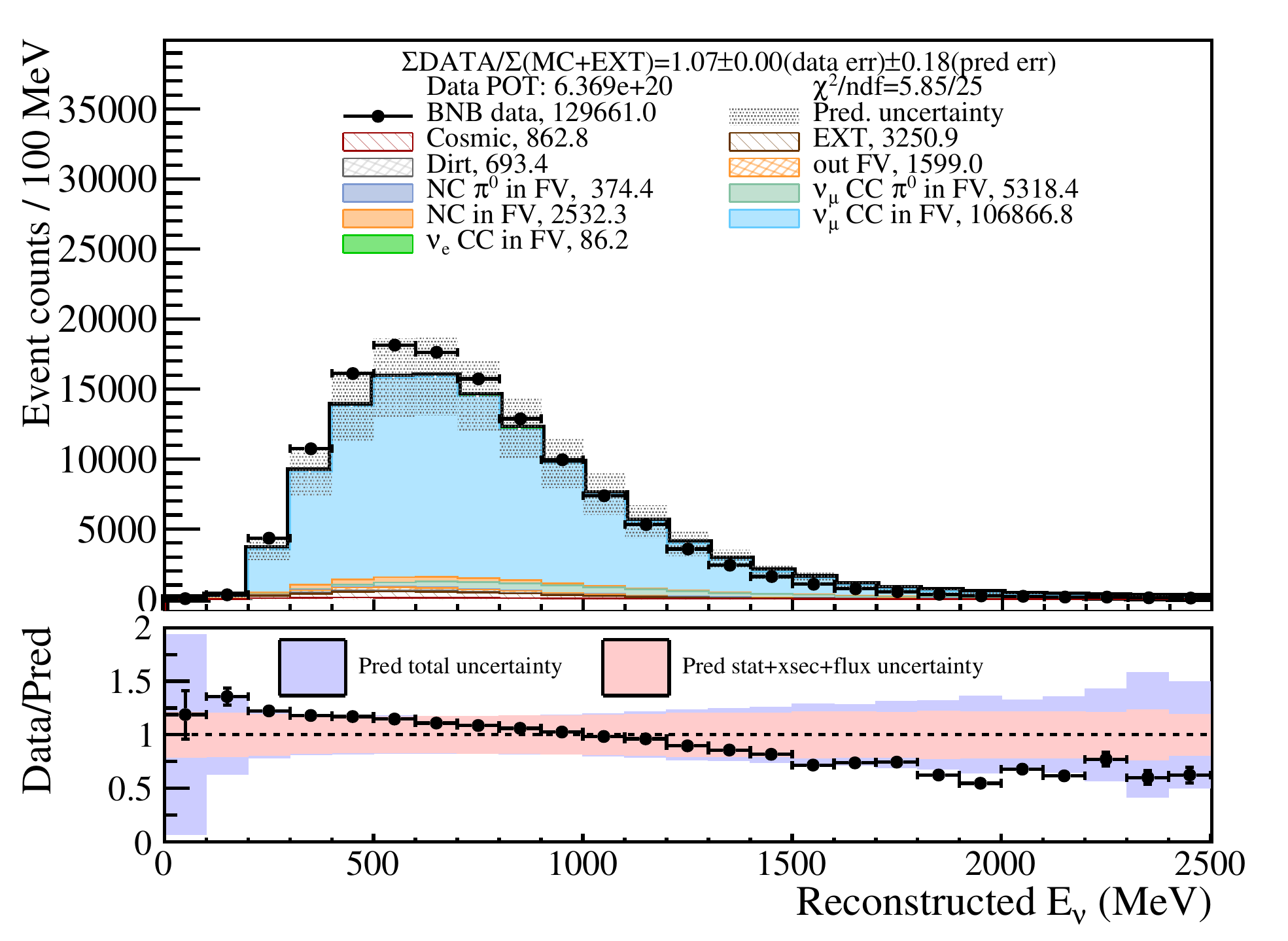}
    \put(-100,115){MicroBooNE}
    \put(-80,100){FC+PC}
    \caption{Neutrino energy, broken down with event types}
    \label{fig:numu_recoEnu}
  \end{subfigure}
  \begin{subfigure}[]{0.48\textwidth}
    \includegraphics[width=\textwidth]{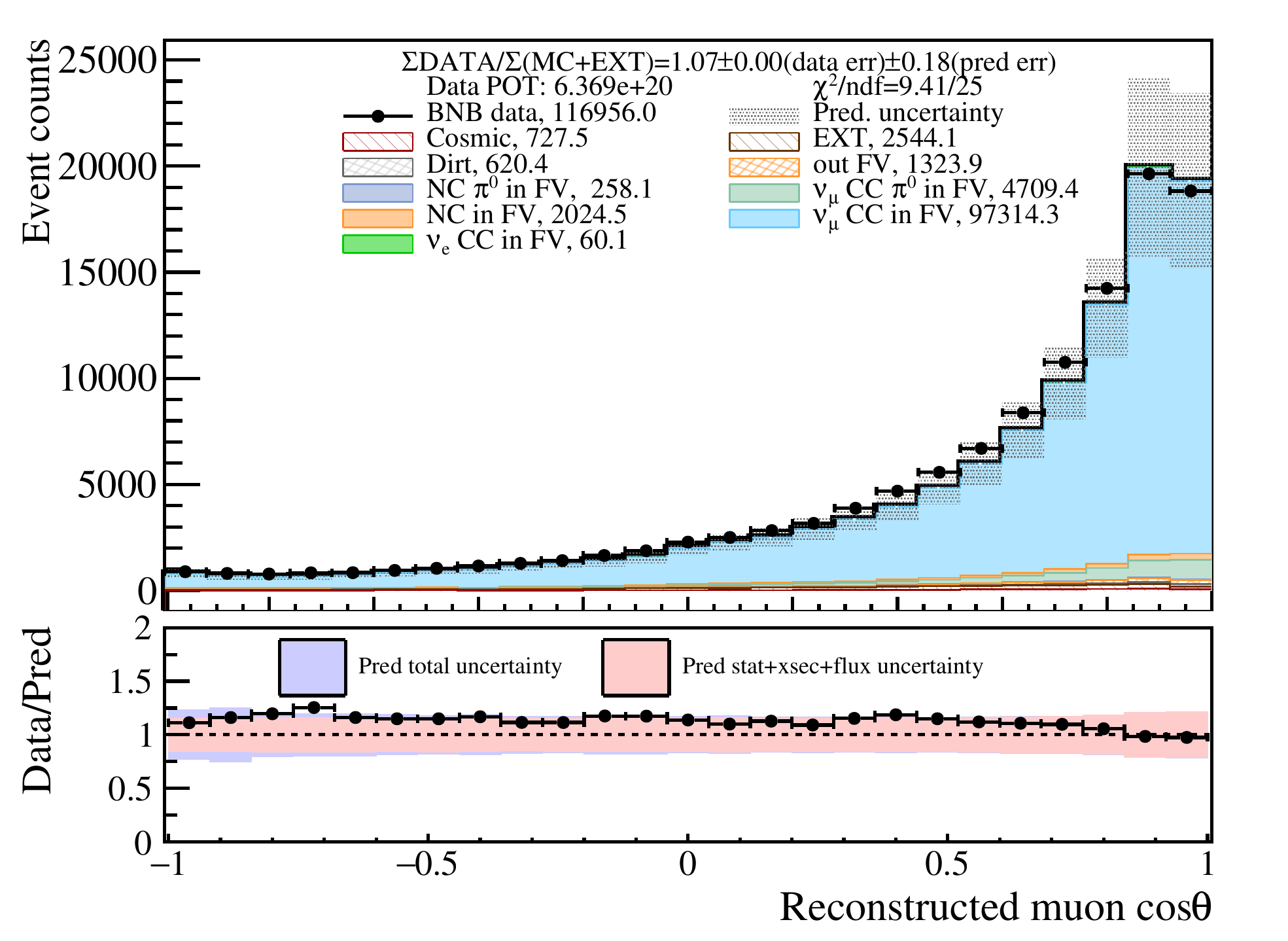}
    \put(-180,115){MicroBooNE}
    \put(-180,100){FC+PC}
    \caption{Muon cos$\theta$, broken down with event types}
    \label{fig:numu_costheta}
  \end{subfigure}
  \begin{subfigure}[]{0.48\textwidth}
    \includegraphics[width=\textwidth]{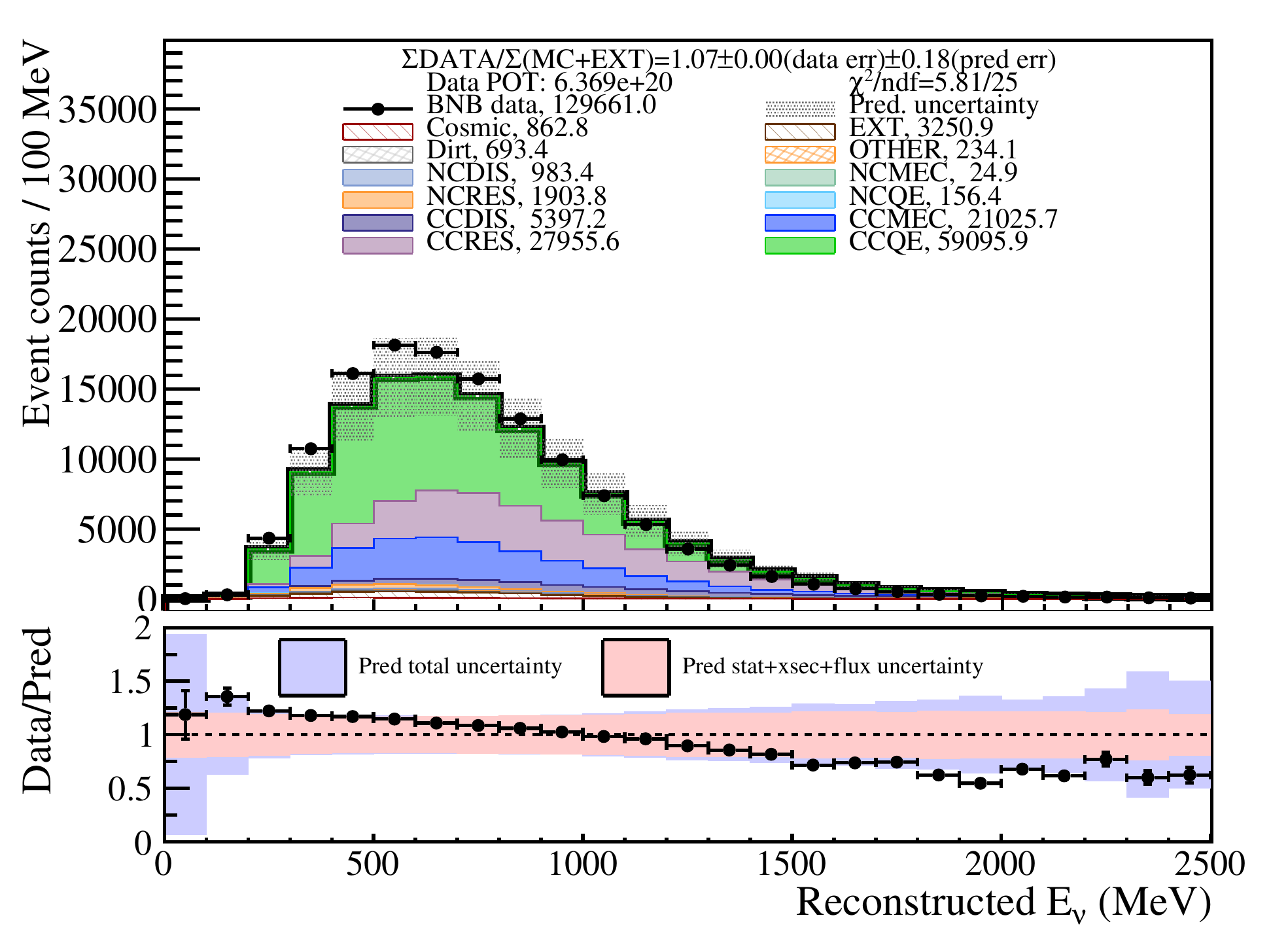}
    \put(-100,115){MicroBooNE}
    \put(-80,100){FC+PC}
    \caption{Neutrino energy, broken down with interaction types}
    \label{fig:numu_recoEnu_interactiontype}
  \end{subfigure}
  \begin{subfigure}[]{0.48\textwidth}
    \includegraphics[width=\textwidth]{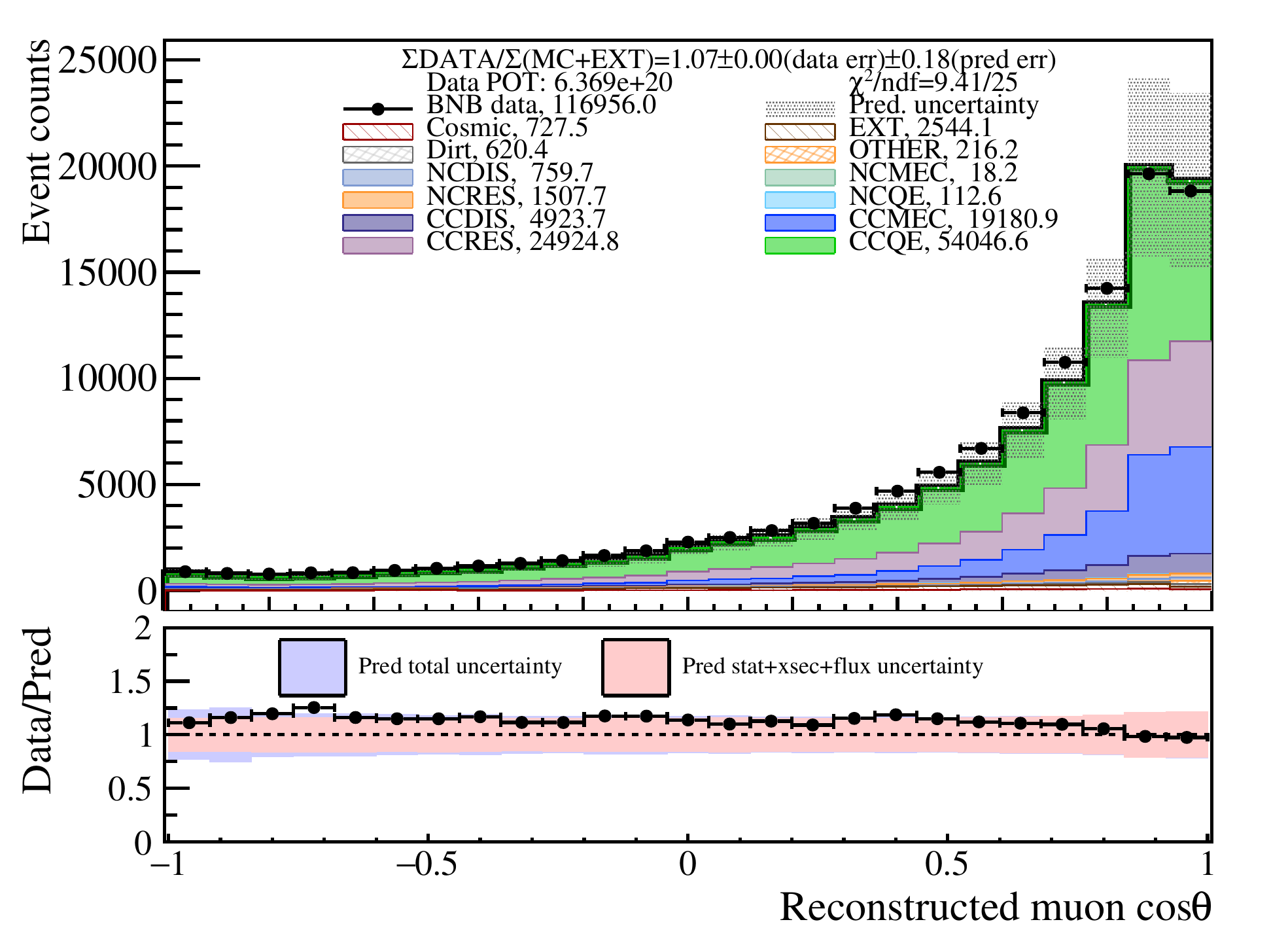}
    \put(-180,115){MicroBooNE}
    \put(-180,100){FC+PC}
    \caption{Muon cos$\theta$, broken down with interaction types}
    \label{fig:numu_costheta_interactiontype}
  \end{subfigure}
  \begin{subfigure}[]{0.48\textwidth}
    \includegraphics[width=\textwidth]{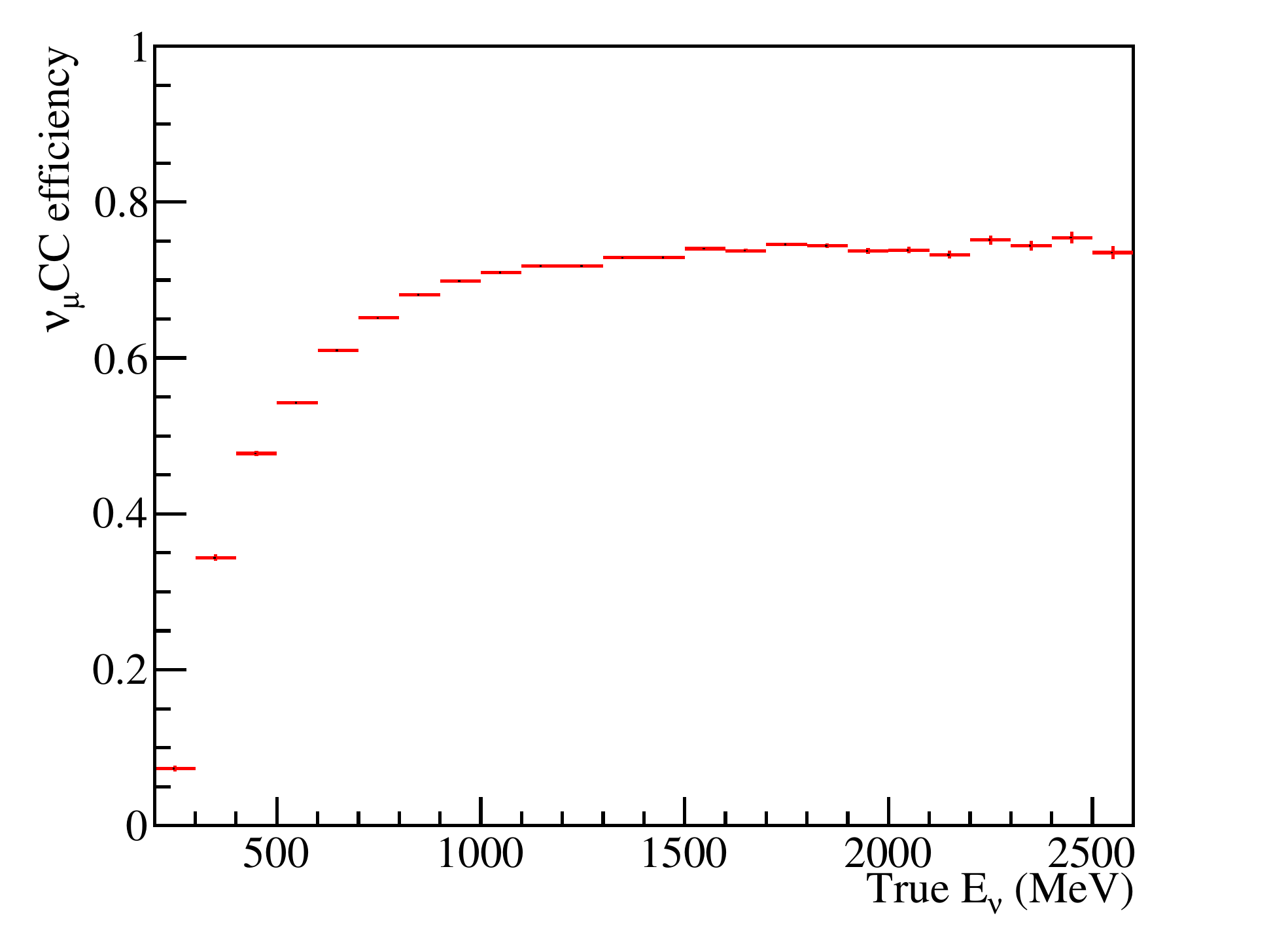}
    \put(-130,160){MicroBooNE Simulation}
    \caption{Efficiency as a function of true neutrino energy}
    \label{fig:numu_eff_recoEnu}
  \end{subfigure}
  \begin{subfigure}[]{0.48\textwidth}
    \includegraphics[width=\textwidth]{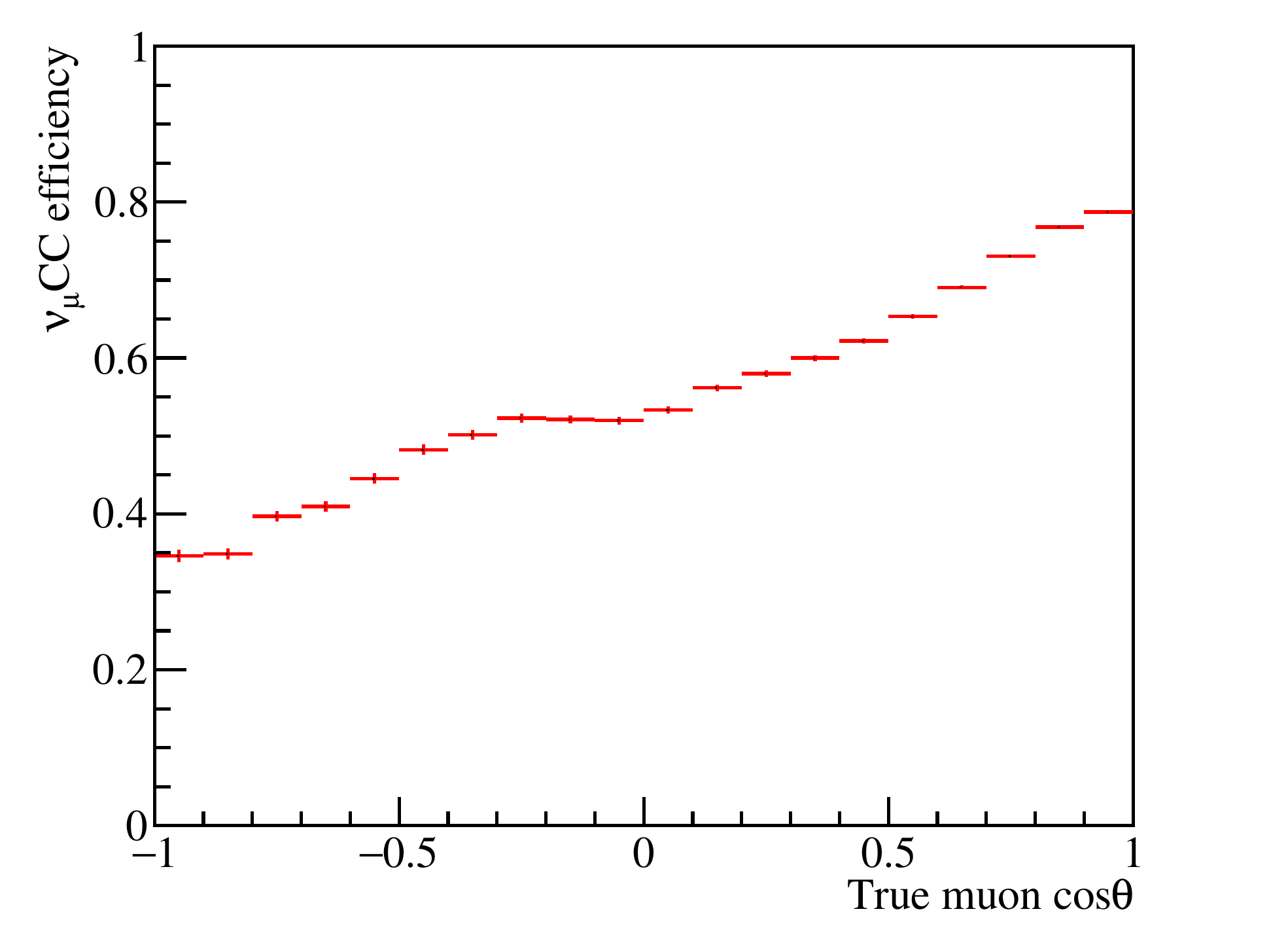}
    \put(-130,160){MicroBooNE Simulation}
    \caption{Efficiency as a function of true muon cos$\theta$}
    \label{fig:numu_eff_costheta}
  \end{subfigure}

  \caption{The final \numuCC\ selections as a function of reconstructed neutrino energy [(a) and (c)] and reconstructed muon cos$\theta$ [(b) and (d)]. (a) and (b) are categorized by event types, and (c) and (d) by interaction types. The number of events correspond to the range shown in the plot. The bottom sub-panels present both the statistical and systematic uncertainties. The pink band includes the statistical, cross section, and flux uncertainties. The purple band corresponds to the full uncertainty with the addition of the detector systematic uncertainty. The selection efficiencies are shown as a function of (e) true neutrino energy and (f) true muon cos$\theta$ with only statistical uncertainty considered. The other dimensions are integrated in calculating these efficiencies.}
  \label{fig:numu_selection_recoEnu_costheta}
  
\end{figure*}
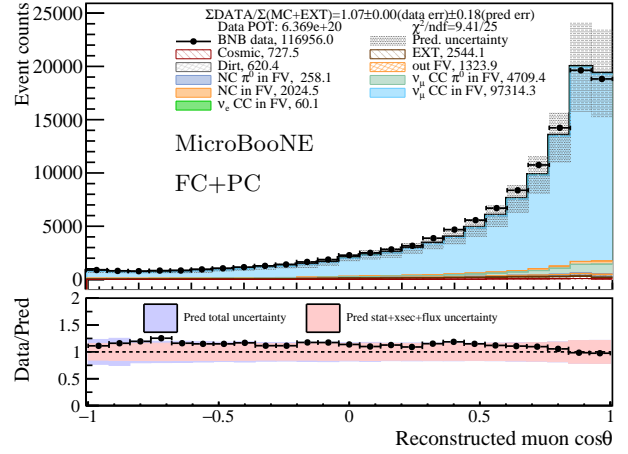
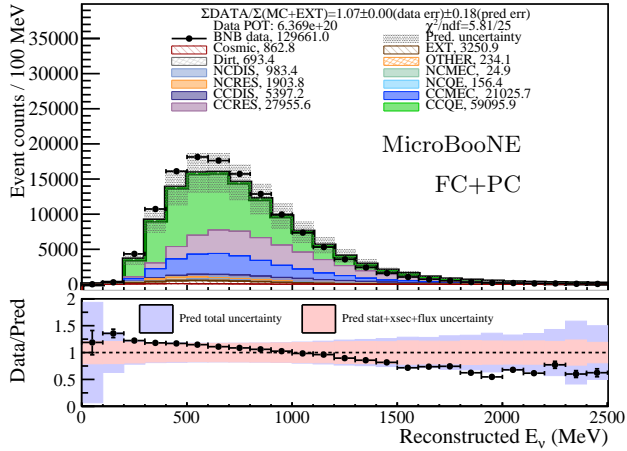
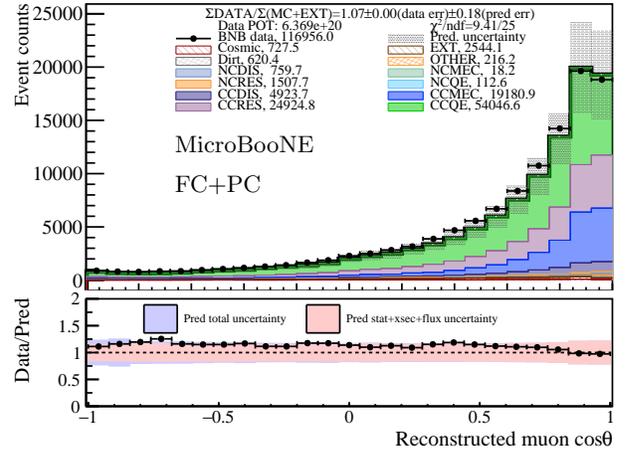
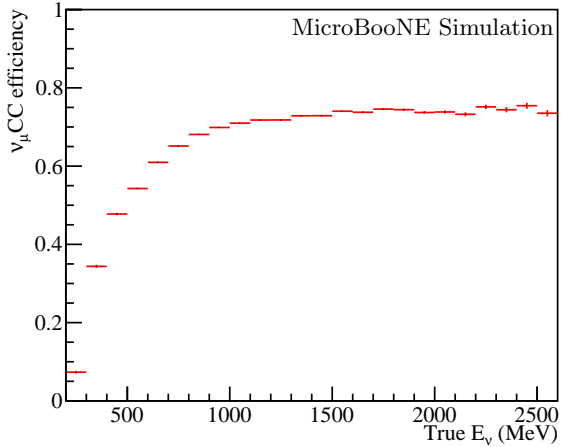
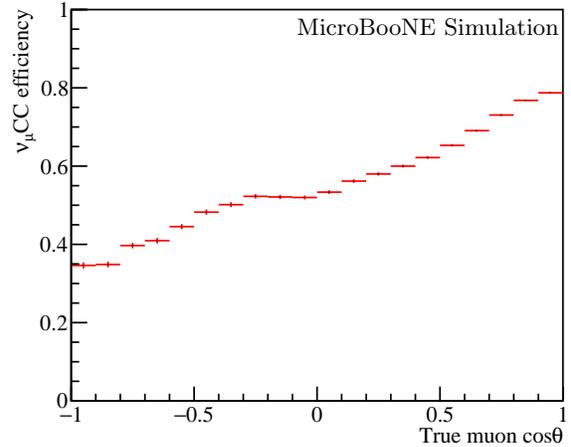

% NC rejection
For NC neutrino interaction background inside the fiducial volume, the main difference from \numuCC\ events is the absence of a primary muon at the neutrino vertex. However, the separation of \numuCC\ and this NC background, which mainly relies on the discrimination of charged pions and muons, is very difficult if only the \dqdx\ information is used. To further reject such NC background, the activities associated with charged pions, e.g. proton scattering, and the relatively large-angle deflection ($\sim$10 degrees) of the trajectory of charged pions can be used to provide additional separation power. 

% BDT selection
With the identification of the major features of the residual cosmic-ray background, neutrino-induced background originating outside the fiducial volume, and NC events inside the fiducial volume, the BDT was trained and applied to improve the \numuCC\ selection. A similar training strategy as discussed in 
Sec.~\ref{sec:nueCC} is used with a signal definition switched to the \numuCC\ events.
Figure~\ref{fig:numu_bdt} shows the \numuCC\ selection efficiency and purity as a function of the \numuCC\ BDT score as well as the distribution of the \numuCC\ BDT score. The final cut value of 0.9 was chosen for the \numuCC\ selection with a 68\% efficiency and 92\% purity. Figure~\ref{fig:numu_selection_recoEnu_costheta} shows the selected \numuCC\ events and selection efficiency as a function of neutrino energy and muon cos$\theta$. The efficiency is generally higher for more forward-going angles as events with forward-going angles are more likely to have a typical topology of a \numuCC\ event to which the BDT input variables are tuned. The ``slope'' of data-prediction ratios present in the bottom panel of Fig.~\ref{fig:numu_recoEnu} will be discussed in Sec.~\ref{sec:energy_validation}.
For some specific backgrounds, the BDT could select a small number of poorly reconstructed events. This is because the \numuCC\ BDT was trained on a data set without an explicit request of a muon being present. Therefore, some \numuCC\ candidates may not have a primary muon identified. This results in a non-zero number of entries in the first bin (less than 100 MeV) in Figs.~\ref{fig:numu_recoEnu}~and~\ref{fig:numu_recoEnu_interactiontype} in which case the muon rest mass is not considered in the neutrino energy reconstruction and, also, in an absence of some events that have no reconstructed muon cos$\theta$ in Figs.~\ref{fig:numu_costheta}~and~\ref{fig:numu_costheta_interactiontype}. We can do this because of the high initial signal-to-background ratio of the events.
%Such plotting style applies to the same type of figures in this paper.

\subsection{\pizero\ Selection}\label{sec:pi0}
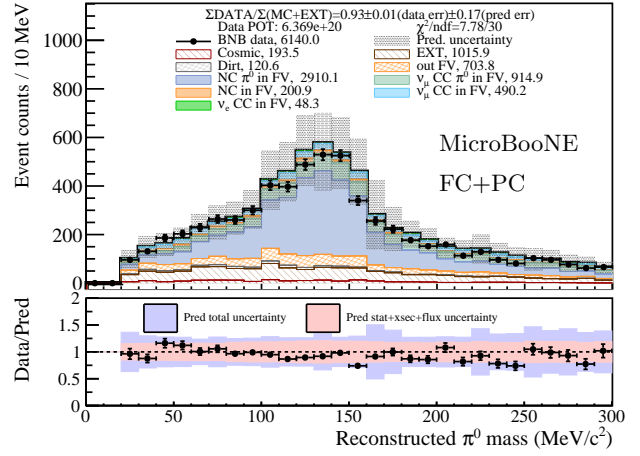
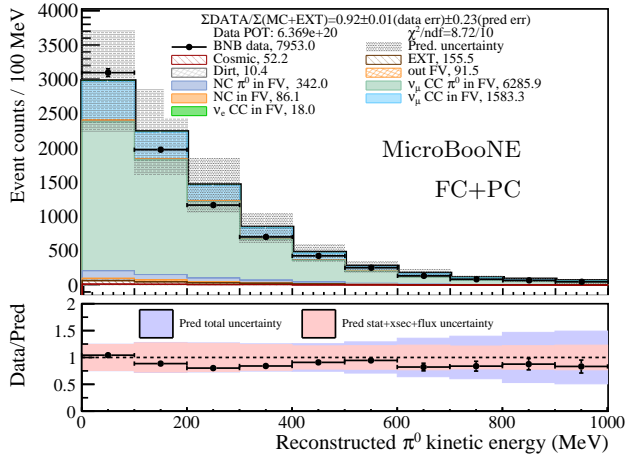
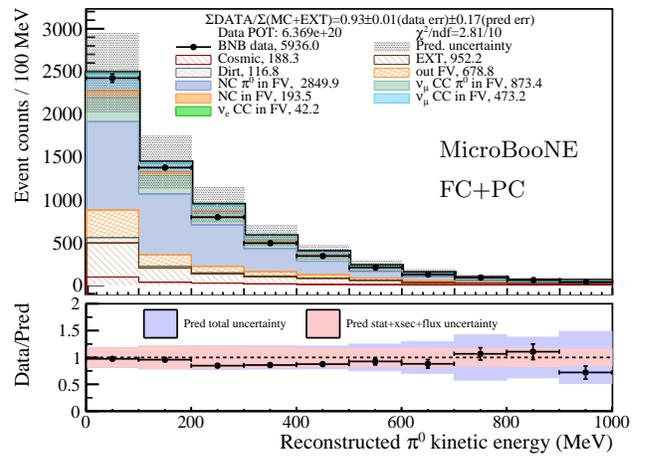
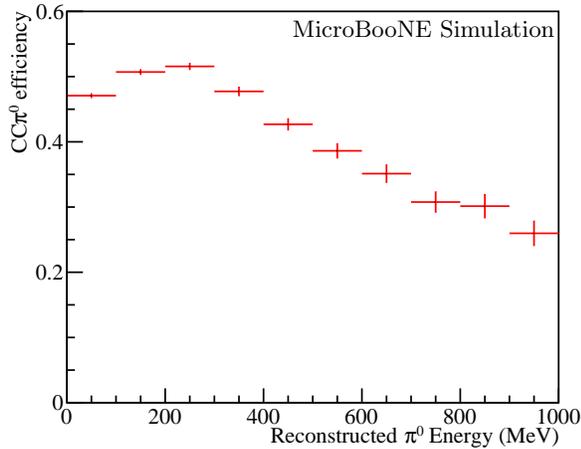
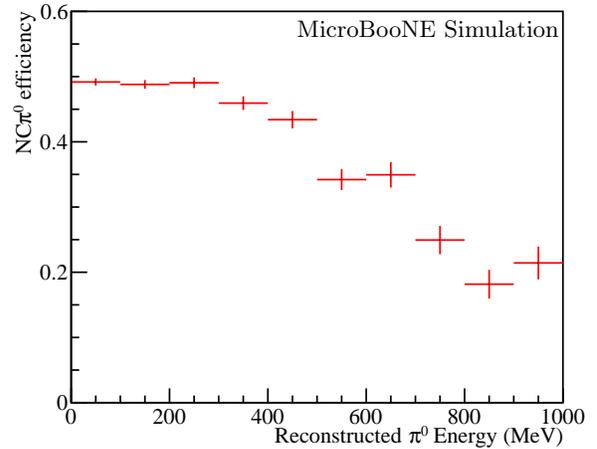
\begin{figure*}[htp!]
  \captionsetup[subfigure]{justification=centering}
  \centering
  \begin{subfigure}[]{0.48\textwidth}
    \includegraphics[width=\textwidth]{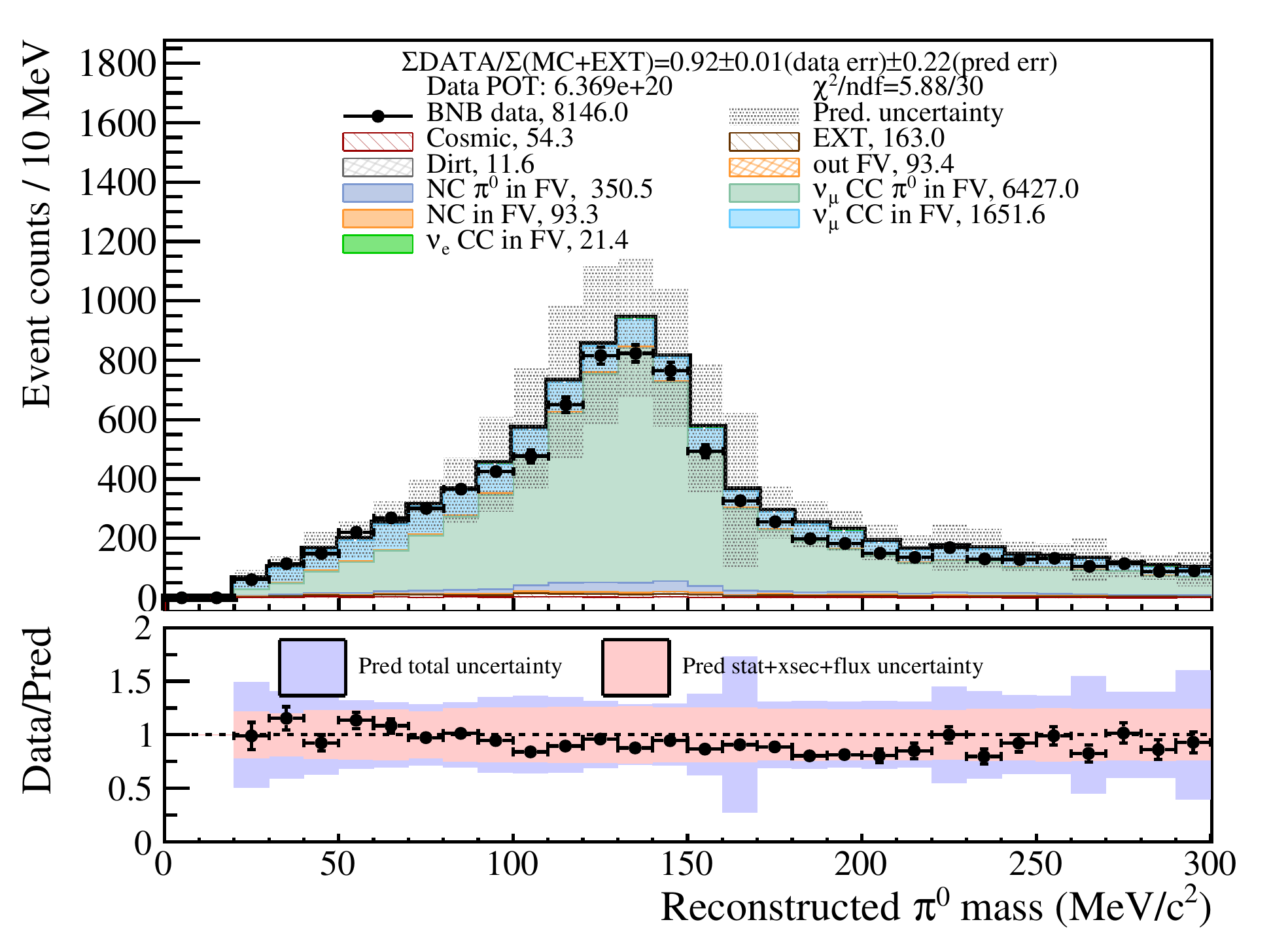}
    \put(-100,115){MicroBooNE}
    \put(-80,100){FC+PC}
    \caption{CC\pizero reconstructed \pizero\ mass}
  \end{subfigure}
  \begin{subfigure}[]{0.48\textwidth}
    \includegraphics[width=\textwidth]{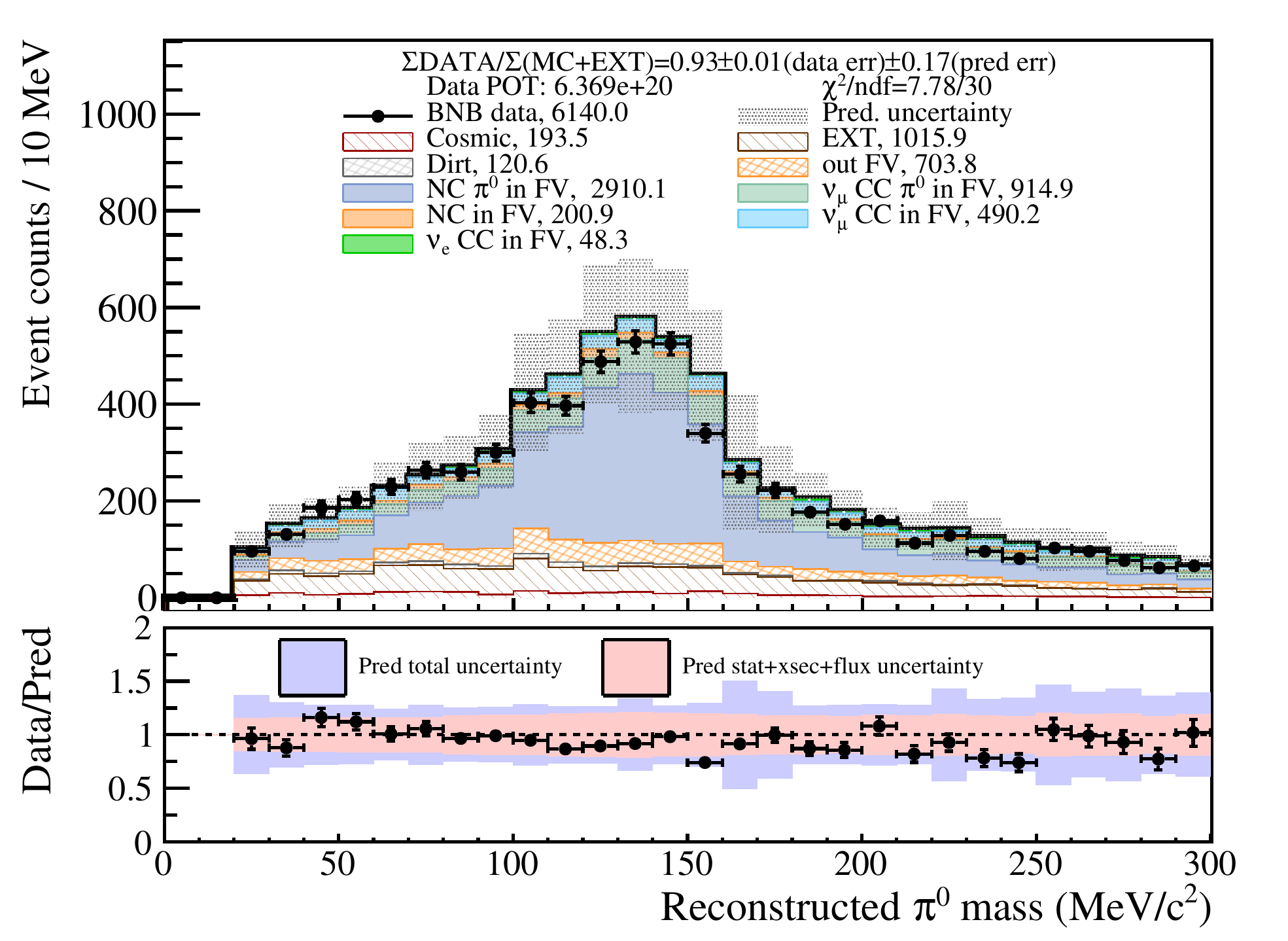}
    \put(-80,115){MicroBooNE}
    \put(-80,100){FC+PC}
    \caption{NC\pizero reconstructed \pizero\ mass}
  \end{subfigure}   
  \begin{subfigure}[]{0.48\textwidth}
    \includegraphics[width=\textwidth]{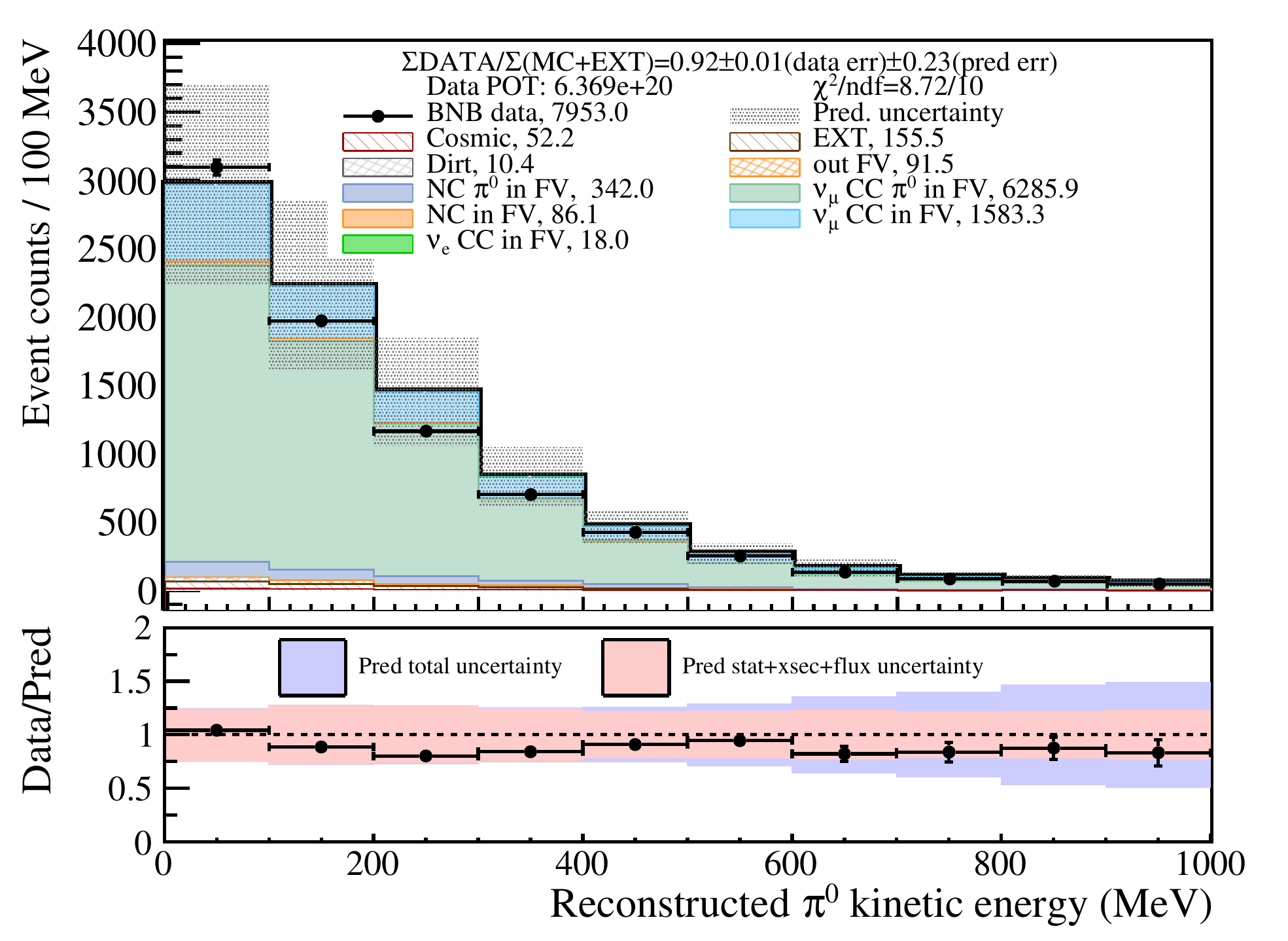}
    \put(-100,115){MicroBooNE}
    \put(-80,100){FC+PC}
    \caption{CC\pizero reconstructed \pizero\ kinetic energy}
  \end{subfigure}
  \begin{subfigure}[]{0.48\textwidth}
    \includegraphics[width=\textwidth]{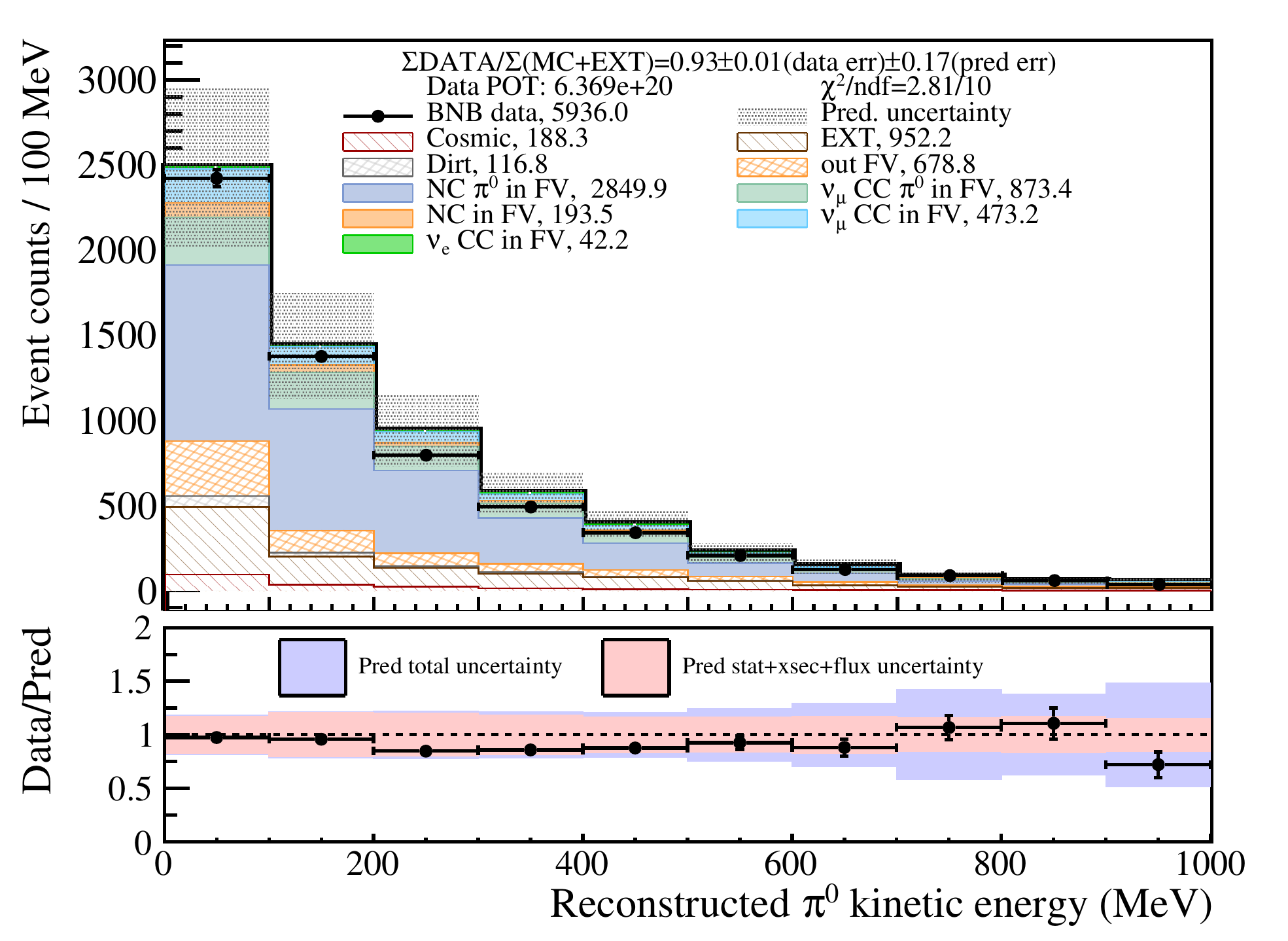}
    \put(-80,115){MicroBooNE}
    \put(-80,100){FC+PC}
    \caption{NC\pizero reconstructed \pizero\ kinetic energy}
  \end{subfigure}
  \begin{subfigure}[]{0.48\textwidth}
    \includegraphics[width=\textwidth]{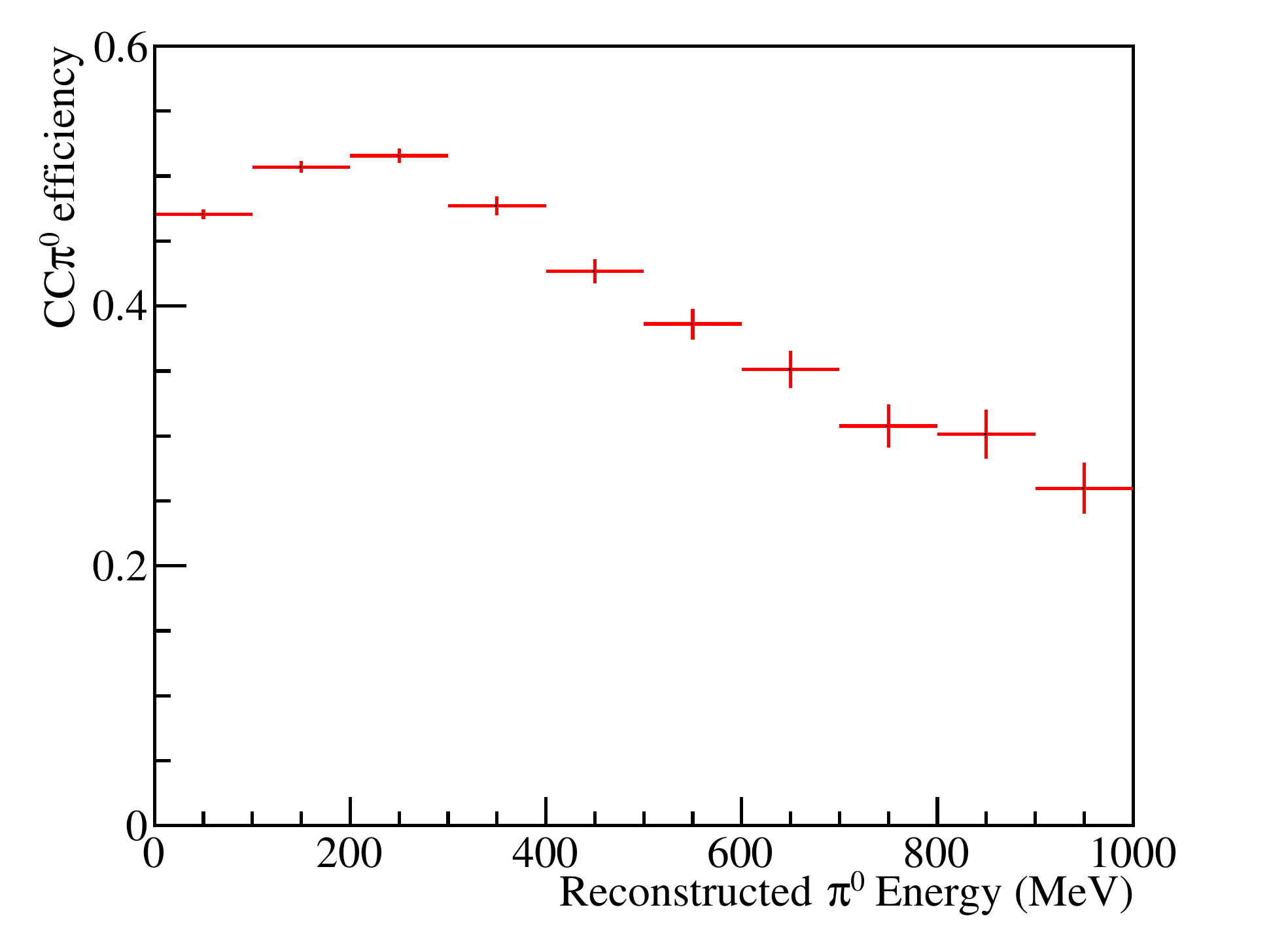}
    \put(-130,160){MicroBooNE Simulation}
    \caption{CC\pizero\ selection efficiency as a function of reconstructed \pizero\ kinetic energy}
  \end{subfigure}
  \begin{subfigure}[]{0.48\textwidth}
    \includegraphics[width=\textwidth]{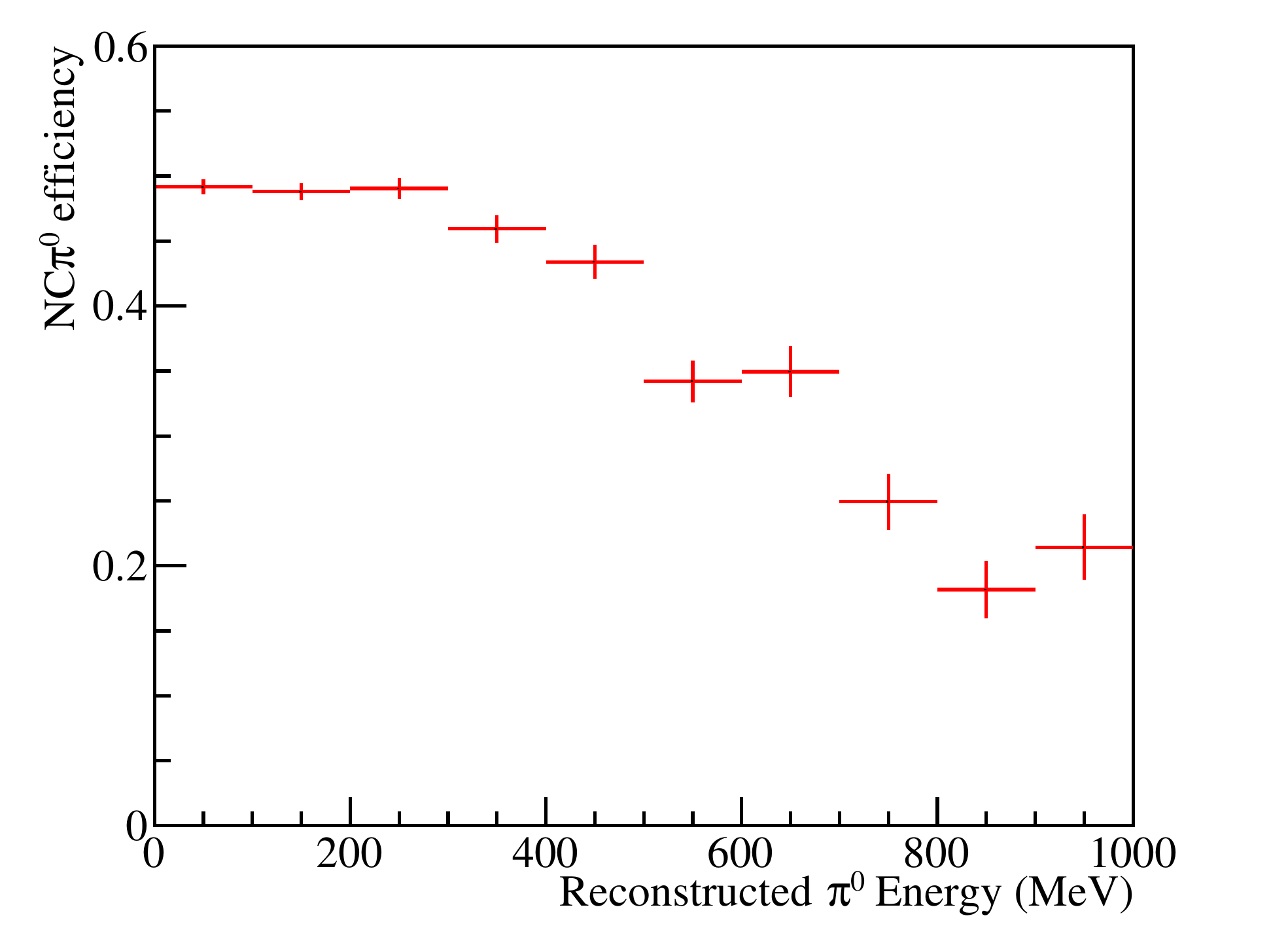}
    \put(-130,160){MicroBooNE Simulation}
    \caption{NC\pizero\ selection efficiency as a function of reconstructed \pizero\ kinetic energy}
  \end{subfigure}
  \caption{
  Distributions of the reconstructed \pizero\ invariant mass [(a) and (b)], reconstructed \pizero\ kinematic energy [(c) and (d)] and selection efficiency [(e) and (f)] as a function of reconstructed \pizero\ kinetic energy for CC\pizero\ [(a), (c), and (e)] and NC\pizero\ [(b), (d), and (f)]. Only the statistical uncertainty was considered in the efficiency plots (e) and (f). The bottom sub-panels of plots [(a)-(d)] present both the statistical and systematic uncertainties. The pink band includes the statistical, cross section, and flux uncertainties. The purple band corresponds to the full uncertainty with the addition of the detector systematic uncertainty. A consistency is observed between the data and simulation validating the energy scale reconstruction for EM showers.}
  \label{fig:pio_mass}
\end{figure*}

The \numuCC\ selection described in the previous section is employed to select CC\pizero\ events from \numuCC\ interactions. 
Additionally, a NC\pizero\ selection is constructed by considering the non-cosmic events that fail the \numuCC\ 
selection ($\nu_\mu$ BDT score smaller than zero). 
In the reconstruction of \pizero\ events, the pair of EM showers with highest energies, supposedly from \pizero\ two-$\gamma$ decay, is chosen to be the ones pointing to the same vertex. 
\pizero\ particles are identified as primary particles by placing a maximum distance cut between the neutrino vertex and the 
\pizero\ vertex. 
More details on the vertexing, clustering, and  pattern recognition of \pizero\ events can be found in Ref.~\cite{wire-cell-pr}. Further selection 
cuts use the $\gamma$ energies, the distances between the 
neutrino vertex and the $\gamma$ vertices, the opening angle between the two $\gamma$'s, and the reconstructed \pizero\ invariant mass. The distribution of the reconstructed \pizero\ mass of the selected CC and NC \pizero\ events can be found in Fig.~\ref{fig:pio_mass} where a Gaussian fit over the peak region of data returns a mean$\pm$width of 131.2$\pm$22.1~MeV/c$^{2}$ for the CC\pizero\ selection and 130.4$\pm$19.3~MeV/c$^{2}$ for the NC\pizero\ selection. 
The best-fit mass values have $<$1 MeV differences compared to the MC reconstructed mass values and are consistent with the expected \pizero\ invariant mass of 135~MeV/c$^{2}$. The best-fit peak width does not include the contribution from the long tails, 
which is the result of imperfect event reconstruction. 
Table~\ref{tab:pizero_effpur} lists the efficiency and purity of the \pizero\ selections. 
Here, the efficiency is defined with respect to all CC or NC \pizero\ events with their neutrino interaction vertex inside the fiducial volume.
The dominant background component for the CC\pizero\ selection is \numuCC\ events without a \pizero\ in the final state, and the dominant background for the NC\pizero\ selection is external events originating from cosmic-ray muons or neutrino interactions outside the fiducial volume.

\begin{table}[t]
    \centering
    \begin{tabular}{llrr}
        \hline
        \hline
        Event selection & Containment  & Efficiency & Purity \\
        \hline 
        CC\pizero\      & FC+PC        & 32\%       & 72\% \\
                        & FC           & 12\%       & 75\% \\
                        & PC           & 20\%       & 71\% \\
        \hline
        NC\pizero\      & FC+PC        & 25\%       & 44\% \\
        \hline
        \hline
    \end{tabular}
\caption[]{Selection efficiency and purity for CC\pizero\ FC+PC, CC\pizero\ FC, CC\pizero\ PC, and NC\pizero\ FC+PC samples.}
\label{tab:pizero_effpur}
\end{table}

\section{Systematic Uncertainties}\label{sec:systematics}
\begin{figure*}[!htp]
  \centering
  \includegraphics[width=0.8\textwidth]{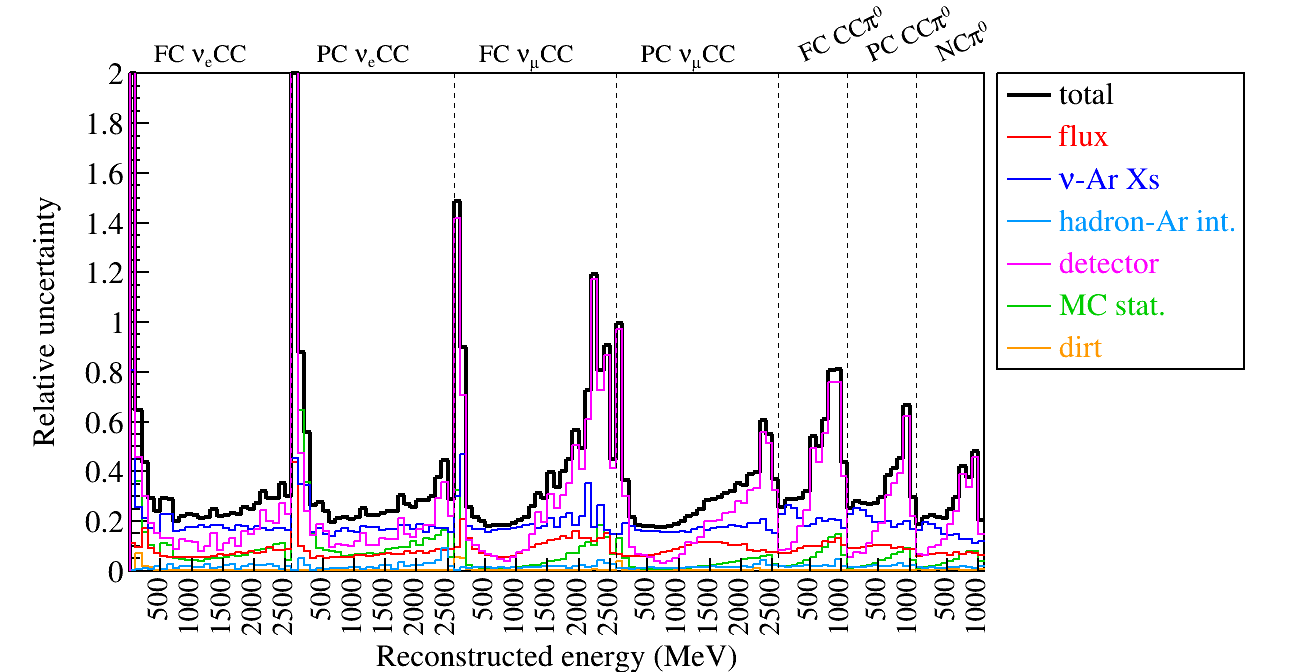}
  \put(-360, 205){MicroBooNE Simulation}
  \caption{Summary of relative uncertainties ($\frac{\mathrm{absolute\ error}}{\mathrm{prediction}}$) of all systematic uncertainty sources for the seven channels. Spikes at very low or high energy regions for detector systematic uncertainty are largely attributed to statistical errors stemming from the finite size of related MC samples.}
  \label{fig:canv_h2_relerr_total}
\end{figure*}

In this analysis, we consider sources of systematic uncertainties from i) the neutrino flux of the BNB, ii) neutrino-argon cross sections of the \textsc{Genie} event generator, iii) hadron-argon interactions of the \textsc{Geant4} simulation, iv) detector response resulting from imperfect 
calibration, and v) the finite statistics of MC samples used for prediction, as well as vi) additional uncertainty for dirt events, which originate from neutrino interactions outside the cryostat. The various sources of systematic uncertainty each have different impacts on the reconstruction or selection efficiency (for both signal and background) as well as the reconstruction of kinematic variables of the predicted events. The uncertainty due to limited statistics of MC samples and beam-off data is particularly important for estimating backgrounds of rare event searches, e.g. the selection of BNB \nueCC, which is $\sim$0.5\% of the total flux. Figure~\ref{fig:canv_h2_relerr_total} summarizes the relative uncertainties from all systematic uncertainty sources for the seven channels. In the following, we describe each uncertainty in detail. 

\subsection{Uncertainties from the Model of Neutrino Beam Flux}~\label{sec:syst_flux}
The calculation of the BNB neutrino flux and its uncertainties follows earlier work from the MiniBooNE collaboration, which includes a well constrained beamline simulation based on the \textsc{Geant4} framework~\cite{Agostinelli:2002hh} as well as techniques to handle systematic uncertainties~\cite{AguilarArevalo:2008yp}. Our flux prediction uses the updated flux calculation that takes into account the SciBooNE measurement of  $p + Be \rightarrow K^+$ production in the BNB~\cite{Cheng:2011wq,Mariani:2011zd}, which provided a better constraint on kaons produced in the BNB. In addition, our flux prediction evaluates $\pi^+$ and $\pi^-$ production uncertainties directly from HARP pion production data~\cite{HARP:2007dqt} rather than using a fit parameterization. This technique allows the HARP measurement uncertainties to be more properly propagated to the calculated neutrino flux. 

\begin{table*}[htp!]
\begin{center}
\begin{tabular}{c|c}
\hline \hline
 Tuning parameter name & Parameter type \\\hline 
 $\pi^+$ hadron production & FLUX \\
 $\pi^-$ hadron production & FLUX  \\
 $K^+$ hadron production & FLUX  \\
 $K^-$ hadron production & FLUX  \\
 $K^0_L$ hadron production & FLUX  \\
 horn current distribution & FLUX  \\
 horn current calibration & FLUX \\
 nucleon total scattering Xs & FLUX \\
 nucleon inelastic scattering Xs & FLUX \\
 nucleon quasi-elastic scattering Xs & FLUX \\
 pion total scattering Xs & FLUX \\
 pion inelastic scattering Xs & FLUX  \\
 pion quasi-elastic scattering Xs& FLUX  \\ \hline
 MicroBooNE \textsc{Genie} All& \textsc{Genie} Xs ($\mu$B tune) \\
 Strength of the CCQE RPA correction & \textsc{Genie} Xs ($\mu$B tune) \\
 Parameterization of the CCQE nucleon axial form factor & \textsc{Genie} Xs \\
 Parameterization of the CCQE nucleon vector form factors & \textsc{Genie} Xs  \\
 Changes angular distribution of nucleon cluster in MEC & \textsc{Genie} Xs ($\mu$B tune) \\
 CCMEC Cross-section Shape & \textsc{Genie} Xs ($\mu$B tune)\\
 Angular distribution for RES $\Delta\rightarrow N + \pi$ & \textsc{Genie} Xs \\
 Angular distribution for RES $\Delta\rightarrow N + \gamma$ & \textsc{Genie} Xs ($\mu$B tune) \\
 Scaling factor for CC coherent $\pi$ production & \textsc{Genie} Xs  ($\mu$B tune) \\
 Scaling factor for NC coherent $\pi$ production & \textsc{Genie} Xs ($\mu$B tune) \\\hline
 Second-class vector current & Xs \\
 Second-class axial current & Xs  \\\hline
 $\pi^-$ interactions & \textsc{Geant4} \\
 $\pi^+$ interactions & \textsc{Geant4}  \\
 proton interactions & \textsc{Geant4}  \\\hline
 \hline
\end{tabular}
\end{center}
\caption[]{Summary of tuning parameters used in generating the covariance matrix for systematic uncertainties. For each tuning parameter, a covariance matrix is generated according to its number of universes.  The final covariance matrix is the summation of all individual covariance matrices.
%For MicroBooNE \textsc{Genie} All, see Table~\ref{tab:system2} for more details. (JLR removed, since this is not in the table now.)
For \textsc{Genie} Xs, the label ``$\mu$B tune" indicates that the tuning parameter and/or its uncertainty is from MicroBooNE \textsc{Genie} tune other than the default treatment from \textsc{Genie} v3.0.6.}
\label{tab:system1}
\end{table*}

\begin{figure*}[htp!]
  \centering
  \includegraphics[width=0.8\textwidth]{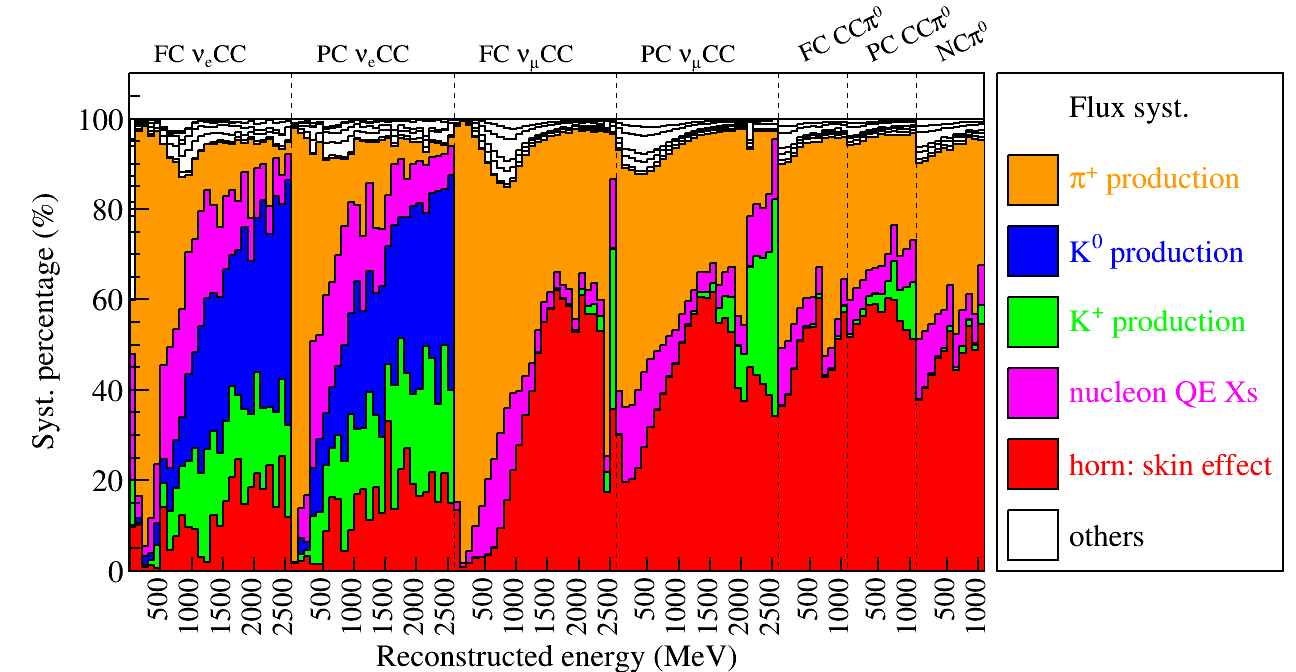}
  \put(-360, 205){MicroBooNE Simulation}
  \caption{Fraction ($\frac{\sigma_i^2}{\sigma^2_{\mathrm{flux}}}\times100$) of uncertainties of total flux systematics for the seven channels.}
  \label{fig:canv_h2_basic_fraction_flux}
\end{figure*}

\begin{figure}[thp!]
  \centering
  \includegraphics[width=0.5\textwidth]{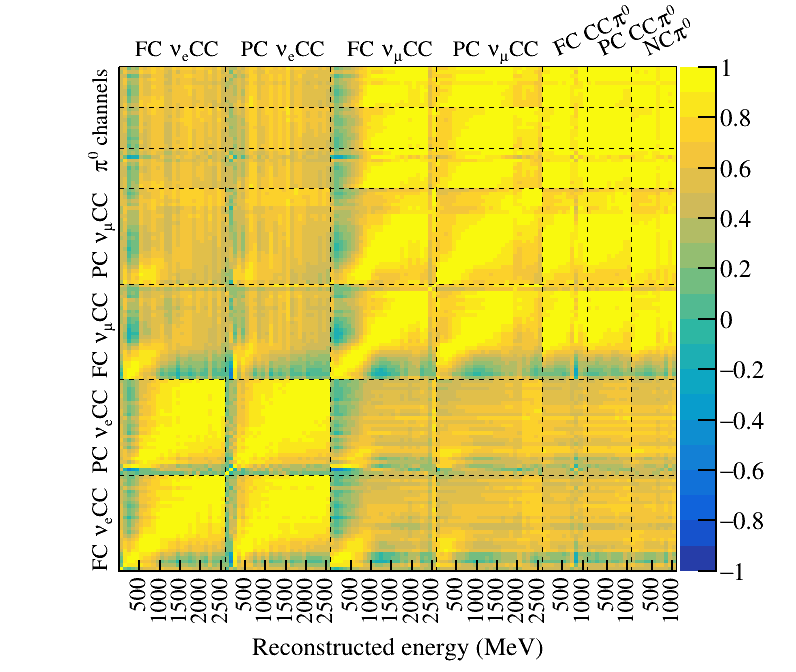}
  \put(-213, 210){MicroBooNE Simulation}
  \caption{Correlations of flux systematics for the seven channels. }
  \label{fig:canv_h2_correlation_flux}
\end{figure}

As summarized in Table~\ref{tab:system1}, the systematic uncertainties in the predicted flux include effects from i) hadron production of $\pi^{+}$, $\pi^{-}$, $K^{+}$, $K^{-}$, and $K^0_L$ and ii) non-hadron production: modeling of the horn current distribution, horn current calibration, and pion and nucleon total, inelastic, and quasielastic scattering cross-sections on beryllium and aluminum. There is also an overall 2\% normalization
uncertainty associated with the POT counting.
Figure~\ref{fig:canv_h2_basic_fraction_flux} shows the fractional contributions to the overall flux systematic uncertainty from each source, for the seven channels of the analysis. In the low-energy region, e.g. $E^{\text{rec}}_{\nu}<500$ MeV, which is relevant to the search for a $\nu_e$ low-energy excess, the flux systematic is limited by the hadron production of $\pi^+$, which, with its decay to muons, produces most of the $\nu_e$'s and $\nu_{\mu}$'s in this energy range.
Figure~\ref{fig:canv_h2_correlation_flux} shows the correlations of flux systematics for the seven channels. There are obviously strong correlations between the low energy ranges of $\nu_e$ events and $\nu_\mu$ events given that they all originate from $\pi^+$ decays. There are strong correlations in the high energy range between $\nu_e$ and $\nu_\mu$ given that these largely originate from $\pi^+$ and $K^+$ decays. There are also strong correlations between the high energy $\nu_\mu$ and the entire energy range of the $\pi^0$ events since the neutrino interactions that enter into the $\pi^0$ channels are almost all from $\nu_\mu$ interactions (negligible $\nu_e$) which pass the energy threshold of $\pi^0$ resonance interactions.

\subsection{Uncertainties from the Neutrino-Argon Interaction Cross Sections}\label{sec:syst_xs}
As introduced in Sec.~\ref{sec:uboone_mc}, \textsc{Genie} v3.0.6~\cite{GENIE:2021npt}, with parameters
governing the CCQE and CC2p2h models adjusted~\cite{uboone_genie_tune} by MicroBooNE, is used to generate exclusive neutrino-argon interactions. 

\begin{figure}[htp!]
  \centering
  \includegraphics[width=0.5\textwidth]{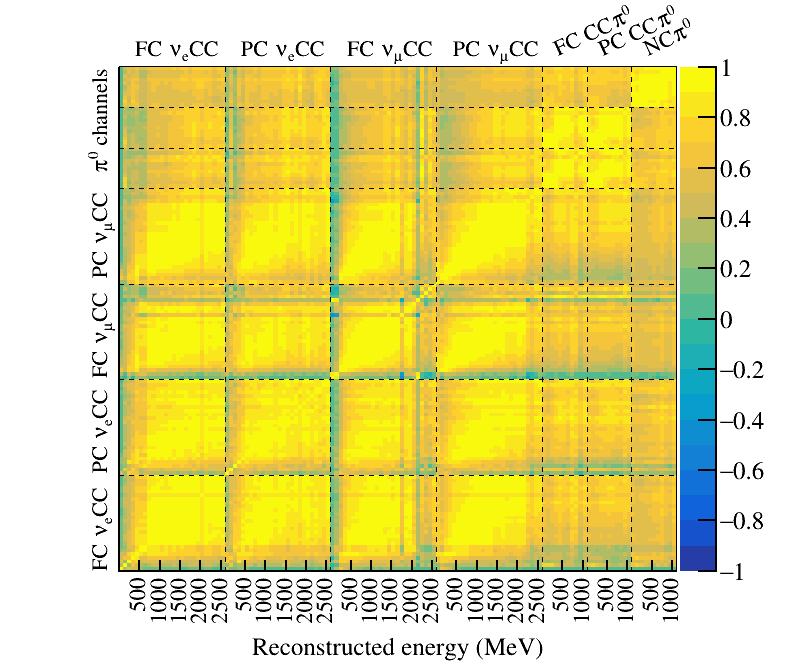}
  \put(-213, 210){MicroBooNE Simulation}
  \caption{Correlations of cross section systematics for the seven channels.}
  \label{fig:canv_h2_correlation_Xs}
\end{figure}

Table~\ref{tab:system1} summarizes the uncertainties considered in the reweighting procedure for neutrino-argon interaction cross sections (Xs). In particular, the contribution labeled ``MicroBooNE \textsc{Genie} All'' contains 46 tuning parameters that are simultaneously varied in generating hundreds of universes. The variations in these universes are used to construct the covariance matrix. 
%The details of the 46 knobs can be found in Table.~\ref{tab:system2}. 
These tuning parameters cover a wide range of models including CCQE, CC resonance (CCRES), CC non-resonance, CC transition, CC deep-inelastic scattering (CCDIS), NC interactions, and final-state interactions. In addition to ``MicroBooNE \textsc{Genie} All'', as listed in Table~\ref{tab:system1}, there are additional 9 ``\textsc{Genie} Xs'' parameters taking into account other cross-section uncertainties in \textsc{Genie}, each of which have only two universes corresponding to the nominal value and the 1$\sigma$ bound of the tuning parameter, respectively. The 1$\sigma$ uncertainty is taken to be the difference. There are two other tuning parameters labeled ``Xs'' focusing on second-class currents~\cite{Weinberg:1958ut} that may contribute to $\nu_e/\nu_\mu$ CCQE cross section differences as suggested by Ref.~\cite{Day:2012gb}. %Figure~\ref{fig:canv_h2_relerr_total} summarizes the relative uncertainties of all sources of systematic uncertainty for the seven channels.  <== JLR: I don't think you need to repeat this.

Figure~\ref{fig:canv_h2_correlation_Xs} shows the correlations of cross section systematics for the seven channels. There are generally strong correlations between $\nu_e$CC and $\nu_\mu$CC across the entire energy range, which is a natural result of the lepton universality assumption. In the low-energy region, e.g. $E^{\text{rec}}_{\nu}<500$ MeV which is relevant to the search for a $\nu_e$ low-energy excess, the neutrino cross section systematic is limited by the level of suppression 
of the CCQE cross section behavior at low $Q^2$ (four-momentum transfer from the lepton) because of 
long-range nucleon-nucleon correlations, which is poorly constrained by existing data.
We should further note that some inconsistencies were identified in the \textsc{Genie} v3.0.6 reweighing code used to 
evaluate FSI-related systematics as also discussed in Ref.~\cite{uboone_genie_tune}, but the effect of these were found to have a negligible impact on the overall 
analysis sensitivity and have been ignored.

\subsection{Uncertainties from the Hadron-Argon Interaction}\label{sec:syst_geant}
Charged hadrons can scatter, both elastically and inelastically, with external argon nuclei via hadronic interactions. These interactions can lead to the production of additional particles or can cause large angle changes in particle trajectories that may affect the reconstructed neutrino energy. Therefore, it is important to include the uncertainties of these secondary hadron-argon interactions. Studies were performed separately for elastic and inelastic interactions for protons and charged pions. %Elastic-scattering uncertainties were found to have negligible impact, and are therefore not included in this analysis.

\begin{figure}[t]
  \centering
  \includegraphics[width=0.5\textwidth]{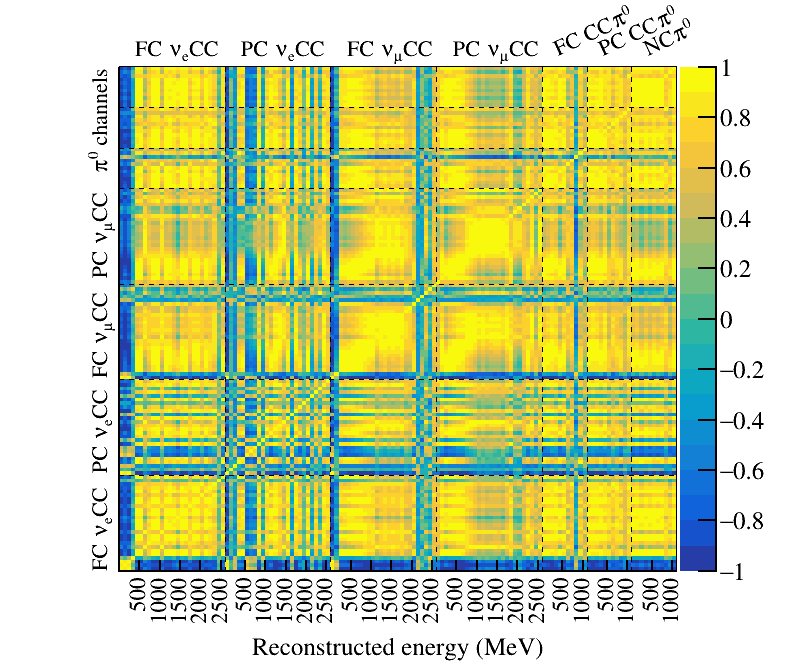}
  \put(-213, 210){MicroBooNE Simulation}
  \caption{Correlations of hadron-argon interaction systematics for the seven channels.}
  \label{fig:canv_h2_correlation_geant}
\end{figure}

\textsc{Geant4} is used to propagate all hadrons through the detector medium based on a semi-classical cascade model~\cite{Wright:2015xia}. Using the \textsc{Geant4reweight}~\cite{Calcutt:2021zck} package, events with inelastic hadronic interactions are reweighted independently for interactions containing protons, positive pions, and negative pions. 
For each particle type and independent of the particle's energy, the inelastic cross section is varied around its mean by $\mathcal{O}$(20\%), based on the uncertainties of world data. 
Figure~\ref{fig:canv_h2_correlation_geant} shows the correlations of hadron-argon interaction cross section systematics for the seven channels. 
The contribution (square of the uncertainty) from this source of systematic uncertainty to the total systematic uncertainty is up to 1\% as indicated in Fig.~\ref{fig:canv_h2_relerr_total}. 

\subsection{Uncertainties from the Model of Detector Effects}\label{sec:syst_det}
There are four major categories of detector systematic uncertainties each targeting one area of the detector
calibration effort:  i) variations related to the light yield (LY) and propagation simulation, ii) variation in the space 
charge effect, iii) variation in the recombination model, and iv) variations on the amplitude and width of the deconvolved ionization charge waveforms for other discrepancies between the detector response used in simulation and that in data. 
We vary the magnitude of the space charge effect based upon the measurements of the spatial distortions at the edge of the TPC, extrapolated to the bulk field~\cite{Adams:2019qrr, Abratenko:2020bbx}. A different recombination model, which provides slightly better agreement to the data, is used to estimate the data/MC difference in the $dE/dx$ to $dQ/dx$ conversion. The variations in the wire waveform include effects as a function of $x$ (drift direction), $\left(y,z\right)$ (vertical/beam directions), $\theta_{xz}$ and $\theta_{yz}$ (the angular orientation of the particle’s trajectory with respect to the global coordinate system); these variations are constructed by comparing the waveforms in data and simulation. 
More details on the variations of the wire waveform can be found in Ref.~\cite{uBpublic_wiremod}.

For each source of detector systematic uncertainty, the same set of MC simulation events are re-simulated with a change to the detector modeling parameter of interest. The systematic uncertainty is estimated by comparing efficiency and reconstructed kinematic variables in the new and old simulation.
The change of each detector modeling parameter is treated as 1$\sigma$, meaning that there is, in principle, only one degree of freedom in constructing each detector covariance matrix after factoring in the statistical uncertainties. The usage of the same set of events in the old and new detector simulation aims to reduce statistical fluctuation, which is estimated using the bootstrapping method~\cite{bootstrap}. 

The central idea of bootstrapping is to re-sample existing MC events in order to estimate the uncertainties on the central value of the prediction. The basic procedure is illustrated in the following:
\begin{itemize}
    \item Choose a common set of events in the central value (CV) and 1$\sigma$ variation simulation samples. 
    \item %Use the bootstrapping method to <== JLR: don't use bootstrapping in this description, since it's the thing you're trying to describe.
    Re-sample MC events to form distributions with statistical uncertainty corresponding to that of the expected POT exposure. During this process, one can naturally take into account the event weight in the re-sampling process. 
    \item Apply the event selection requirements to the re-sampled events in both the CV and the 1$\sigma$ samples and calculate the difference in the spectra of different channels. Each such vector is considered one ``universe.'' By repeating this procedure many times, many universes are simulated to form the covariance matrix, $M_R$. This covariance matrix essentially represents the uncertainty on the difference between the CV and 1$\sigma$ sample (nominal difference vector, $\vec{V}^{\text{nominal}}_D$). In other words, these are uncertainties of the uncertainties. 
    \item In order to generate the detector covariance matrix ($M_D$), a 2-step procedure is adopted. In the first step, based on $M_R$, generate a random set of vectors (based on decomposition of the symmetric matrix) $\delta \vec{V}_D$ to be applied to the nominal difference vector ($\vec{V}^{\text{nominal}}_D$) in order to obtain a new vector $\vec{V}_D = \vec{V}^{\text{nominal}}_D + \delta \vec{V}_D$. 
    \item In the second step, generate a single random number $r$ using the normal distribution. Then multiply $\vec{V}_D$ by this random number (bin-to-bin fully correlated) to obtain $r \cdot \vec{V}_D$, which is treated as one universe. 
    \item By multiple repetitions of the above two steps, many universes are simulated, which can the be used to construct the detector covariance matrix.
\end{itemize}
In the aforementioned approach, the statistical uncertainties in the detector variation samples are naturally taken into account in the $M_R$ matrix. Correlation of the detector variation samples is ensured by using the single random number $r$. The relative strengths of the correlated and uncorrelated uncertainties are properly dealt with via this procedure. 

\begin{figure}[t]
  \centering
  \includegraphics[width=0.5\textwidth]{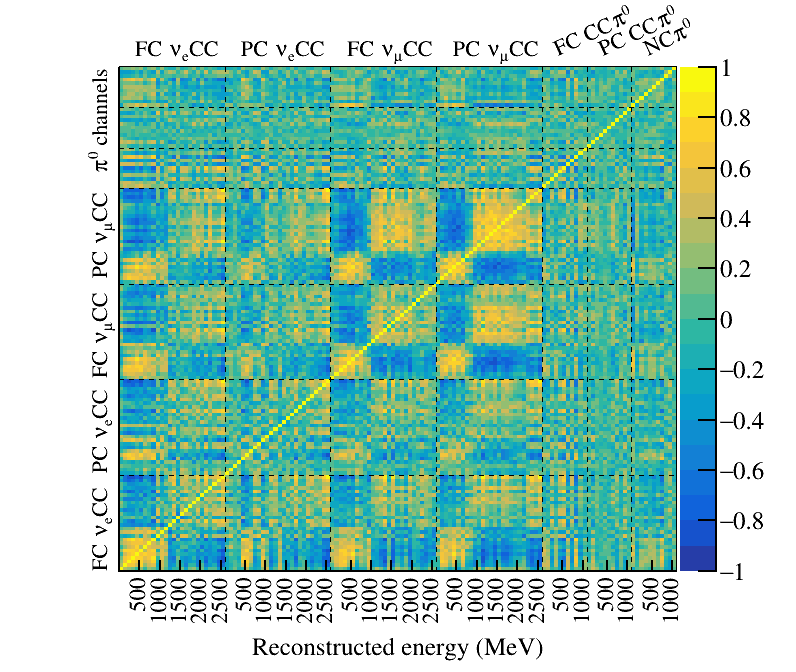}
  \put(-213, 210){MicroBooNE Simulation}
  \caption{Correlations of total detector systematics for the seven channels. The correlations between the FC and PC $\nu_{\mu}$CC channels are dominated by the components of light yield and recombination.}
  \label{fig:canv_h2_correlation_detector}
\end{figure}

\begin{figure*}[htp!]
  \centering
  \includegraphics[width=0.8\textwidth]{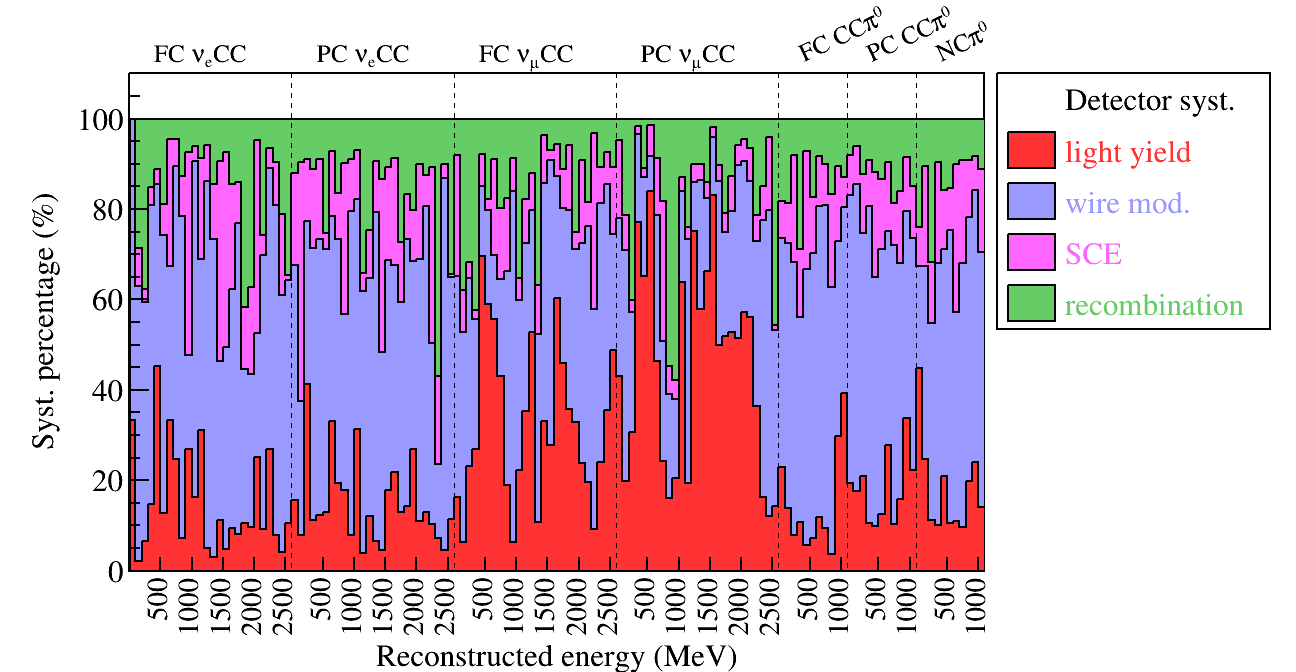}
  \put(-360, 205){MicroBooNE Simulation}
  \caption{Fractional contribution  ($\frac{\sigma_i^2}{\sigma^2_{\mathrm{det}}}\times100$) to the total detector systematic uncertainty for the seven channels.}
  \label{fig:canv_h2_basic_fraction_detector}
\end{figure*}

Figure~\ref{fig:canv_h2_correlation_detector} shows correlations of all detector systematics for the seven channels, including statistical uncertainties. The correlations seen between the FC and PC \numuCC\ channels and between the FC \numuCC\ and \nueCC\ channels mostly come from the variations in light yield and recombination. Figure~\ref{fig:canv_h2_basic_fraction_detector} shows 
the fractional contributions to the overall detector systematic uncertainty from each type of variation for the seven channels of the analysis.

\subsection{Uncertainties of Finite Statistics in Making Predictions}\label{sec:syst_mcstat}
For a rare event search with low background predictions such as that in the MicroBooNE eLEE
analysis, statistical uncertainties of the MC simulation and dedicated background data
can be important. For example,
when the predicted background is zero, quoting a zero uncertainty on this prediction following
the general error propagation would lead to an underestimation of statistical uncertainties. 
On the other hand, if we use an upper limit to quote the statistical uncertainty, 
the size of statistical uncertainties would increase rapidly through the general error propagation procedure when there are multiple components in estimating the background. In order to 
solve this problem and obtain an optimal estimation of MC statistical uncertainties, we adopt
a new approach by combining the Bayesian approach with the covariance matrix. For the Poisson distribution assuming a unit step function as the prior (1 for $\mu \ge 0$ and 0 for $\mu<0$), an accurate 
approximation of the posterior distribution of the expectation $\mu$ given the observation of 
$N$ is given by
\begin{equation}
    P\left(\mu | N \right) \approx e^{N-\mu + N\cdot log\left(\frac{\mu}{N}\right)},
    \label{eq:Poisson_posterior}
\end{equation}
where the high-order asymptotic expansions (based on Stirling's approximation) of the Gamma function, $\Gamma(\mu+1)$, are ignored. For example, in the case of $N=0$, we have $P\left(\mu | 0\right) = e^{-\mu}$. If we use the covariance matrix formalism to estimate the RMS of $\mu$ given a central value of zero, we would have an uncertainty of 1.41 (standard deviation relative to the central value). Similarly, we can estimate the uncertainties for other values of $N$. 

In MC simulation, different events may be associated with a different weight value, which can lead to deviations from the simple Poisson distribution. For example, if a MC event is associated with a weight
value of two, it would present two events in making a prediction.
In the literature, there are many studies of the treatment of a likelihood function given this situation. For a recent review, see Ref.~\cite{Glusenkamp:2019uir}. In this section, we give a prescription for how to deal with this situation. The actual implementation follows Ref.~\cite{Arguelles:2019izp}. Given events (labeled by $i$) with different weights ($w_i$), we define the mean and variance of the prediction as:
\begin{equation}
    m := \sum_i w_i ~~~ \text{and} ~~~ \sigma^2 := \sum_i w_i^2,
\end{equation}
which can be written using effective uniform weights as
\begin{equation}
    m = w_{\textrm{Eff}} \cdot m_{\textrm{Eff}} ~~~ \text{and} ~~~ \sigma^2 = w_{\textrm{Eff}}^2 \cdot m_{\textrm{Eff}}.
\end{equation}
Define 
\begin{equation}
     \overline{m_{\textrm{Eff}}} := \mu\cdot m/\sigma^2,
\end{equation}
so that $m_{\textrm{Eff}} = m^2/\sigma^2$ approximately follows a Poisson distribution with an expectation value of $\overline{m_{\textrm{Eff}}}$, and then the likelihood function (or distribution) can be written as
\begin{equation}
    P(\mu | m, \sigma^2) = \frac{e^{-\mu \cdot m / \sigma^2} \left( \mu \cdot m/\sigma^2\right)^{m^2/\sigma^2}}{\Gamma\left(m^2/\sigma^2+1\right)},
\end{equation}
where $\mu$ is the expected number of events in data with non-uniform weights. Accordingly, Eq.~\ref{eq:Poisson_posterior} can be modified by replacing ``$\mu$'' and ``$N$'' with $\overline{m_{\textrm{Eff}}}$ and $m_{\textrm{Eff}}$.

The above estimation can easily be extended to the case where several estimations are added together (e.g., a situation with multiple backgrounds). In practice, we first divide all distributions into two categories: one with non-zero event count predictions and one with predictions of exactly zero event count. These two distributions are then convolved. Finally, we correct for the Bayesian prior. For implementation, we give prior distributions when we convolve multiple ($n$) distributions together. If the prior information of each distribution is flat, the overall prior of the summation would be max($\mu$,0)$^{n-1}$, or $\mu^{n-1}$ for $\mu \ge 0$ and 0 for $\mu < 0$. In this case, it is corrected (divided) after the convolution of multiple distributions.

\subsection{Additional Systematics for Dirt Events}\label{sec:syst_dirt}
The dirt events are neutrino interactions originating outside the cryostat. The biggest uncertainty associated with the dirt events is the modeling of the geometry and materials of the foam insulation covering the cryostat outer surfaces, the nozzle penetrations for cryogenic and electronic services, the supporting structures, and the dirt around the detector facility. In addition to the systematic uncertainties associated with neutrino flux, cross section, and detector, we conservatively assign an additional 50\% bin-to-bin uncorrelated uncertainty to dirt events. 

\begin{figure}[htp!]
  \centering
  \includegraphics[width=0.5\textwidth]{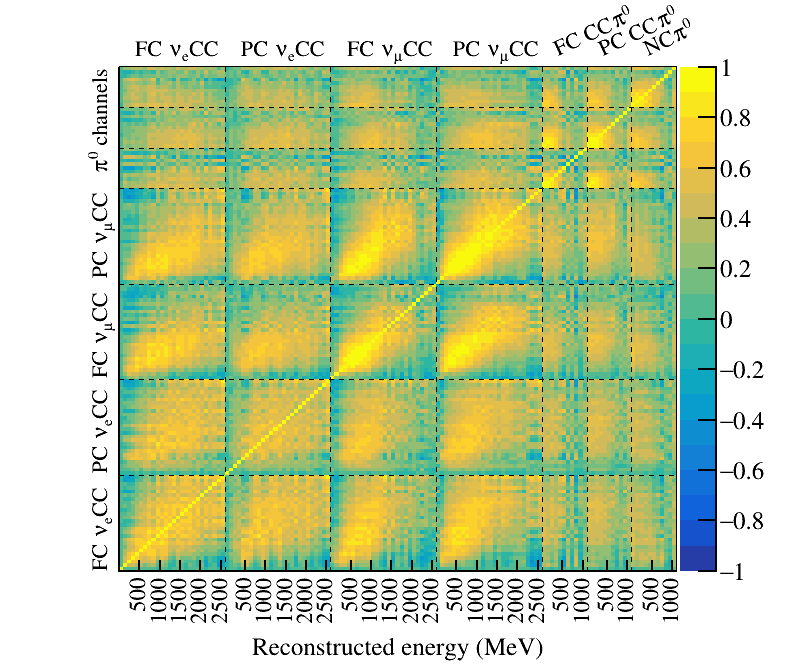}
  \put(-213, 210){MicroBooNE Simulation}
  \caption{Correlations of all systematic uncertainties for the seven channels.}
  \label{fig:canv_h2_correlation_total}
\end{figure}

\begin{figure*}[htp!]
  \centering
  \includegraphics[width=0.8\textwidth]{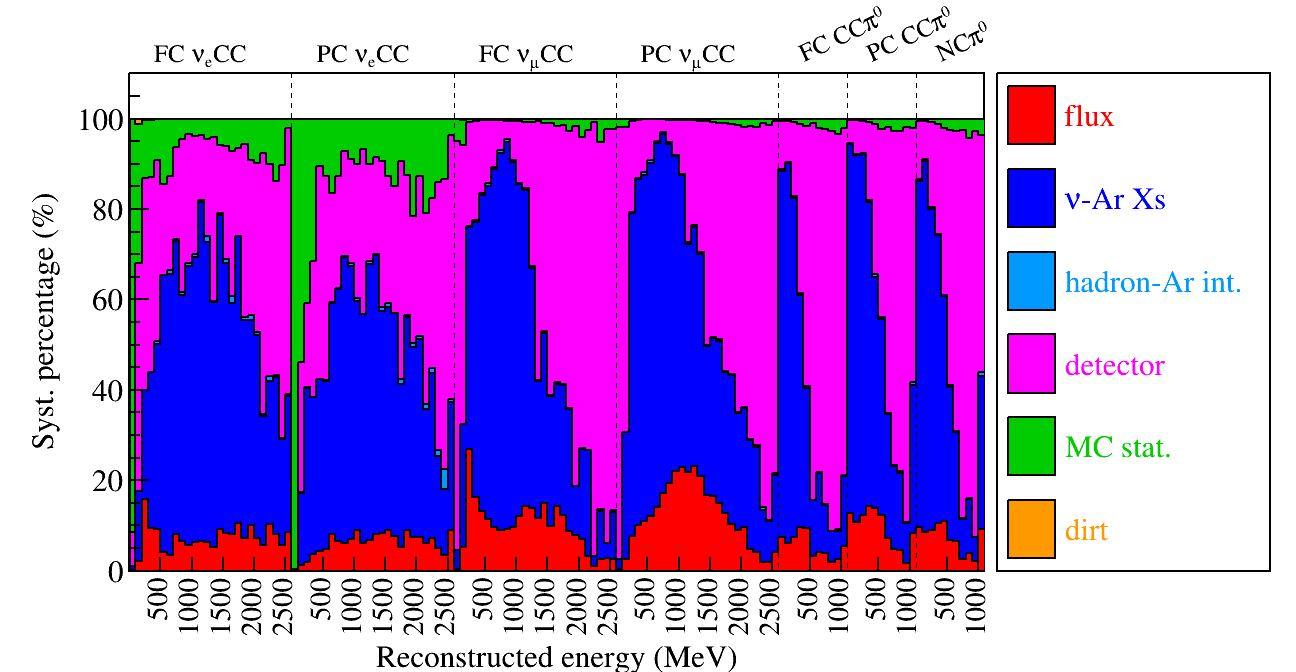}
  \put(-360, 205){MicroBooNE Simulation}
  \caption{Summary of fractional contributions to the overall systematic uncertainty ($\frac{\sigma_i^2}{\sigma^2_{\mathrm{total}}}\times100$) for the seven channels. }
  \label{fig:canv_h2_basic_fraction}
\end{figure*}

\subsection{Summary of Systematic Uncertainties}\label{sec:syst_curr}
Figure~\ref{fig:canv_h2_correlation_total} shows correlations of all sources of systematic uncertainty across the seven channels.  Figure~\ref{fig:canv_h2_basic_fraction} summarizes the fractional contribution of each source of uncertainty to the total systematic uncertainty, for each of the seven channels. In general, without any constraint, the cross section systematic uncertainty dominates in the region where neutrino interaction rate is relatively high, followed by the flux systematic uncertainty. With constraint, as discussed in later sections, cross section and flux uncertainties will be significantly reduced while the data statistical uncertainty becomes predominant in this analysis. 
For \nueCC\ channels, the relative contribution of MC statistical uncertainty, which includes statistical
uncertainty from both beam-off data and MC simulation in making predictions, is significantly higher than that of the \numuCC\ and $\pi^0$ channels mainly because of the limited statistics of the samples in estimating the backgrounds. The equivalent exposure in POT of the beam-off data sample to estimate the beam-off (EXT) background is only 2.5 times of the exposure of the BNB data sample, and the POT of the MC samples to estimate the beam-on background is only 3.7 times of the exposure of the BNB data sample. The situation will be improved with more MC production or data taking. For the \numuCC\ and $\pi^0$ channels, the detector systematic uncertainties become the dominating uncertainty at high energies where the number of events in the distribution become smaller (See. Fig.~\ref{fig:7-channel-numuCC-pi0}). This effect is expected with the bootstrapping method and is the result of the limited MC statistics of the detector systematic CV and 1$\sigma$ detector
variation samples in these regions.

\section{Model Validations}\label{sec:validation}
This section summarizes the validations of the overall model, which includes i) the prediction of $\nu_e$CC signal through the examination of data and model consistency in \numuCC\ events assuming lepton universality with the difference in 
lepton masses included in the neutrino-argon interaction cross section model, ii) constraints on the \nueCC\ backgrounds through the examination of data and model consistency in the various $\pi^0$ channels, and iii) modeling in the neutrino energy reconstruction.

\subsection{Methodology}\label{sec:validation_method}
The covariance matrix formalism is adopted to construct the $\chi^2$ test statistic:
\begin{equation}
    \chi^2 = \left(M-P\right)^T\times Cov_{\text{full}}^{-1}\left(M,P\right) \times \left(M-P\right),
    \label{eq:test-stat-chi2}
\end{equation}
where $M$ and $P$ are vectors of measurement and prediction, respectively. The $Cov\left(M,P\right)$ is the full covariance matrix:
%\begin{widetext}
\begin{dmath}\label{eq:test-stat-1}
    Cov_{\text{full}} = Cov^{\text{stat}}_{\text{CNP}} + Cov^{\text{sys}}_{\text{MC stat}} + Cov^{\text{sys}}_{\text{xs}} + Cov^{\text{sys}}_{\text{flux}} + Cov^{\text{sys}}_{\text{det}} + Cov^{\text{sys}}_{\text{add}}. 
\end{dmath}
%\end{widetext}
The $Cov^{\text{stat}}_{\text{CNP}}$ is the diagonal covariance matrix constructed based on the combined-Neyman-Pearson (CNP) method~\cite{Ji:2019yca} with the statistical uncertainty square being $3/\left(1/M_i + 2/P_i\right)$ for the $i$th bin. The $Cov^{\text{sys}}_{\text{MC stat}}$ is the diagonal covariance matrix containing the statistical uncertainties corresponding to finite statistics in making predictions as described in Sec.~\ref{sec:syst_mcstat}. The other four covariance matrices $Cov^{\text{sys}}_{\text{xs}}$, $Cov^{\text{sys}}_{\text{flux}}$, $Cov^{\text{sys}}_{\text{det}}$, $Cov^{\text{sys}}_{\text{add}}$, are the covariance matrices corresponding to uncertainties from cross section (Sec.~\ref{sec:syst_xs}), neutrino flux (Sec.~\ref{sec:syst_flux}), detector performance (Sec.~\ref{sec:syst_det}), and dirt (Sec.~\ref{sec:syst_dirt}), respectively. 
The dependence on the LEE strength $x$ is considered in calculating the covariance. It is based on dedicated systematic and statistical covariance matrices for the \LEE\ component and correlations between the \LEE\ component and the other channels.

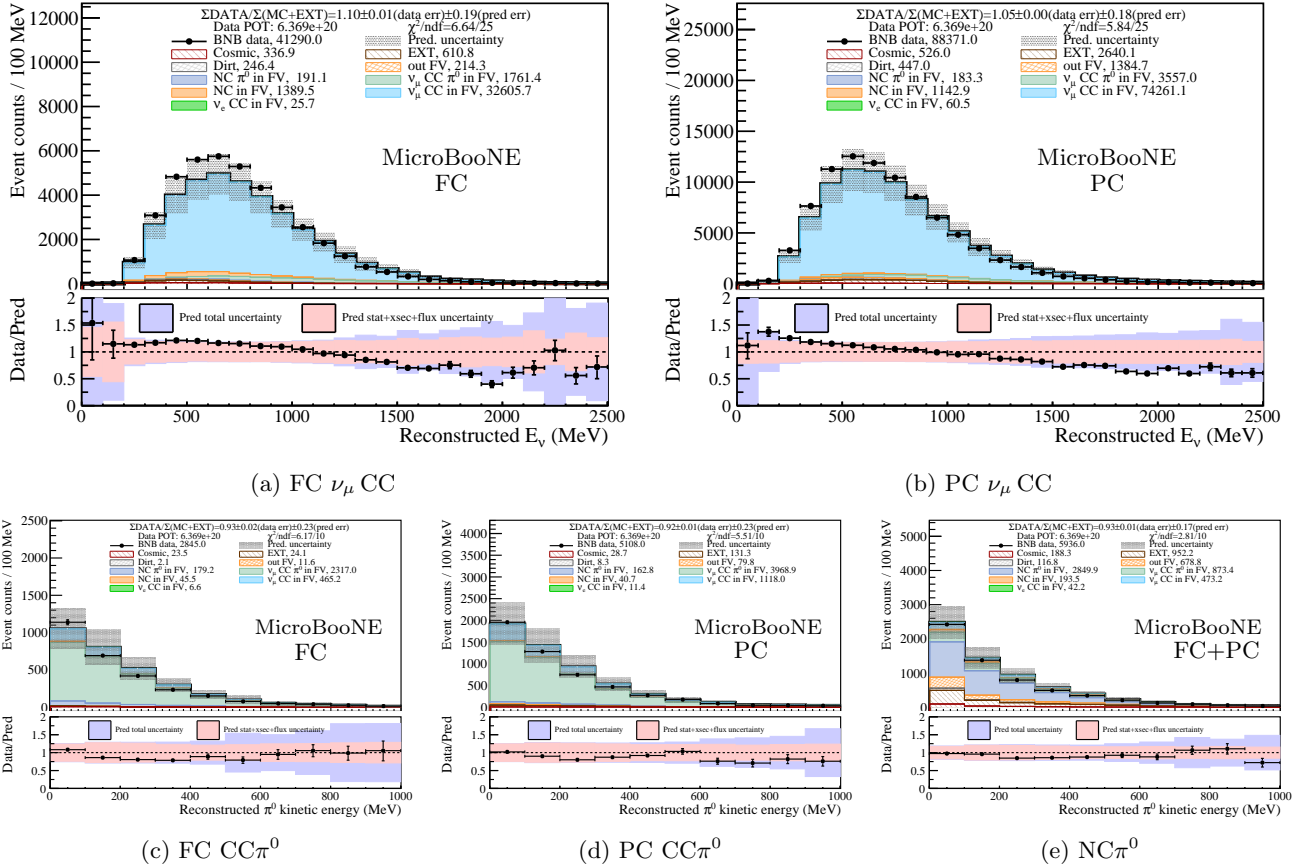
\begin{figure*}[!htp]
  \captionsetup[subfigure]{justification=centering}
  \centering
  \begin{subfigure}[]{0.48\textwidth}
    \includegraphics[width=\textwidth]{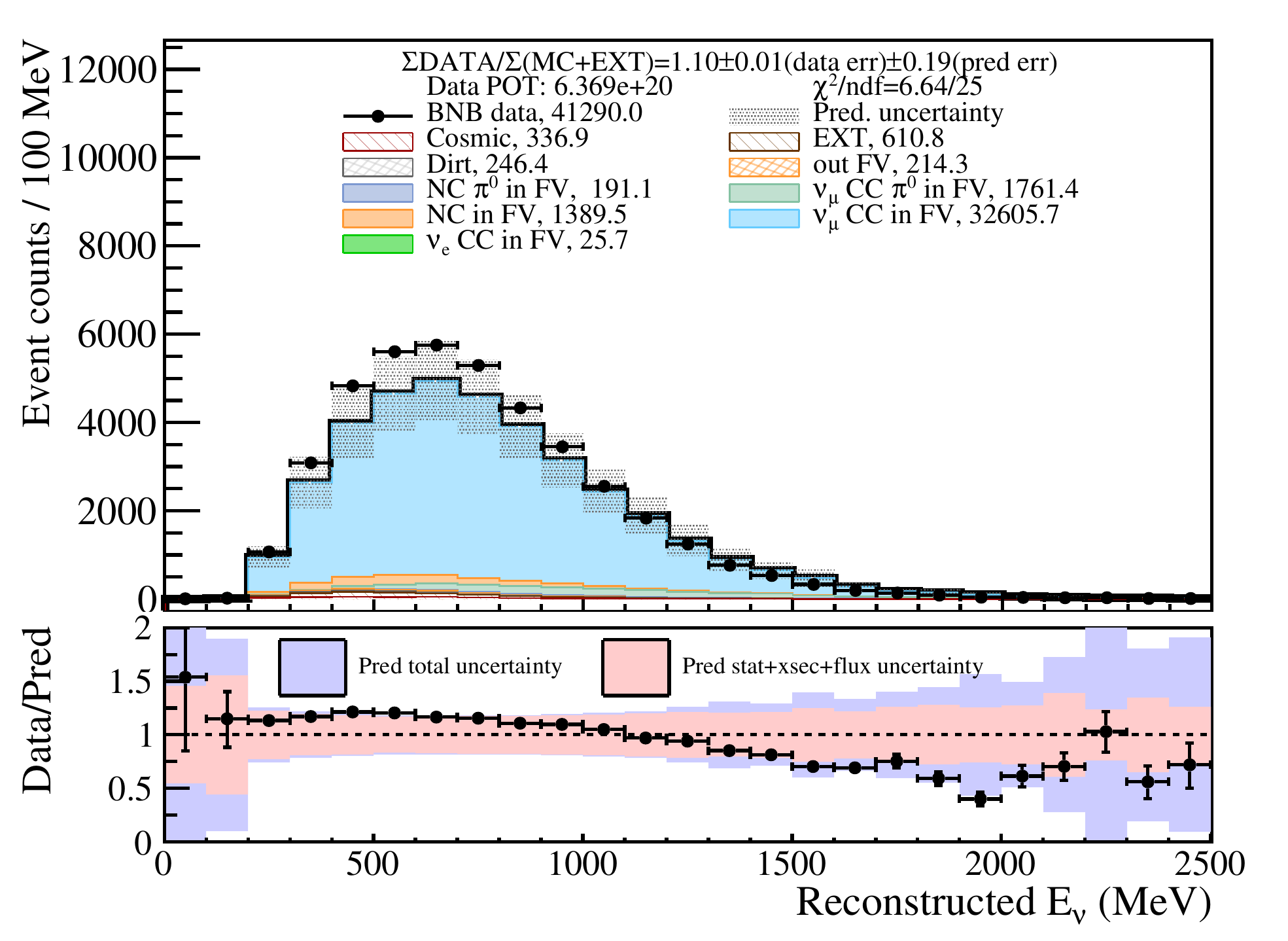}
    \put(-100,110){MicroBooNE}
    \put(-80,100){FC}
    \caption{FC \numuCC}
  \end{subfigure}
  \begin{subfigure}[]{0.48\textwidth}
    \includegraphics[width=\textwidth]{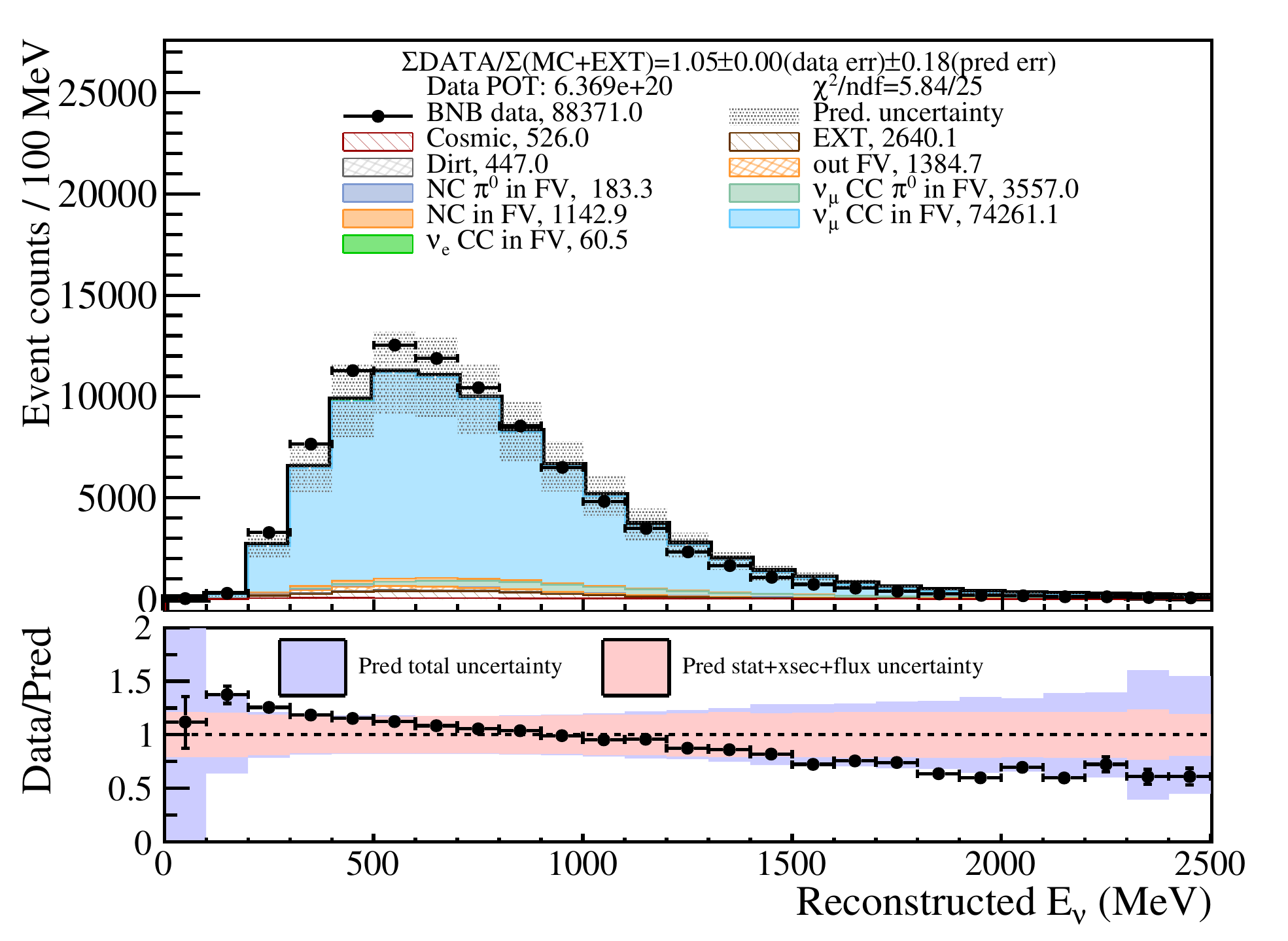}
    \put(-100,110){MicroBooNE}
    \put(-80,100){PC}
    \caption{PC \numuCC}
  \end{subfigure}
  \begin{subfigure}[]{0.32\textwidth}
    \includegraphics[width=\textwidth]{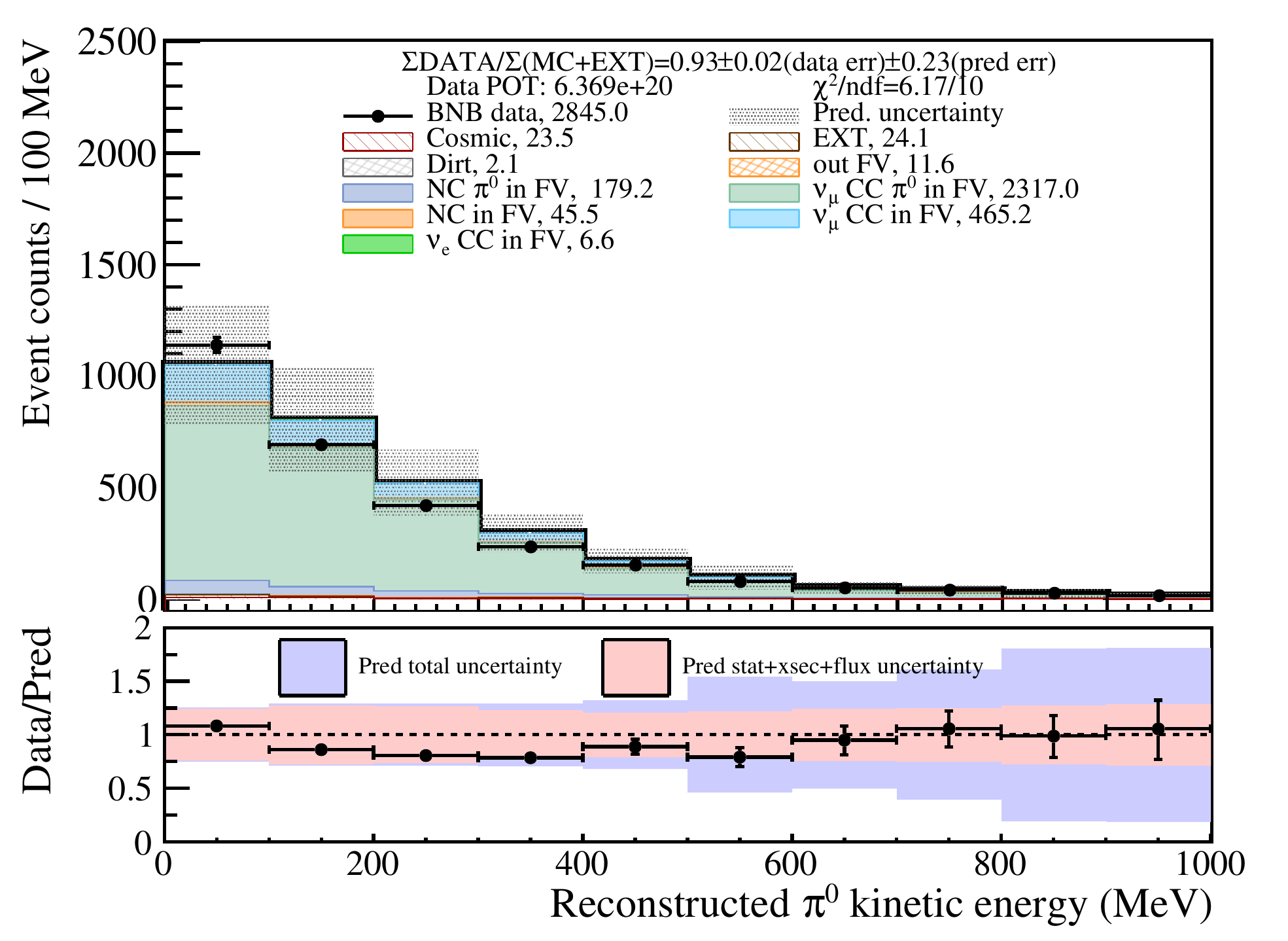}
    \put(-65,71){\footnotesize MicroBooNE}
    \put(-50,61){FC}
    \caption{FC CC$\pi^0$}
  \end{subfigure}
  \begin{subfigure}[]{0.32\textwidth}
    \includegraphics[width=\textwidth]{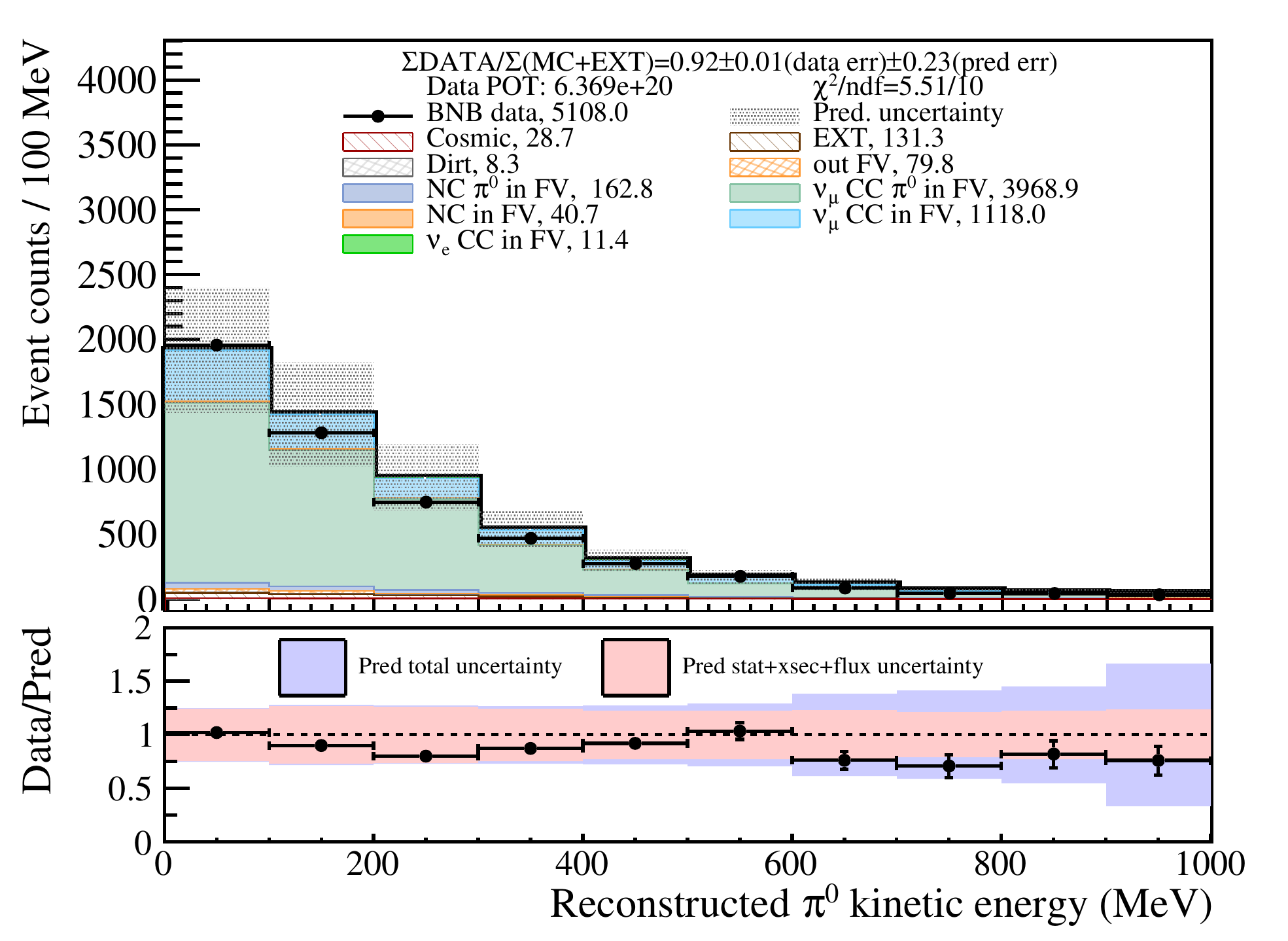}
    \put(-65,71){\footnotesize MicroBooNE}
    \put(-50,62){PC}
    \caption{PC CC$\pi^0$}
  \end{subfigure}
  \begin{subfigure}[]{0.32\textwidth}
    \includegraphics[width=\textwidth]{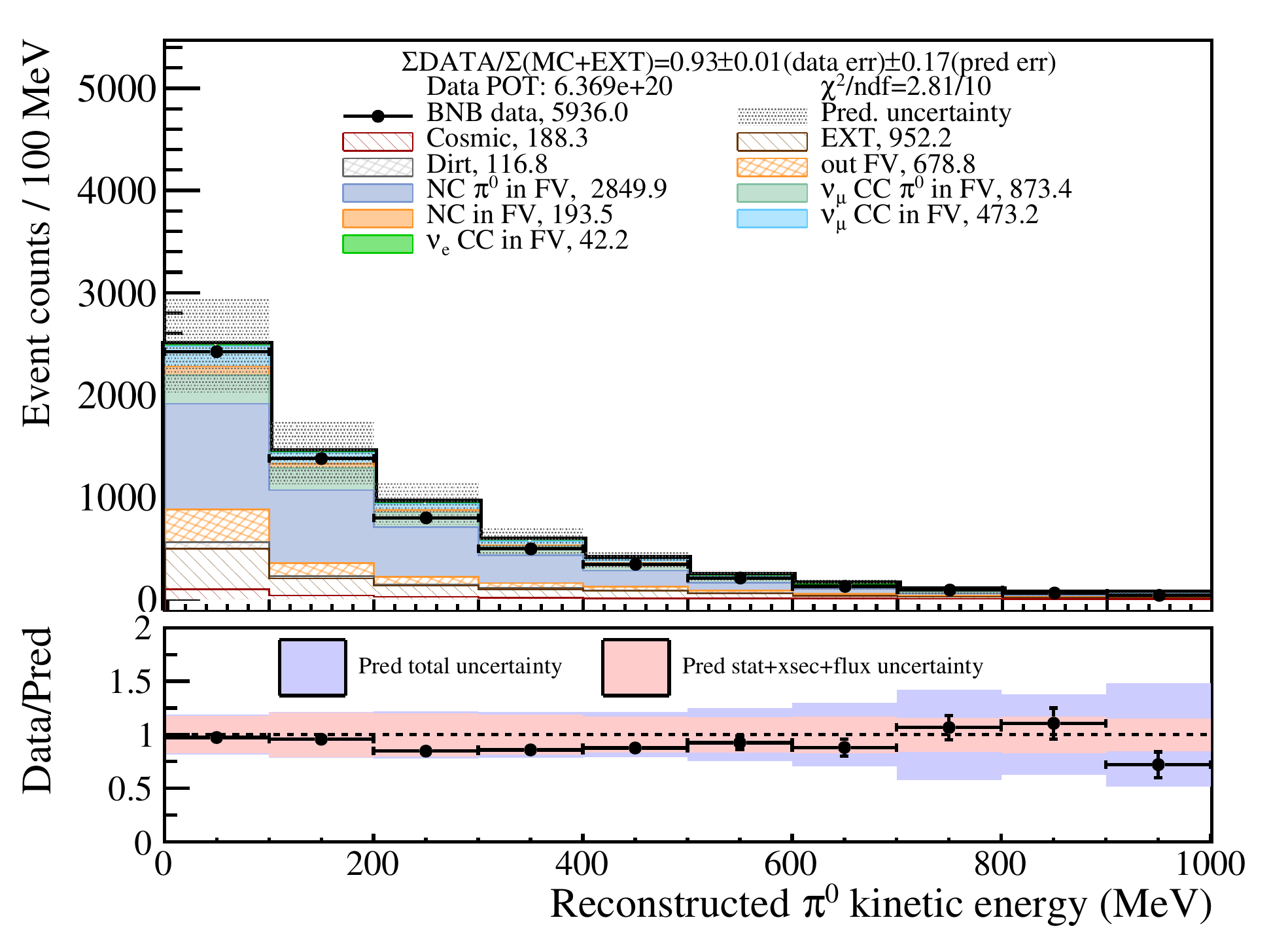}
    \put(-65,71){\footnotesize MicroBooNE}
    \put(-50,62){FC+PC}
    \caption{NC$\pi^0$}
  \end{subfigure}

\caption{Event distributions of FC \numuCC, PC \numuCC, FC CC$\pi^0$, PC CC$\pi^0$, and NC$\pi^0$ in (a)-(e), respectively. These five selections serve as constraints to reduce the systematic uncertainties in searching for eLEE, thus maximizing the physics sensitivity. The breakdown of each component for different final states for both signal and background events is shown in the legend. The bottom sub-panels present the data-to-prediction ratios as well as the statistical and systematic uncertainties. The pink band includes the statistical, cross section, and flux uncertainties. The purple band corresponds to the full uncertainty with an addition of detector systematic uncertainty.}
  \label{fig:7-channel-numuCC-pi0}
\end{figure*}

%gof test
The goodness-of-fit (GoF) test can be performed to test the compatibility between the data and the overall model. Following the recommendation of Ref.~\cite{comp_teststat}, we adopt the Pearson $\chi^2$ construction (instead of the CNP construction) for the
data statistical uncertainty:
\begin{dmath}
    Cov_{\text{GoF}} = Cov^{\text{stat}}_{\text{Pearson}} + Cov^{\text{sys}}_{\text{MC stat}} + Cov^{\text{sys}}_{\text{xs}} + Cov^{\text{sys}}_{\text{flux}} + Cov^{\text{sys}}_{\text{det}} + Cov^{\text{sys}}_{\text{add}},
\end{dmath}
with the statistical uncertainty square being $P_i$ for the $i$th bin. Given the null hypothesis (i.e. eLEEx=0), the $\chi^2$ value can be used to perform the GoF test by comparing to the $\chi^2$ distribution with the associated number of degrees of freedom ($ndf$), which is the total number of bins of the measurement.

\begin{table*}[htp!]
    \centering
    \begin{tabular}{l|c|c|l}
        \hline
        \hline
        Channel & $\chi^2/ndf$ w/o constr. & $\chi^2/ndf$ w/ constr. & Notes \\
        \hline
        FC $\nu_\mu$CC  & 6.64/25 & N/A & No constraint, see other checks \\
        \hline
        PC $\nu_\mu$CC  & 5.84/25 & 6.94/25 & Constrained by FC $\nu_\mu$CC \\
        \hline
        FC CC$\pi^0$     & 6.17/10 & 7.39/10 & \multirow{3}{20em}{Constrained by both FC and PC $\nu_\mu$CC}  \\
        PC CC$\pi^0$     & 5.51/10 & 6.80/10 &   \\
        NC$\pi^0$        & 2.81/10 & 5.33/10 &  \\
        \hline
        PC $\nu_e$CC  & 24.93/25 & 24.19/25 & See Sec.~\ref{sec:results_nueCC}; constrained by the above five channels; \SM~hypothesis \\
        \hline
        FC $\nu_e$CC & 12.55/25 & 17.86/25 & See Sec.~\ref{sec:results_nueCC}; constrained by the other six channels; \SM~hypothesis \\
        \hline
        \hline
    \end{tabular}
\caption[]{$\chi^2/ndf$ without and with constraints, for various selection channels. The central values and uncertainties (covariance) of the MC prediction are both changed with the constraint. Pearson construction of $\chi^2$, instead of CNP construction of $\chi^2$, is used for these goodness-of-fit tests. }
\label{tab:GoFvalues_constraint}
\end{table*}

\begin{figure*}[htp!]
\captionsetup[subfigure]{justification=centering}
  \centering
  \begin{subfigure}[]{0.92\columnwidth}
    \includegraphics[width=\textwidth]{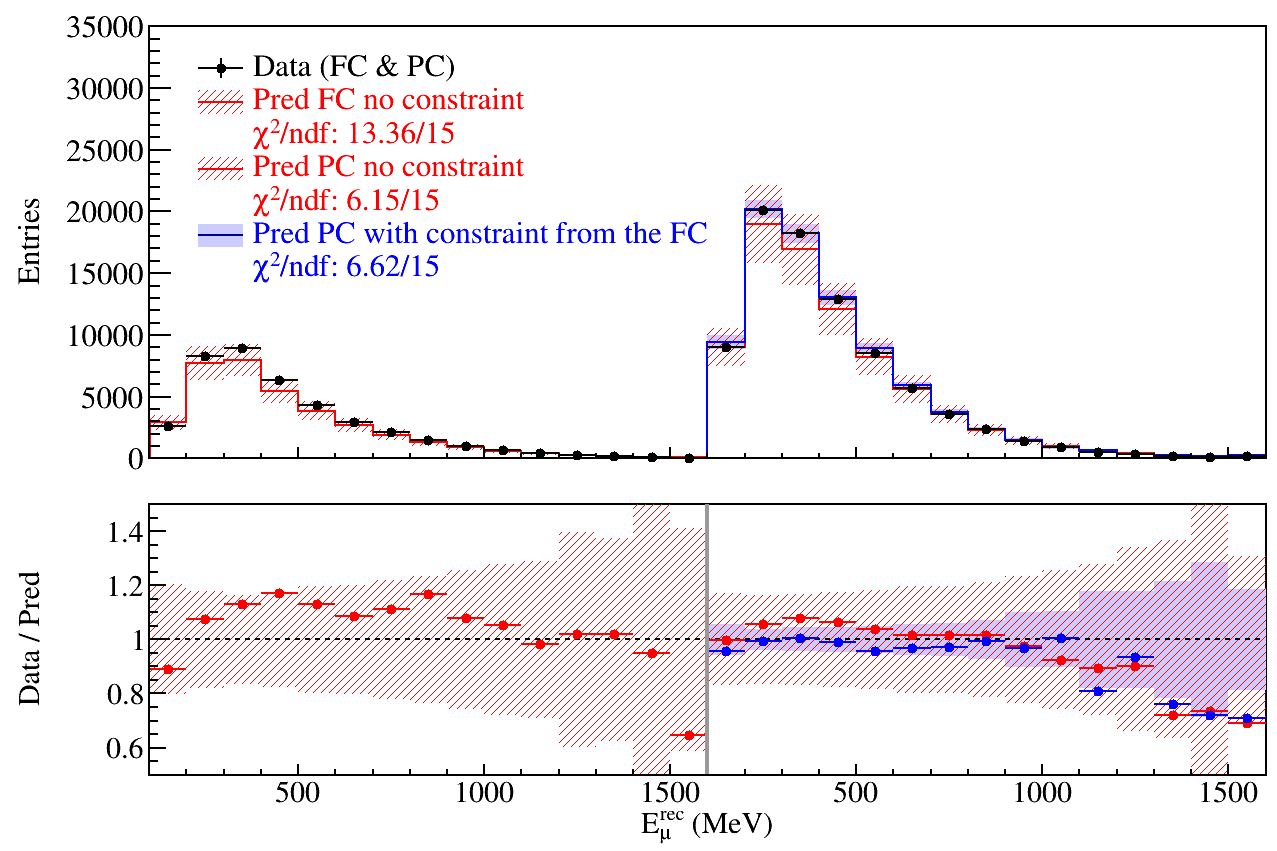}
    \put(-128, 138){\footnotesize{MicroBooNE 6.369$\times$10$^{20}$ POT}}
    \put(-150,90){FC}
    \put(-50, 90){PC}
    \caption{\numuCC\ primary muon energy.}
    \label{fig:constraint_Emuon}
  \end{subfigure}
  \begin{subfigure}[]{0.92\columnwidth}
    \includegraphics[width=\textwidth]{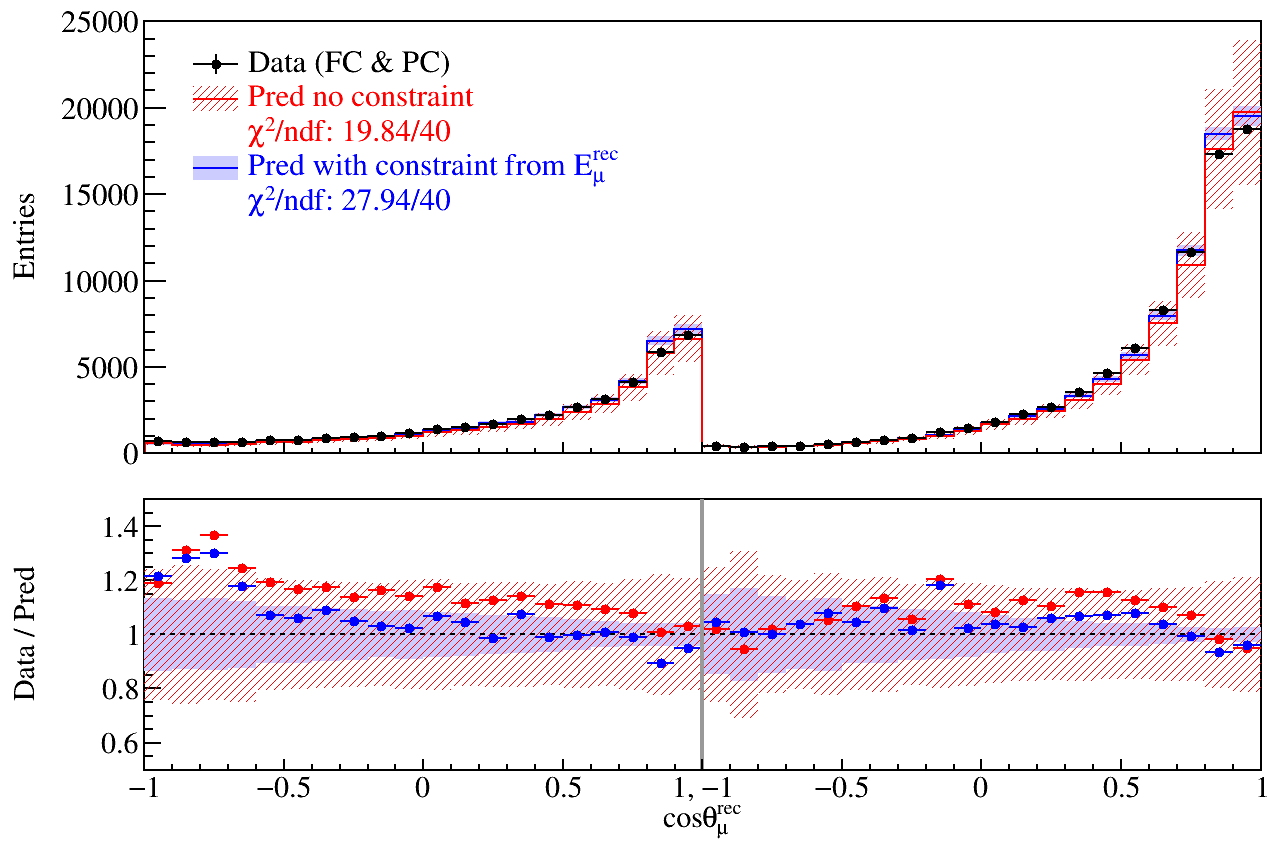}
    \put(-128, 136){\footnotesize{MicroBooNE 6.369$\times$10$^{20}$ POT}}
    \put(-150,90){FC}
    \put(-50, 90){PC}
    \caption{\numuCC\ primary muon polar angle.}
    \label{fig:constraint_costheta}
  \end{subfigure}
  \begin{subfigure}[]{0.92\columnwidth}
    \includegraphics[width=\textwidth]{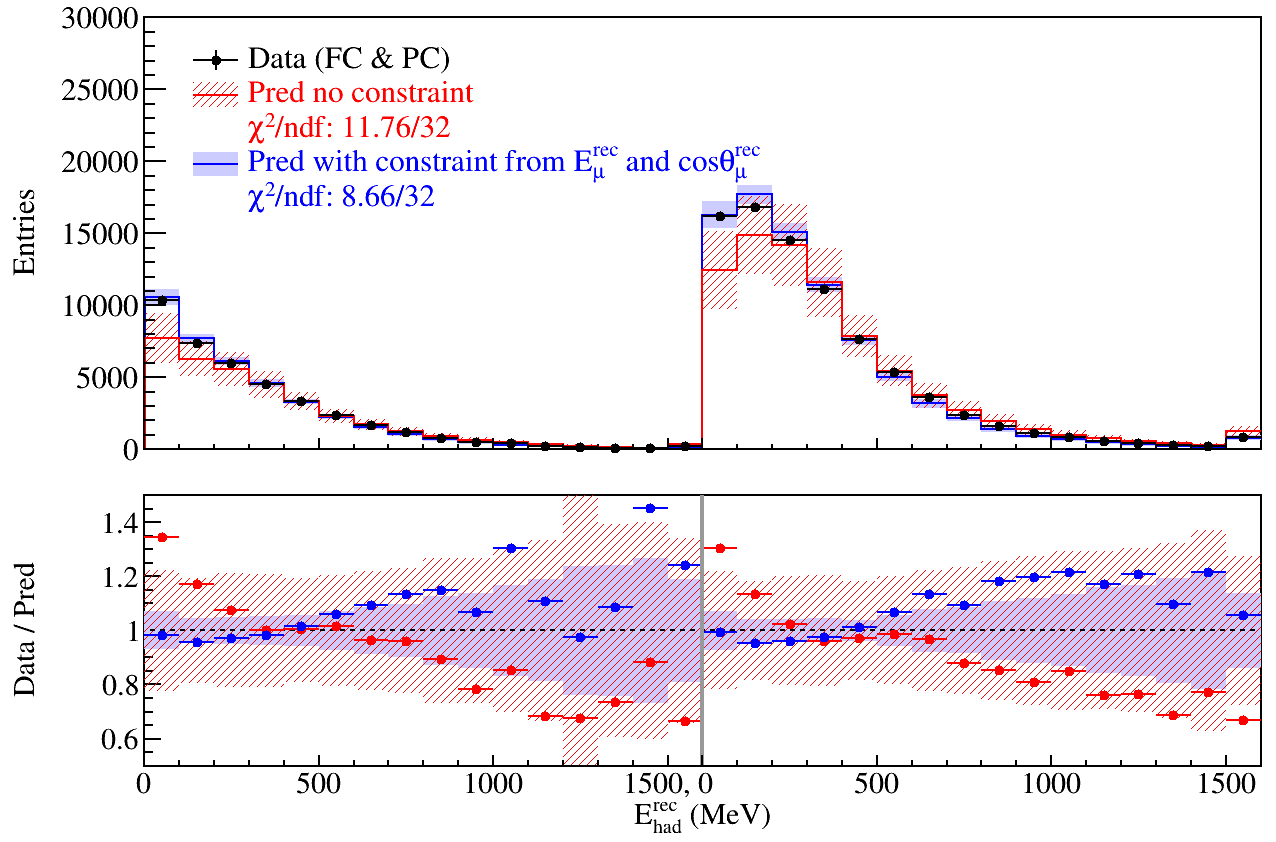}
    \put(-128, 138){\footnotesize{MicroBooNE 6.369$\times$10$^{20}$ POT}}
    \put(-150,90){FC}
    \put(-50, 90){PC}
    \caption{\numuCC\ hadronic energy ($E_{\rm had}=E_{\nu}$-$E_{\mu}$).}
    \label{fig:constraint_Ehad}
  \end{subfigure}
  \caption{Event distributions of reconstructed muon energy (a), muon angle (b), hadronic energy (c), for selected $\nu_\mu$CC events with fully contained (FC) events on the left half and partially contained (PC) events on the right half in each sub-figure. Black points are from data measurements. Red (blue) histograms and error bands are for the MC prediction before (after) constraint. The last bin is the overflow bin for FC or PC muon energy or hadronic energy distributions.}
  \label{fig:validation_constraint}
\end{figure*}

% conditional gof test
The above GoF test using a Pearson $\chi^2$ construction provides an overall evaluation of the model and the null hypothesis compared to the data. This evaluation can be used to study different parts of the model using the conditional covariance matrix formalism~\cite{cond_cov}. For example, given the full covariance (stat + sys) containing two channels (X, Y):
\begin{equation}
    \Sigma = \begin{pmatrix}
\Sigma^{XX} & \Sigma^{XY} \\
\Sigma^{YX} & \Sigma^{YY} 
\end{pmatrix},~~~ n: {\rm measurement}, ~~~ \mu:~{\rm prediction},
\end{equation}
we can derive the conditional mean and conditional covariance of the prediction of $X$ given the constraints from the measurement of $Y$:
\begin{eqnarray}
\mu^{X,\text{const.}} &=& \mu^{X} + \Sigma^{XY} \cdot \left(\Sigma^{YY} \right)^{-1} \cdot \left( n^Y - \mu^Y \right) \\
\Sigma^{XX, \text{const.}} &=& \Sigma^{XX} - \Sigma^{XY} \cdot \left(\Sigma^{YY} \right)^{-1} \cdot \Sigma^{YX}.
\end{eqnarray}
Thus, a goodness-of-fit test can be performed on $Y$ first, and then performed on $X$ after the constraints from $Y$. This allows the examination of the model on $X$ and $Y$ individually, which provides more information about the compatibility between the model and data. 

The current cross-section model we use has conservative uncertainties in general, and along with correlated systematics, the 
reduced $\chi^2$ values, which is the ratio between $\chi^2$ and number of degrees of freedom, are generally low suggesting that 
the model describes the data well within the given uncertainty.

\subsection{Validation of Monte Carlo Models with Goodness-of-fit Tests}\label{sec:model_validation}
As previously introduced, we use a seven-channel fit strategy in searching for a low-energy excess. The seven channels are FC \nueCC, PC \nueCC,  FC \numuCC, PC \numuCC, FC CC$\pi^0$, PC CC$\pi^0$, and NC$\pi^0$. The selections for these seven channels are exclusive from each other. In particular, the candidates in the CC$\pi^0$  excludes the selected \nueCC\ candidates, and the candidates in the \numuCC\ channel excludes \nueCC\ and CC$\pi^0$ candidates. Figure~\ref{fig:7-channel-numuCC-pi0} shows the five non-\nueCC\ channels in the eLEE search that serve as constraints to reduce the systematic effects in the measurement of \nueCC. The $\chi^2/ndf$ value in each plot is calculated based on the data-MC difference and the full covariance matrix taking into account the bin-to-bin correlations from the off-diagonal terms. The error bands in the plot correspond to only diagonal terms in the covariance matrix.

Figure~\ref{fig:7-channel-numuCC-pi0} shows that the data and MC simulation agree within uncertainties, and the $\chi^2/ndf$ values without any constraint suggest consistency albeit with conservative uncertainties. To further validate the MC models, i.e. data-MC consistency, GoF tests using the conditional mean prediction and conditional covariance of the prediction, as described in Sec.~\ref{sec:validation_method}, are performed on these channels before applying them in the final eLEE search. 
After constraints, the $\chi^2/ndf$ values based on both of the reduced systematic uncertainty and adjusted central values of the MC prediction are summarized in Table~\ref{tab:GoFvalues_constraint}. The $\chi^2/ndf$ values after constraints are generally increased to a certain degree mostly because of the reduced systematic uncertainty. For the PC \numuCC, the uncertainty is reduced by about 75\% with constraints from the FC $\nu_\mu$CC. For the three $\pi^0$ channels, the uncertainties are reduced by about 30\% with constraints from the FC and PC \numuCC\ channels. All the $\chi^2/ndf$ values after constraints are still less than 1 indicating consistency between data and MC within the estimated systematic uncertainties of the MC prediction.

\subsection{Validation of energy reconstruction}
\label{sec:energy_validation}
One may note that the data-to-prediction ratios demonstrate a slope for $\nu_\mu$CC channels, which motivates further energy reconstruction validation, particularly on inclusive hadronic final states, in addition to the validations on $dE/dx$ and EM shower energy scale using $dE/dx$ profiles from various reconstructed particles and $\pi^0$ invariant mass. 

As discussed in Appendix, this ``slope'' can be decomposed using different final state topologies, such as via 0$p$X$\pi$/N$p$X$\pi$ (N$\ge$1 and X$\ge$0) separation, where $p$ represents reconstructed protons with kinetic energy greater than 35~MeV, and there is no requirement regarding whether charged pions are reconstructed. 
The enhancement in the data measurement is found to be concentrated in the 0$p$X$\pi$ channels which essentially correspond to the events with low hadronic energy while N$p$X$\pi$ channels have an overall good agreement between data and MC.
To quantitatively assess this issue, a conditional constraint study was performed with the high-statistics \numuCC\ events in this analysis following the methodology described in Sec.~\ref{sec:validation_method}.
The more precise measurements of leptonic kinematics, such as muon energy ($E_{\mu}$) and muon polar angle relative to the BNB direction (cos$\theta_{\mu}$), are utilized to constrain/adjust the MC prediction on the hadronic kinematics, such as the reconstructed hadronic energy (reconstructed $E_{\nu}$-$E_{\mu}$), which is expected to contain significant missing energy from invisible neutral particles. 

The measurements of muon kinematics and hadronic energy constraint results are shown in Fig.~\ref{fig:validation_constraint}.
The measurements of $E_{\mu}$ and cos$\theta_{\mu}$ show good data-MC consistency for the MC predictions with or without certain constraints. The measurement of hadronic energy shows an excess for low hadronic energy events. With constraints from the measurements of $E_{\mu}$ and cos$\theta_{\mu}$, the measurement of hadronic energy, as in Fig.~\ref{fig:constraint_Ehad}, shows a good agreement ($\chi^2/ndf$=8.66/32) with the constrained MC prediction even though the common systematic effect between leptonic and hadronic systems (e.g. neutrino flux uncertainties) is significantly reduced resulting in a much smaller residual systematic uncertainty (blue error band). This is an important evidence that the origin of this issue is essentially from the common systematics, most likely cross-section modeling, shared by both leptonic and hadronic final states and can be adequately described/allowed by the current models and their variations in MicroBooNE MC simulation. More discussion can be found in Ref.~\cite{WC_numuXS}.
This constraint study is sensitive to the potential missing energy in the hadronic system. For example, 5-10\% (relative to the true energy transfer) additional invisible energy would result in a noticeable increase of the $\chi^2/ndf$ values in this test. See the supplementary material of Ref.~\cite{WC_numuXS} for more details. This stringent test validates the MC model as well as our neutrino energy reconstruction treatment (leptonic and hadronic energy) and gives us confidence using the overall MC model in the seven-channel fit.

\begin{figure*}[htp!]
\captionsetup[subfigure]{justification=centering}
  \centering
  \begin{subfigure}[]{0.48\textwidth}
    \includegraphics[width=\textwidth]{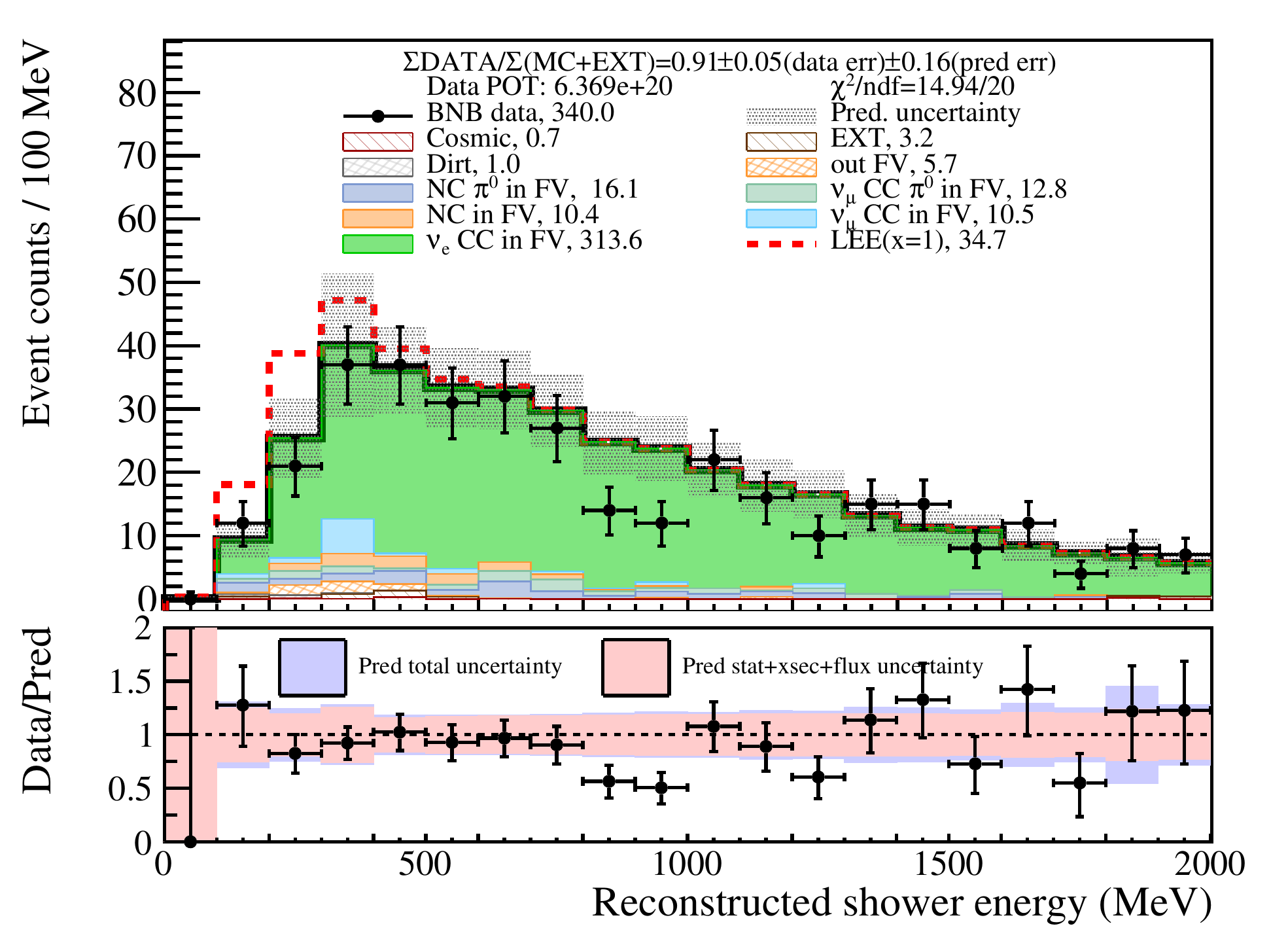}
    \put(-95,110){MicroBooNE}
    \put(-95,100){FC, unconstrained}
    \caption{\nueCC\ shower energy, fully contained events.}
    \label{fig:nueCC_FC_showerenergy}
  \end{subfigure}
  \begin{subfigure}[]{0.48\textwidth}
    \includegraphics[width=\textwidth]{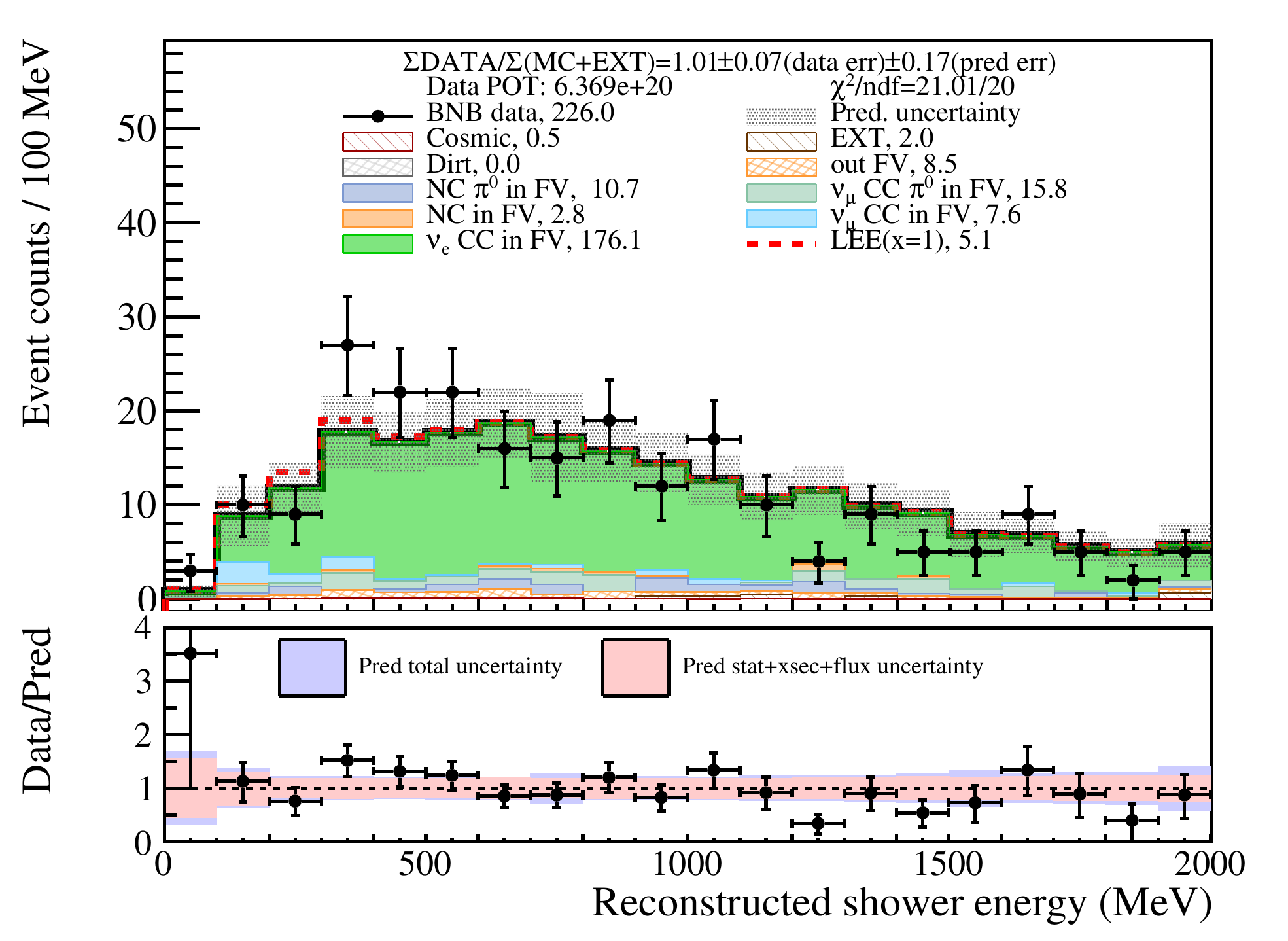}
    \put(-95,110){MicroBooNE}
    \put(-95,100){PC, unconstrained}
    \caption{\nueCC\ shower energy, partially contained events.}
    \label{fig:nueCC_PC_showerenergy}
  \end{subfigure}
  \begin{subfigure}[]{0.48\textwidth}
    \includegraphics[width=\textwidth]{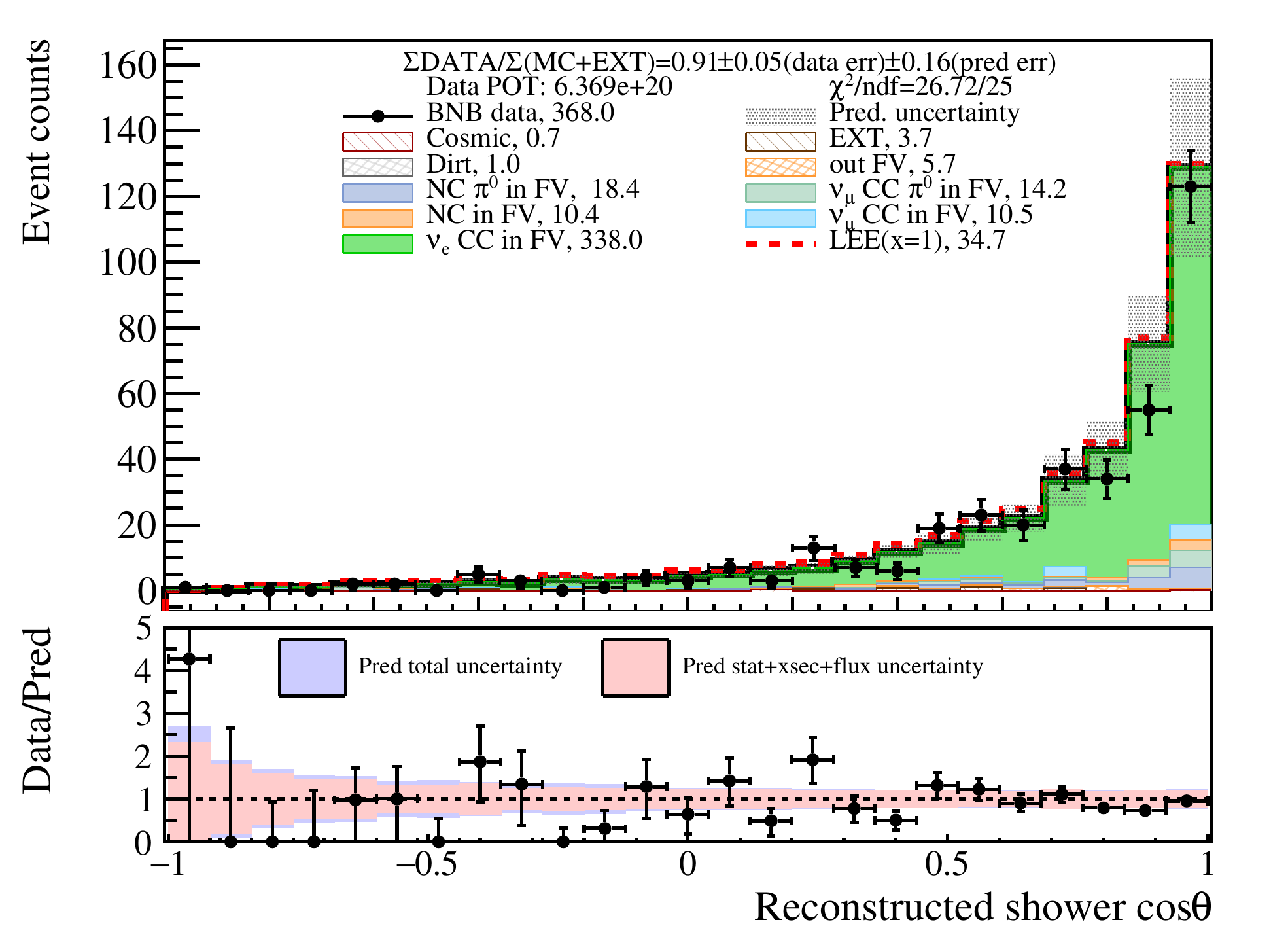}
    \put(-195,110){MicroBooNE}
    \put(-195,100){FC, unconstrained}
    \caption{\nueCC\ shower cos$\theta$, fully contained events.}
    \label{fig:nueCC_FC_costheta}
  \end{subfigure}
  \begin{subfigure}[]{0.48\textwidth}
    \includegraphics[width=\textwidth]{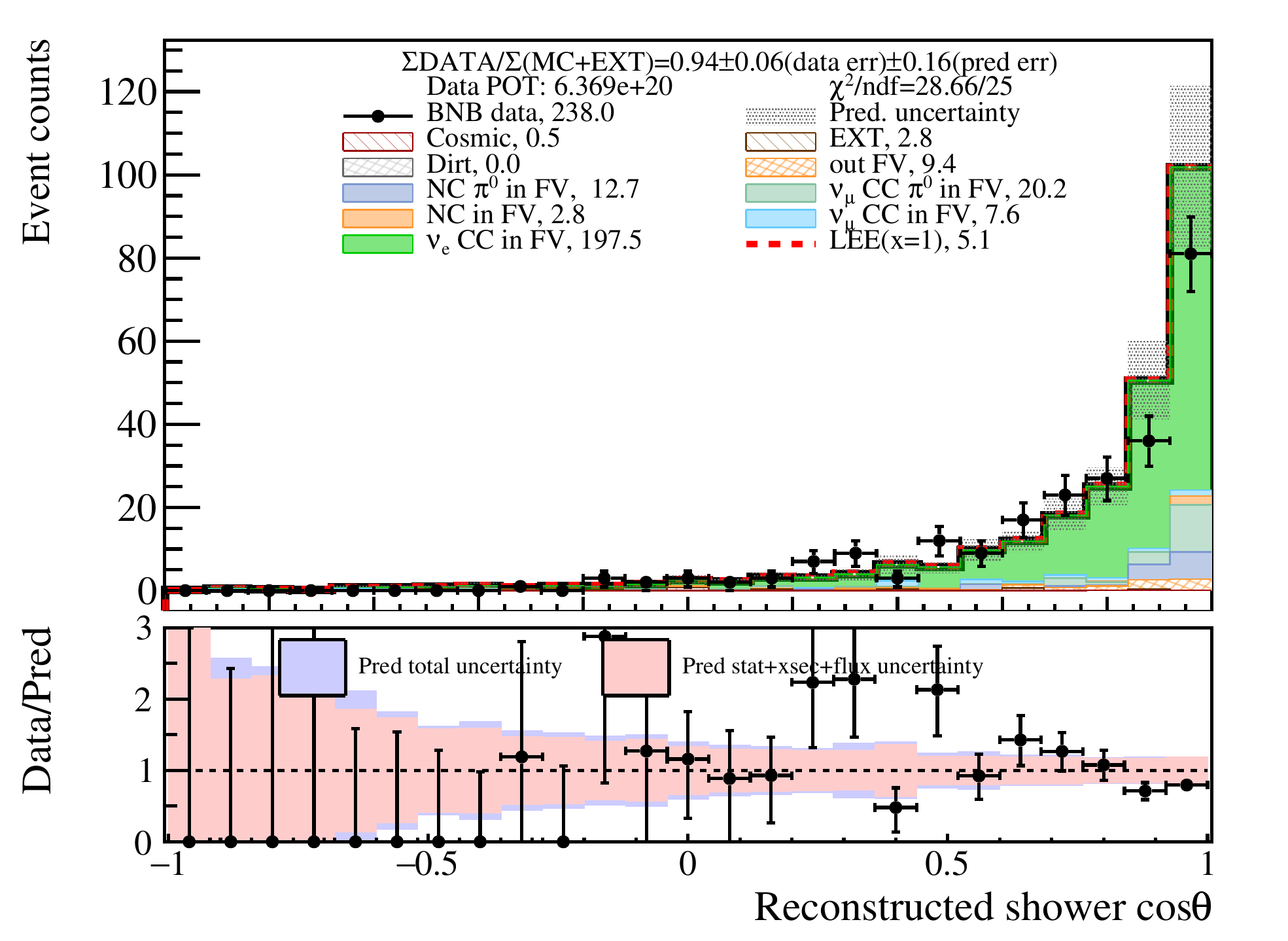}
    \put(-195,110){MicroBooNE}
    \put(-195,100){PC, unconstrained}
    \caption{\nueCC\ shower cos$\theta$, partially contained events.}
    \label{fig:nueCC_PC_costheta}
  \end{subfigure}
  \begin{subfigure}[]{0.48\textwidth}
    \includegraphics[width=\textwidth]{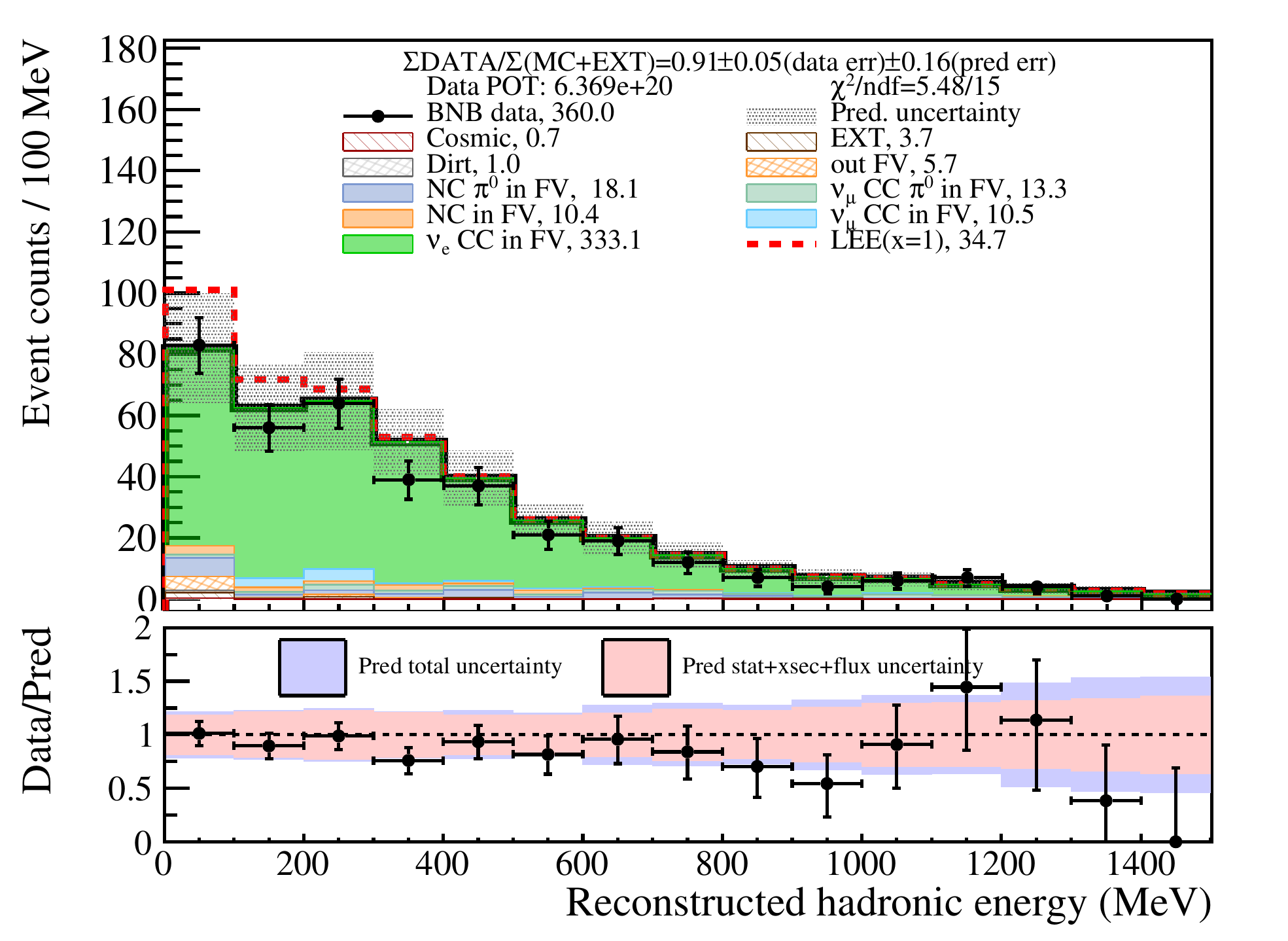}
    \put(-95,110){MicroBooNE}
    \put(-95,100){FC, unconstrained}
    \caption{\nueCC\ hadronic energy, fully contained events.}
    \label{fig:nueCC_FC_Ehadron}
  \end{subfigure}
  \begin{subfigure}[]{0.48\textwidth}
    \includegraphics[width=\textwidth]{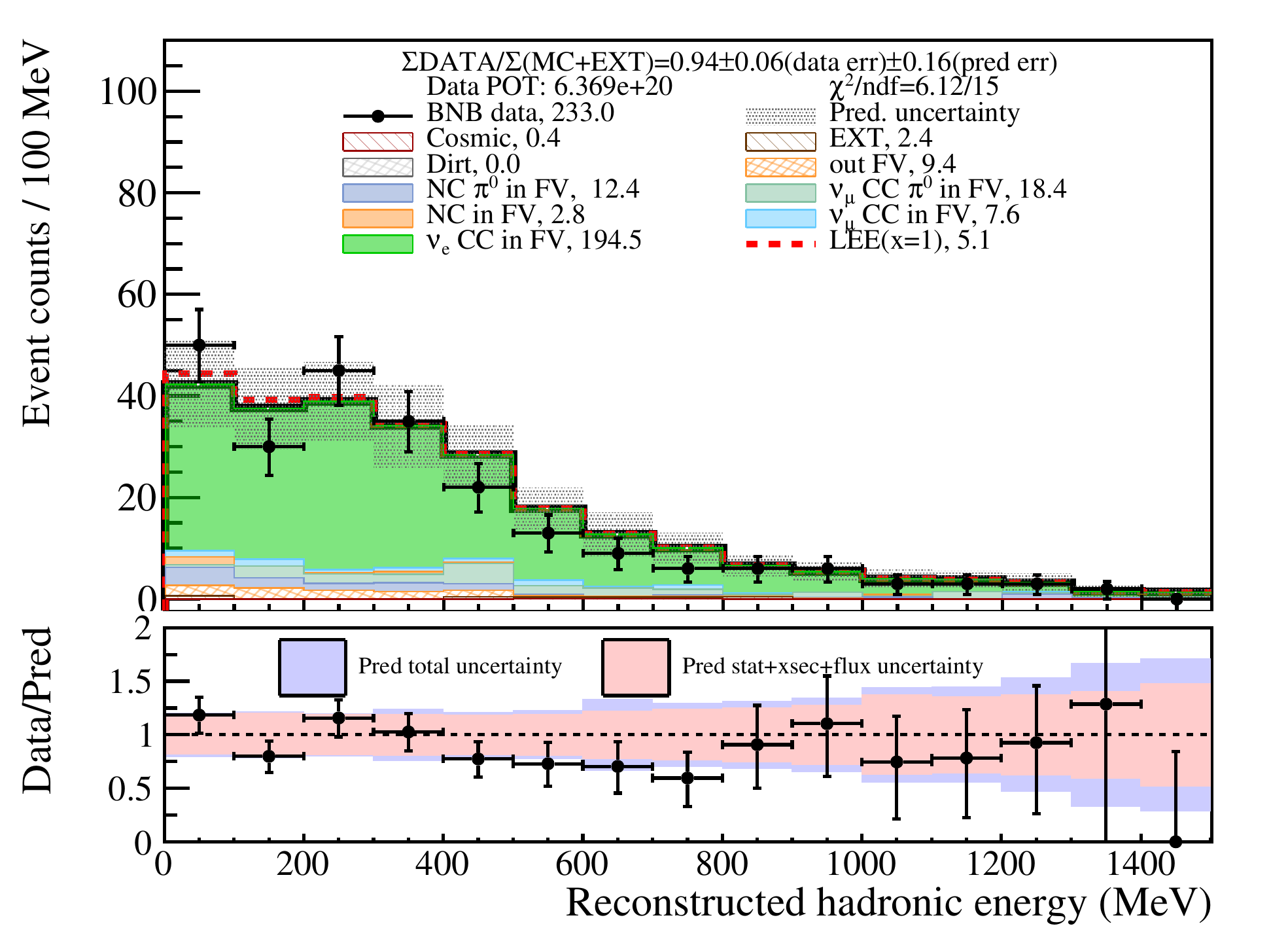}
    \put(-95,110){MicroBooNE}
    \put(-95,100){PC, unconstrained}
    \caption{\nueCC\ hadronic energy, partially contained events.}
    \label{fig:nueCC_PC_Ehadron}
  \end{subfigure}
\caption{Event distributions of FC \nueCC\ [(a), (c), and (e)] and PC \nueCC\ [(b), (d), and (f)] candidates as a function reconstructed shower energy [(a) and (b)], shower cos $\theta$ [(c) and (d)], and hadronic energy [(e) and (f)]. No constraint is used. Data-to-prediction ratios, $\chi^2$, and error bands are calculated based on the \SM~hypothesis. The MC expectation of the \LEE~component is added on top of the energy spectrum as represented by the dashed red curve.  The breakdown of each component for different final states for both signal and background events is shown in the legend. The bottom sub-panels present the data-to-prediction ratios as well as the statistical and systematic uncertainties. The pink band includes the statistical, cross section, and flux uncertainties. The purple band corresponds to the full uncertainty with an addition of detector systematic uncertainty.}
\label{fig:nueCC_variables}
\end{figure*}
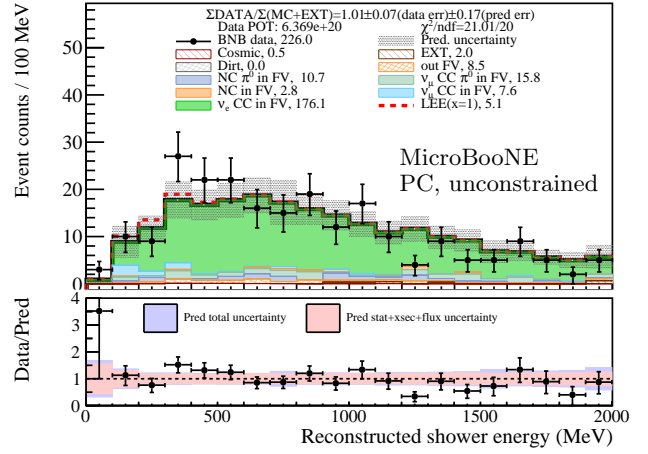
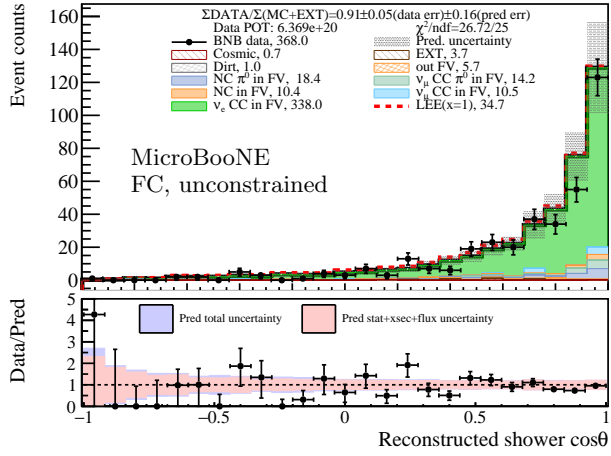
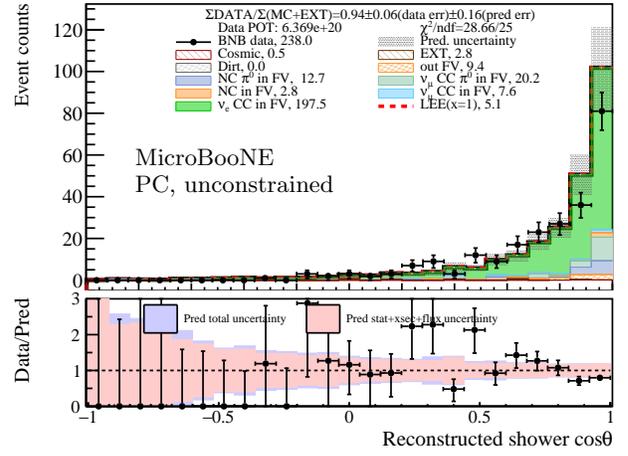
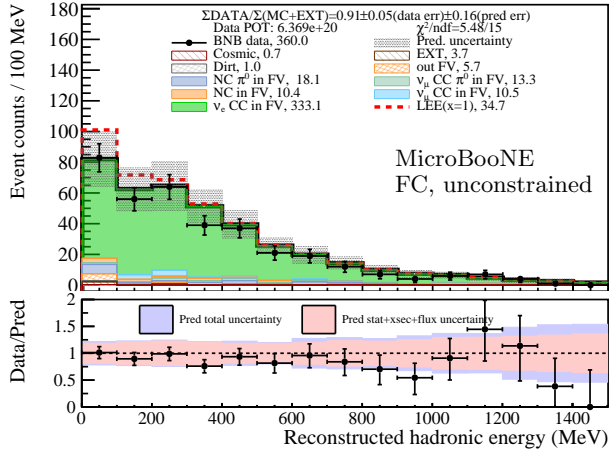
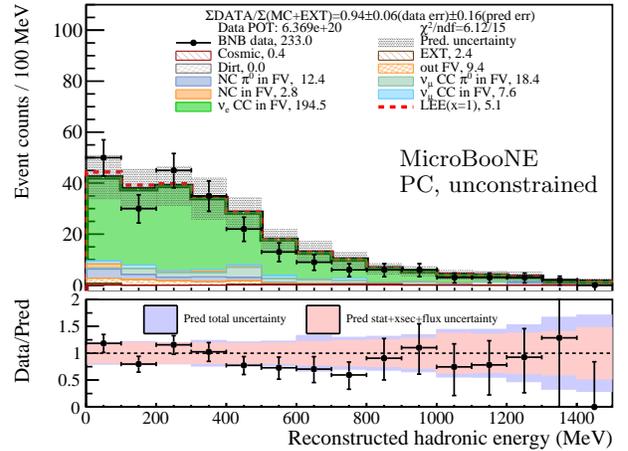

\section{Results}~\label{sec:results}
In this section, 
eLEE sensitivity, nue CC selected data and MC comparisons (Sec.~\ref{sec:results_nueCC}), and various statistical analyses
(Sec.~\ref{sec:simple_likelihood} and Sec.~\ref{sec:nested_likelihood}) 
including the best-fit of the eLEE strength and Feldman-Cousins confidence intervals are reported. 
The eLEE model used here is generated by unfolding the MiniBooNE's observed excess to true neutrino energy under a CCQE hypothesis and scaling the flux of intrinsic \nue\ with the excess-to-intrinsic \nue\ ratios.

As explained in Sec.~\ref{sec:overview}, a blind analysis strategy was implemented for the MicroBooNE LEE analysis which sequesters $\nu_e$CC candidates in BNB data below a reconstructed neutrino energy of 600~MeV until the analysis had been finalized. This procedure ensured that there was no bias in the reconstruction or event selection while allowing for cross-checks of data selection and simulation performance using sideband events with similar kinematic and/or topological characteristics to those of \nueCC\ signal. 

\subsection{\nueCC\ Selection Results}\label{sec:results_nueCC}
\begin{figure*}[!htp]
\captionsetup[subfigure]{justification=centering}
  \centering
  \begin{subfigure}[]{0.95\columnwidth}
    \includegraphics[width=\textwidth]{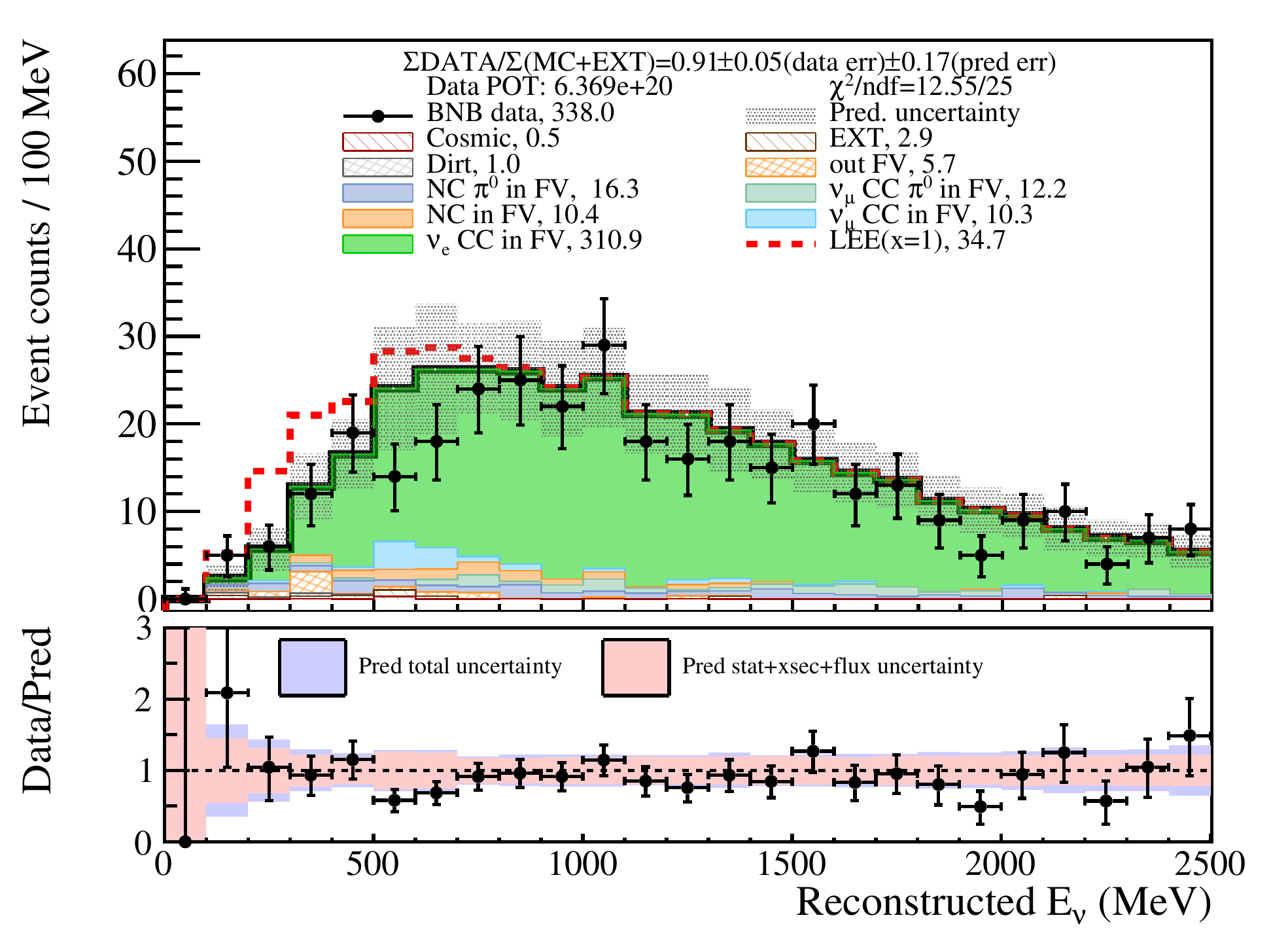}
    \put(-92,113){MicroBooNE}
    \put(-92,103){FC, unconstrained}
    \caption{\nueCC\ fully contained events.}
    \label{fig:7-channel-nueCC-FC}
 \end{subfigure}
 \begin{subfigure}[]{0.95\columnwidth}
    \includegraphics[width=\textwidth]{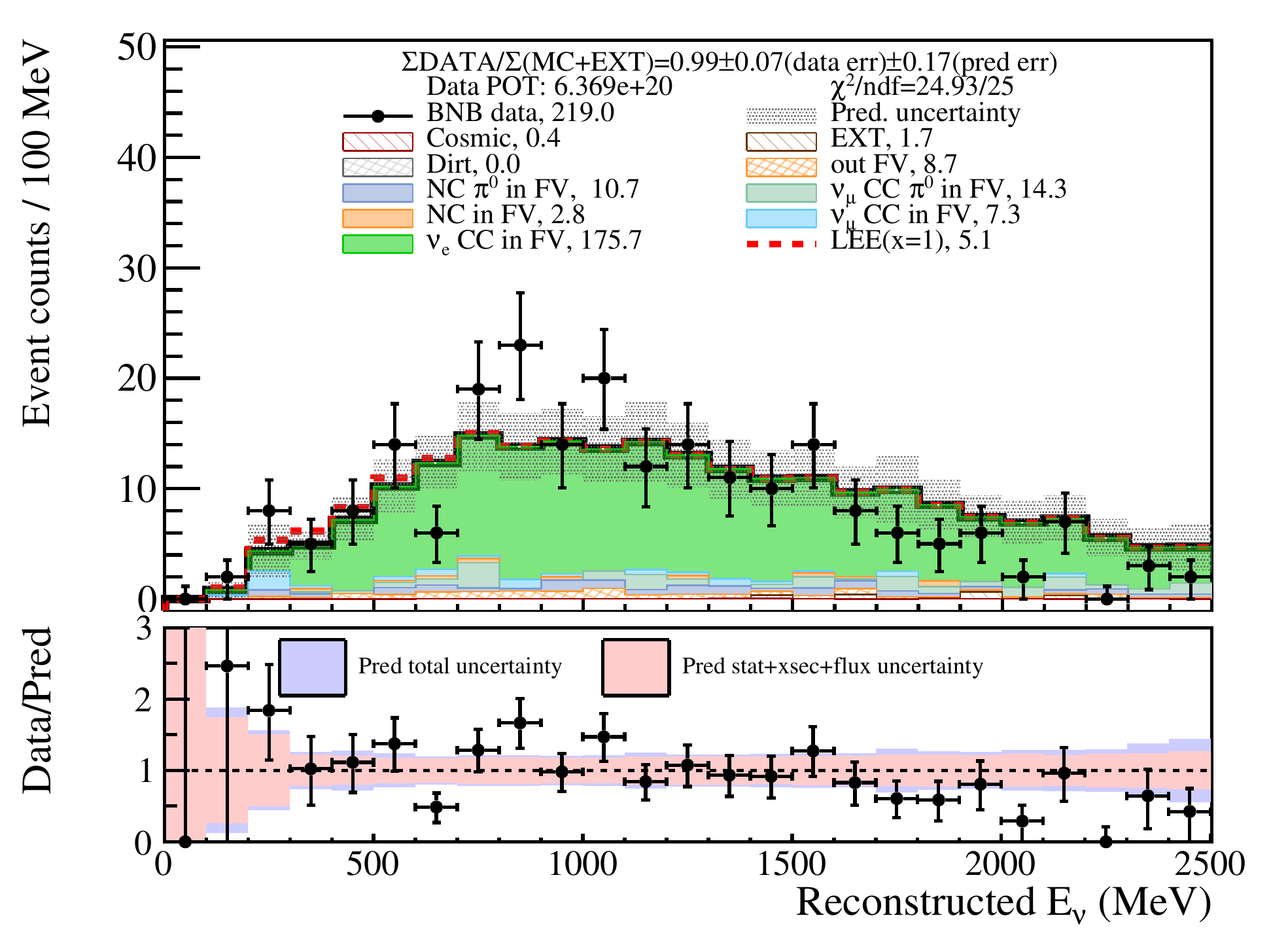}
    \put(-92,113){MicroBooNE}
    \put(-92,103){PC, unconstrained}
    \caption{\nueCC\ partially contained events.}
    \label{fig:7-channel-nueCC-PC}
 \end{subfigure}
\caption{Event distributions of (a) FC \nueCC\ and (b) PC \nueCC\ candidates as a function reconstructed $E_{\nu}$. No constraint is used. Data-to-prediction ratios, $\chi^2$, and error bands are calculated based on the \SM~hypothesis. The MC expectation of the \LEE~component is added on top of the energy spectrum as represented by the dashed red curve.  The breakdown of each component for different final states for both signal and background events is shown in the legend. The bottom sub-panels present the data-to-prediction ratios as well as the statistical and systematic uncertainties. The pink band includes the statistical, cross section, and flux uncertainties. The purple band corresponds to the full uncertainty with an addition of detector systematic uncertainty.}
\label{fig:7-channel-nueCC}
\end{figure*}

$\nu_e$CC events are identified as described in Sec.~\ref{sec:nueCC} starting from a generic neutrino selection that is followed by various dedicated selection criteria based on Wire-Cell event reconstruction and BDTs. The \nueCC\ events are selected with a high purity as shown in Fig.~\ref{fig:nue_selection_recoEnu_theta}. To further maximize the eLEE sensitivity, we divide the $\nu_e$CC candidates into two sub-samples:  a fully contained (FC) and a partially contained (PC) sample, both of which follow the aforementioned seven-channel fit strategy.
The FC and PC $\nu_e$CC selection results are shown in Fig.~\ref{fig:nueCC_variables} and Fig.~\ref{fig:7-channel-nueCC}. The green histogram represents the MC predicted \nueCC\ signal expectation obtained from the beam simulation and the \textsc{Genie} interaction models tuned to MicroBooNE (\SM\ hypothesis). The figures also contain the data-to-prediction ratios, $\chi^2$ per number of degrees-of-freedom, and error bands calculated based on the simulation of the expected \nueCC\ signal and background.
The MC model of the \nueCC\ signal with the \LEE~component added to the beam expectation is indicated by the red dashed histogram.
The $\chi^2/ndf$ values (assuming \SM~hypothesis) for FC and PC \nueCC\ channels without and with the constraints from the \numuCC\ and \pizero\ channels are summarized in Table~\ref{tab:GoFvalues_constraint}.

\begin{figure}[htp!]
    \includegraphics[width=0.99\columnwidth]{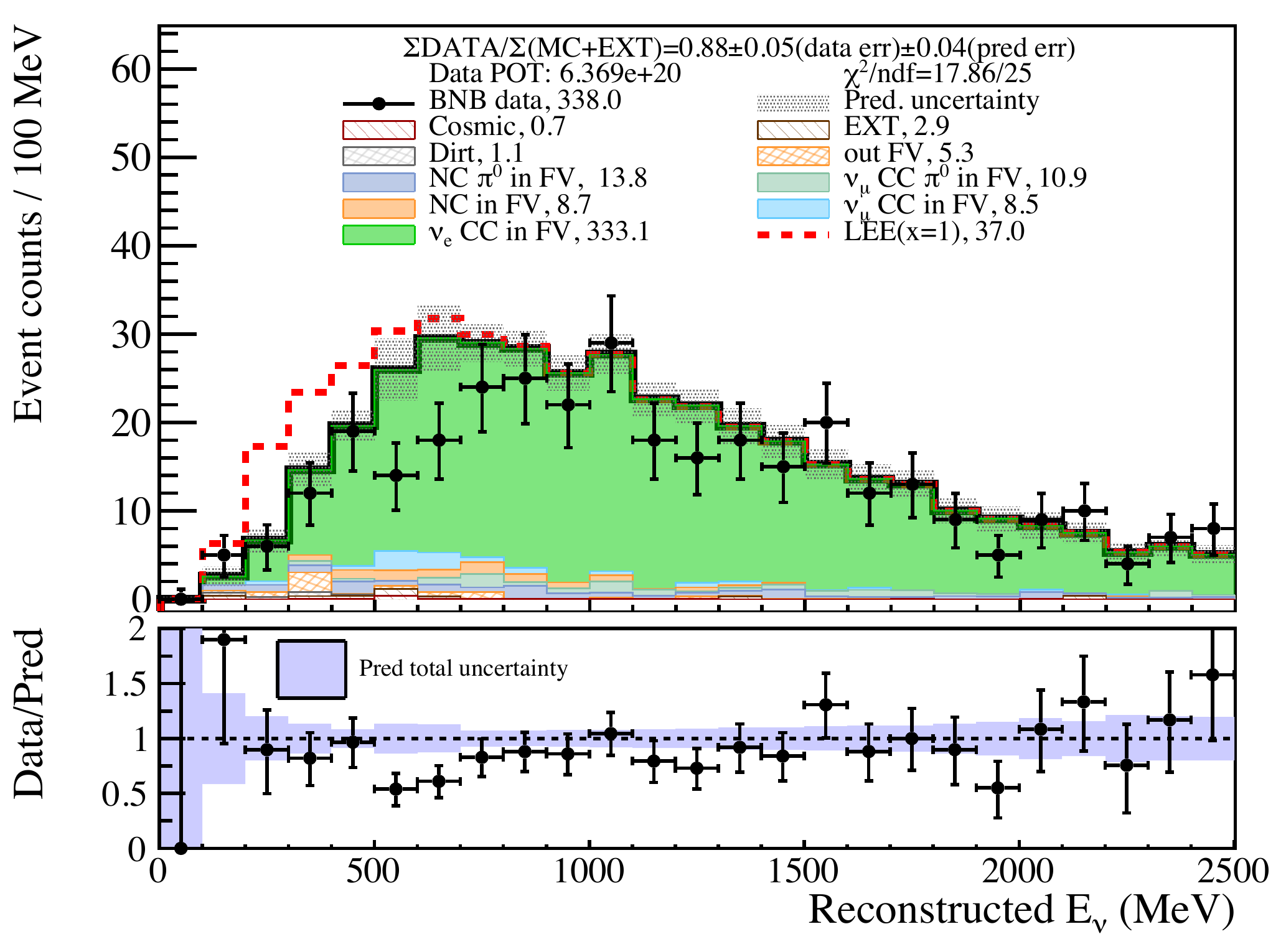}
   \put(-85,122){MicroBooNE}
    \put(-85,112){FC, constrained}
\caption{Event distribution of FC \nueCC\ candidates as a function reconstructed $E_{\nu}$. Same set-up as that of Fig.~\ref{fig:7-channel-nueCC} except that the constraint from PC \nueCC, FC \numuCC, PC \numuCC, FC CC$\pi^0$, PC CC$\pi^0$, 
and NC$\pi^0$ is used. }
\label{fig:nueCC_FC_constraint_breakdown}
\end{figure}

The neutrino energy spectrum of the data compared to the constrained MC prediction for the FC \nueCC\ channel, 
of which the low energy region is most sensitive to the eLEE search is shown in Fig.~\ref{fig:nueCC_FC_constraint_breakdown}. 
The applied constraints include the PC \nueCC\ channel\footnote{The LEE component is included in predicting the PC \nueCC\ distribution for the corresponding hypothesis.} and various \numu\ and \pizero\ channels. The constrained MC prediction with the eLEE component added is indicated by the red dashed histogram in the figure. 
The number of predicted events for each category of background, beam intrinsic \nueCC, and eLEE \nueCC\ in the energy region of reconstructed $E_\nu<$ 600 MeV are summarized in Table~\ref{tab:nueCC_FC_constraint_eventcounts_LEEregion}. Definitions of each category can be found in Sec.~\ref{sec:numuCC}. The main background comes from NC \pizero\ events in which one of the \pizero\ decay $\gamma$'s (the other one may exit the active volume) is mis-identified as a primary electron EM shower. The \nueCC\ selection is also contaminated by CC or NC interactions with no \pizero\ in the final state in which case the typical failure mode is that the decay electrons from muons or charged pions are mis-identified as primary electron EM showers alongside bad reconstruction of the neutrino vertex.  
Note that the ``Cosmic'' category corresponds to the cosmic-ray background estimated from the overlay MC. This background is essentially a mismatch between charge and light signals which accounts for the mis-selected cosmic-ray background from neutrino activity in addition to the cosmic-ray background estimated from beam-off data with no neutrino activity (``EXT'' category). Since this ``Cosmic'' background is associated with neutrino interactions in the overlay MC, the predicted number will change accordingly in the constraint.

In Table.~\ref{tab:nueCC_FC_constraint_eventcounts_LEEregion},  expected number of events for each type of background, beam 
intrinsic \nueCC, and \LEE~\nueCC\ in the energy region of reconstructed $E_\nu<$600 MeV for FC \nueCC\ are shown. 
The uncertainties for each category of the predicted background or \nueCC\ signal 
are systematic uncertainties originating from the flux, cross section (including both $\nu$-Argon and hadron-Argon interactions), detector, and the limited statistics of the samples used in making predictions (e.g. MC statistical uncertainty). ``Dirt'' background has an additional 50\% systematic uncertainty. The statistical uncertainties are presented only for beam-on and beam-off data results.
With the constraints, the systematic uncertainty is reduced significantly for the \nueCC\ signal. However, the uncertainties of the predictions of various backgrounds change only slightly because the uncertainty of each background is dominated by the MC statistical uncertainty. With the constraints, the contributions from flux and cross section systematic uncertainties are largely reduced and the central values of the predictions are adjusted as well. The constraining power is mostly from \numuCC\ channels, while \pizero\ channels help constrain the residual \pizero\ background in \nueCC\ candidate events. The PC \nueCC\ channel has an insignificant effect in the constraint or in the eLEE search, and it serves as a good validation check on the data and simulation. After such constraints, the uncertainty of the result is dominated by statistical uncertainty followed by the systematic uncertainties from the MC statistics, detector response, cross section, and neutrino flux. 

\begin{table}[htp!]
    \centering
    \begin{tabular}{l|cc}
        \hline
        \hline
        Category & w/o constraint &  w/ constraint \\
        \hline
 %       Beam $\nu_e$CC   & 42.55 $\pm$ 10.57 & 51.50 $\pm$ 2.57 \\
 Beam $\nu_e$CC   & 42.6 $\pm$ 10.6 & 51.5 $\pm$ 2.6 \\
  %      $\nu_{\mu}$CC $\pi^0$   & 0.62 $\pm$ 0.80 & 0.79 $\pm$ 0.75 \\
  $\nu_{\mu}$CC $\pi^0$   & 0.6 $\pm$ 0.8 & 0.8 $\pm$ 0.8 \\
  %      $\nu_{\mu}$CC (non-$\pi^0$) & 3.92 $\pm$ 4.16 & 3.08 $\pm$ 3.07 \\
  $\nu_{\mu}$CC (non-$\pi^0$) & 3.9 $\pm$ 4.2 & 3.1 $\pm$ 3.1 \\
  %      NC $\pi^0$       & 4.51 $\pm$ 2.34 & 4.25 $\pm$ 1.59 \\
  NC $\pi^0$       & 4.5 $\pm$ 2.3 & 4.3 $\pm$ 1.6 \\
  %      NC (non-$\pi^0$) & 2.96 $\pm$ 1.41 & 2.91 $\pm$ 1.23 \\
  NC (non-$\pi^0$) & 3.0 $\pm$ 1.4 & 2.9 $\pm$ 1.2 \\
  %      Out of FV        & 3.76 $\pm$ 1.97 & 3.41 $\pm$ 1.55 \\
   Out of FV        & 3.8 $\pm$ 2.0 & 3.4 $\pm$ 1.6 \\
  %      Dirt             & 1.03 $\pm$ 0.99 & 1.15 $\pm$ 0.92 \\
  Dirt             & 1.0 $\pm$ 1.0 & 1.2 $\pm$ 0.9 \\
  %      Cosmic           & 0.32 $\pm$ 0.57 & 0.45 $\pm$ 0.55 \\
  Cosmic           & 0.3 $\pm$ 0.6 & 0.5 $\pm$ 0.6 \\
  %      EXT (beam-off data)    & \multicolumn{2}{c}{1.87 $\pm$ 1.68}  \\ \hline
  EXT (beam-off data)    & \multicolumn{2}{c}{1.9 $\pm$ 1.7}  \\ \hline
  %      Pred. total (\SM)   & 61.54$\pm$15.32$\pm$7.7 & 69.58$\pm$4.99$\pm$8.0 \\
  Predicted total (\SM)   & 61.5 $\pm$ 15.3 $\pm$ 7.7 & 69.6 $\pm$ 5.0 $\pm$ 8.0 \\
  %      Pred. total (\LEE)  & 91.78$\pm$23.44$\pm$8.7 & 103.82$\pm$7.39$\pm$9.0\\
  Predicted total (\LEE)  & 91.8 $\pm$ 23.4 $\pm$ 8.7 & 103.8 $\pm$ 7.4 $\pm$ 9.0\\
        BNB data  & \multicolumn{2}{c}{56} \\
        \hline
        \hline
    \end{tabular}
\caption[]{Expected number of events for each type of background, beam intrinsic \nueCC, and \LEE~\nueCC\ in the energy region of reconstructed $E_\nu<$600 MeV for FC \nueCC, without and with all available constraints. All the expected numbers correspond to 6.37$\times$10$^{20}$ POT. The uncertainties are systematic uncertainties for each category of the background and \nueCC\ signal except for the predicted total where the second uncertainty is the statistical uncertainty using the CNP formalism. See text in
Sec.~\ref{sec:results_nueCC} for more detailed discussion.}
\label{tab:nueCC_FC_constraint_eventcounts_LEEregion}
\end{table}

\begin{figure}[htp!]
\captionsetup[subfigure]{justification=centering}
  \centering
  \begin{subfigure}[]{0.92\columnwidth}
    \includegraphics[width=\textwidth]{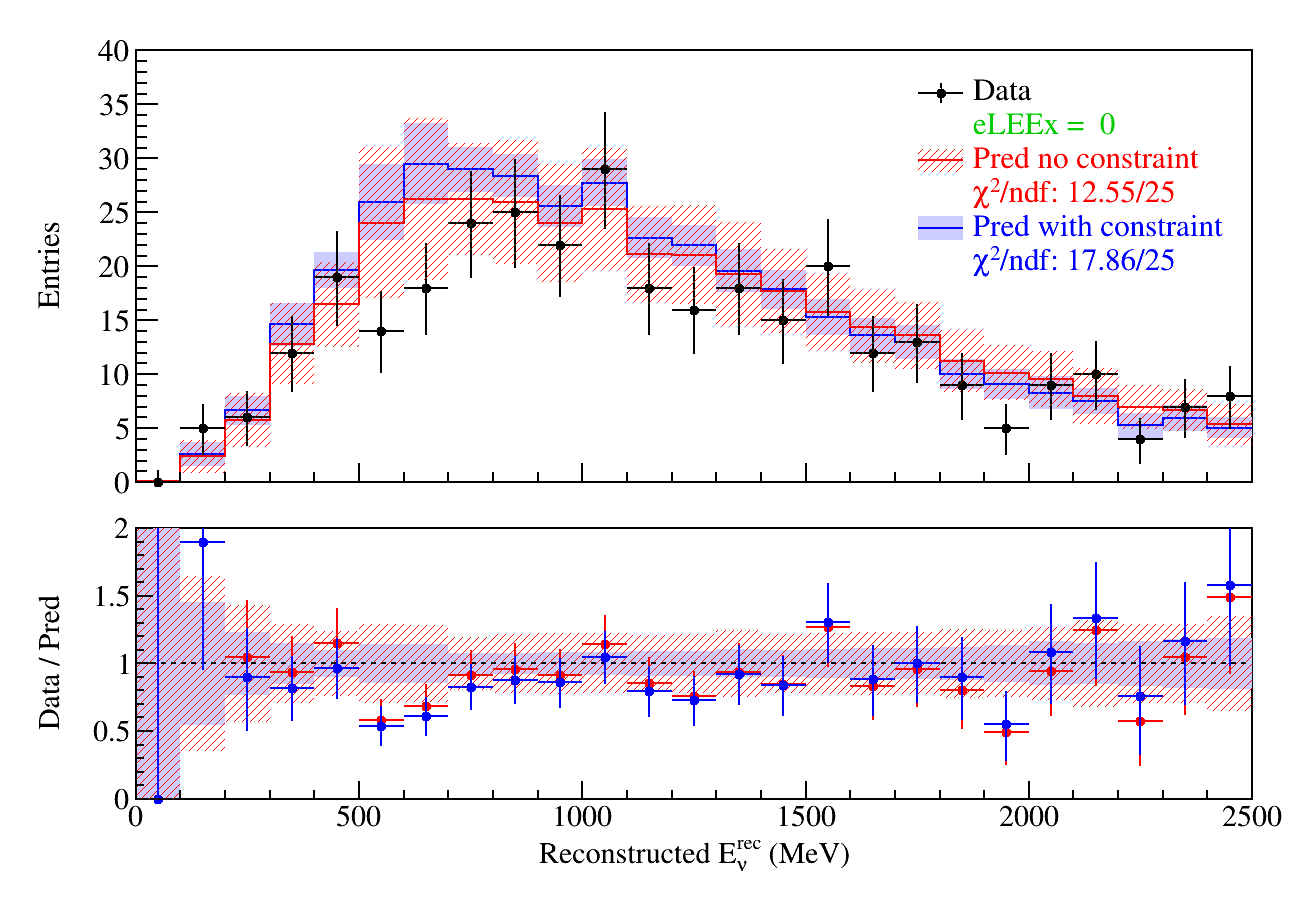}
    \put(-135, 155){MicroBooNE 6.369$\times$10$^{20}$ POT}
    \put(-200,140){FC}
    \caption{\SM~hypothesis.}
    \label{fig:nueCC_FC_constraint_LEEx0}
  \end{subfigure}
  \begin{subfigure}[]{0.92\columnwidth}
    \includegraphics[width=\textwidth]{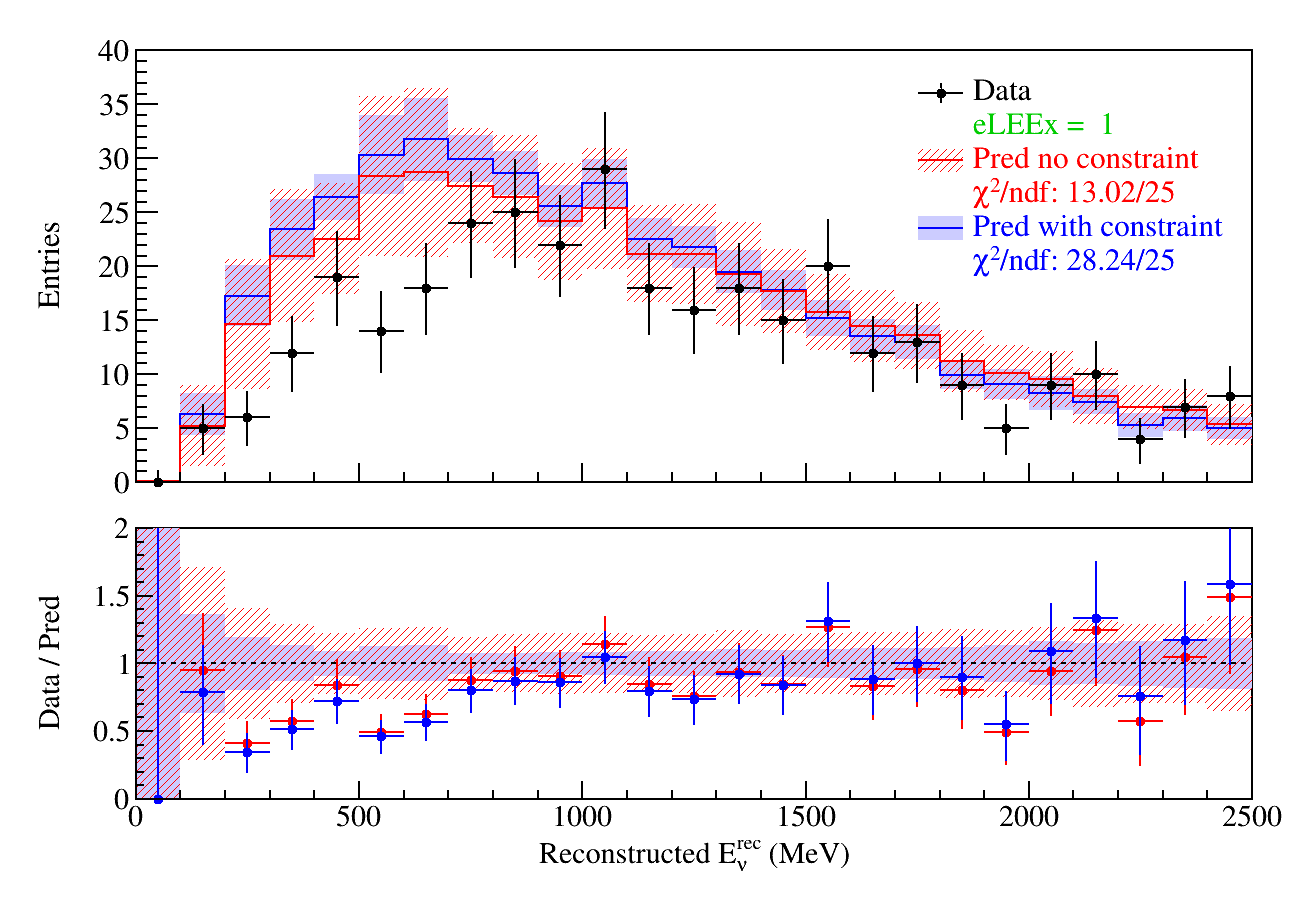}
    \put(-135, 155){MicroBooNE 6.369$\times$10$^{20}$ POT}
    \put(-200,140){FC}
    \caption{\LEE~hypothesis}
    \label{fig:nueCC_FC_constraint_LEEx1}
  \end{subfigure}
  \caption{Event distributions of FC \nueCC\ candidates as a function of reconstructed $E_{\nu}$. (a) An \SM~hypothesis and (b) an \LEE~hypothesis are assumed in MC prediction and uncertainty calculation, respectively. Black points are from data measurements. Red (blue) histograms and error bands are for MC prediction before (after) constraint from the other six channels.}
  \label{fig:nueCC_FC_constraint}
\end{figure}

The data comparisons to the energy spectra of the MC predictions before and after constraint, without and with
\LEE\ \nueCC, for the FC \nueCC\ channel are shown in Fig.~\ref{fig:nueCC_FC_constraint}.
No significant discrepancy between data and the MC \SM~hypothesis (obtained from the BNB beam simulation with 
the MicroBooNE tuned \textsc{Genie} interaction models) was observed after applying all available constraints.
The two data points between 500 MeV and 700 MeV show deficits compared to the constrained prediction of the \SM\ hypothesis. A quantitative examination considering full uncertainties yields a local p-value of 0.039 (2.1$\sigma$),  which is consistent with the hypothesis of
statistical fluctuation and/or systematic variations.
The goodness-of-fit $\chi^2/ndf$ (Pearson format) and $p$-values for the entire energy region as well as the 
low energy region (less than 600 MeV) are summarized in Table~\ref{tab:summary_LEE_chi2}. Comparing the 
$\chi^2/ndf$ values, one can see the impact from the constraint on the sensitivity to the eLEE search as well 
as the level at which the current data measurement excludes the \LEE~hypothesis. More detailed statistical
analyses are reported in the following sections.

\begin{table}[htp!]
    \centering
    \begin{tabular}{l|c|c}
    \hline\hline
    \multicolumn{3}{c}{$\chi^2/ndf$, \SM}\\\hline
    Energy region & \makecell[c]{w/o constraint} & \makecell[c]{w/ constraint }\\\hline
     (0, 2500) MeV & \makecell[c]{12.55/25\\$p_{\text{val}}=0.982$} & \makecell[c]{17.86/25\\$p_{\text{val}}=0.848$}\\\hline
     (0, 600) MeV & \makecell[c]{4.25/6\\$p_{\text{val}}=0.643$} & \makecell[c]{5.78/6\\$p_{\text{val}}=0.448$}\\\hline\hline
    \multicolumn{3}{c}{$\chi^2/ndf$, \LEE}\\\hline
     Energy region & \makecell[c]{w/o constraint} & \makecell[c]{w/ constraint}\\\hline
     (0, 2500) MeV & \makecell[c]{13.02/25\\$p_{\text{val}}=0.976$} & \makecell[c]{28.24/25\\$p_{\text{val}}=0.297$}\\\hline
     (0, 600) MeV & \makecell[c]{4.23/6\\$p_{\text{val}}=0.646$} & \makecell[c]{15.73/6\\$p_{\text{val}}=0.015$}\\\hline\hline
    \end{tabular}    
\caption{Summary of the goodness-of-fit $\chi^2/ndf$ values of FC $\nu_e$CC energy distributions without or with the constraint from the other six channels for \SM\ and \LEE\ hypotheses, respectively. In all model goodness-of-fit tests, Pearson construction of $\chi^2$, instead of the CNP construction of $\chi^2$, is used. The $p$-value for each goodness-of-fit test is also shown.}
\label{tab:summary_LEE_chi2}
\end{table}

\subsection{Simple vs. Simple Likelihood Ratio Test}\label{sec:simple_likelihood}
Using the $\chi^2$ (CNP format) as defined in Eq.~\eqref{eq:test-stat-chi2}, we construct a simple-vs-simple test statistic:
\begin{equation}
    \Delta \chi_{\text{simple}}^2 = \chi^2 | _{\text{eLEEx=1}} - \chi^2 | _{\text{eLEEx=0}}
\end{equation}
where $x$ represents the expected eLEE strength in the prediction. This test statistic allows one to calculate the p-value using a frequentist approach (i.e. with pseudo experiments) assuming the MC prediction with the \SM\ hypothesis or with the \LEE\ hypothesis is true. 
In generating the pseudo experiments, the events in all channels are randomized according to their associated systematic 
uncertainties (covariance matrix) and statistical uncertainties (Poisson distribution). In the case the expected counts
in a given bin is negative, which can happen for large systematic uncertainties, the corresponding pseudo experiment is discarded
and a new psuedo experiment is generated. 
This test statistic tests the compatibility between the data from all seven channels and the given hypothesis. Note that as opposed to Table~\ref{tab:summary_LEE_chi2} which focuses on the low energy FC \nueCC\ events, both the FC \nueCC\ channel and the PC \nueCC\ channel that have eLEE sensitivity and the other \numuCC\ and \pizero\ channels are simultaneously employed in the hypothesis test here and in the following section, in which case the \numuCC\ and \pizero\ channels effectively serve as constraints.

\begin{figure}[!htp]
  \captionsetup[subfigure]{justification=centering}
  \centering
   \begin{subfigure}[]{0.95\columnwidth}
    \includegraphics[width=\textwidth]{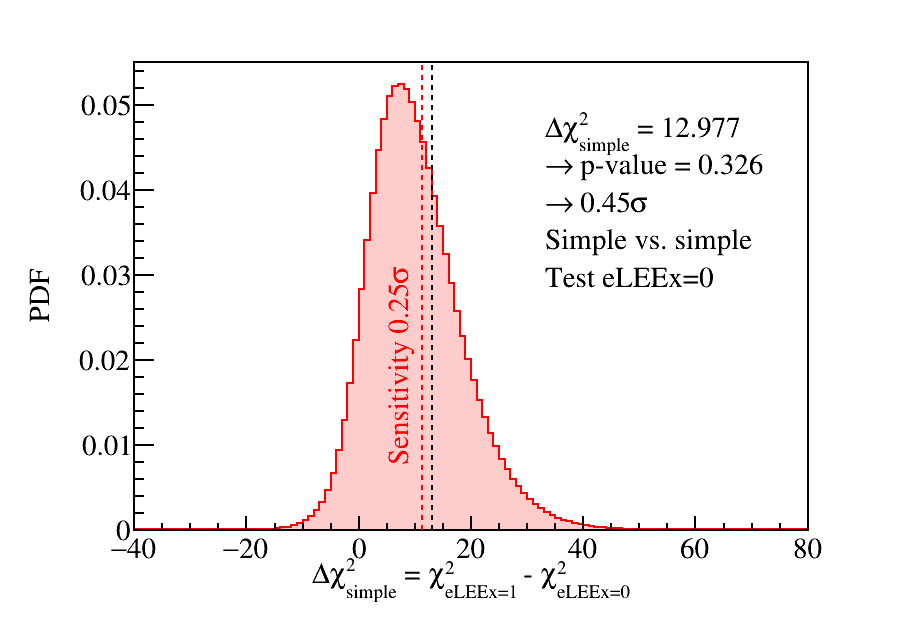}
    \put(-150, 150){MicroBooNE 6.369$\times$10$^{20}$ POT}
    \caption{\SM\ hypothesis.}
    \label{fig:canv_stat_simple_LEEx0}
  \end{subfigure}
  \begin{subfigure}[]{0.95\columnwidth}
    \includegraphics[width=\textwidth]{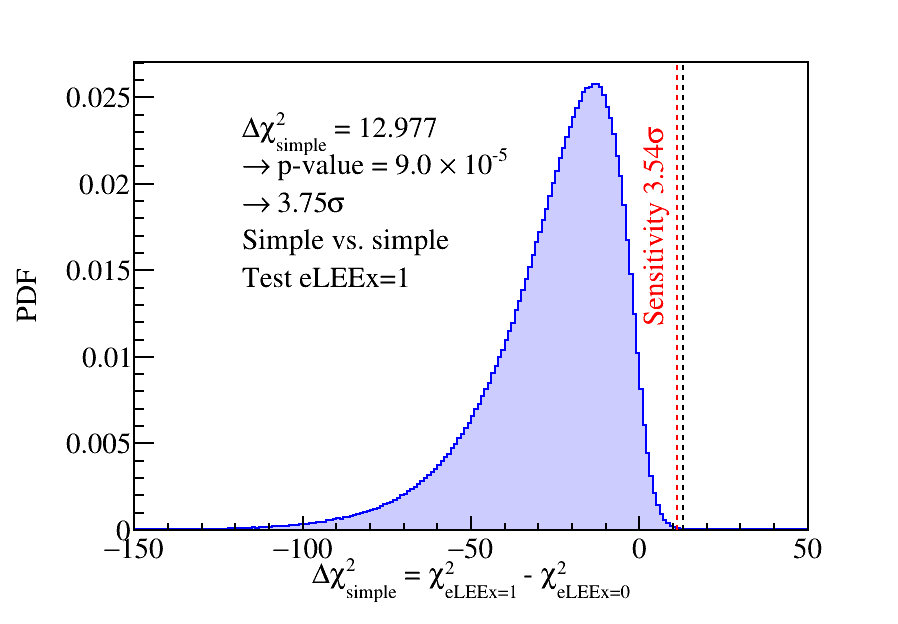}
    \put(-150, 150){MicroBooNE 6.369$\times$10$^{20}$ POT}
    \caption{\LEE\ hypothesis.}
    \label{fig:canv_stat_simple_LEEx1}
  \end{subfigure}
  \caption{Distributions of simple-vs-simple $\Delta \chi^2_{\text{simple}}$ assuming (a) the \SM\ hypothesis is true and (b) the \LEE\ hypothesis is true. Both distributions are obtained from 4 million pseudo experiments. Data $\Delta \chi^2_{\text{simple}}$ values are indicated by dashed vertical lines.}
  \label{fig:canv_stat_simple}
\end{figure}

Figure~\ref{fig:canv_stat_simple} shows the $\Delta \chi^2_{\text{simple}}$ values and distributions of data and pseudo experiments for \SM\ and \LEE\ hypotheses separately. Assuming the \SM\ hypothesis is true, the p-value of the data measurement is derived to be 0.326 (one-sided), which corresponds to a 0.45$\sigma$ significance level. This result demonstrates good agreement between the data measurement and the \SM\ hypothesis. Assuming the \LEE\ hypothesis is true, the p-value of the data measurement is derived to be $9.0\times10^{-5}$ (one-sided), which disfavors the \LEE\ hypothesis at a 3.75$\sigma$ significance level.

\subsection{Nested Likelihood Ratio Test and Best-fit of eLEE strength}\label{sec:nested_likelihood}
Besides the simple-vs-simple test statistic, another test statistic is constructed to further test the compatibility between observation in all seven channels and various hypotheses and to find the best fit of an eLEE model where the strength of the eLEE signal, $x$, is allowed to vary such that $x=1$ would be the eLEE model strength based on the extrapolation from MiniBooNE and values less than or greater than that indicate a smaller or larger excess compared to MiniBooNE. The test statistic is defined as follows:
%MB: Is the eLEE strength defined somewhere? Should we add a sentence here to better define what that means? My attempt is above
\begin{equation}\label{eq:nested}
    \Delta \chi_{\text{nested}}^2 = \chi^2 | _{\text{eLEEx=x}_{0}} - \chi_{\text{min}}^2 | _{\text{eLEEx=x}_{\text{min}}},~~x_{\text{min}}\ge0.
\end{equation}
The value of $x_0$ represents the null hypothesis: $x_0=0$ represents the \SM\ hypothesis (expectation from the beam simulation and the tuned \textsc{Genie} interaction models), and $x_0=1$ represents the \LEE\ hypothesis.  $x_{\text{min}}$ is the best-fit value of $x$ in the allowed region ($x_{\text{min}}\ge0$) after minimization.
This test statistic is a nested likelihood ratio test statistic comparing the null hypothesis ($x=x_0$) with an alternative hypothesis corresponding to $x_{min}$, which can in principle take any value within the allowed region, and is included to quantify the search for an eLEE along with the Feldman-Cousins approach~\cite{Feldman:1997qc}.

\begin{figure}[htp!]
  \centering
  \includegraphics[width=0.5\textwidth]{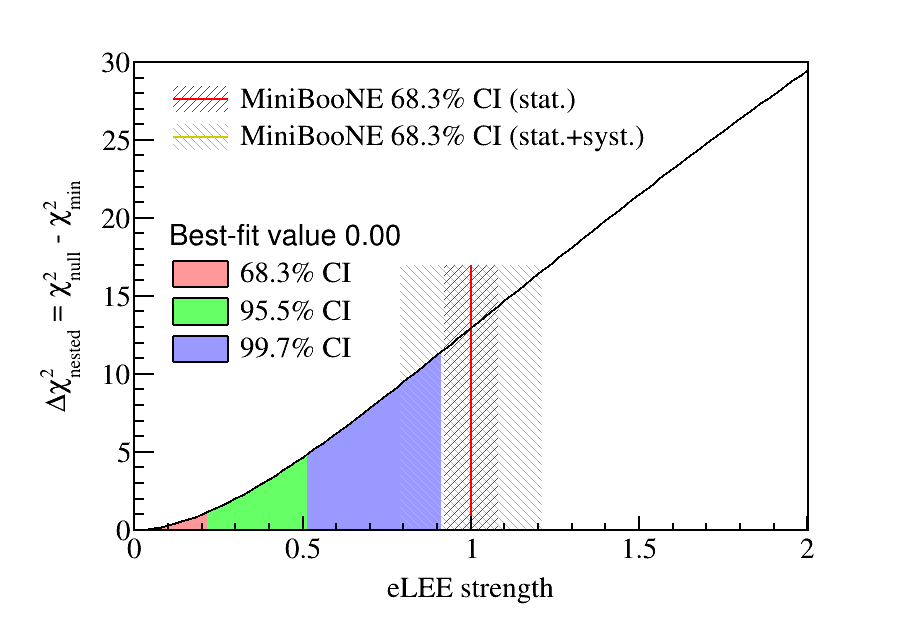}
  \put(-150, 162){MicroBooNE 6.369$\times$10$^{20}$ POT}
  \caption{The $\Delta \chi^2_{\text{nested}}$ with the nested likelihood ratio test. The best-fit eLEE strength is at $x_{\text{min}}=0$. The Feldman-Cousins 68.3\%, 95.5\% and 99.7\% confidence intervals (CI) are displayed with the red, green and blue bands, respectively. The 68.3\% confidence interval of the eLEE strength extracted from the MiniBooNE 2020 result
  are also shown for comparison purpose.}
  \label{fig:canv_stat_bestFit}
\end{figure}

The best-fit value of the eLEE strength $x_{min}$ is determined by minimizing $\chi^2_{\text{eLEEx}}$ with the covariance matrix as defined in Eq.~\eqref{eq:test-stat-1} that accounts for the statistical and systematic uncertainties varying with different eLEE strengths. As shown in Fig.~\ref{fig:canv_stat_bestFit}, the best-fit eLEE strength is at the boundary $x_{\text{min}}=0$ which corresponds to the \SM\ hypothesis. The Feldman-Cousins 68.3\%, 95.5\% and 99.7\% confidence level upper limits from the data measurement are calculated to be 0.217, 0.513, and 0.911, with the expected upper limits for the \SM\ hypothesis at 0.243, 0.563, and 0.958, respectively.
% data CI vs sensitivity CI
% 68.3% 0.217 vs 0.243
% 90.0% 0.396 vs 0.424
% 95.5% 0.513 vs 0.563
% 99.7% 0.911 vs 0.958

In comparison, the 68\% confidence interval (CI) of the eLEE strength estimated from the MiniBooNE 2020 result~\cite{Aguilar-Arevalo:2020nvw} with full and statistical-only uncertainties are
shown in Fig.~\ref{fig:canv_stat_bestFit} as well.
The lower limits of the 68\% MiniBooNE full and statistical-only CIs are disfavored at significance levels 
of more than 2.6$\sigma$ and 3.0$\sigma$, respectively.
Note that some of the systematic uncertainties are correlated between the MiniBooNE and MicroBooNE experiments and a direct comparison of the two experiments taking into account correlated systematic uncertainties
is unavailable. Therefore two separate significance levels derived based on the published information 
are used to indicate the bounds of this comparison.

\begin{figure}[htp!]
  \centering
  \includegraphics[width=0.5\textwidth]{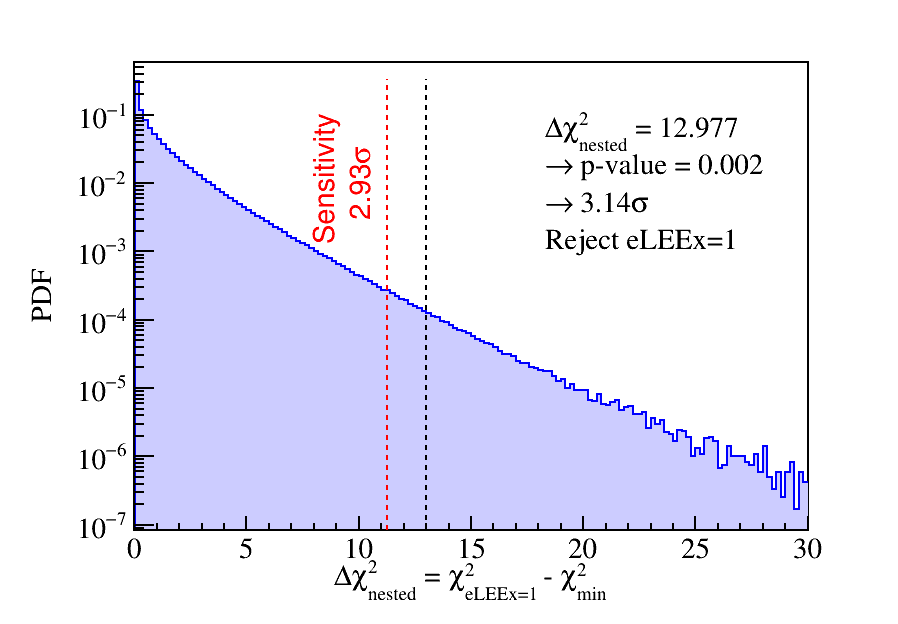}
  \put(-150, 162){MicroBooNE 6.369$\times$10$^{20}$ POT}
  \caption{Distribution of $\Delta \chi^2_{\text{nested}}$ assuming \LEE\ is true. The sensitivity of rejecting the \LEE\ hypothesis is at 2.93$\sigma$ following the Feldman-Cousins procedure. Using the same procedure, the data measurement rejects \LEE\ hypothesis at the p-value of 0.002, which corresponds to 3.14$\sigma$. The results are obtained from 10 million pseudo experiments.}
  \label{fig:canv_stat_exLEE}
\end{figure}

Based on this nested likelihood ratio test statistic, the significance level of rejecting the \LEE\ hypothesis 
is estimated following the Feldman-Cousins procedure.
As indicated by the black dashed line in Fig.~\ref{fig:canv_stat_exLEE}, the $\Delta \chi^2_{\text{nested}}$ value of the data measurement is 12.977. Comparing to the $\Delta \chi^2_{\text{nested}}$ distribution from the pseudo experiments assuming the \LEE\ hypothesis is true, the \LEE\ hypothesis is disfavored at a p-value of 0.002, which corresponds to a 3.14$\sigma$ significance level. In comparison with the data result, the sensitivity of rejecting the \LEE\ hypothesis is also calculated using the Asimov data 
set of the \SM\ hypothesis (best-fit of eLEE strength, $x_{\text{min}}=0$) following the procedure in Ref.~\cite{Cowan:2010js}. The constraint on FC and PC $\nu_e$CC channels from the other five channels are taken into account in the Asimov data set.
The resulting sensitivity, as indicated by the red dashed line in Fig.~\ref{fig:canv_stat_exLEE}, is that the \LEE~hypothesis, 
which represents the median of the MiniBooNE result, is disfavored at 2.93$\sigma$ with the $\Delta \chi^2_{\text{nested}}$ value 
of 11.26. Various hypothesis test results as well as sensitivity values (using the Asimov data set) are summarized in Table~\ref{tab:stat_test_summary}. This nested likelihood ratio hypothesis 
test obtains a similar level of rejection of the \LEE\ hypothesis as the aforementioned
simple-vs-simple likelihood ratio hypothesis test, and provides a rigorous cross-check of 
the result. 
%This sensitivity result is consistent with the \LEE\ rejection significance level from data.

\begin{table}[htp!]
\centering
\begin{tabular}{c|c|c} 
\hline\hline
Scenario & Sensitivity & Data result\\\hline 
\makecell[c]{$\Delta \chi^2_{\text{simple}}$}$\begin{cases}\text{null hypothesis \SM}\\ \text{null hypothesis \LEE}\end{cases}$ & \makecell[c]{0.25$\sigma$\\3.54$\sigma$} & \makecell[c]{0.45$\sigma$\\3.75$\sigma$}\\\hline
$\Delta \chi^2_{\text{nested}}$ with null hypothesis \LEE & 2.93$\sigma$ & 3.14$\sigma$\\
\hline\hline
\end{tabular}
\caption{Summary of the hypothesis test results. The simple vs. simple $\Delta \chi^2_{\text{simple}}$ results are from Fig.~\ref{fig:canv_stat_simple} and the nested likelihood ratio $\Delta \chi^2_{\text{nested}}$ results are from Fig.~\ref{fig:canv_stat_exLEE}. The significance level is calculated based on the pseudo experiments corresponding to each null hypothesis. The sensitivity is calculated using an Asimov data set assuming the \SM\ hypothesis is true.}
\label{tab:stat_test_summary}
\end{table}

\section{Summary}\label{sec:summary}
In this article, we report a search for a low-energy excess in \nueCC\ interactions using BNB data in the MicroBooNE experiment. 
A data-driven LEE model motivated by the previous observation of an electron-like low-energy excess (eLEE) from the MiniBooNE neutrino experiment is built to quantify the search result. With the single MicroBooNE detector, measurements of \numuCC\ interactions and CC or NC interactions with a $\pi^0$ in the final state are used to constrain the prediction of \nueCC\ interactions as well as other neutrino backgrounds reducing the systematic uncertainty and thus maximizing the eLEE search sensitivity. 

A blind analysis scheme was adopted in analyzing the data with a BNB exposure of 6.369$\times$10$^{20}$ POT, which corresponds to the first three years of MicroBooNE data taking.
After unblinding the $\nu_e$CC events with reconstructed neutrino energy below 600 MeV and using a nested likelihood ratio test statistic using all available information, the best-fit LEE strength is determined to be 0 (\SM) with the Feldman-Cousins 68.3\%, 95.5\%, and 99.7\% C.L. upper limits at 0.217, 0.513, and 0.911, respectively. Using a simple-vs-simple hypothesis test, the \LEE\ hypothesis is rejected at 3.75$\sigma$, while the \SM\ hypothesis is shown to be consistent with the observation at 0.45$\sigma$. Various hypothesis test results can be found in Table~\ref{tab:stat_test_summary}.

Regarding the $\nu_e$CC hypothesis to explain the electron-like low-energy 
excess observed in the MiniBooNE experiment, we compare the confidence 
intervals obtained from the MicroBooNE and MiniBooNE experiments in the context 
of the eLEE model. The MiniBooNE 1$\sigma$ statistical uncertainty band is entirely outside the 3$\sigma$ allowed range of the LEE strength parameter $x$ derived from this analysis in MicroBooNE. 
Even in the absence of any consideration of cross-experiment correlations, the MiniBooNE 1$\sigma$ full uncertainty
band is well outside the 2$\sigma$ allowed $x$ parameter range from 
MicroBooNE. We should note that the current eLEE model only takes into account the excess as a function of true neutrino energy 
with other kinematics modeled as for the BNB intrinsic \nueCC\ events. 
If we consider the lepton kinematic distributions reported by MiniBooNE~\cite{Aguilar-Arevalo:2020nvw} with an enhanced excess in the forward direction, the achieved \nueCC\ selection efficiency in this work, which is better at forward angles, would yield a stronger exclusion of the eLEE signal. 

While we observe a null result in searching for a $\nu_e$ LEE signal using the current available data set, further tests of alternative hypotheses explaining the MiniBooNE low-energy excess using single-photon-like events due to processes either within or beyond the standard model~\cite{MiniBooNE:2008yuf, Wang:2013wva, Bertuzzo:2018itn, Ballett:2018ynz, Abdallah:2020vgg} will be carried out and reported in future MicroBooNE publications. While this analysis did not perform a fit using a sterile-neutrino oscillation model, future analyses in MicroBooNE
with a full data set at 12.25$\times$10$^{20}$ POT exposure, as well as the 
short-baseline neutrino program~\cite{Antonello:2015lea}, will explicitly test short-baseline oscillation models in both appearance and disappearance channels.

%\clearpage
\begin{acknowledgments}
This document was prepared by the MicroBooNE collaboration using the resources of the Fermi National Accelerator Laboratory (Fermilab), a U.S. Department of Energy, Office of Science, HEP User Facility. Fermilab is managed by Fermi Research Alliance, LLC (FRA), acting under Contract No. DE-AC02-07CH11359.  MicroBooNE is supported by the following: the U.S. Department of Energy, Office of Science, Offices of High Energy Physics and Nuclear Physics; the U.S. National Science Foundation; the Swiss National Science Foundation; the Science and Technology Facilities Council (STFC), part of the United Kingdom Research and Innovation; the Royal Society (United Kingdom); and The European Union’s Horizon 2020 Marie Sklodowska-Curie Actions. Additional support for the laser calibration system and cosmic ray tagger was provided by the Albert Einstein Center for Fundamental Physics, Bern, Switzerland. We also acknowledge the contributions of technical and scientific staff to the design, construction, and operation of the MicroBooNE detector as well as the contributions of past collaborators to the development of MicroBooNE analyses, without whom this work would not have been possible.
\end{acknowledgments}

\newpage

\appendix*

\section{Separation of 0$p$X$\pi$ and N$p$X$\pi$ samples} \label{sec:appendix}
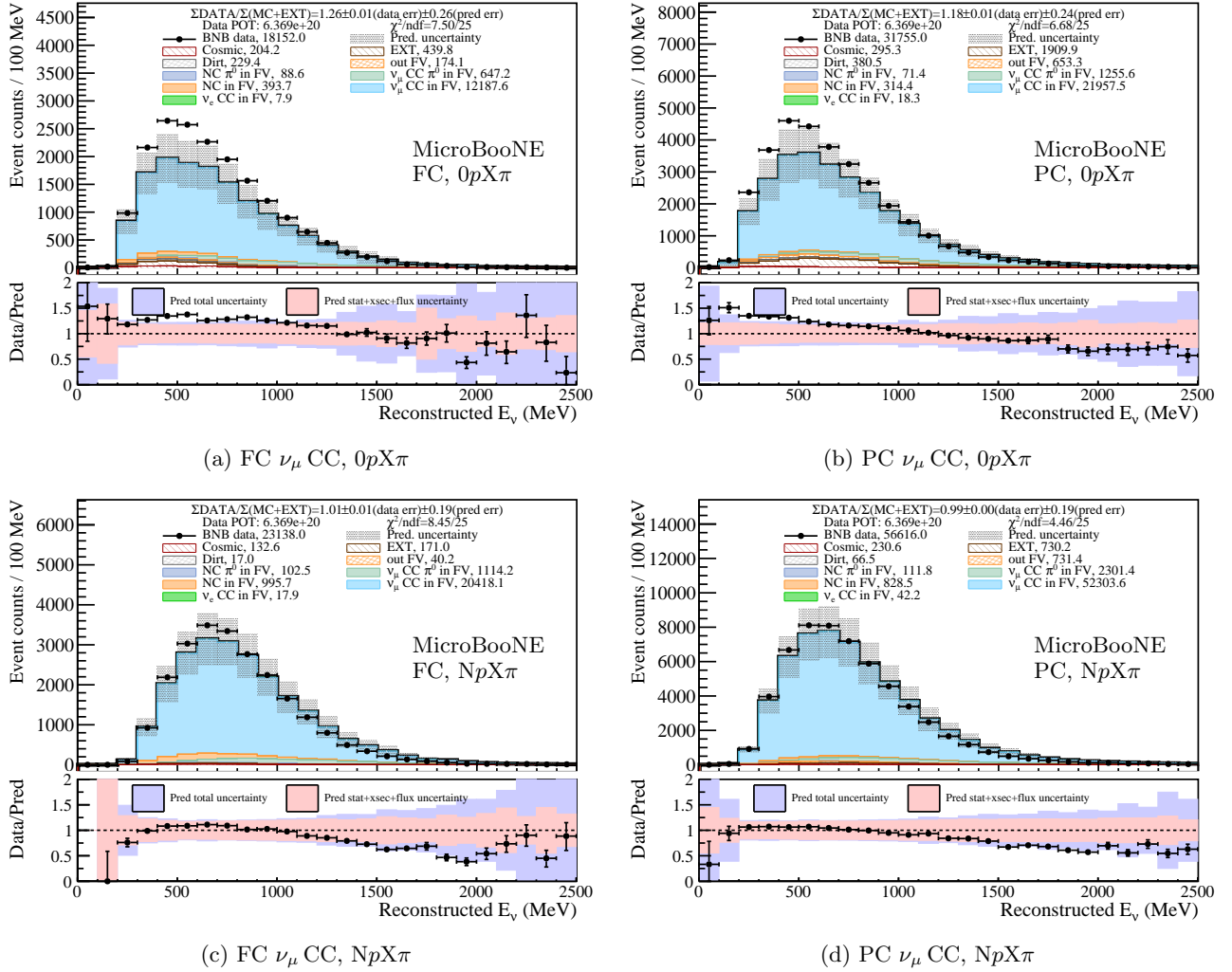
\begin{figure*}[!htp]
  \captionsetup[subfigure]{justification=centering}
  \centering
  \begin{subfigure}[]{0.48\textwidth}
    \includegraphics[width=\textwidth]{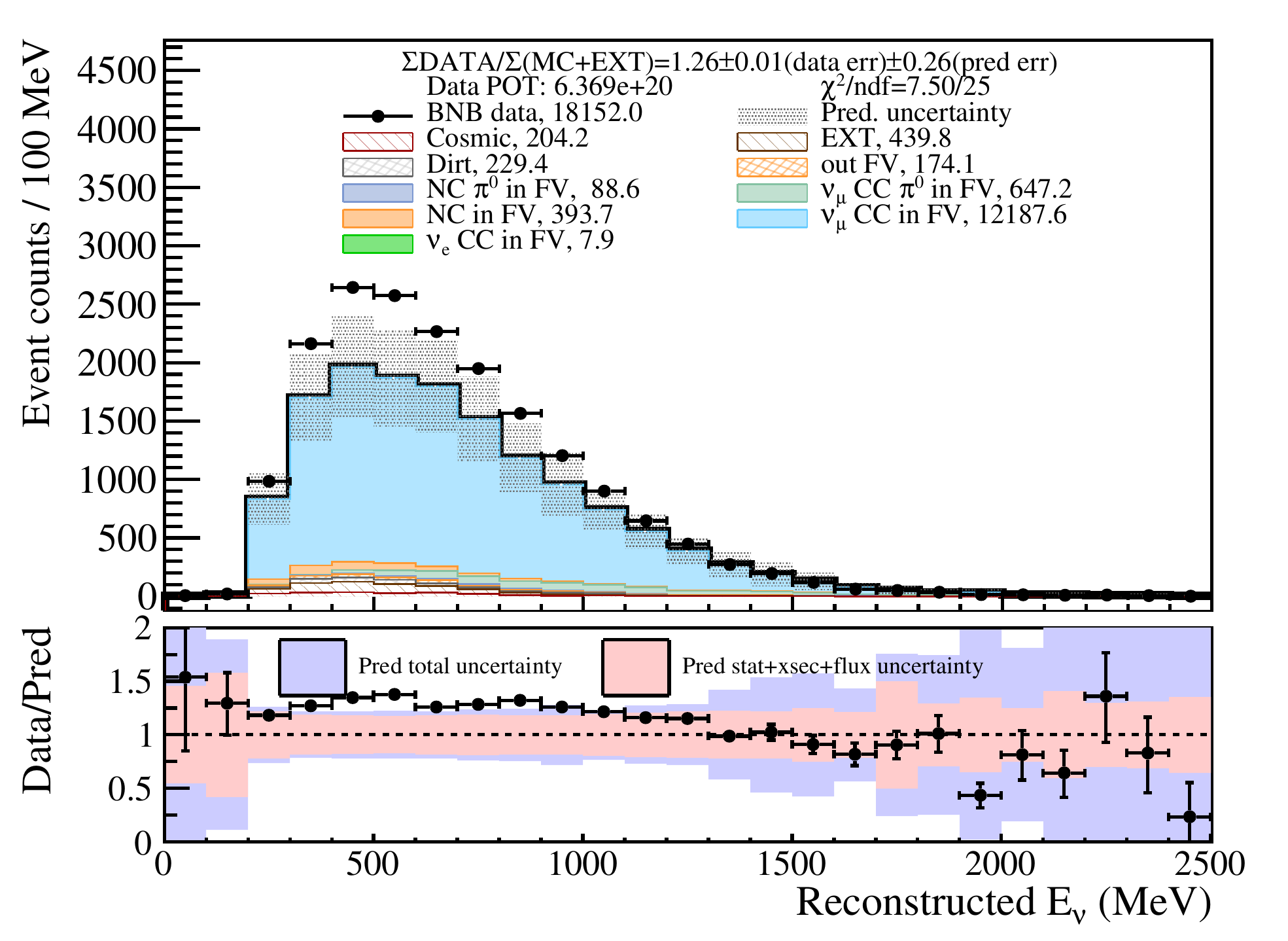}
    \put(-80, 110){MicroBooNE}
    \put(-80, 100){FC, 0$p$X$\pi$}
    \caption{FC \numuCC, 0$p$X$\pi$}
  \end{subfigure}
  \begin{subfigure}[]{0.48\textwidth}
    \includegraphics[width=\textwidth]{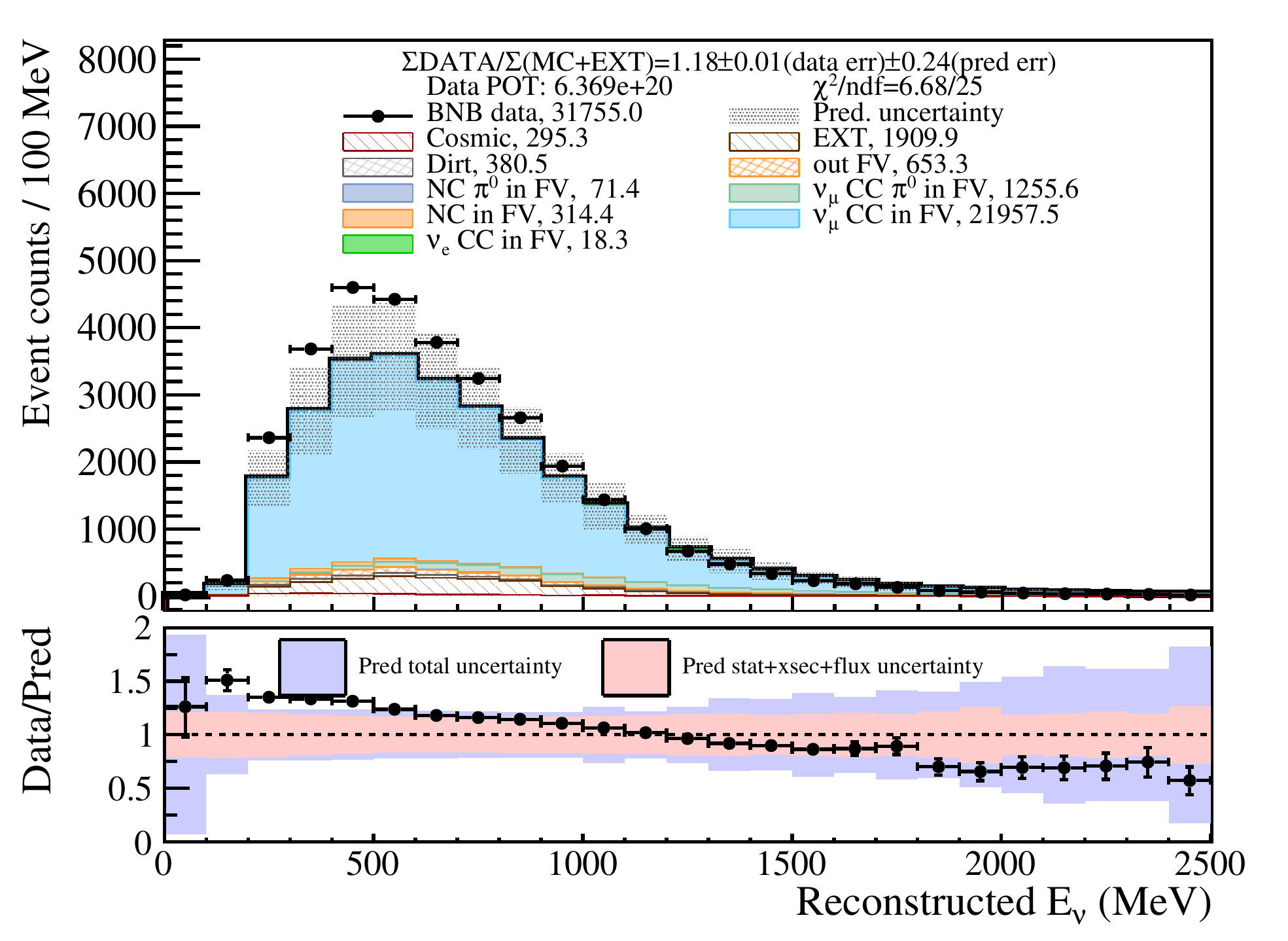}
    \put(-80, 110){MicroBooNE}
    \put(-80, 100){PC, 0$p$X$\pi$}
    \caption{PC \numuCC, 0$p$X$\pi$}
  \end{subfigure}
  \begin{subfigure}[]{0.48\textwidth}
    \includegraphics[width=\textwidth]{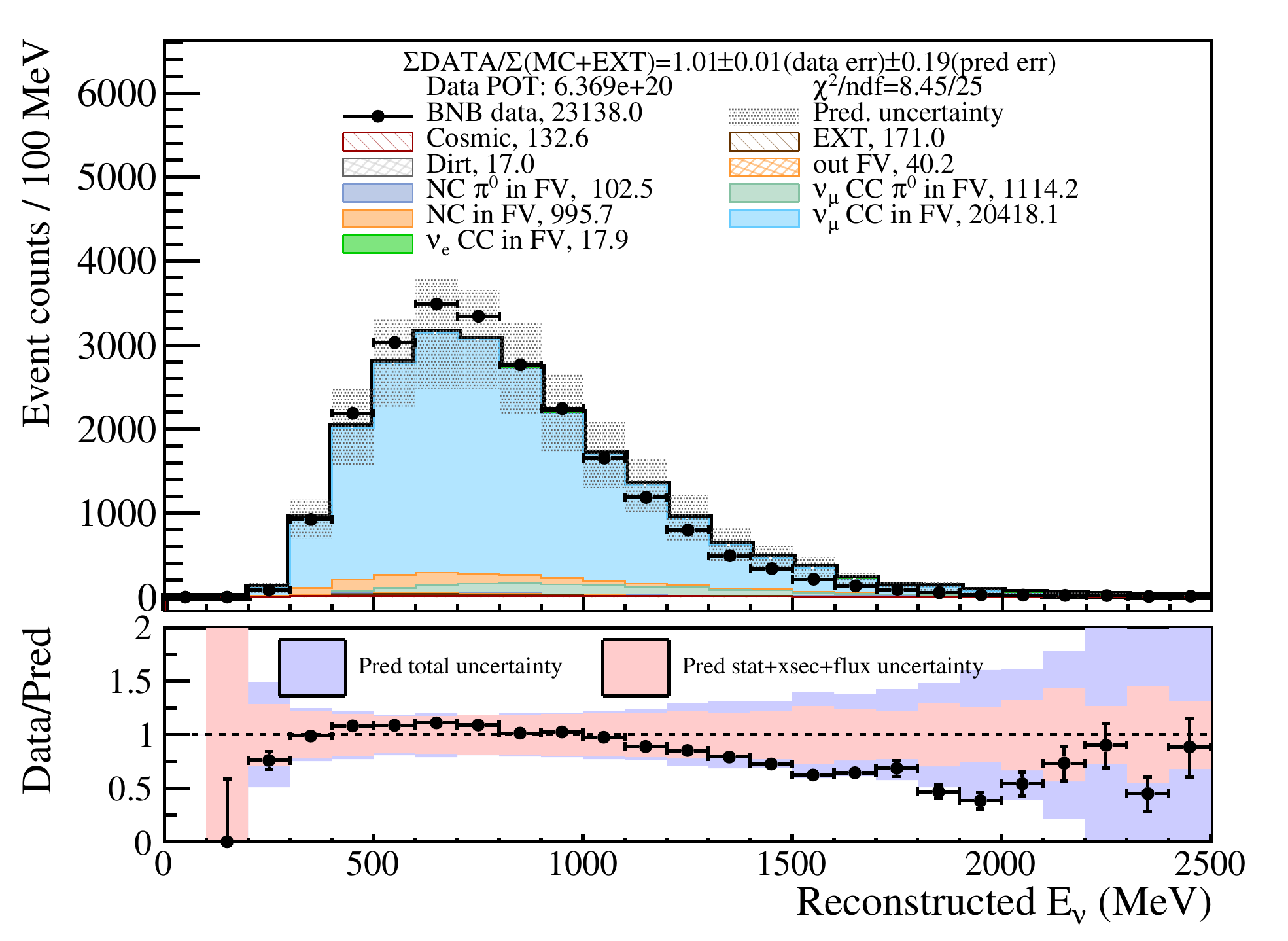}
    \put(-80, 110){MicroBooNE}
    \put(-80, 100){FC, N$p$X$\pi$}
    \caption{FC \numuCC, N$p$X$\pi$}
  \end{subfigure}
  \begin{subfigure}[]{0.48\textwidth}
    \includegraphics[width=\textwidth]{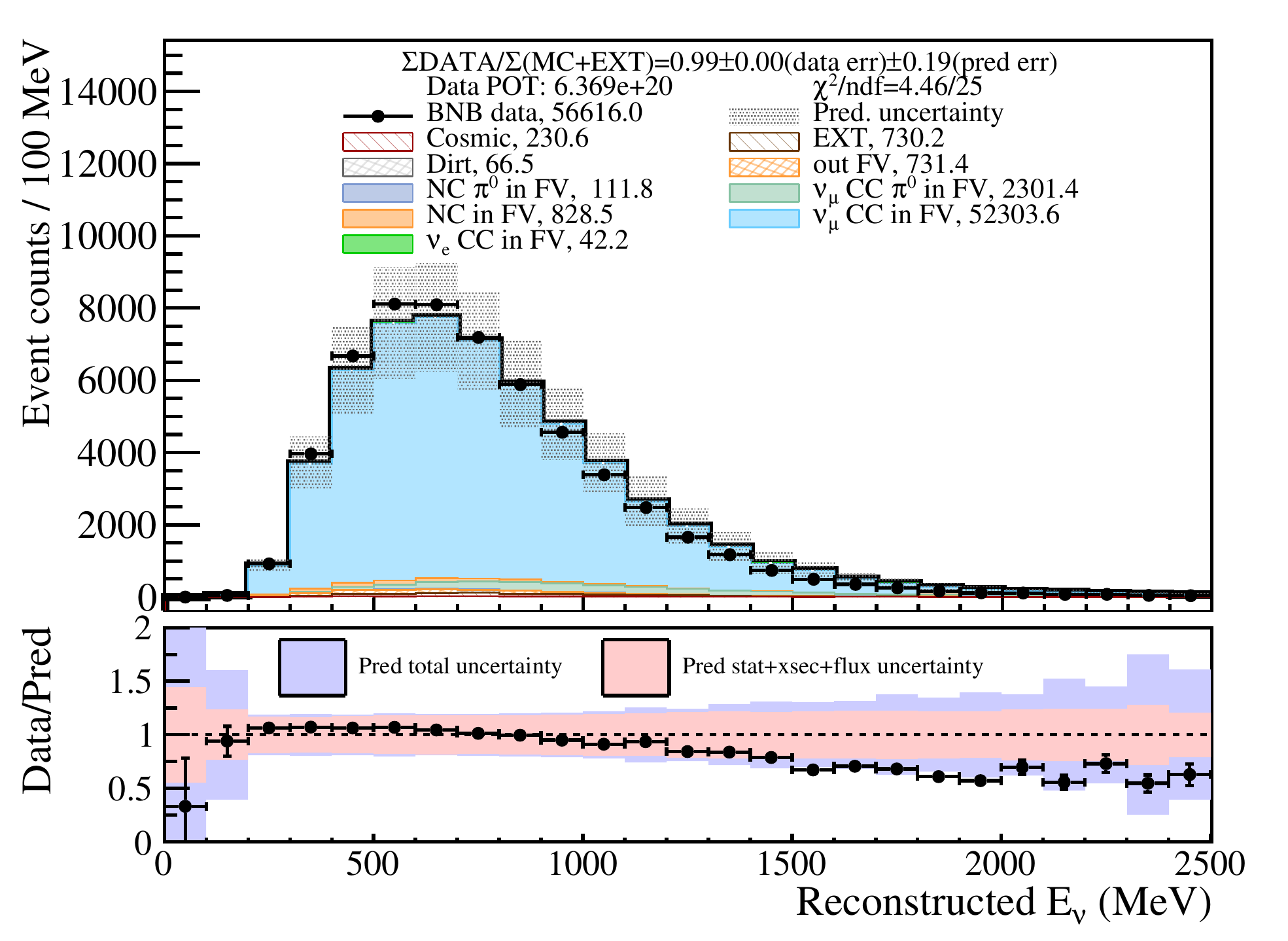}
    \put(-80, 110){MicroBooNE}
    \put(-80, 100){PC, N$p$X$\pi$}
    \caption{PC \numuCC, N$p$X$\pi$}
  \end{subfigure}
\caption{Event distributions of FC \numuCC\ 0$p$X$\pi$, PC \numuCC\ 0$p$X$\pi$, FC \numuCC\ N$p$X$\pi$, and PC \numuCC\ N$p$X$\pi$ samples in figures (a)-(d), respectively. The breakdown of each component for different final states for both signal and background events is shown in the legend (see definitions in Sec.~\ref{sec:nueCC}). The bottom sub-panels present the data-prediction ratios as well as the statistical and systematic uncertainties. The pink band includes the MC statistical, cross section, and flux uncertainties. The purple band corresponds to the full uncertainty with an addition of detector systematic uncertainty. No constraint is applied.}
  \label{fig:11-channel-numuCC}
\end{figure*}

\begin{figure*}[!htp]
  \captionsetup[subfigure]{justification=centering}
  \centering
  \begin{subfigure}[]{0.48\textwidth}
    \includegraphics[width=\textwidth]{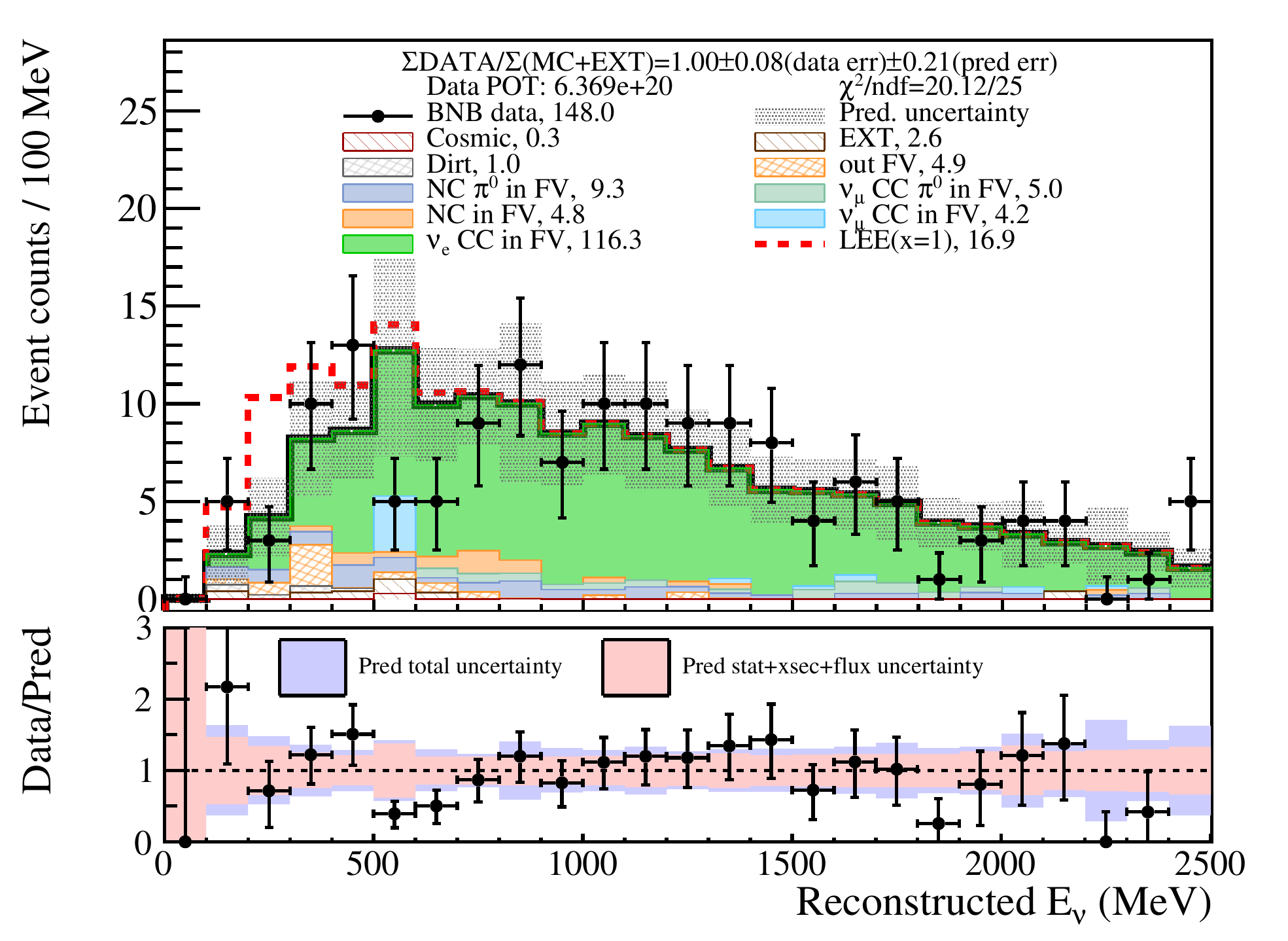}
    \put(-80, 118){MicroBooNE}
    \put(-80, 108){FC, 0$p$X$\pi$}
    \caption{FC \nueCC, 0$p$X$\pi$}
  \end{subfigure}
  \begin{subfigure}[]{0.48\textwidth}
    \includegraphics[width=\textwidth]{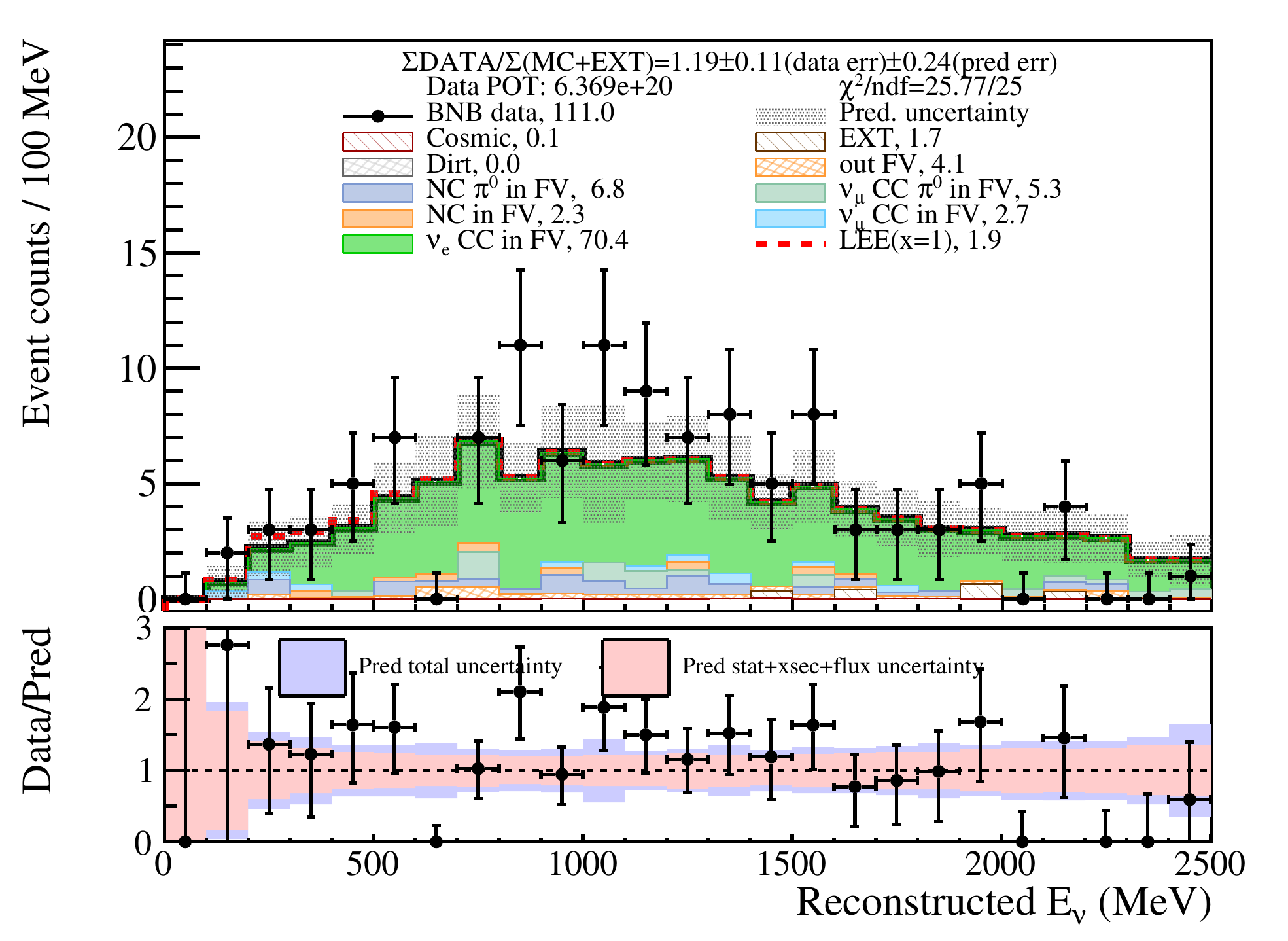}
    \put(-80, 118){MicroBooNE}
    \put(-80, 108){PC, 0$p$X$\pi$}
    \caption{PC \nueCC, 0$p$X$\pi$}
  \end{subfigure}
  \begin{subfigure}[]{0.48\textwidth}
    \includegraphics[width=\textwidth]{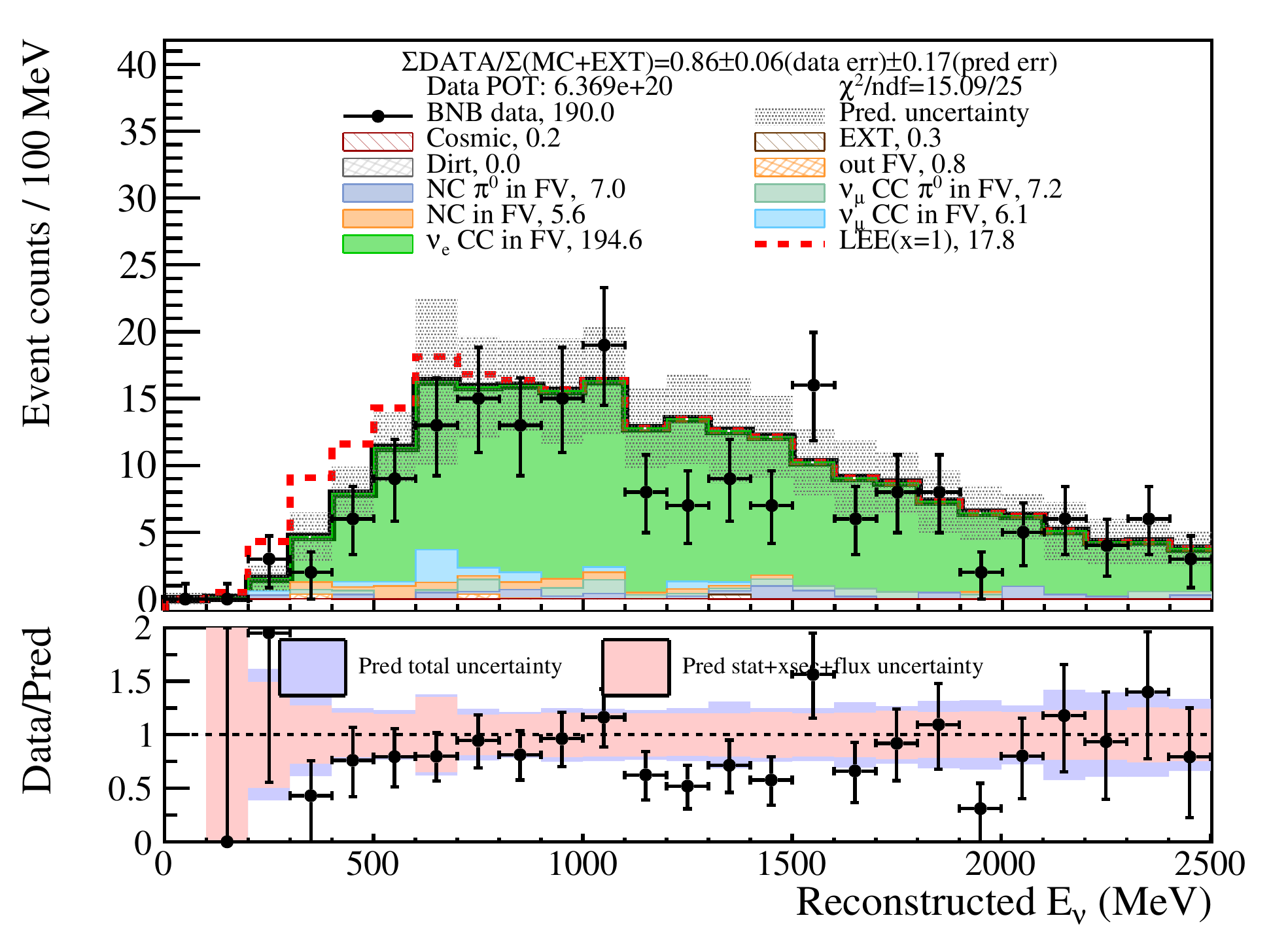}
    \put(-80, 118){MicroBooNE}
    \put(-80, 108){FC, N$p$X$\pi$}
    \caption{FC \nueCC, N$p$X$\pi$}
  \end{subfigure}
  \begin{subfigure}[]{0.48\textwidth}
    \includegraphics[width=\textwidth]{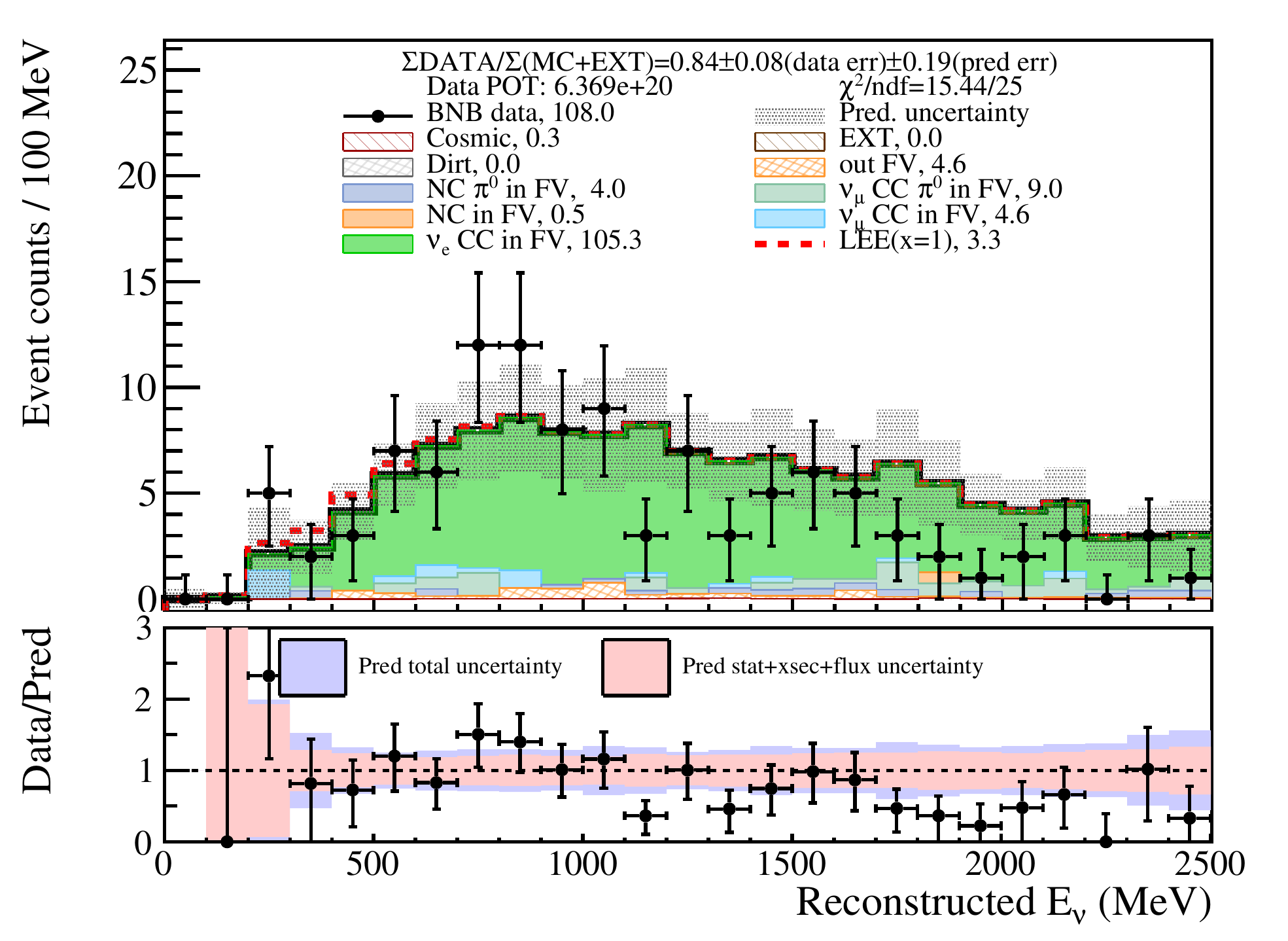}
    \put(-80, 118){MicroBooNE}
    \put(-80, 108){PC, N$p$X$\pi$}
    \caption{PC \nueCC, N$p$X$\pi$}
  \end{subfigure}
\caption{Event distributions of FC \nueCC\ 0$p$X$\pi$, PC \nueCC\ 0$p$X$\pi$, FC \nueCC\ N$p$X$\pi$, and PC \nueCC\ N$p$X$\pi$ samples in figures (a)-(d), respectively. The breakdown of each component for different final states for both signal and background events is shown in the legend (see definitions in Sec.~\ref{sec:nueCC}). The bottom sub-panels present the data-prediction ratios as well as the statistical and systematic uncertainties. The pink band includes the MC statistical, cross section, and flux uncertainties. The purple band corresponds to the full uncertainty with an addition of detector systematic uncertainty. No constraint is applied.}
  \label{fig:11-channel-nueCC}
\end{figure*}

As shown in Fig.~\ref{fig:validation_constraint} and discussed in Sec.~\ref{sec:energy_validation}, the enhancement of the \numuCC\ events with low hadronic energies can be adequately described by the simulation, in particular by the cross-section model and its allowed variation which is most relevant to the potential missing energy in the hadronic system, employed in the MicroBooNE MC simulation.

To further validate the reconstruction of interactions with inclusive hadronic final states and cross-check its impact on the analysis results, the inclusive \numuCC\ events (Fig.~\ref{fig:7-channel-numuCC-pi0}(a) and \ref{fig:7-channel-numuCC-pi0}(b)) are further divided into 0$p$X$\pi$ and N$p$X$\pi$ (N$\ge$1) channels where $p$ is defined as reconstructed protons with kinetic energy greater than 35~MeV (corresponding to a length of 1~cm) and X is the number of reconstructed pions (X$\ge$0).
The separated 0$p$X$\pi$ and N$p$X$\pi$ channels for the selected \numuCC\ events are shown in Fig.~\ref{fig:11-channel-numuCC}. 
Interestingly, the enhancement of \numuCC\ events is found to show up in the 0$p$X$\pi$ channel, whereas the N$p$X$\pi$ channel has an overall good agreement between the data and MC prediction. 
This suggests that the enhancement of the inclusive \numuCC\ events in the relatively low energy region is less likely to be caused by potential bias in the neutrino flux prediction.
Based on the nominal MC simulation, the \numuCC\ 0$p$X$\pi$ FC+PC events, excluding true \nueCC\ and NC events, have 41\% (46\%) true 0$p$X$\pi$ events, and the N$p$X$\pi$ FC+PC events have 96\% (97\%) true N$p$X$\pi$ events in the full energy region (low energy region of $E^{\rm rec}_{\nu}<$ 600~MeV). 
The kinetic energy threshold of the protons in the truth-level study is 40~MeV.
To complement the above truth-level studies which indicate contamination of N$p$X$\pi$ events in the 0$p$X$\pi$ \numuCC\ selections, we assess from hand-scans of both data and MC simulation that protons which contaminate the selection are generally low energy. It supports the statement that the \numuCC\ channels successfully split interactions by the amount of proton hadronic activity, consistent with the goal of this study.
This indicates the excess 0$p$X$\pi$ events are not due to the migration of N$p$X$\pi$ events with high energy protons.
Finally, the hypothesis of migration of high energy events into the low energy region is disfavored because the deficit in the high energy region for both 0$p$X$\pi$ and N$p$X$\pi$ \numuCC\ channels has insufficient events to account for the excess in the low energy region. 

The inclusive \nueCC\ events (Fig.~\ref{fig:7-channel-nueCC}) are divided into 0$p$X$\pi$ and N$p$X$\pi$ channels as well in order to investigate the consistency between \numuCC\ and \nueCC\ selections assuming the enhancement of 0$p$X$\pi$ events is largely attributed to the cross section modeling. 
The separated 0$p$X$\pi$ and N$p$X$\pi$ channels for the selected \nueCC\ events are shown in Fig.~\ref{fig:11-channel-nueCC}, with no constraints applied in the MC predictions. 
Based on a truth-level study similar to that for \numuCC\ events, the \nueCC\ 0$p$X$\pi$ FC+PC events, excluding true \numuCC\ and NC events, have 39\% (49\%) true 0$p$X$\pi$ events, and the N$p$X$\pi$ FC+PC events have 94\% (98\%) true N$p$X$\pi$ events in the full energy region (low energy region of $E^{\rm rec}_{\nu}<$ 600~MeV). 
The data-prediction ratio in the relatively low energy region of the \nueCC\ 0$p$X$\pi$ channel appears higher than that in the \nueCC\ N$p$X$\pi$ channel, in analogy to the \numuCC\ results.

\begin{figure}[!htp]
  \centering
   \includegraphics[width=0.99\columnwidth]{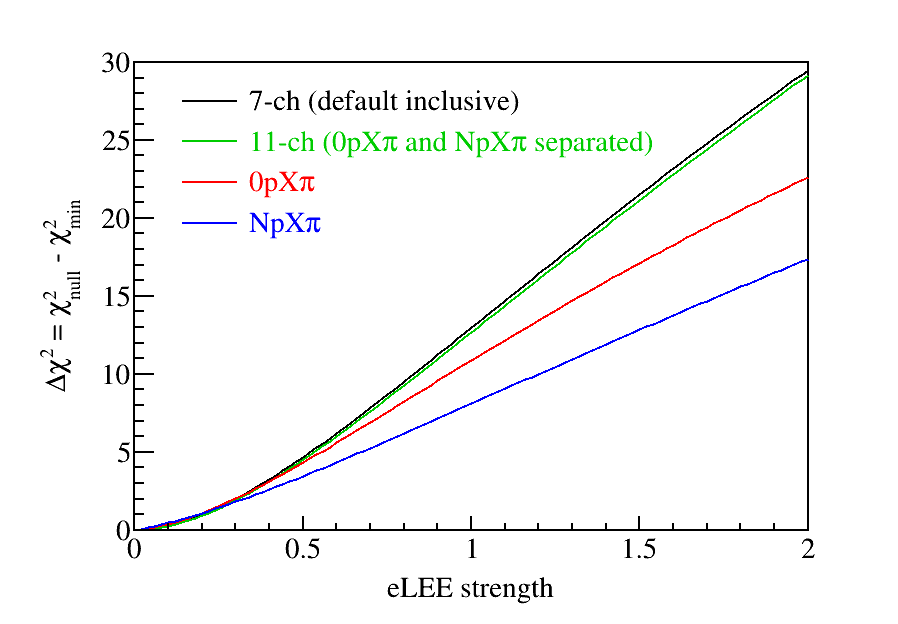}   
   \put(-150, 156){MicroBooNE 6.369$\times$10$^{20}$ POT}
  \caption{
  $\Delta\chi^2_{\rm nested}$ value as a function of eLEE strength. ``7-ch'' (black curve) corresponds to the result from the default 7-channel simultaneous fit which is the same as the one in Fig.~\ref{fig:canv_stat_bestFit}. ``11-ch'' (green curve) corresponds to the result from a 11-channel simultaneous fit combining separated 0$p$X$\pi$ and 
  N$p$X$\pi$ channels for \nueCC\ and \numuCC\ FC and PC events together with the dedicated \pizero\ channels. 
  ``0pX$\pi$'' (red curve) corresponds to the result from the 0$p$X$\pi$ channels for \nueCC\ and \numuCC\ FC and PC events as well as the dedicated \pizero\ channels. ``NpX$\pi$'' (blue curve) corresponds to the result from the 
  N$p$X$\pi$ channels for \nueCC\ and \numuCC\ FC and PC events as well as the dedicated \pizero\ channels.
  }
  \label{fig:nueCC_LEEfit_0pNp}
\end{figure}

%We also notice that the data statistics of \nueCC\ channels is insufficient with respect to the \numuCC\ channels. 
In order to understand the impact of, and account for the observed behavior in the data-prediction ratios in the 0$p$X$\pi$ and N$p$X$\pi$ channels on the eLEE analysis, the eLEE strength fit is repeated separating the inclusive \numuCC\ and \nueCC\ channels by these exclusive final-state topologies.
The default 7 channels used in the eLEE analysis are expanded to 11 channels where each \numuCC\ or \nueCC\ channel is separated into two (0$p$X$\pi$ and N$p$X$\pi$).
The results can be found in Fig.~\ref{fig:nueCC_LEEfit_0pNp}. The 11-channel result combining separated 0$p$X$\pi$ and N$p$X$\pi$ channels is well consistent with the default 7-channel result. Furthermore, the individual 0$p$X$\pi$ or N$p$X$\pi$ results also have best-fit eLEE strengths at zero albeit the uncertainties and sensitivities are different. Therefore, the consistency between \numuCC\ and \nueCC\ selections with respect to the different hadronic final states is validated, and it supports the inclusive 7-channel fit strategy as well as the overall conclusion of this paper.
Additional work studying \numuCC\ interaction cross-sections in exclusive final states is ongoing within the collaboration with the goal of further improving our ability to model these processes.

\clearpage
\bibliography{wire-cell-lee}% Produces the bibliography via BibTeX.

\end{document}